\documentclass[a4paper]{aa}
\usepackage{lscape}

\usepackage{graphicx}
\usepackage{latexsym,amsmath,amssymb}
\usepackage{natbib}
\bibpunct{(}{)}{;}{a}{}{,}

\def \xmm {XMM-Newton}
\def \chandra {$Chandra$}
\def \suzaku {$Suzaku$}

\def \nh {N${\rm _H}$}
\def \ergsec{\hbox{erg s$^{-1}$}}
\def \hcm {\hbox {\ifmmode $ atom cm$^{-2}\else atom cm$^{-2}$\fi}}
\def \arcmin {\hbox{$^\prime$}}
\def \arcsec {\hbox{$^{\prime\prime}$}}
\def \deg {$^{\circ}$}
\def \chisq {$\chi ^{2}$}
\def \rchisq {$\chi_{\nu} ^{2}$}
\def \approxgt{\mathrel{\hbox{\rlap{\lower.55ex \hbox {$\sim$}}
        \kern-.3em \raise.4ex \hbox{$>$}}}}
\def \approxlt{\mathrel{\hbox{\rlap{\lower.55ex \hbox {$\sim$}}
        \kern-.3em \raise.4ex \hbox{$<$}}}}

\newcommand{\mc}{\multicolumn}

\newcommand {\fetfour} {\ion{Fe}{xxiv}}
\newcommand {\fetfive} {\ion{Fe}{xxv}}
\newcommand {\fetsix} {\ion{Fe}{xxvi}}
\newcommand {\ka} {K$\alpha$}

\newcommand {\fettwenty} {\ion{Fe}{xx}}

\newcommand {\fettwo} {\ion{Fe}{xxii}}
\newcommand {\fetthree} {\ion{Fe}{xxiii}}
\newcommand {\canineteen} {\ion{Ca}{xix}}

\newcommand {\areighteen} {\ion{Ar}{xviii}}

\newcommand {\ssixteen} {\ion{S}{xvi}}

\newcommand {\oeight} {\ion{O}{viii}}

\newcommand {\neten} {\ion{Ne}{x}}

\def\ergsec{\hbox{erg s$^{-1}$}}
\def\ergcmsec{\hbox{erg cm$^{-2}$ s$^{-1}$}}
\def\countsec{\hbox{count s$^{-1}$}}

\def \ser {Ser\,X$-$1}
\def \sixteen {4U\,1636$-$536}

\def \seventeenof {4U\,1705$-$44}
\def \gxtfn {GX\,349$+$2}
\def \gxtfz {GX\,340$+$0}

\def \sax {SAX\,J1808.4$-$3658}
\def \fifteen {4U\,1543$-$62}
\def \seventeenth {4U\,1735$-$44}
\def \seventeentw {4U\,1728$-$34}
\def \seventeensix {GX\,9$+$9}
\def \aql {Aql\,X$-$1}
\def \osix {4U\,0614$+$09}
\def \sixo {4U\,1608$-$52}
\def \cenxf {Cen\,X$-$4}
\def \xtee {XTE\,J1807$-$294}
\def \igr {IGR\,J00291$+$5934}

\def \countsec{\hbox{counts s$^{-1}$}}

\newcommand {\egau} {$E_{\rm gau}$}

\newcommand {\ktdbb} {$kT_{\rm dbb}$}
\newcommand {\ktbb} {$kT_{\rm bb}$}

\newcommand {\ew} {$EW$}
\newcommand {\ews} {$EW$s}

\def \nh {$N{\rm _H}$}
\def \nhabs {$N{\rm _H^{abs}}$}

\newcommand {\ttnh} {$\times~$10$^{22}$~atom~cm$^{-2}$}

\def \xil {$\xi$}

\def \kpl {$k_{\rm pl}$}
\def \kbb {$k_{\rm bb}$}
\def \kdbb {$k_{\rm dbb}$}
\def \kgau {$k_{\rm gau}$}

\begin{document}

\title{A systematic analysis of the broad Fe K$\alpha$ line in neutron star LMXBs with XMM-Newton}

\author{C. Ng\inst{1} \and M. D{\'i}az Trigo\inst{2} \and M. Cadolle Bel\inst{3} \and S. Migliari\inst{1}}
\institute{
  ESAC, P.O. Box 78, E-28691 Villanueva de la Ca\~nada, Madrid, Spain
  \and
  XMM-Newton Science Operations Centre, 
  Science Operations Department, ESAC, P.O. Box 78, E-28691 Villanueva de la Ca\~nada, Madrid, Spain   \and 
  Integral Science Operations Centre, 
  Science Operations Department, ESAC, P.O. Box 78, E-28691 Villanueva de la Ca\~nada, Madrid, Spain
}

\date{Received ; Accepted:}

\authorrunning{Ng et al.}

\titlerunning{Broad Fe lines in neutron star LMXBs}

\abstract{We analysed the \xmm\ archival observations of 16 neutron star (NS) low-mass X-ray binaries (LMXBs) to study the Fe~K emission in these objects. The sample includes all the observations of NS LMXBs performed in EPIC pn Timing mode with \xmm\ publicly available until September 30, 2009. We performed a detailed data analysis considering pile-up and background effects. The properties of the iron lines differed from previous published analyses due to either incorrect pile-up corrections or different continuum parameterization. 80\% of the observations for which a spectrum can be extracted showed significant Fe line emission. We found an average line centroid of 6.67\,$\pm$\,0.02~keV and a finite width, $\sigma$, of 0.33\,$\pm$\,0.02~keV. The equivalent width of the lines varied between 17 and 189 eV, with an average weighted value of 42\,$\pm$\,3~eV. For sources where several observations were available the Fe line parameters changed between observations whenever the continuum changed significantly. The line parameters did not show any correlation with luminosity. Most important, we could fit the Fe line with a simple Gaussian component for all the sources. The lines did not show the asymmetric profiles that were interpreted as an indication of relativistic effects in previous analyses of these LMXBs.
}\maketitle

\section{Introduction}
\label{sect:intro}

Accreting binaries often show iron line emission in their X-ray
spectra. The ability of \chandra, \xmm, \suzaku\ and {\it Swift} to obtain medium
to high resolution spectra covering the iron K energy band has opened
a new era in the observations of stellar-mass black holes (BHs) and NS
binaries. The large effective area of these observatories is
crucial for the detection of iron line emission while the high
resolution enables us to discriminate between narrow and intrinsically
broad features. Broad iron lines are also present in Active Galactic
Nuclei \citep{tanaka95nat, nandra97apj} though studies of large samples of sources show
that the average fraction of broad Fe lines is never higher than 40\%
\citep{guainazzi06an, nandra07mnras}.

The origin of broad iron lines has been extensively
discussed in the literature \citep[e.g.][]{sunyaev80aa, cygx1:fabian89mnras, 
done07mnras, ross07mnras, titarchuk09apj}.
However, even after the advent of 
the current powerful X-ray observatories, the exact determination of the line width and
the mechanisms responsible for it are still controversial. Undoubtedly the
most exciting possibility is that such lines originate in the
disc close to the BH event horizon or to the NS by
fluorescent emission following illumination (and photoinisation) of
the accretion disc by an external source of X-rays \citep[][]{reynolds03ps, 
fabian05, matt06an}. 
In this model, the combination of relativistic Doppler
effects arising from the high orbital velocities and gravitational
effects due to the strong gravitational field in the vicinity of the compact object
smear the reflected features.
In this case,
detailed X-ray spectroscopy of iron line features can be used to study
Doppler and gravitational redshifts, thereby providing key information
on the location and kinematics of the material in the vicinity of the
compact object. Most interesting of all is the potential for
establishing BH spin using relativistic iron \ka\ lines.
The spin value is constrained mainly by the lower boundary of the
broad line, which depends on the inner boundary of the disc emission
(identified with the marginally stable orbit) where the gravitational
redshift is maximal. Therefore, lines which are strongly skewed toward
lower energies can indicate black hole spin \citep[][]{miller09apj, 
1753:hiemstra09mnras, 1655:reis09mnras}.

The recent claim of broad skewed iron lines from NS binaries has
opened the exciting possibility of determining an upper limit to
the radius of the NS, the most difficult parameter to obtain in order to
constrain the equation of state of NSs \citep{serx1:bhattacharyya07apjl}.
The advantage of using 
iron K emission lines as a probe of the NS
radius is that they only require short observations to clearly reveal
the relativistic lines, and do not require any knowledge of the
distance to the object.

An alternative location of the line emitting region could be the inner
part of the so-called Accretion Disc Corona (ADC), formed by
evaporation of the outer layers of the disc illuminated by the
emission of the central object \citep[e.g.,][]{white82apj}. In the
corona, where the gas is highly ionised, the iron line emission is
likely to come from recombination onto hydrogenic and helium-like
iron. In the region below the transition region the iron will be at
most a few times ionised and fluorescence will be the dominant process
\citep{kallman89apj}.  In the corona, Compton scattering will broaden
and shift the centroid of the line. The line broadening and shift are
due to Doppler and/or recoil effects, depending on the temperature of
the plasma compared to the energy of the photons
\citep{pozdnyakov79aa, sunyaev80aa}.  The final line width is
determined by the simultaneous effects of blending, Compton scattering
and rotation. For a value of {\it f}\,$\sim$\,0.1 ($f$ being the
fraction of the incident X-ray flux which is assumed to penetrate the
base of the hot medium or corona), \citet{kallman89apj} calculated a
centroid energy of 6.6--6.7~keV for the iron line and a width of
$\sim$0.6~keV at small radii (R~$\approxlt$~10$^8$~cm).  For {\it
f}\,$\sim$\,1, the width can be as large as $\sim$1~keV
(FWHM). Alternatively, widths of $\sim$1~keV can also be attained via
emission from a photoionised gas with large Thomson depth ($\tau
\approxgt$3). Such large Compton depths could arise if the corona is
heated by some mechanism other than X-rays, or if the line of sight of
the observer through the corona were very different from that of the
compact X-ray source \citep{kallman89apj}.

A third scenario for the formation of broad iron lines in LMXBs could
be one where extensive red wings form by recoil of line photons in an
optically thick medium expanding, or converging, at relativistic
velocities \citep{titarchuk03apj, laming04apj, laurent07apj}. 
It is assumed that the wide-open wind is launched
at some disc radius where presumably the local radiation force exceeds
the local disc gravity. The wind should be illuminated by the emission
of X-rays formed in the innermost part of the source. The \ka\ line is
generated in a narrow wind shell and the line profile is formed in the
partly ionised wind when \ka\ line photons are scattered off the
diverging flow (wind) electrons \citep{titarchuk09apj}.  The
red-skewed part of the spectrum is formed by photons undergoing
multiple scatterings while the primary peak is formed by photons
escaping directly to the observer.

The line properties of an ADC are in general expected to be different
from those of an accretion disc \citep{brandt94mnras}. In the former case
the equivalent width (\ew) is nearly independent of inclination except at
very high angles \citep{vrtilek93apj}, while the dependence on
inclination of the case of an accretion disc is significant
\citep{brandt94mnras}. Therefore iron lines appear to be a powerful tool
for distinguishing between an ADC, an accretion disc, or any other origin of
the iron line in LMXBs. In any case, a proper modelling of the iron line
is mandatory to determine uniquely the origin of the line and, in the
case of relativistic broadened iron lines, to derive the spin of the
BH or an upper limit to the radius of the NS.

\citet{white86mnras} performed a systematic analysis of six NS LMXBs 
with EXOSAT and found line emission in 83\% of the objects. The lines were
broad with FWHM of 0.8-1.3~keV and had centroid energies of 6.6-6.9~keV. Further 
the line properties did not show obvious correlation
with luminosity. This led \citet{white86mnras} to conclude that the only
plausible broadening mechanism was Comptonization in a cloud with a Thomson depth of
a few and an electron temperature close to 7~keV. \citet{hirano87pasj} analysed 
ten NS LMXBs observed with {\it Tenma}
and found line emission from six of them. The line energy had a weighted average of
6.66\,$\pm$0.05~keV with \ews\ in the range 20-60~eV. A significant
line width of 0.55~keV was only obtained for one source, though a width of $\sim$1~keV
could not be rejected for the other cases. They attributed the line emission
to recombination processes in the accretion disc corona and explained its width
as Compoton scattering in the region with an optical depth of 1 and an average
temperature $\sim$1~keV.     
\citet{asai00apjs} extended the analysis of the iron K
emission lines to twenty NS LMXBs observed with ASCA. They detected
significant iron lines from about half of the sources. The average
properties of the lines were a line centre of 6.56~keV and a finite
width of $\sim$0.5~keV (FWHM) in six of the sources. They also found a
large scatter in the \ew\ of the detected lines, ranging between 10 and
170~eV. They concluded that the iron K lines are likely produced
through the radiative recombination of photoionised plasma and explained
the line width as a combination of line blending, Doppler broadening
and Compton scatterings. 

The picture outlined by the analyses based on EXOSAT, {\it Tenma} and ASCA
data changed with recent analyses of observations of NS
LMXB observations by \xmm, \chandra\ and \suzaku. A number of skewed
iron lines have been claimed and their origin attributed to
relativistic effects \citep[e.g.,][]{serx1:bhattacharyya07apjl,
1808:papitto09aa, 1705:disalvo09mnras, gx340:dai09apjl} or to Compton
scattering in an optically thick medium expanding, or converging, at
relativistic velocities \citep[e.g.,][]{laurent07apj, titarchuk09apj}.
However only some of the NS LMXBs showed such asymmetric lines while
others showed symmetric lines \citep[e.g.,][]{cackett08apj}.  A
systematic analysis of the iron~K lines is at this stage fundamental
in order to establish why some sources exhibit skewed lines while
others show symmetric lines or why lines are only present in some of
the sources. Having an answer to these questions would help to clarify
the currently controversial origin of the lines.

With this aim we performed a systematic analysis of all the NS
LMXBs observed by \xmm\ since the beginning of the mission and
publicly available up to the 30th of September 2009. We excluded from this sample
observations of dipping or ADC sources, i.e. with known inclinations
above 70\deg, since their analyses at the Fe band are complicated due
to strong absorption in the line of sight.
We chose \xmm\ to
perform this study for the following reasons: firstly, its high
effective area both below and above the iron~K band allows a good
determination of the shape and width of the iron line as well as
the continuum. Secondly, \xmm\
has provided most of the detections and well determined profiles of
broad iron lines. Finally, \xmm, after 10 years of operation, has
an ample archive of public data which can be used to determine class
properties in the least unbiased way.

In total we analysed 26 observations of 16 sources from the \xmm\
archive. We restricted our analysis to sources observed in the EPIC~pn
Timing Mode. This mode is specially suited for the observation of
bright sources. Therefore, by selecting
all the observations performed in this mode we obtained a sample of
spectra with the best possible statistics.
 
In what follows we first present the properties of the selected sample
and our analysis method. Special attention was payed to eliminate
pile-up effects and to the treatment of background. Then, we performed spectral
fitting accounting for the excess emission at the Fe~K energy band. We
finally discuss the implications of our analysis regarding the characteristics 
of the lines and their possible origin. In an appendix we compare the
characteristics of the lines in this work with previous analyses of the
same observations and discuss the reason for the discrepancies whenever
found.

\section{Observations and data reduction}
\label{sec:observations}

The XMM-Newton Observatory \citep{xmm:jansen01aa} includes three
1500~cm$^2$ X-ray telescopes each with an EPIC (0.1--15~keV) at the
focus. Two of the EPIC imaging spectrometers use MOS CCDs
\citep{xmm:turner01aa} and one uses pn CCDs
\citep{xmm:struder01aa}. The RGSs
\citep[0.35--2.5~keV,][]{xmm:denherder01aa} are located behind two of
the telescopes. In addition, there is a co-aligned 30~cm diameter
Optical/UV Monitor telescope \citep[OM,][]{xmm:mason01aa}, providing
simultaneously coverage with the X-ray instruments.  Data products
were reduced using the Science Analysis Software (SAS) version
9.0. The EPIC MOS cameras were not used in most of the observations
analysed in this paper because their telemetry was allocated to the
EPIC pn camera to avoid Full Scientific Buffer in the latter. Since
the pn has an effective area $\sim$5 times higher at 7~keV than the
MOS CCDs and the latter were not available for most of the
observations, we present in this work only the analysis of the EPIC pn data. 
We did not analyse the RGS data since they do not cover the Fe~K energy band in
which we are interested.

\begin{table*}
\begin{center}
\caption[]{XMM-Newton archival observations of NS LMXBs performed in EPIC pn Timing mode ordered by Right Ascension. $T$ is the total effective EPIC pn exposure time. $C$ is the pn 0.7--10~keV persistent
emission count rate after dead time correction, calculated as the mean of the Gaussian function used to fit the count rate distribution ($\sigma$ represents the standard deviation of the distribution). The source class is A (atoll), Z (Z source), UCXB (ultra-compact X-ray binary) or AMXP (accreting millisecond X-ray pulsar). T means transient system. The last column shows the number of columns excised from the centre of the PSF to remove the pile-up effects in spectral analysis.
}
\begin{tabular}{ccclcccccc}
\hline \noalign {\smallskip}
\hline \noalign {\smallskip}
Source  & Class & Observation & \mc{3}{c}{Observation Times (UTC)} & $T$  & $C$ [$\sigma$] & Pile-up &
Columns \\
 & & ID   & \mc{2}{c}{Start}  & End & (ks) & (s$^{-1}$) & & removed \\
        & &        & (year~mon~day)& (hr:mn) & (hr:mn) & & \\
\hline \noalign {\smallskip}
\osix\ & A,UCXB & 0111040101 & 2001 Mar 13 & 12:27& 17:11 & 10 & 240 [8] & N & 0 \\
\cenxf\ & T & 0144900101 & 2003 Mar 01 & 15:11 & 14:48 & - & - & - & - \\
\fifteen\ & UCXB & 0061140201 & 2001 Feb 04 &13:16 &03:13 & 46 & 203 [5] & N & 0 \\
\sixo\ & A, T & 0074140101 & 2002 Feb 13 & 16:04 & 20:57 & - & - & - & -\\
& & 0074140201 & 2002 Feb 15 & 01:34 & 06:22 & - & - & - & - \\
\sixteen\ & A & 0303250201 & 2005 Aug 29 & 17:47 & 02:47 & 29 & 243 [15] & N & 0 \\
& & 0500350301 & 2007 Sep 28 & 15:07&00:17 & 19 & 507 [11] & Y & 1 \\
& & 0500350401 & 2008 Feb 27 &03:38 & 15:01& 37 & 652 [24] & Y & 3 \\
\gxtfz\ & Z & 0505950101 & 2007 Sep 02 & 12:40 & 02:34 & 40 & 801 [50] & Y & 8 \\
\gxtfn\ & Z & 0506110101 & 2008 Mar 19 & 15:07 & 23:00& 7 & 2043 [69] & Y & 8 \\
\seventeenof\ & A & 0402300201 & 2006 Aug 26 & 04:27 & 14:57 & 35 & 30 [1] & N & 0 \\
& & 0551270201 & 2008 Aug 24 & 01:41 & 17:15 & 45 & 742 [31]& Y & 7 \\
\seventeensix\ & A &  0090340101 & 2001 Sep 04 & 09:37 & 15:17 & 1 & 1380 [11] & Y & 4 \\
& & 0090340601 & 2002 Sep 25 &09:15 & 16:10& 5 & 1458 [25] & Y & 5\\
\seventeentw\ & A & 0149810101 & 2002 Oct 03 & 21:48& 05:55& 26 & 88 [1] & N & 0 \\
\seventeenth\ & A & 0090340201 & 2001 Sep 03 &02:57  &09:06 & 5 & 1198 [23] & Y & 3 \\
\ser\ & A & 0084020401 & 2004 Mar 22 &14:58 &21:18 & 6 & 1074 [23] & Y & 4 \\
& & 0084020501 & 2004 Mar 24 & 14:47& 21:10& 7 & 925 [14] & Y & 4 \\
& & 0084020601 &2004 Mar 26  &14:18 &20:41 & 5 & 1014 [32] & Y & 4 \\
\aql\ & A, T & 0112440101 &2002 Oct 27 & 01:03 & 03:30 & - & - & - & - \\
& & 0112440301 & 2002 Oct 15 & 01:55 & 04:17 & - & - & - & - \\
& & 0112440401 & 2002 Oct 17 & 01:42 & 05:52 & - & - & - & -\\
& & 0303220201 & 2005 Apr 07 & 14:30 & 18:58 & 3 & 228 [6] & N & 0 \\
\igr\ & AMXP, T & 0560180201 & 2008 Aug 25 & 04:45 & 14:25 & - & - & - & -\\
\xtee\ & AMXP, T & 0157960101 &2003 Mar 22 & 13:40 & 18:40 & 9 & 41 [1] & N & 0 \\
\sax\ & AMXP, T & 0560180601 & 2008 Sep 30 & 23:15 & 17:19 & 43 & 550 [10] & Y$^{a}$ & 2\\
\noalign {\smallskip} \hline \label{tab:obslog}
\end{tabular}
\end{center}
\footnotetext{}{$^a$For \sax, we found indications for very small pile-up in the two central columns. 
Therefore we present the analysis with both the full PSF and after removal of the 2 central columns in
the following sections.}
\end{table*}

Table~\ref{tab:obslog} is a summary of the XMM-Newton observations. The
EPIC pn was used in Timing Mode for all the observations. In this mode 
only one CCD chip is
operated and the data are collapsed into a one-dimensional row
(4\farcm4) and read out at high speed, the second dimension being
replaced by timing information. This allows a time resolution of
30~$\mu$s. 
We used the SAS task {\tt epfast} on the event files to correct for a 
Charge Transfer Inefficiency (CTI) effect which has been seen in the EPIC 
pn Timing mode when high count rates are present~\footnote[1]{More 
information about the CTI correction can be 
found in the {\it EPIC status of calibration and data analysis} and 
in the Current Calibration File (CCF) release note 
{\it Rate-dependent CTI correction for EPIC-pn Timing Modes}, by
Guainazzi et al. (2008), at 
http:$\slash\slash$xmm.esac.esa.int$\slash$external$\slash$xmm$\_$calibration}.
Ancillary response files 
were generated using the SAS task
{\tt arfgen} following the recommendations of the {\it XMM-Newton SAS
User guide~\footnote[2]{http:$\slash\slash$xmm.esac.esa.int}} for piled-up observations 
in Timing mode whenever applicable. 
Response matrices were
generated using the SAS task {\tt rmfgen}.
We extracted one EPIC pn spectrum per observation, not taking into account
any intra-observational variability (see Sect.~\ref{sec:x-lc}).
Bursts were excluded for the calculation of the total energy
spectra whenever present. 

Light curves were generated with the SAS task
{\tt epiclccorr}, which 
accounts for time dependent corrections within a exposure, like dead
time and GTIs.

\subsection{Pile-up treatment}
\label{sec:pileup}

Since the average count rate in the EPIC pn was close to, or above, the
800~\countsec\ level at which pile-up effects may become significant 
for at least eight observations in the sample, we
investigated in detail the presence of pile-up before extracting the
spectra. We used the SAS task {\tt epatplot}, which utilizes the
relative ratios of single- and double-pixel events deviating from
the standard values as a diagnostic tool in case of significant pile-up 
in the pn camera Timing mode. We found that the spectra of twelve
observations from eight sources were affected by pile-up (see
Table~\ref{tab:obslog}).

Then, we extracted several spectra selecting
single and double Timing Mode events (patterns 0 to 4) 
and different spatial regions for each of the piled-up observations. 
For \seventeenof\ (Obs~0551270201), \sixteen,
\seventeenth\ and \seventeensix\ (Obs~0090340601) the source
coordinates fell in the centre of the central column of the
CCD. Therefore, source events were first extracted from a 64\arcsec\
(15 columns) wide box centred on the source position (Region~1). Then,
we excluded the neighbouring 1, 3, 5 and 7 columns from the centre of Region~1
(Regions~2--5, respectively) and extracted one spectrum for each of the defined
regions. For \gxtfz, \gxtfn, \ser\ and \seventeensix\ (Obs~0090340101)
the source fell between two columns in the CCD. Therefore, source events were
first extracted from a 68\arcsec\ (16 columns) wide box centred on the
source position (Region~1) and Regions~2--5 were defined by excluding
2, 4, 6, and 8 columns respectively from the centre of Region~1. This was 
done to preserve the best symmetry when excluding piled-up
events. 
Table~\ref{tab:obslog} lists the number of columns that had to be
extracted from each source in order to obtain spectra free of pile-up.
It is evident that pile-up starts to be important already at a count
rate of $\approxgt$450~s$^{-1}$. Pile-up depends on the spectral 
shape and the time variation of the source in a complex way.
Therefore, although the average count rate of an
observation generally gives an indication of whether the effects of pile-up are 
important, it is also of {\it outmost} importance to carefully inspect the
{\tt epatplot} to evaluate the PSF radius at which the relative
ratios of single- and double-pixel events do not deviate from the standard
values in the full energy band.

 As an example of the use of {\tt epatplot} to determine the 
amount of pile-up, we show the {\tt epatplot}s in a highly 
piled-up source, \gxtfn, and in a
dim source free of pile-up, \seventeentw, in Appendix ~\ref{sec:app1}.

\subsection{Background treatment}
\label{sec:bkg}

In the EPIC pn Timing mode, there is no source-free background
region, since the PSF of the telescope extends further than the
central CCD boundaries. The central CCD has a field of view of 
13\farcm6 $\times$ 4\farcm4 in the pn. In timing mode, the largest
column is the one in which the data are collapsed into one-dimensional
row. Therefore, the maximum angle for background extraction is 2\arcmin, 
compared to 5\arcmin\ for imaging modes. In
our sample, sixteen out of the nineteen observations (for which source
emission is detected) are very bright, with total count rates
$\approxgt$200~s$^{-1}$ (see Table~\ref{tab:obslog}). Therefore, the
spectra from these sources will not be significantly modified by the
``real'' background which contributes less than 1\% to the total count
rate in most of the bandwidth. Conversely, subtracting the background
extracted from the outer columns of the central CCD will modify the
source spectrum, since the PSF is energy dependent and the source
photons scattered to the outer columns do not show the same energy
dependence as the photons focused on the inner columns.  

For the reasons mentioned above, in this work we chose not to 
subtract the ``background'' extracted from the outer 
regions of the central CCD. This is an appropriate
method for all the sources for which the ``real'' background is
negligible compared to the source count rate at all energies. In
contrast, for sources with \nh\,$\approxgt$\,1~\ttnh,
the background is expected to contribute significantly at energies $\approxlt$1~keV, where
most of the source photons are absorbed in the interstellar medium. 
Therefore for
these sources it is mandatory to remove the bins where the background
is significant before spectral analysis. 

With this aim, we extracted the background from the
outer regions of the central CCD of ``blank fields'',
where we took as examples of ``blank fields'' the observations 
of \aql\ and \sixo\ listed in Table~\ref{tab:obslog} for which the source
was not significantly detected. Then, we inspected the spectra of 
sources with \nh\,$\approxgt$\,0.5~\ttnh,
and compared the ``blank field'' background 
with the
background extracted from the observation for which the source is
being analysed. 
We scaled the former so that it was comparable to the
latter at energies where 
we did not expect a significant flux from the source. We estimated
such energy as the one at which simulated spectra with different values 
of \nh\ started to flatten as a consequence of being dominated by background
events. Then we removed energy bins where we expect a contribution from the background
to the total count rate of more than 1\%. For \seventeenof,
\seventeentw\ and \gxtfz\ we removed bins below 1.5, 1.8 and 2.2~keV,
respectively. We used all the other spectra from 0.7~keV up to
10~keV. 

\begin{figure*}[!ht]
\includegraphics[angle=0,width=0.5\textwidth]{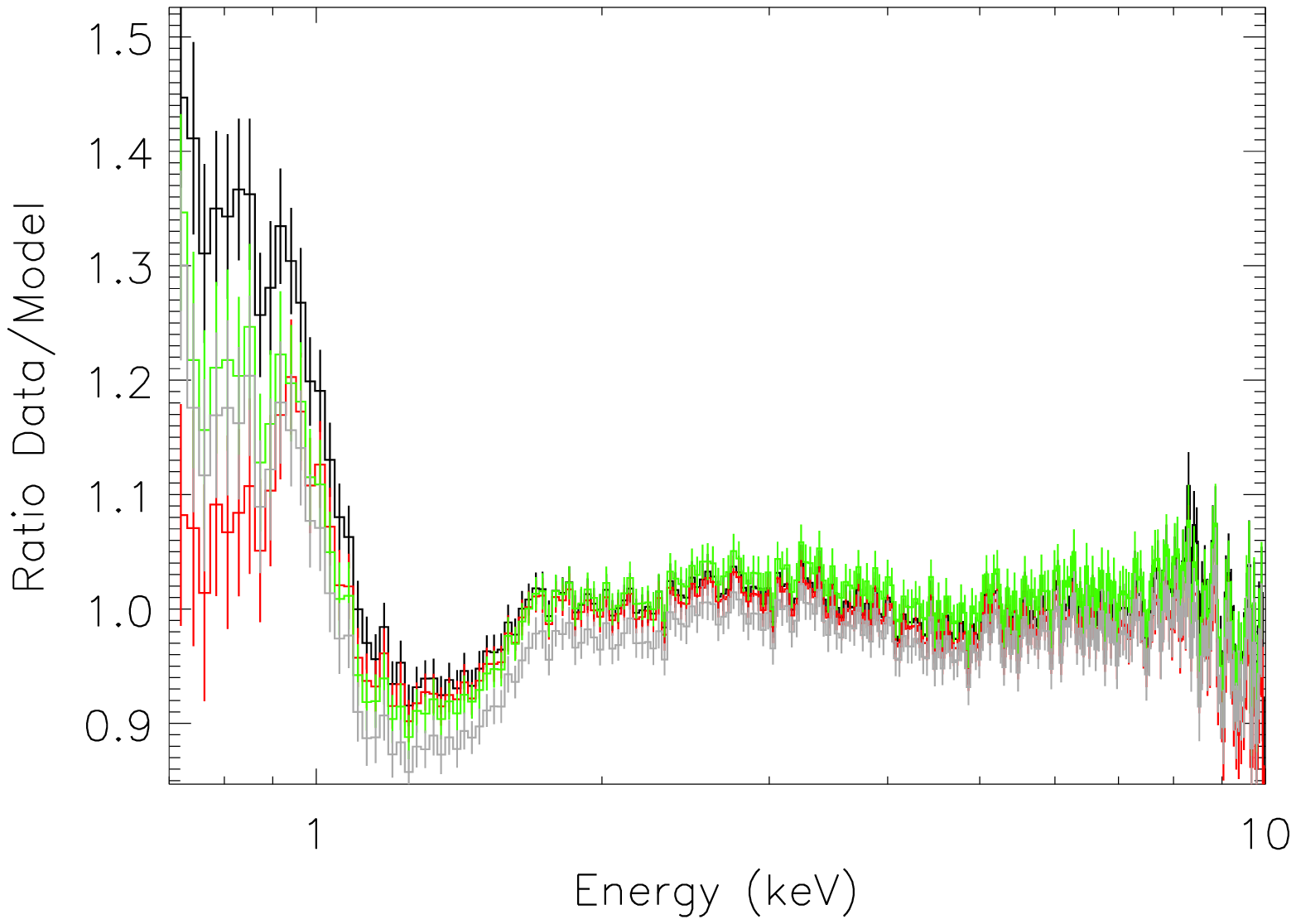}
\includegraphics[angle=0,width=0.5\textwidth]{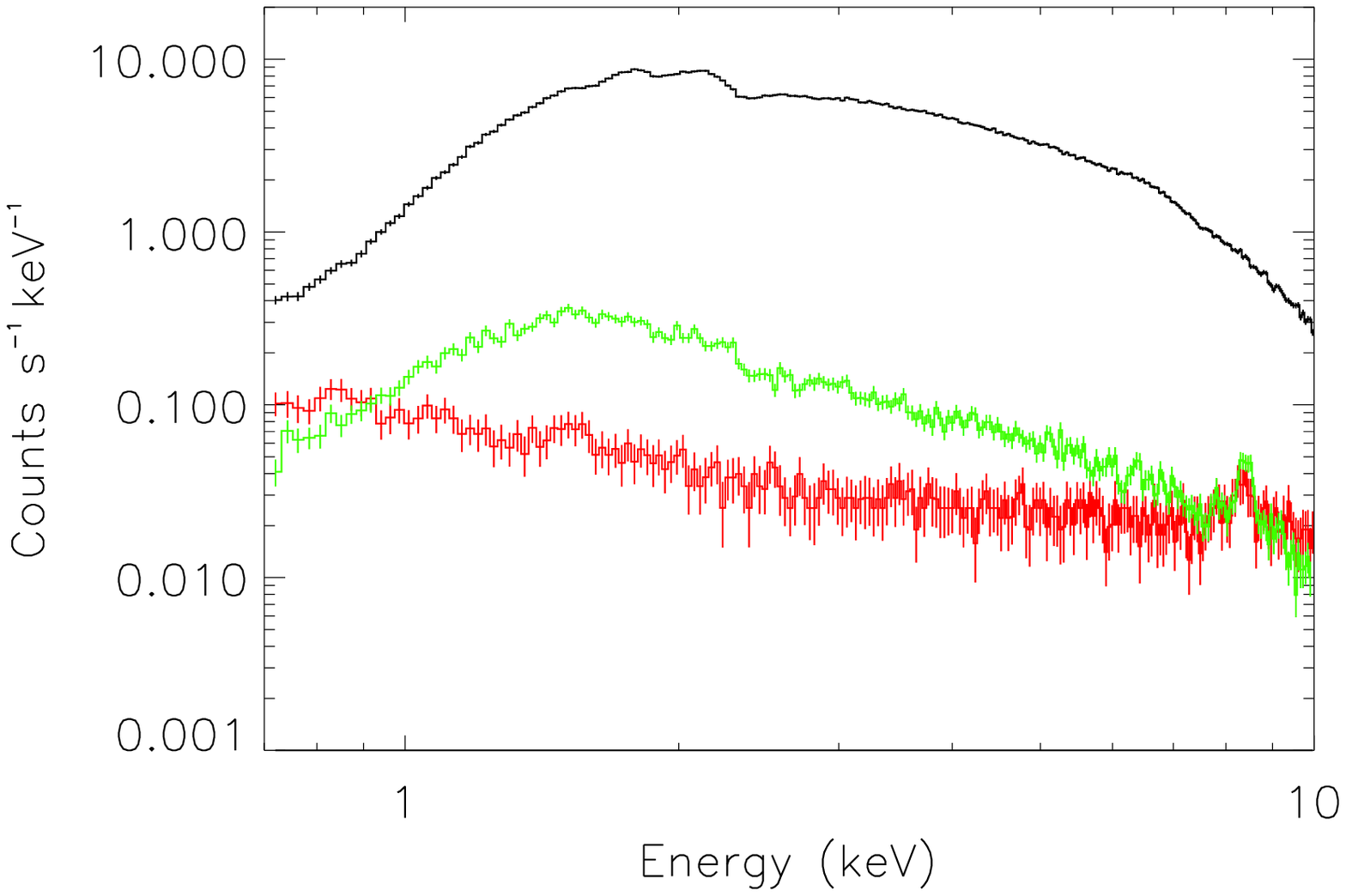}
\includegraphics[angle=0,width=0.5\textwidth]{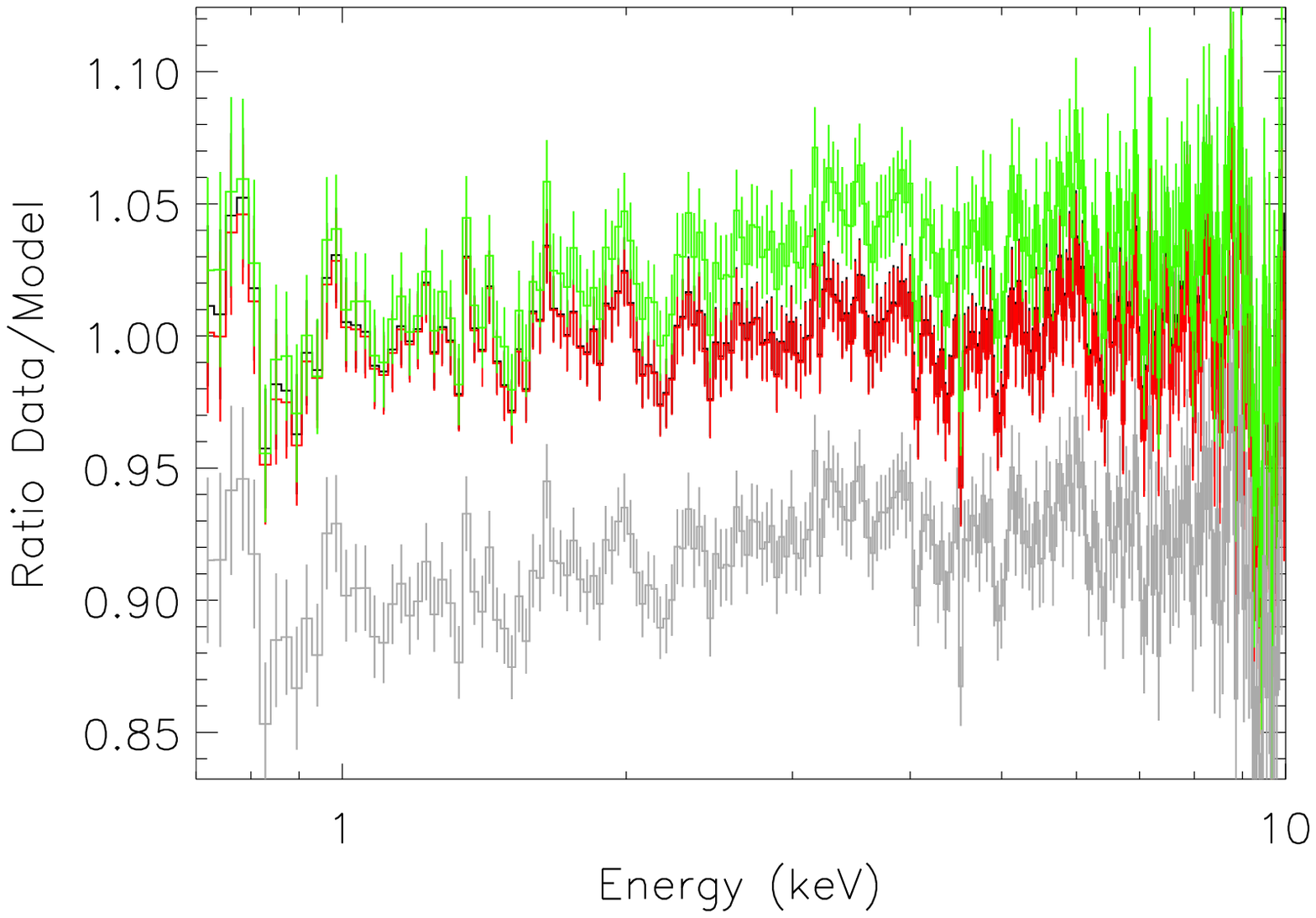}
\includegraphics[angle=0,width=0.5\textwidth]{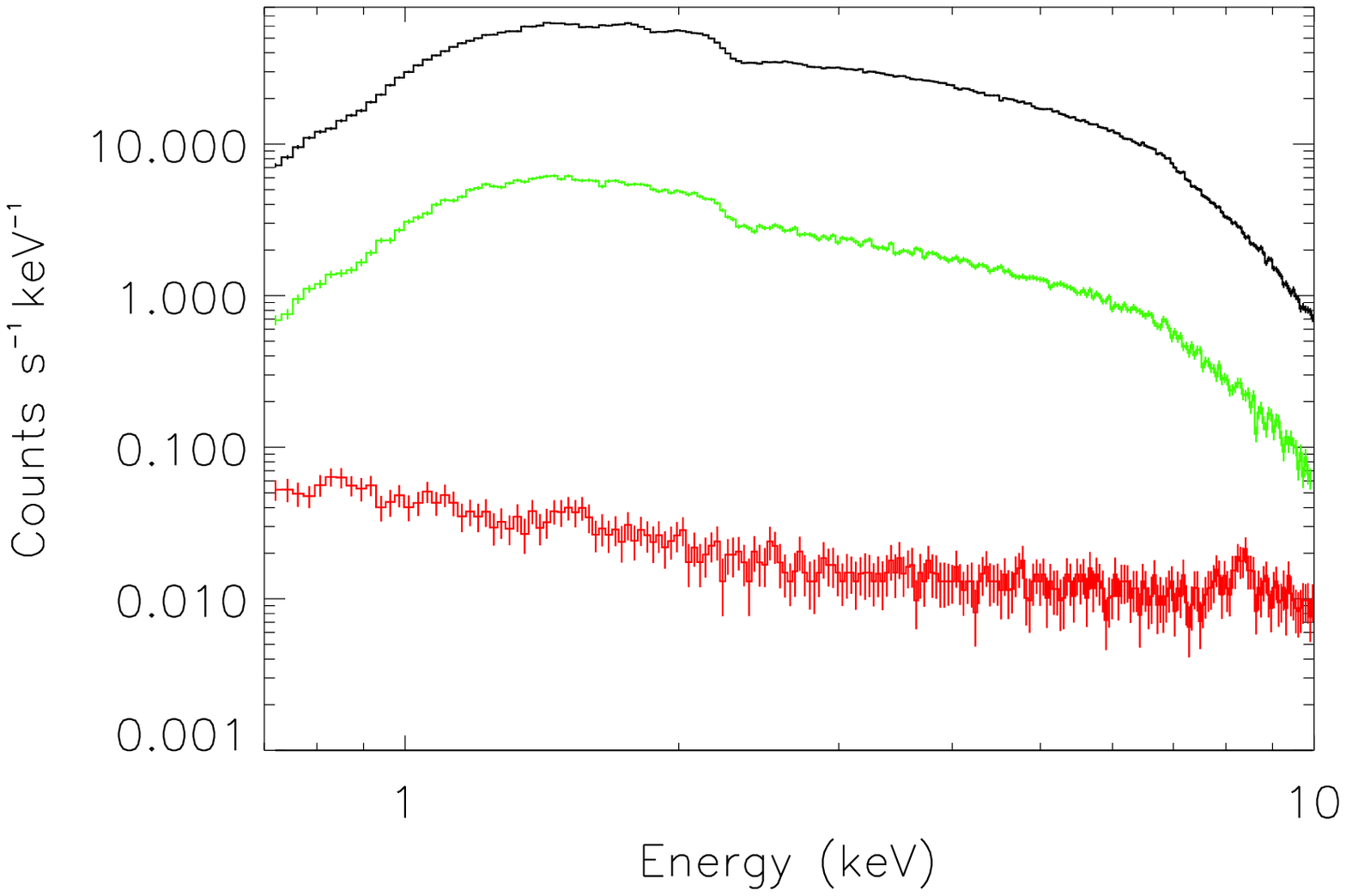}
\caption{{\it Left:} Ratio of Spec~1 (see Sect.~\ref{sec:bkg}) to its
best-fit continuum model (black) and the difference from Spec~2 (red)
and Spec~3 (green) with respect to the same model. Spec~3 has been
scaled so that the energy bin at 10~keV coincides with Spec~1. The
original Spec~3 is shown in grey. {\it Right:} Source spectrum
(black), background from a blank field (red) and background extracted
from the outer columns of the central CCD (green).
The upper and lower panels show the results for \seventeenof\ and \gxtfn\ 
respectively.
}
\label{fig:background}
\end{figure*}

To illustrate the effects of background subtraction in the spectral
fitting and evaluate the validity of our method we chose the weakest and
brightest sources of our sample, \seventeenof\ (Obs~0402300201) and \gxtfn. We
examined the differences in the spectra after (1) not subtracting any
background from the source spectrum (Spec~1), (2) subtracting the
background extracted from a blank field (Spec~2), and (3) subtracting
the background extracted from the outer columns of the central CCD
during the observation (Spec~3). Fig.~\ref{fig:background} (left panels) shows the
ratio of Spec~1 to its best-fit continuum model (black) and the
difference from Spec~2 (red) and Spec~3 (green) with respect to the
same model. The red and black residuals are
consistent within the errors at all energies for \gxtfn\ (lower panels)
and above $\sim$1~keV for \seventeenof\ (upper panels). The deviation
below $\sim$1~keV for \seventeenof\ is expected, since at such
energies the background photons have a significant contribution to the total 
spectrum. In contrast, we
observe an energy dependent discrepancy of Spec~3 with respect to
Spec~1 and~2, as expected due to the energy dependence of the
PSF. These results are consistent with the shape of the background
shown in Fig.~\ref{fig:background} (right panels). The
``background'' extracted from the outer columns of the CCD (green) is
clearly contaminated by the source at all energies (compared to the
source spectrum in black), while the ``background'' extracted from a
blank field (red) shows the shape expected when
compared to Fig.~35 of the {\it XMM-Newton Users Handbook}. 

\section{Light curves and spectral fitting}

\subsection{Light curves}
\label{sec:x-lc}

Fig.~\ref{fig:lc} shows EPIC pn light curves and hardness ratios of all the \xmm\
observations analysed in this paper with a binning of 64~s. The hardness ratio is counts in the
3--10~keV energy range divided by those between 0.7--3~keV, except for \seventeenof,
\seventeentw\ and \gxtfz, for which the soft band is 1.5--3~keV, 1.8--3~keV and 2.2--3~keV,
respectively.
The light curves of observations with count rates 
$\approxgt$~200~\countsec\ suffer from regular telemetry drops which are seen when
the light curves are plotted with a high time resolution,
e.g. 1~s. The SAS task {\tt epiclccorr} accounts for such gaps, as well
as for dead time with a different origin, by
taking into account the count rate in adjacent frames 
(see Sect.~\ref{sec:observations}).

The average count rate changed by more than 2 orders of magnitude
among the studied sources. Observations of the same source taken
within months showed strong variability in some cases (e.g. \sixteen)
and a steady level in other cases (e.g. \ser).

Strong variability of $\approxgt$30\% within one observation is only present in
the light curves of \gxtfz\ and \gxtfn. The variabilities within such
observations were studied by \citet{gx340:dai09apjl} and
\citet{gx349:iaria09aa}, who found that the spectral fits to different
intervals within the observations gave consistent results for the
parameters of the iron~line. 

\begin{figure*}[!ht]
\includegraphics[angle=90,width=0.33\textwidth]{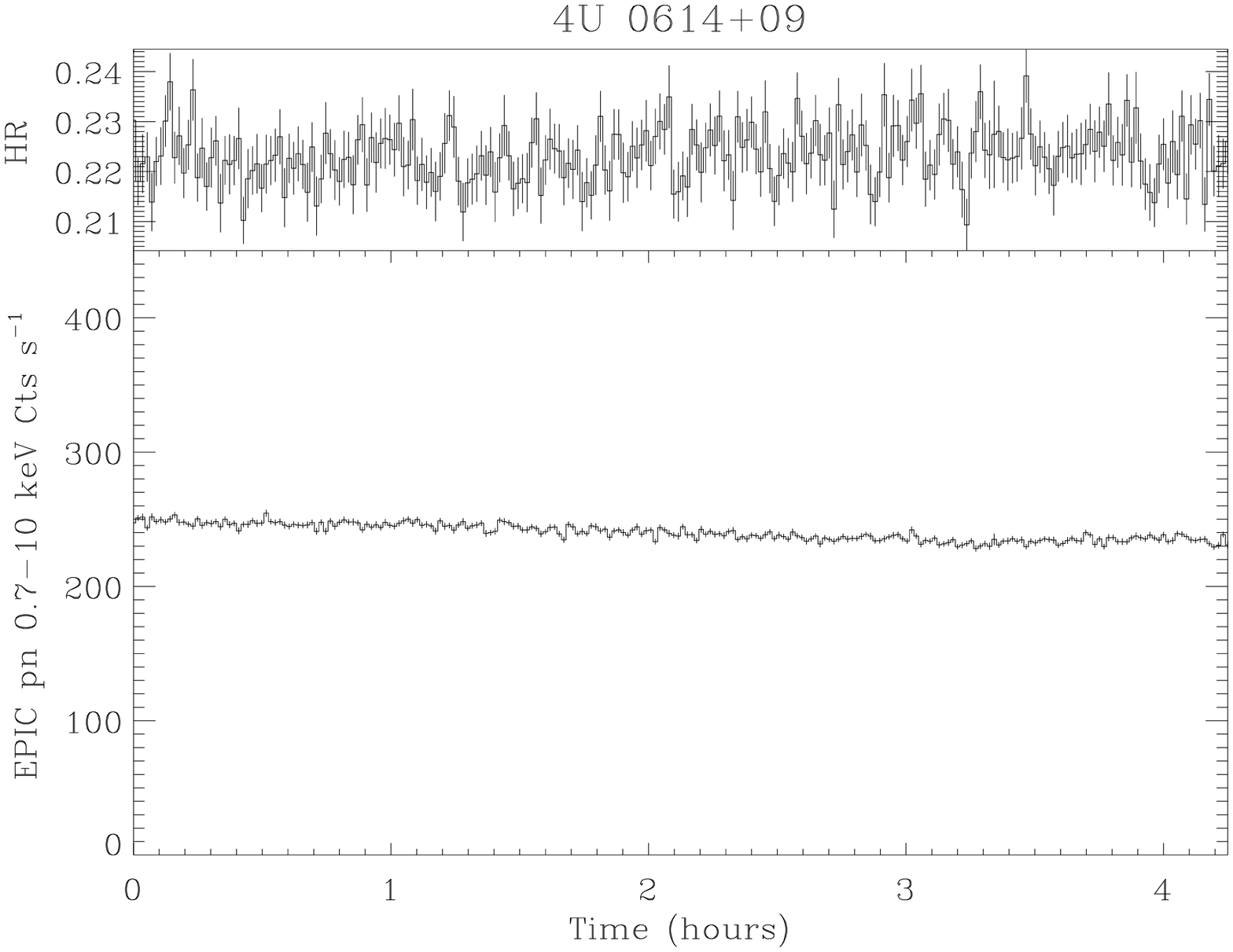}
\includegraphics[angle=90,width=0.33\textwidth]{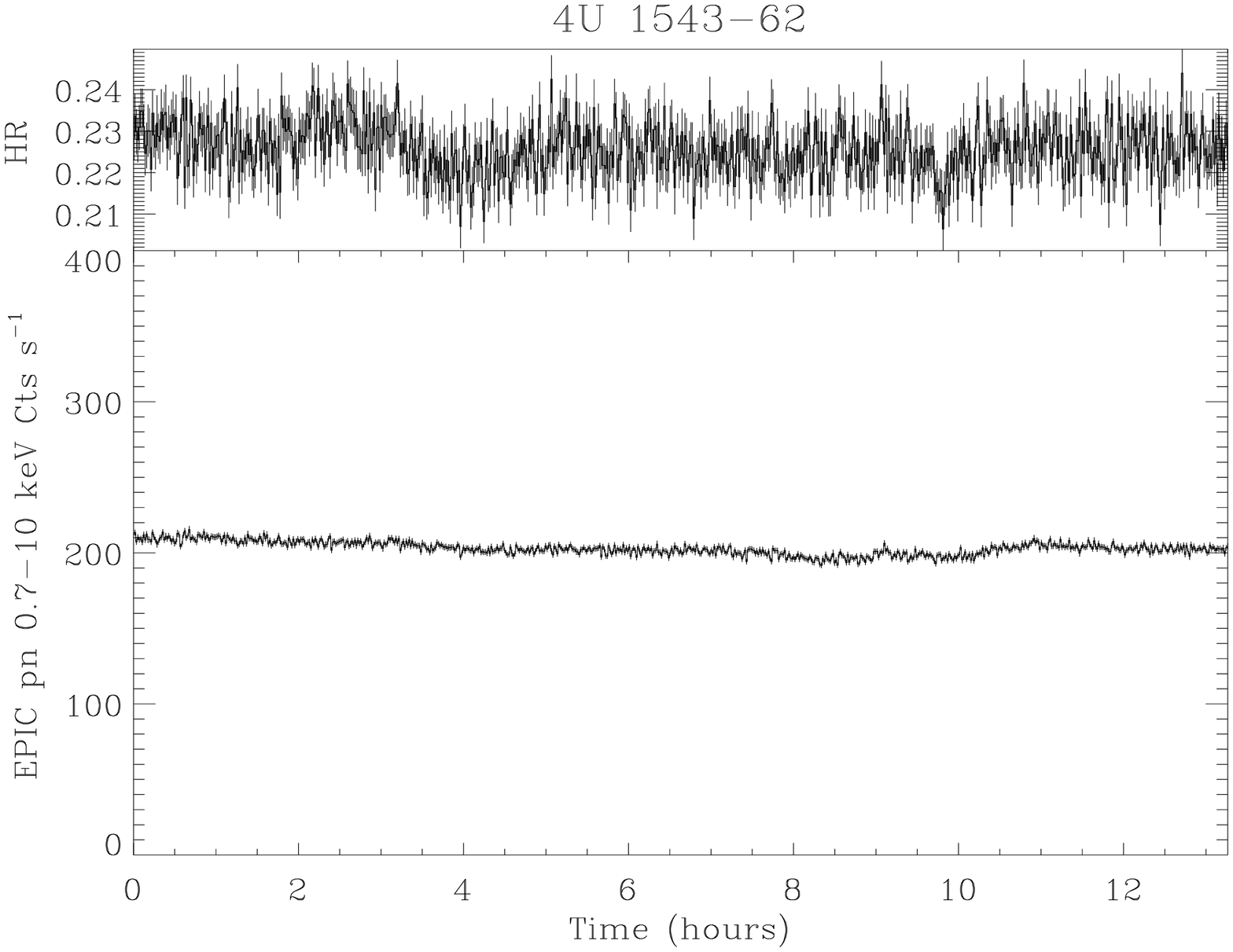}
\includegraphics[angle=90,width=0.33\textwidth]{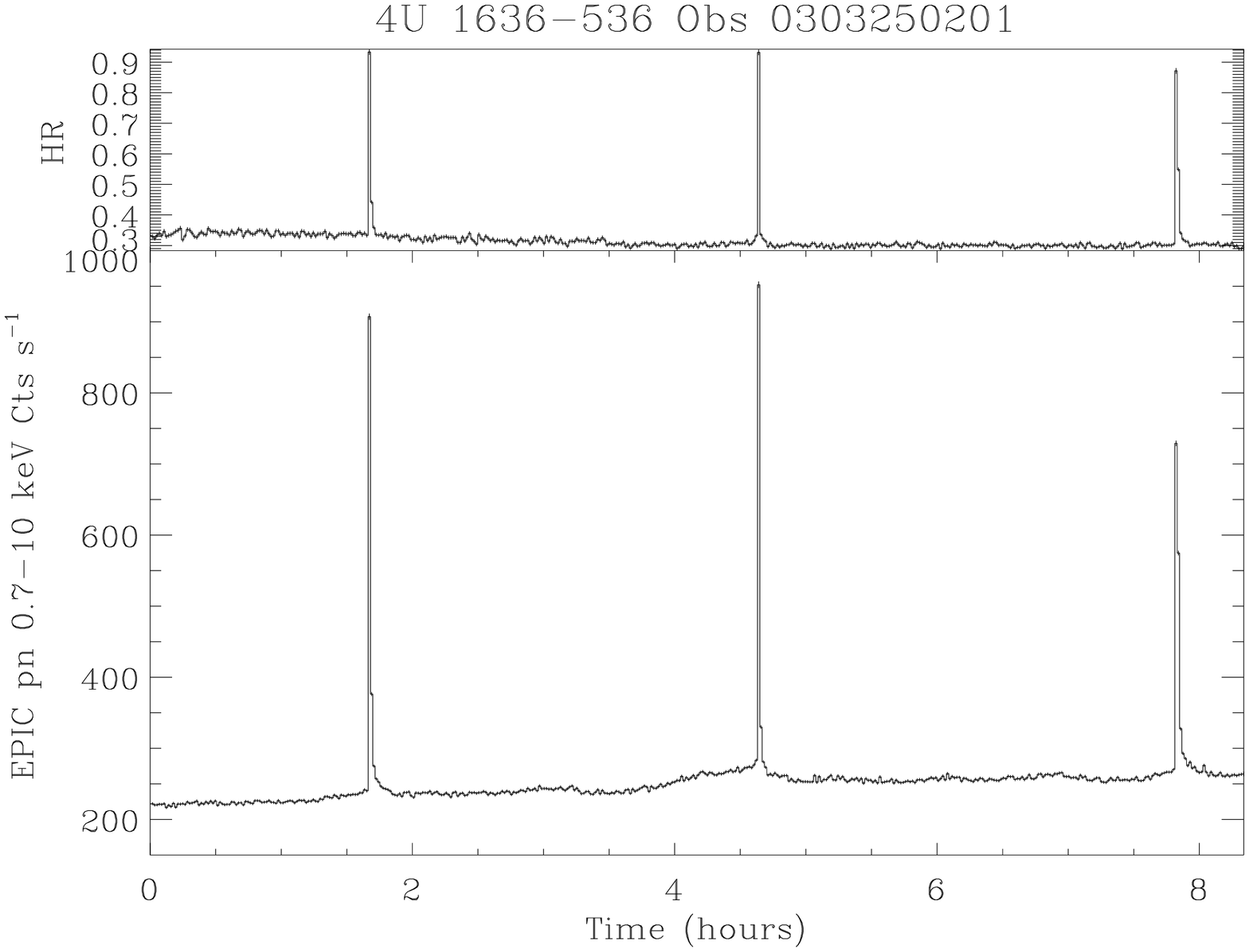}
\vspace{0.1cm}
\includegraphics[angle=90,width=0.33\textwidth]{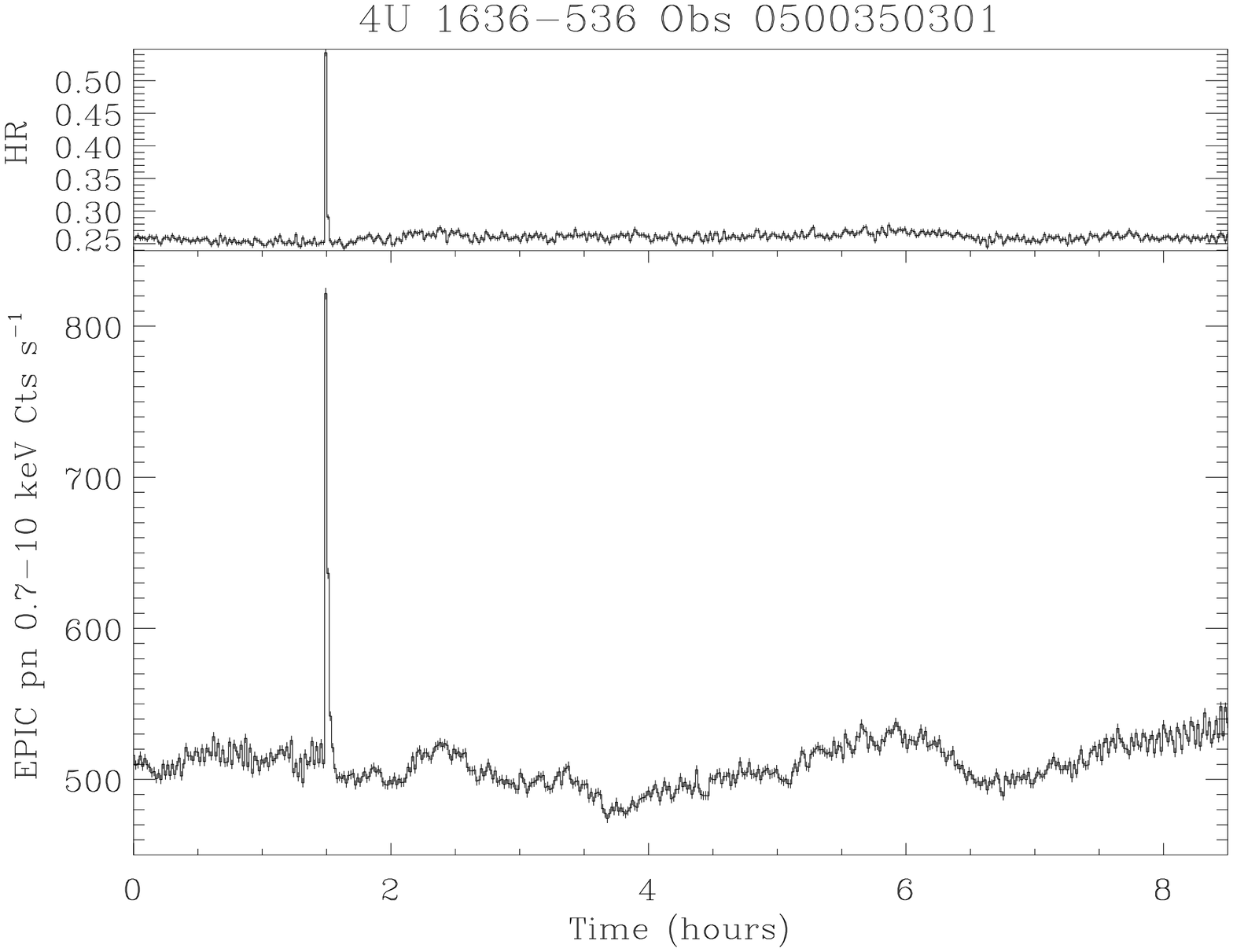}
\includegraphics[angle=90,width=0.33\textwidth]{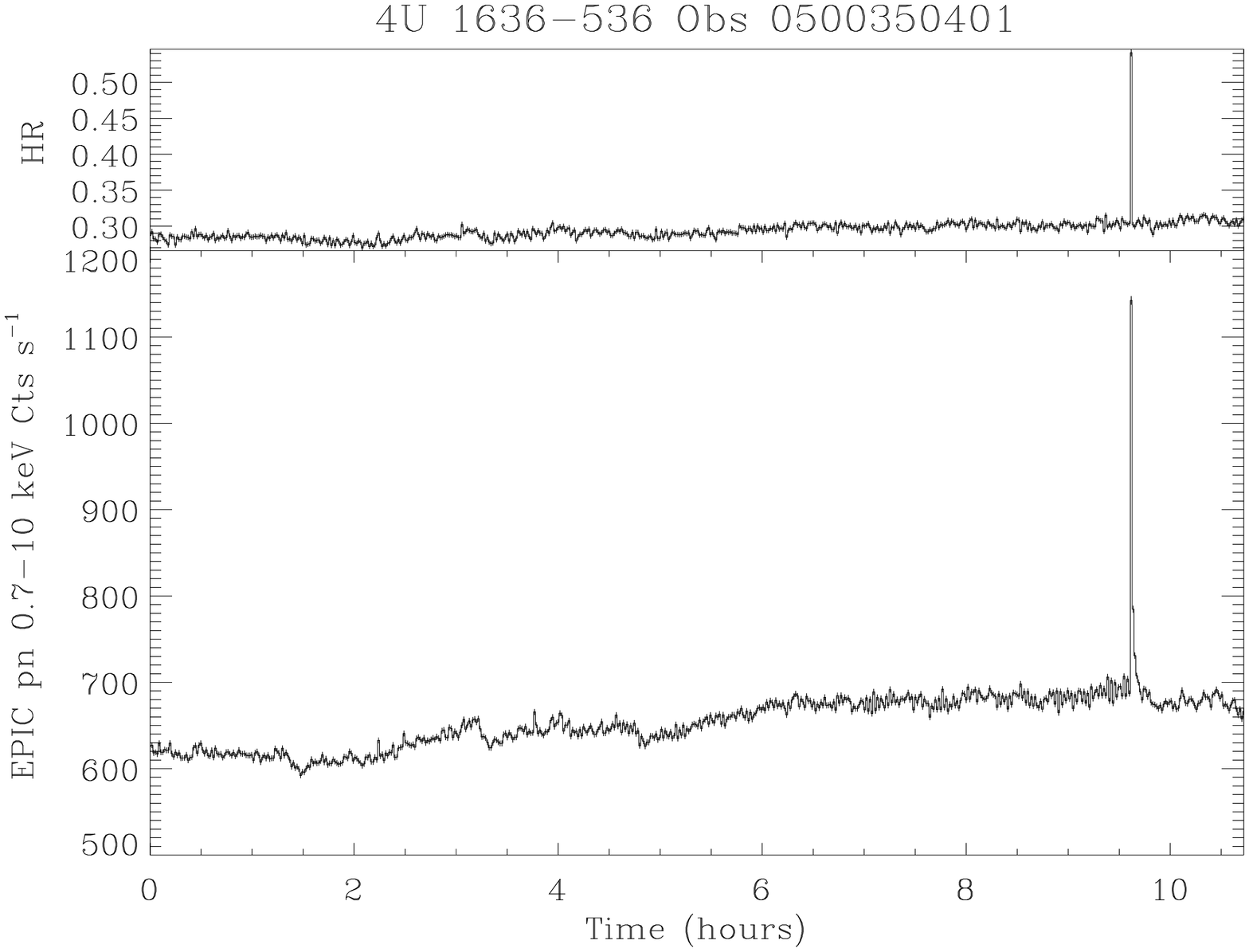}
\includegraphics[angle=90,width=0.33\textwidth]{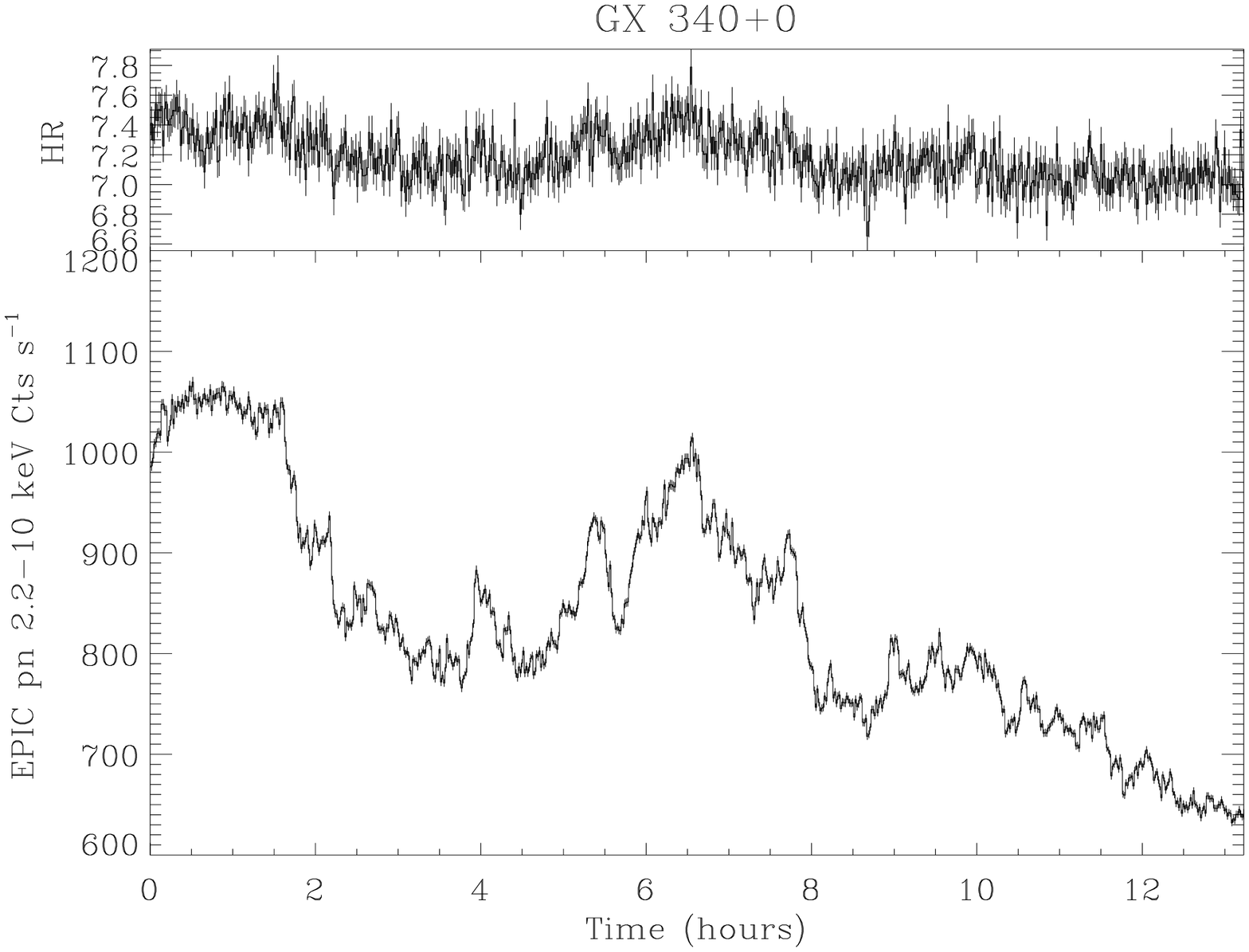}
\vspace{0.1cm}
\includegraphics[angle=90,width=0.33\textwidth]{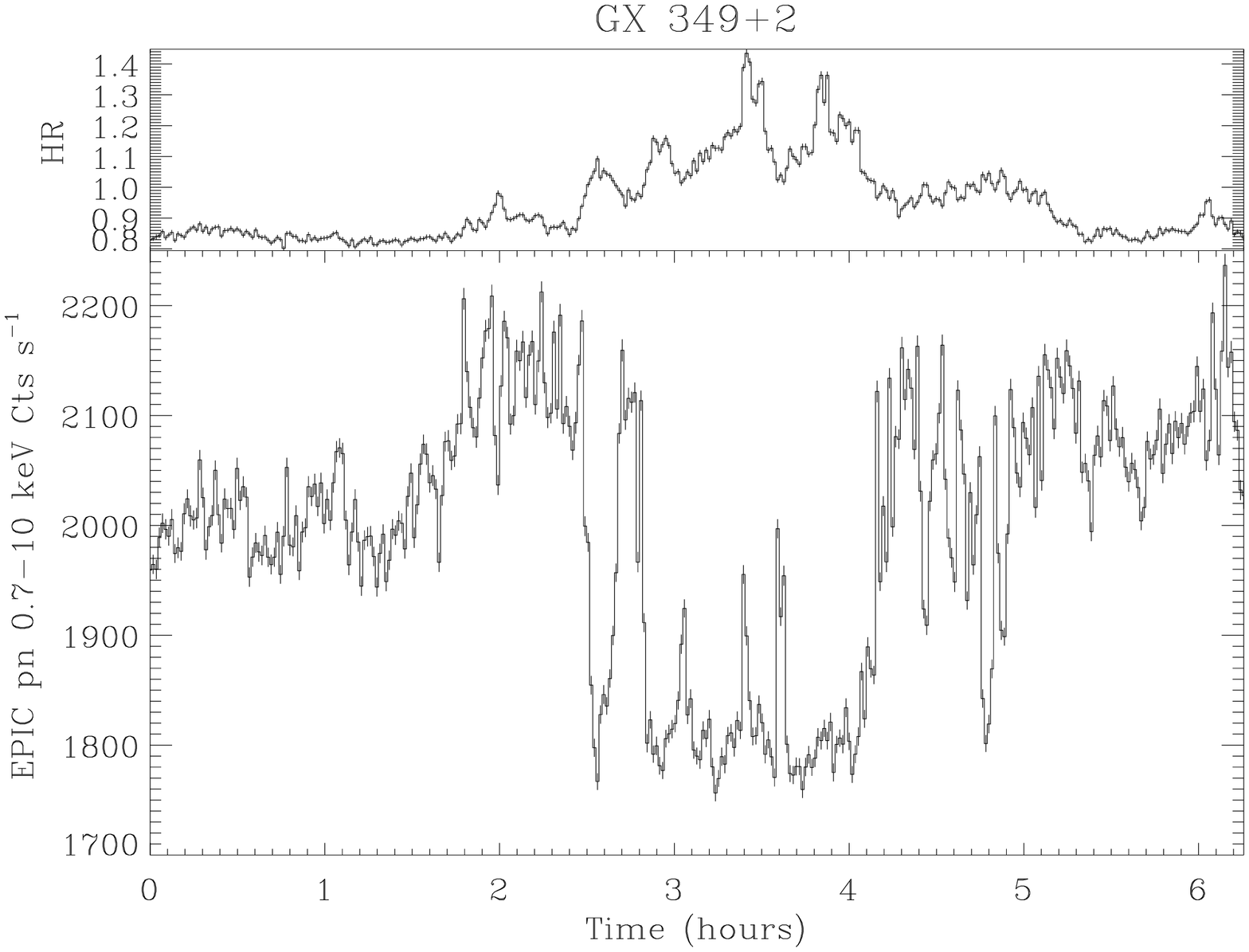}
\includegraphics[angle=90,width=0.33\textwidth]{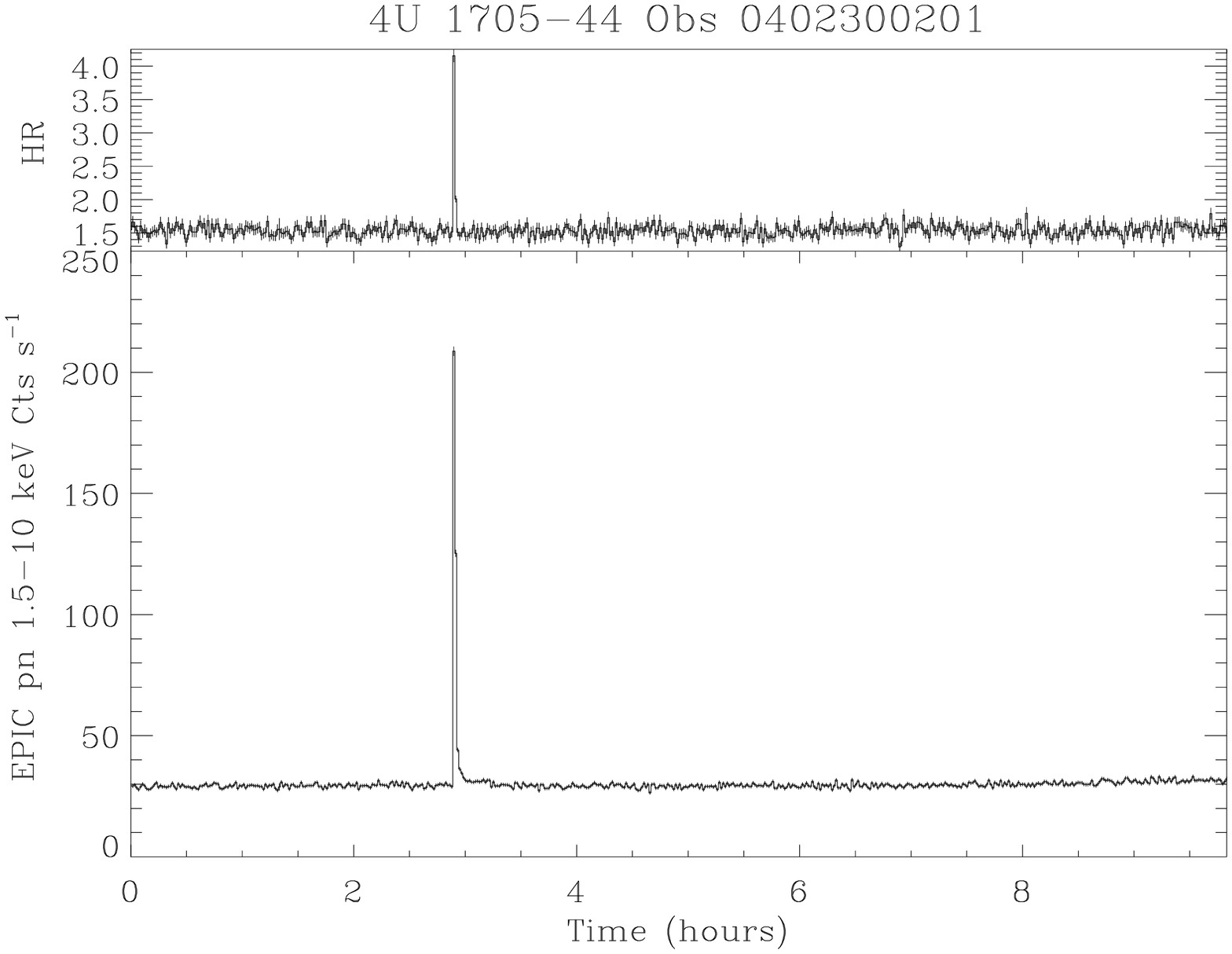}
\includegraphics[angle=90,width=0.33\textwidth]{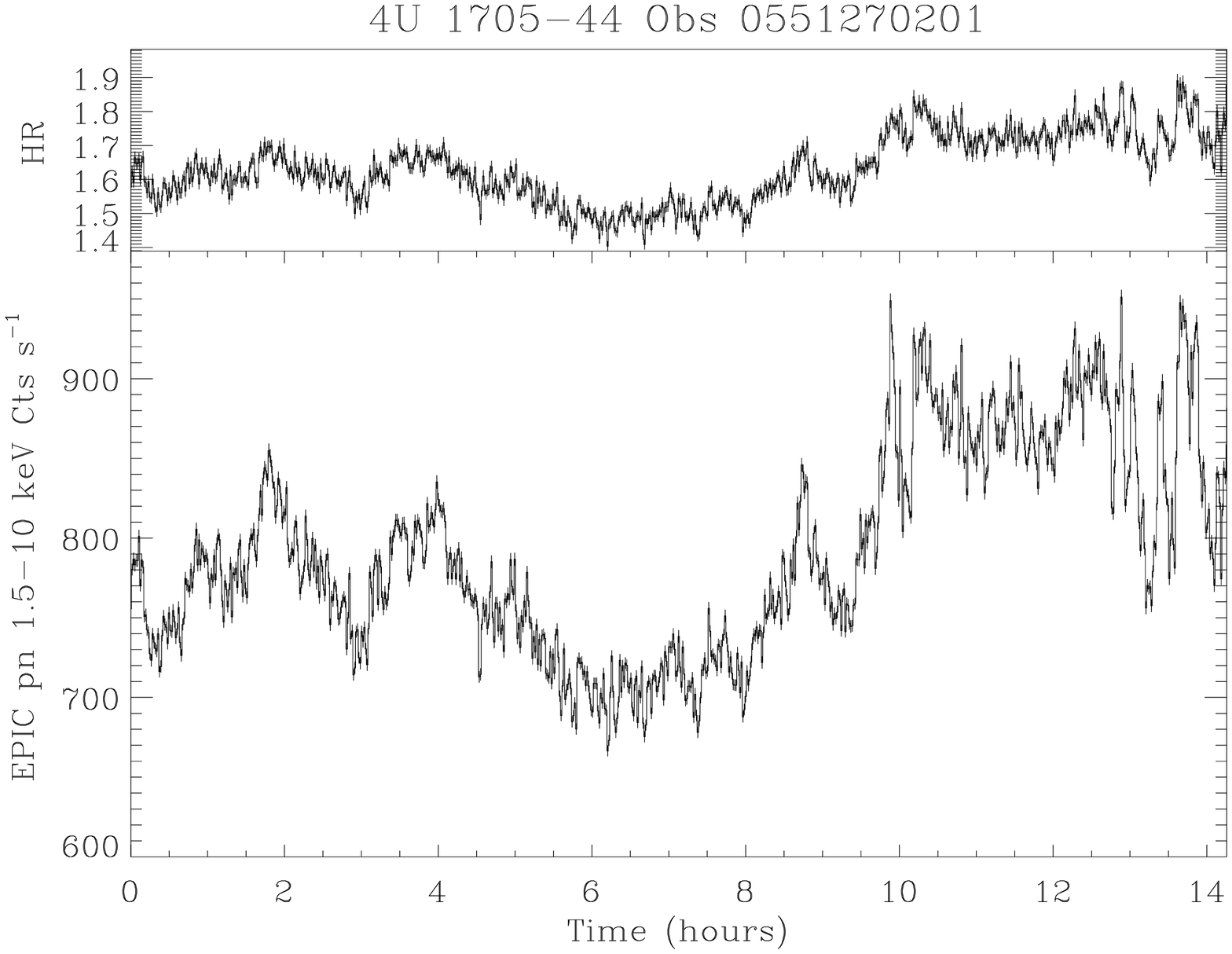}
\vspace{0.1cm}
\includegraphics[angle=90,width=0.33\textwidth]{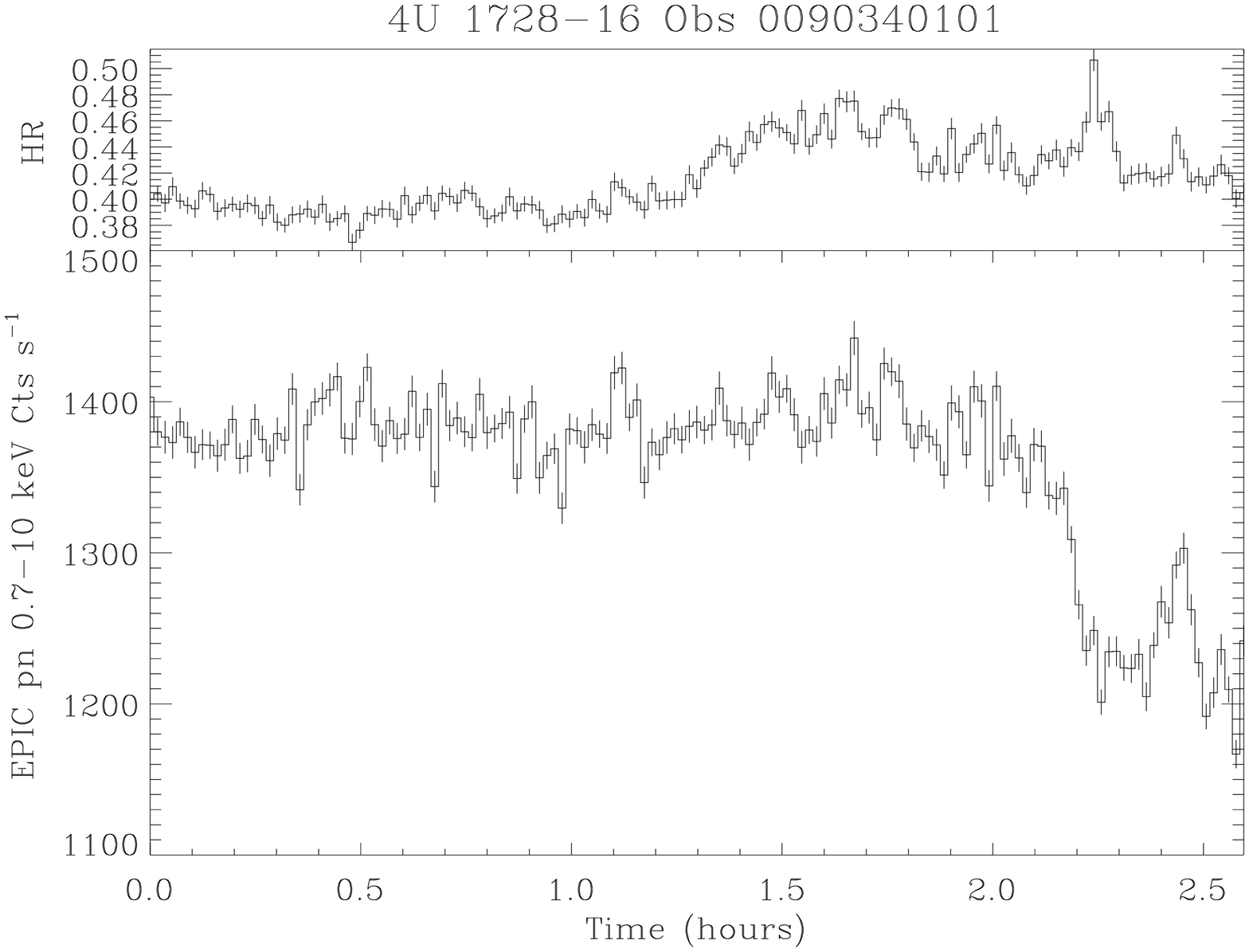}
\includegraphics[angle=90,width=0.33\textwidth]{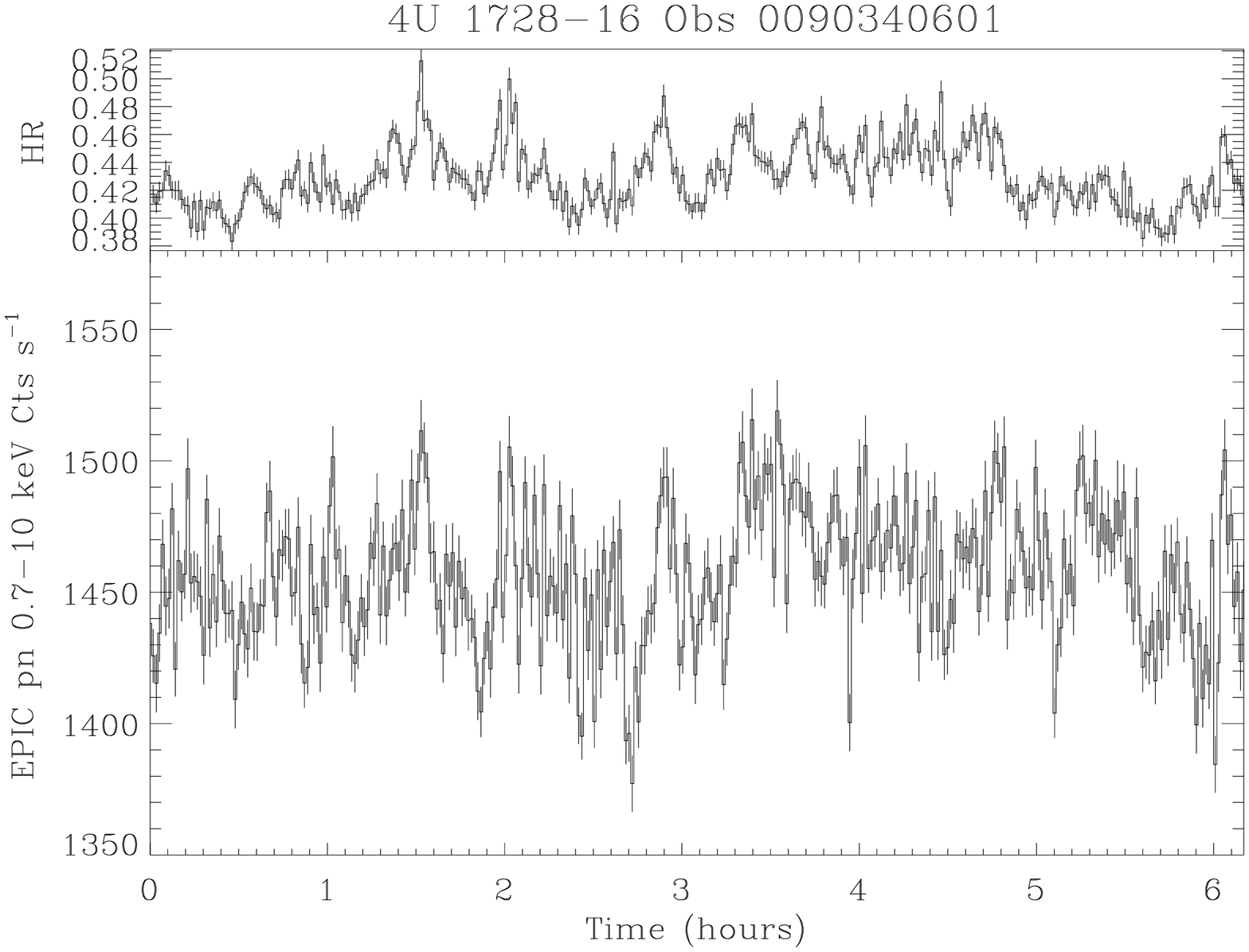}
\includegraphics[angle=90,width=0.33\textwidth]{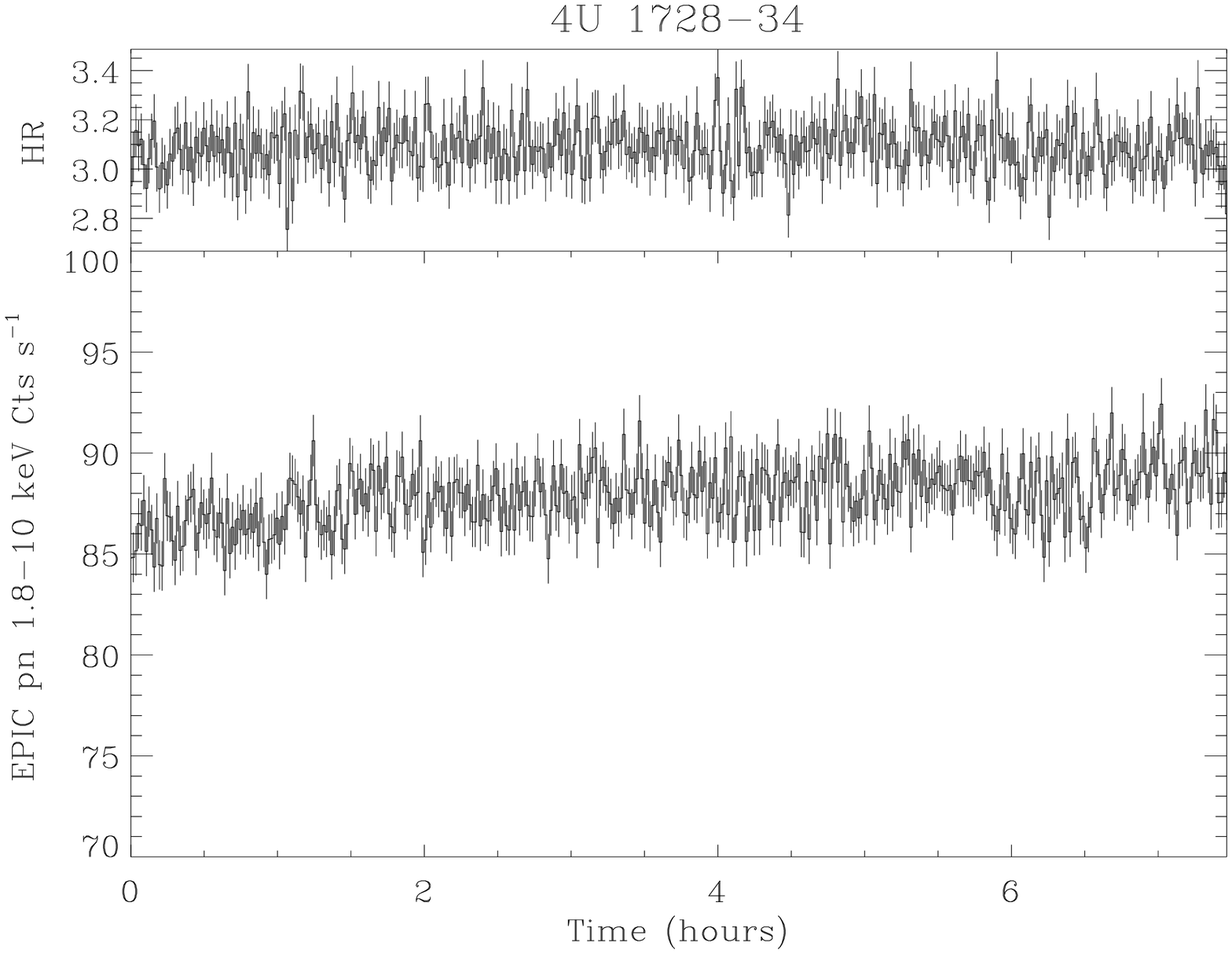}
\vspace{0.1cm}
\includegraphics[angle=90,width=0.33\textwidth]{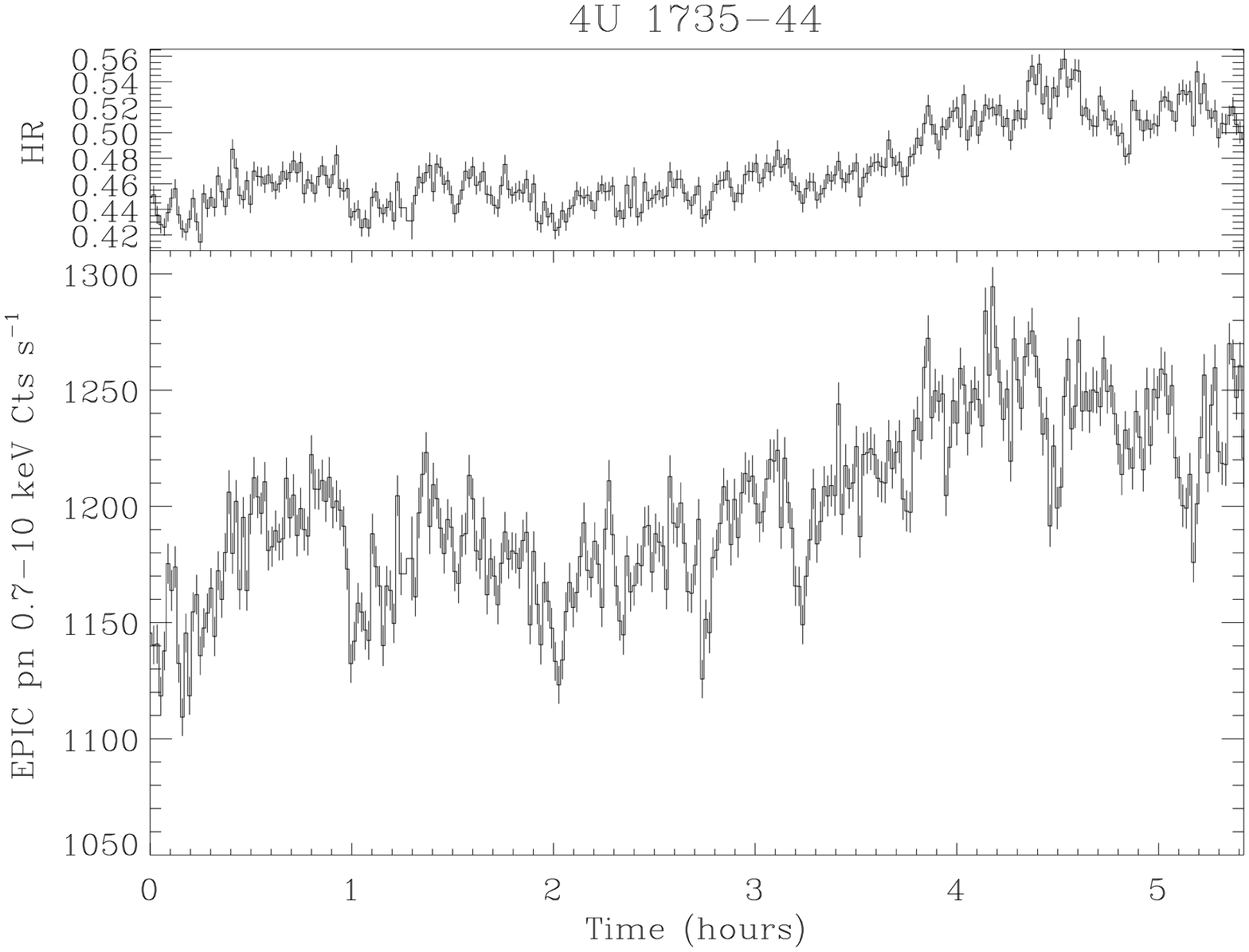}
\includegraphics[angle=90,width=0.33\textwidth]{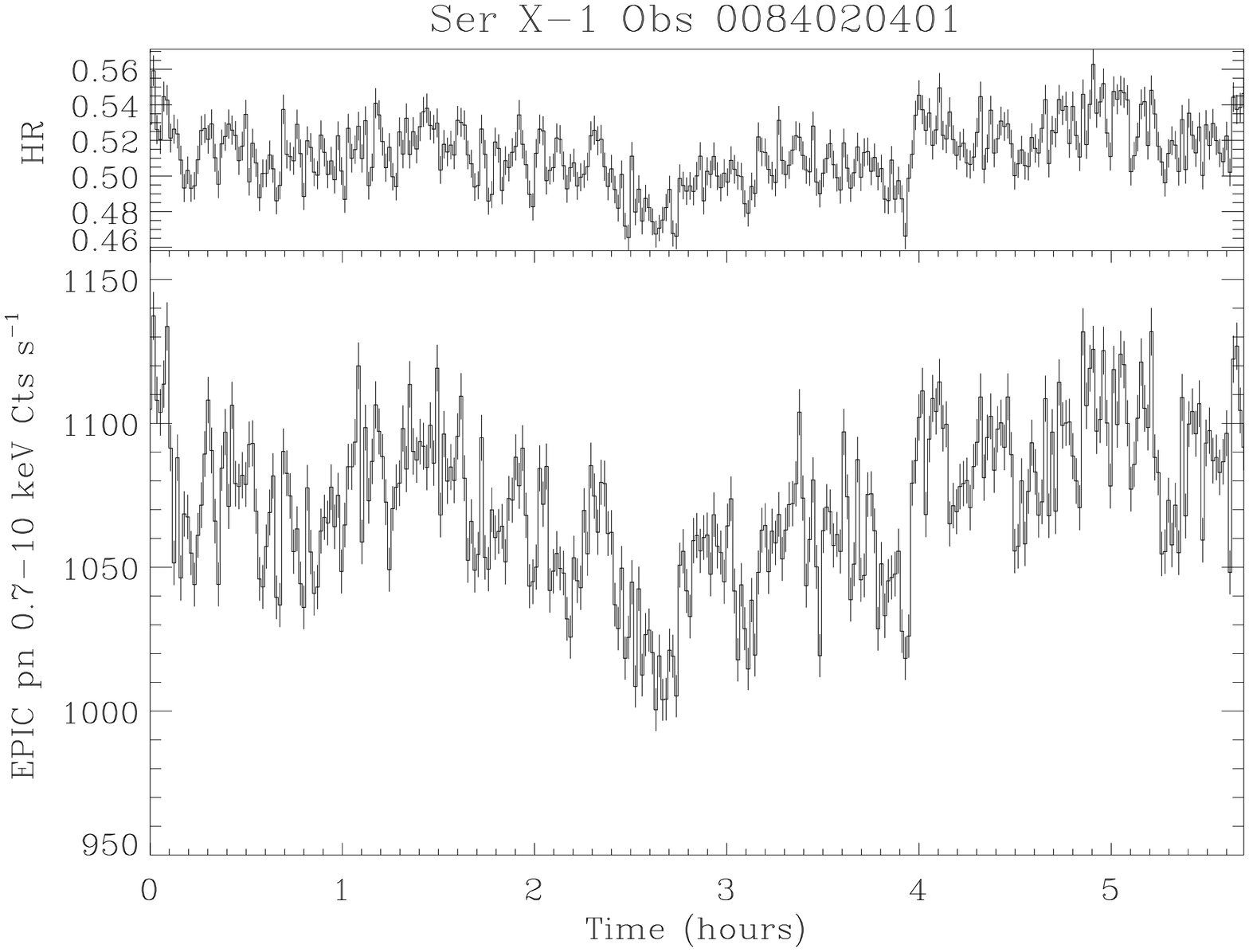}
\includegraphics[angle=90,width=0.33\textwidth]{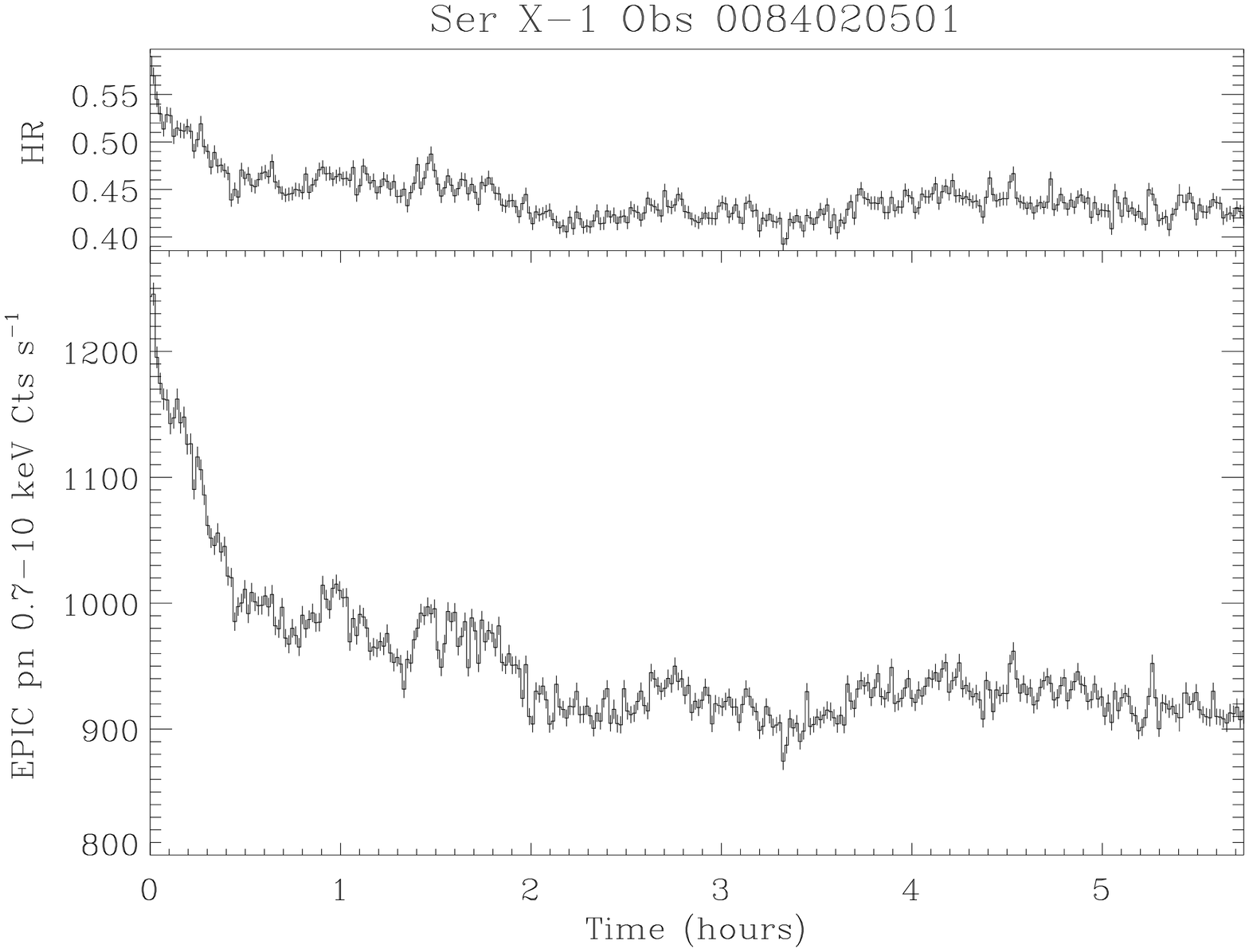}
\caption{EPIC pn 
light curves and hardness ratios (HR) for all the observations of our sample with a
binning of 64~s. Time is shown in hours since the beginning of the
observation.
The hardness ratio is counts in the
3--10~keV energy range divided by those between 0.7--3~keV, except for \seventeenof,
\seventeentw\ and \gxtfz, for which the soft band is 1.5--3~keV, 1.8--3~keV and 2.2--3~keV,
respectively. 
}
\label{fig:lc}
\end{figure*}

\begin{figure*}[!ht]
\includegraphics[angle=90,width=0.33\textwidth]{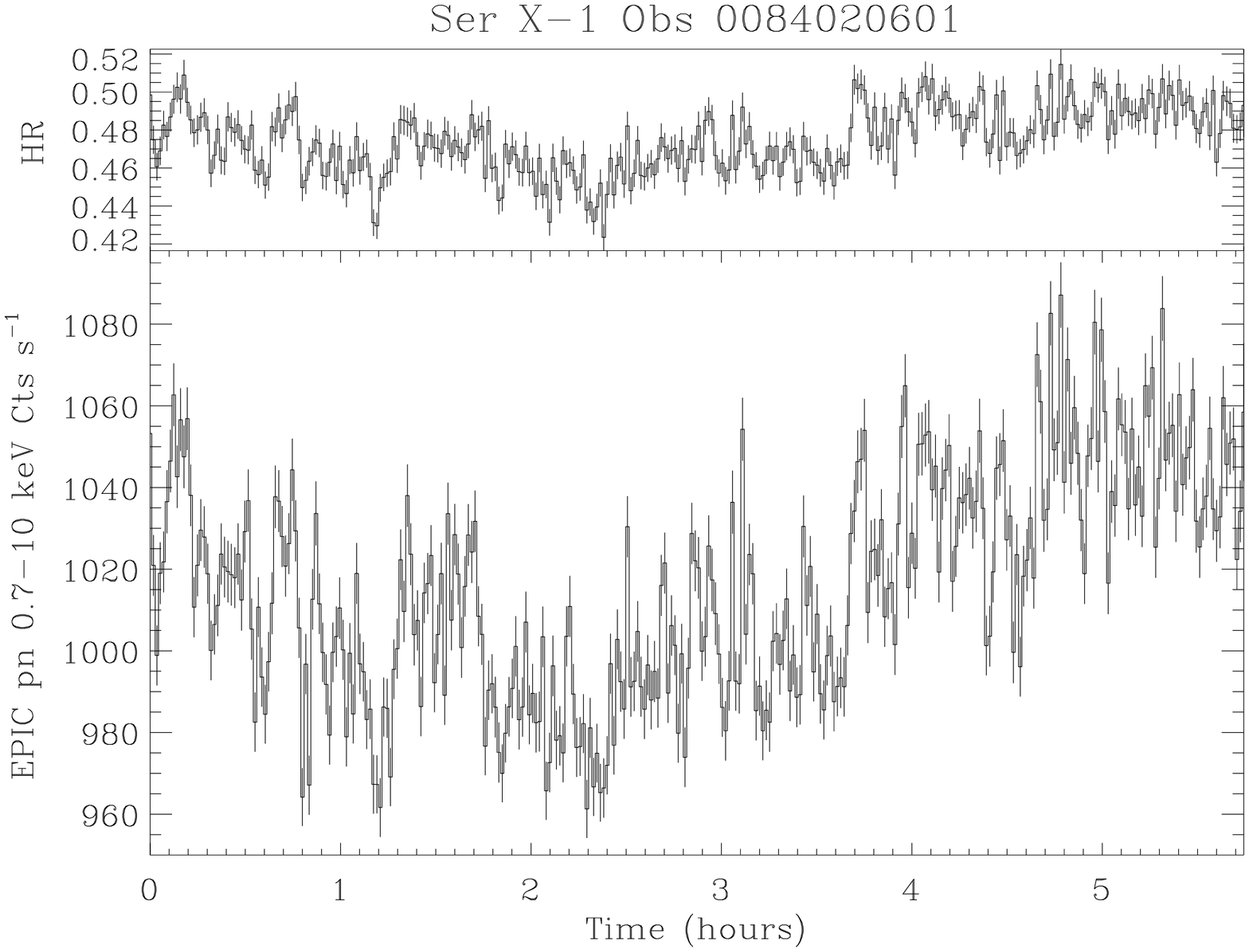}
\includegraphics[angle=90,width=0.33\textwidth]{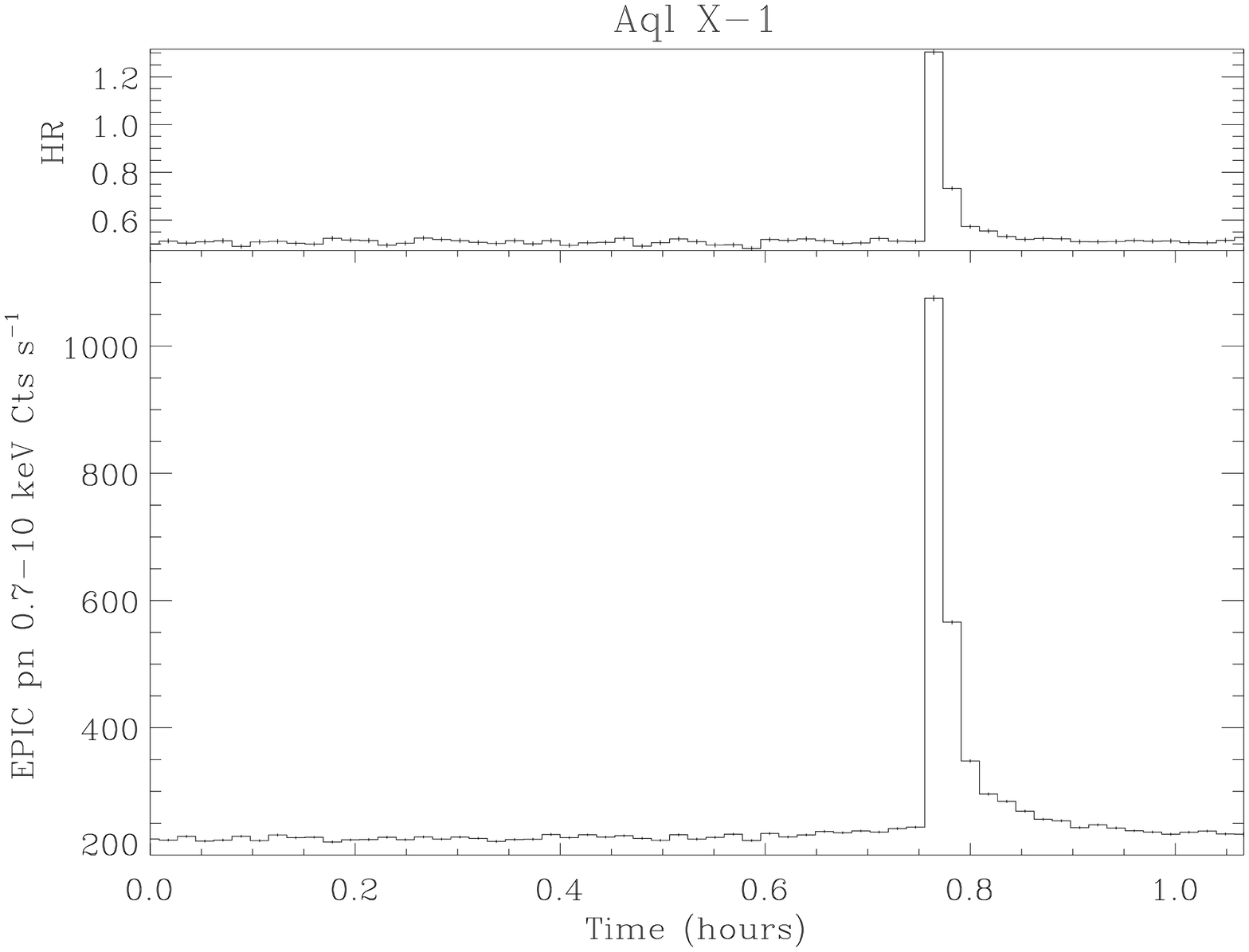}
\includegraphics[angle=90,width=0.33\textwidth]{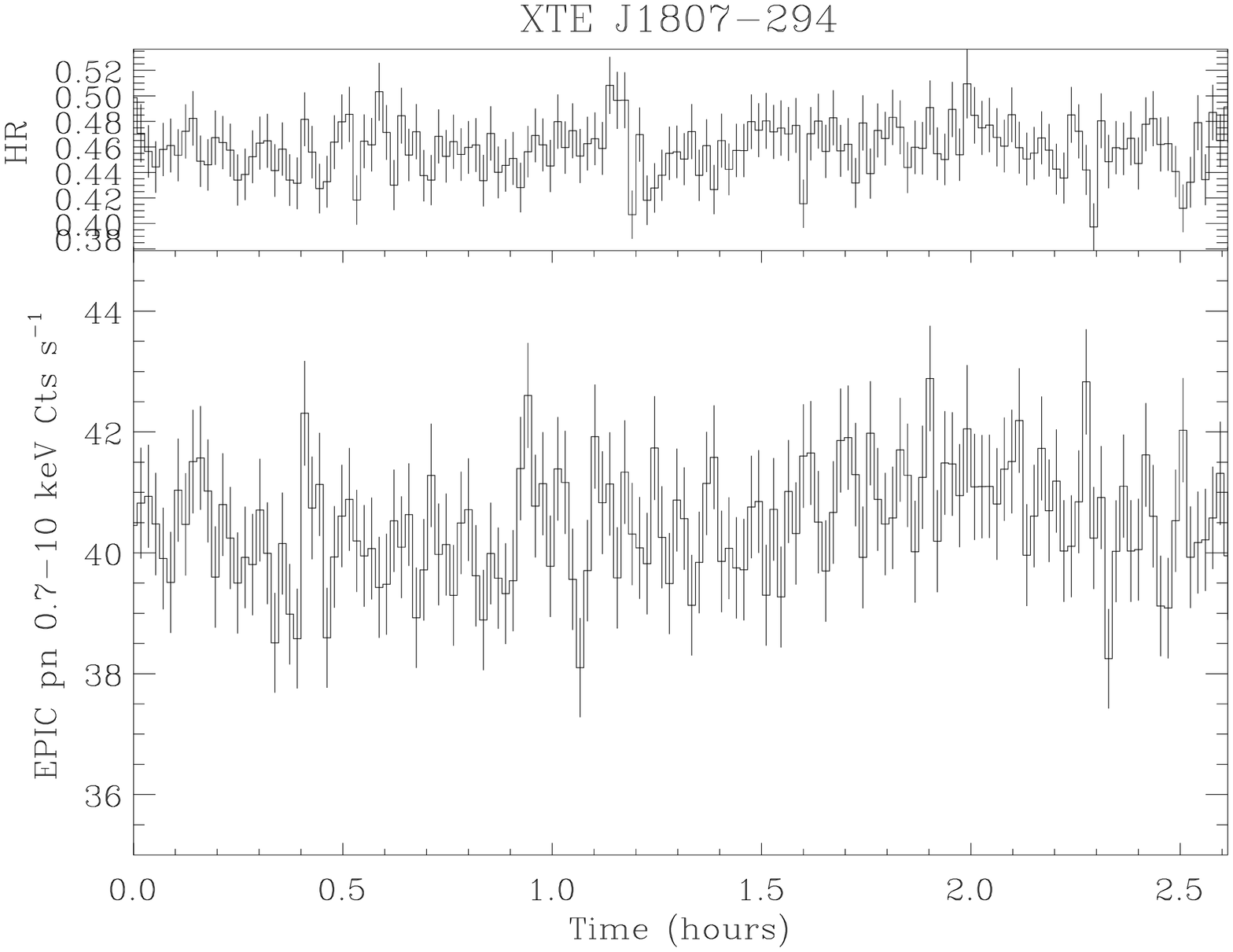}
\vspace{0.1cm}
\includegraphics[angle=90,width=0.33\textwidth]{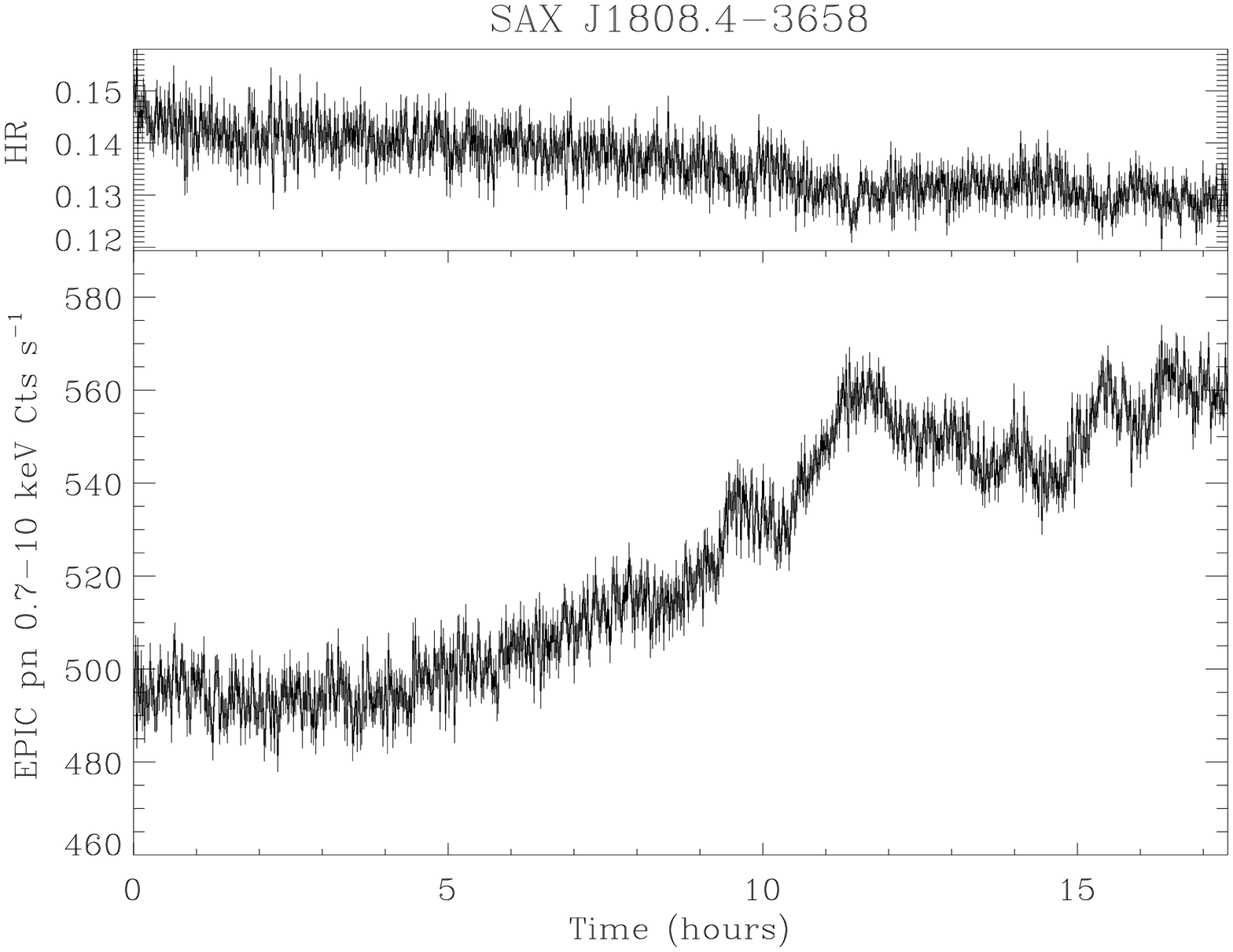}
\caption{Continued from Fig.~\ref{fig:lc}.
}
\label{fig:lc2}
\end{figure*}

\subsection{X-ray spectra}
\label{sec:spectra}

We rebinned the EPIC pn spectra to over-sample the $FWHM$ of the energy
resolution by a factor of 3 and to have a minimum of 25 counts per bin,
to allow the use of the $\chi^2$ statistic. 
To account for systematic effects in the EPIC pn timing mode we added
quadratically a 2\% uncertainty to each pn spectral bin. 
 We performed spectral analysis using XSPEC \citep{arnaud96conf} version 12.3.1. We
used the photo-electric cross-sections of \citet{wilms00apj} to
account for absorption by neutral gas with solar abundances (the so-called 
{\tt tbabs} XSPEC model). Spectral uncertainties are given at 90\%
confidence ($\Delta$\chisq = 2.71), and
upper limits at 95\% confidence.

We first fitted the EPIC~pn spectra with a model consisting of a
blackbody and a disc blackbody, both modified by photo-electric
absorption from neutral material (model {\tt tbabs*(bbodyrad+diskbb)} 
in XSPEC). 

The fits were unacceptable in most of the cases, mainly due to the
presence of a broad emission feature at $\sim$6-7~keV and a
significant excess in the pn spectrum below $\sim$1.5~keV. This
excess consisted of a ``Gaussian-like'' component
centred at $\sim$1~keV. This feature has been previously modelled in a
number of sources either as an emission line or as an edge, and its
nature is unclear \citep[e.g.,][]{1658:sidoli01aa, 1254:boirin03aa,
1916:boirin04aa, 1323:boirin05aa}. If the feature has an astrophysical
origin, its energy is consistent with \neten\
or a blend of \fettwenty-\fetfour\ emission. When the feature is edge-like
its energy is consistent with \oeight. We therefore fitted such a feature with a
Gaussian component or an edge whenever present and discuss its origin 
in Sect.~\ref{sec:discussion}. We fitted the excess at $\sim$6-7~keV with
either a Gaussian or a {\tt laor} component (see below). Finally, 
we added two additional Gaussian absorption/emission features at $\sim$1.84~keV and
$\sim$2.28~keV. Such features are probably due to an incorrect modelling of
the Si and Au absorption in the CCD detectors by the EPIC pn
calibration and are therefore not further discussed.

Summarising, our final model consisted of disc blackbody and blackbody components,
one Gaussian emission feature at $\sim$1~keV (or absorption edge at
$\sim$0.87~keV), and one emission feature at $\sim$6.5~keV (modelled 
with Gaussian or {\tt laor}),
both modified by photo-electric absorption from neutral material, and
two narrow Gaussian features at $\sim$1.84 and $\sim$2.28~keV to account for the 
calibration deficiencies at these energies (Models~1a and 2a, 
see Table~\ref{tab:models}).
\begin{table}
\begin{center}
\caption[]{Description of the models (M) used to fit the spectra in XSPEC notation.
}
\begin{tabular}{ll}
\hline \noalign {\smallskip}
\hline \noalign {\smallskip}
M & Description \\
\hline \noalign {\smallskip}
 1a & {\tt tbabs*(diskbb+bbodyrad+gau$_1$/edge$_1$+gau$_2$)+gau$_3$+gau$_4$} \\
 1b & {\tt tbabs*(diskbb+po+gau$_1$/edge$_1$+gau$_2$)+gau$_3$+gau$_4$} \\
 1c & {\tt tbabs*(bbodyrad+po+gau$_1$/edge$_1$+gau$_2$)+gau$_3$+gau$_4$} \\
 1d & {\tt tbabs*(diskbb+bbodyrad+po+gau$_1$/edge$_1$+gau$_2$)} \\
    & {\tt +gau$_3$+gau$_4$} \\
 2a & {\tt tbabs*(diskbb+bbodyrad+gau$_1$/edge$_1$+laor)+gau$_3$+gau$_4$} \\
 2b & {\tt tbabs*(diskbb+po+gau$_1$/edge$_1$+laor)+gau$_3$+gau$_4$} \\
 2c & {\tt tbabs*(bbodyrad+po+gau$_1$/edge$_1$+laor)+gau$_3$+gau$_4$} \\
 2d & {\tt tbabs*(diskbb+bbodyrad+po+gau$_1$/edge$_1$+laor)} \\
& {\tt +gau$_3$+gau$_4$} \\
\noalign {\smallskip} \hline \label{tab:models}
\end{tabular}
\end{center}
\end{table}

The fits with Model~1a were acceptable for fifteen out of 
the nineteen observations for which we
extracted a spectrum, with \rchisq\ between 0.7 and 1.2
for a number of degrees of freedom (d.o.f.) between 161 and 221. The parameters of the best-fit with
this model are given in Table~\ref{tab:bestfit} and the residuals of
the fits are shown in Fig.~\ref{fig:spectra} (see Appendix~\ref{sec:app2}). 

Substituting the blackbody or disc blackbody components in Model~1a by
a power-law component resulted in fits with similar or worse \rchisq\
for all the observations shown in Table~\ref{tab:bestfit} (see
Table~\ref{tab:chisq}). Substituting the blackbody component
by a cutoff power-law in Model~1a resulted in fits with similar \rchisq.
This is expected, since the blackbody in Model~1a accounts for the emission
of the boundary layer, and could be therefore equally well fitted by a cutoff 
power-law or a {\tt comptt} component, representing saturated comptonization.
Finally, substituting the disc blackbody component by a cutoff power-law in 
Model~1a resulted in fits with similar \rchisq, but with unrealistic values for
the index and the cutoff energy of the cutoff power-law component. For example,
for the bright source \gxtfn, we obtained an index of -0.2\,$\pm$\,0.4 and an energy for the cutoff
of 1.0\,$^{+0.3}_{-0.2}$~keV. Similarly, for the dim source \seventeentw\ we obtained an energy for the cutoff
of 0.51\,$\pm$\,0.02~keV and the index was unconstrained. Fixing the index of the cutoff
power-law to a more realistic value of 1 \citep{cygx2:hasinger90aa} 
yielded fits with significantly worse \rchisq.

We performed a comparison of the \rchisq\
for the different continua before adding the emission line at
$\sim$6.5~keV to avoid the \ew\ of the line being affected by
the possible deficiencies of the fit to the continuum. Whenever
two different continua yielded a similar \rchisq\ before inclusion of
the Fe line, e.g. using a disc blackbody in combination with a blackbody 
or a cutoff power-law components, 
the breadth of the line was similar in both fits.

For one of the remaining four observations, corresponding to \xtee, we
obtained already an acceptable \rchisq\ of 1.20 for 224 d.o.f. when
fitting the continuum with Model~1a. However, a better fit was
obtained substituting the blackbody or disc blackbody components by a
power law (Model~1b or Model~1c, see Table~\ref{tab:models}), 
with a final \rchisq\ of 1.00 (Model~1b) and 1.02 (Model~1c) for 224 d.o.f. None of
the fits showed significant residual emission at $\sim$1 or
$\sim$6.5~keV. We calculated an upper limit of 17~eV for the \ew\ of a
potential Fe~line with an energy of 6.6~keV and a width of 0.3~keV in
this observation. For the other AMXP in our sample, \sax, fitting the
spectrum with Models~1a or~1b gave significantly worse results
(\rchisq\ of 20.7 and 2.2 for 224 d.o.f., respectively) than with
Model 1c (\rchisq\ of 1.1 for 224 d.o.f.). Since \xtee\ was equally
well fitted with Models 1b and 1c and \sax\ was best-fit with Model 1c, we
chose the latter to model both sources and thus allow a better
comparison of the parameters. For \sax, we show two fits,
corresponding to 2 different spectra: one extracted using the full PSF
and one after removal of the 2 central columns of the PSF (marked with
** in Tables~\ref{tab:bestfit2}-\ref{tab:bestfitlines}). The reason is
that it was difficult to decide based on the {\tt epatplot} if the original
spectrum had some residual pile-up. The second spectrum (**) was 
completely free of pile-up. We show the
parameters of the best-fit for \xtee\ and \sax\ in
Table~\ref{tab:bestfit2}.

Similarly, for \osix\ we obtained a better \rchisq\ of
1.19 (220) with Model~1c, compared to 2.25 (220) with Model~1a and 1.21 (220) with 
Model~1b. We show the parameters of the best-fit for this source in
Table~\ref{tab:bestfit2}.

Finally, for Obs~0303250201 of \sixteen\ we needed three continuum components, 
blackbody, disc blackbody and power-law, in order to obtain an acceptable \rchisq. We
note that fits of this observation with a simultaneous RXTE observation require
as well three components and show a power law extending well above 10~keV \citep{1636:pandel08apj}.
We
show the parameters of the best-fit for this observation in
Table~\ref{tab:bestfit2}. 

Next, we examined the asymmetry of the line and evaluated to which
extent was the iron emission better approximated by a model including
relativistic effects. For this, we substituted the second
Gaussian component at $\sim$6.5~keV (gau$_2$) in Models~1a, 1b and 1c
by a {\tt laor} component and re-fitted all the spectra with the new
models (Models 2a, 2b and 2c, see Table~\ref{tab:models}).

We show the parameters of the best-fit with these models in
Tables~\ref{tab:bestfit}-\ref{tab:bestfit2} and the residuals of the
fits in Fig.~\ref{fig:spectra}. The fits with Models~2a, 2b and 2c were
again acceptable, with \rchisq\ between 0.8 and 1.2 for a number of
d.o.f. between 161 and 221, i.e. similar to the \rchisq\ obtained when
a simple Gaussian model was used to fit the lines.

We found that fits to continua different to the one selected 
that yielded significantly higher values of \rchisq\ had a clear influence on the
line breadth for all the sources analysed in this work.

 Narrow absorption features were not visible in any of the spectra, as expected
for sources which are not at high inclination. Similarly, adding an absorption edge
 to the best-fit models shown in Table~\ref{tab:bestfit} did not improve significantly
the goodness of the fits for any source with the exception of \gxtfn. For this source,
the \rchisq\ improved from 1.15 (216) to 1.07 (214) after adding an edge at $\sim$\,9.2\,$\pm$\,0.1~keV, 
that we attribute to absorption from \fetsix. 
The parameters of the Fe line did not change after including this additional edge.
We note that this edge was previously detected in a
 BeppoSAX observation of the source \citep{gx349:iaria04apj}.

For sources with significant emission at $\sim$6.5~keV, the parameters of 
the continuum did not change significantly when fitting the former with
a Gaussian or a {\tt laor} component, except for \seventeentw. Previous
analyses of this source based on \chandra\ and RXTE or BeppoSAX data 
\citep{1728:dai06aa}
showed equally good fits when modelling the emission at $\sim$6.5~keV with
a broad line or with two absorption edges. Therefore, we regard the
\xmm\ fits from \seventeentw\ in this work with caution and discuss 
them in the frame of the
general properties of the iron lines in the following Sections.

\begin{figure*}[!ht]
\includegraphics[angle=0,width=0.33\textwidth]{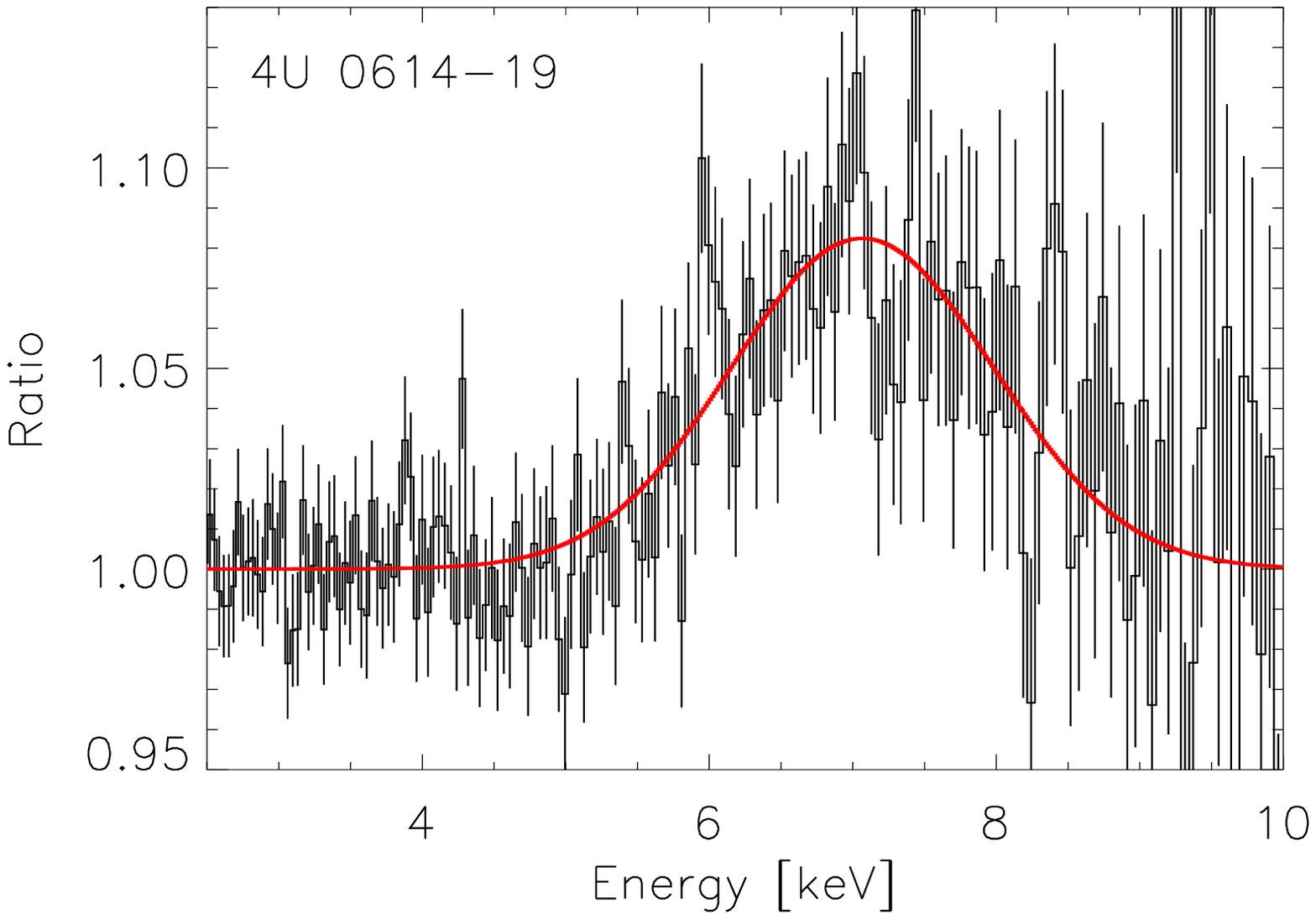}
\includegraphics[angle=0,width=0.33\textwidth]{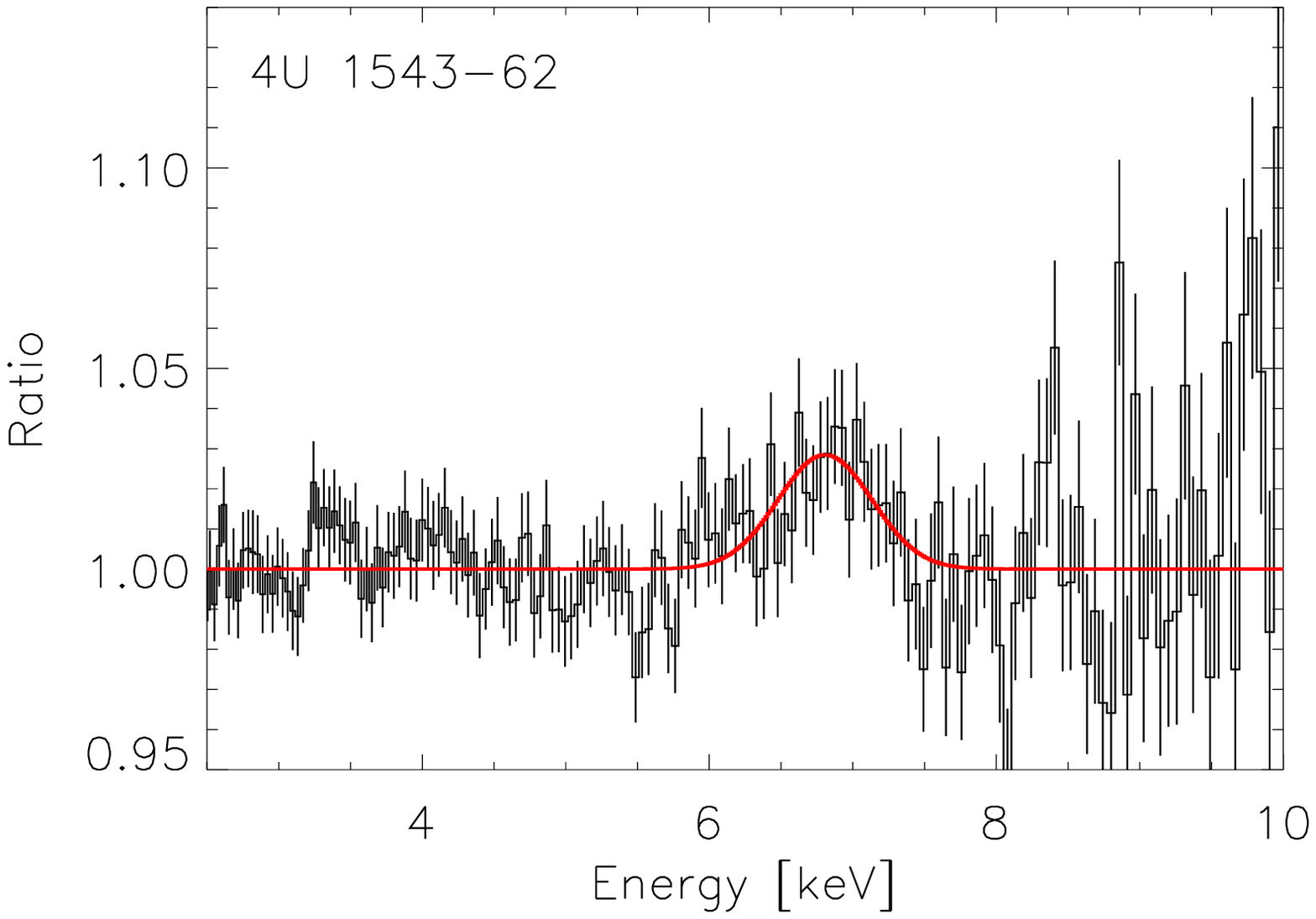}
\includegraphics[angle=0,width=0.33\textwidth]{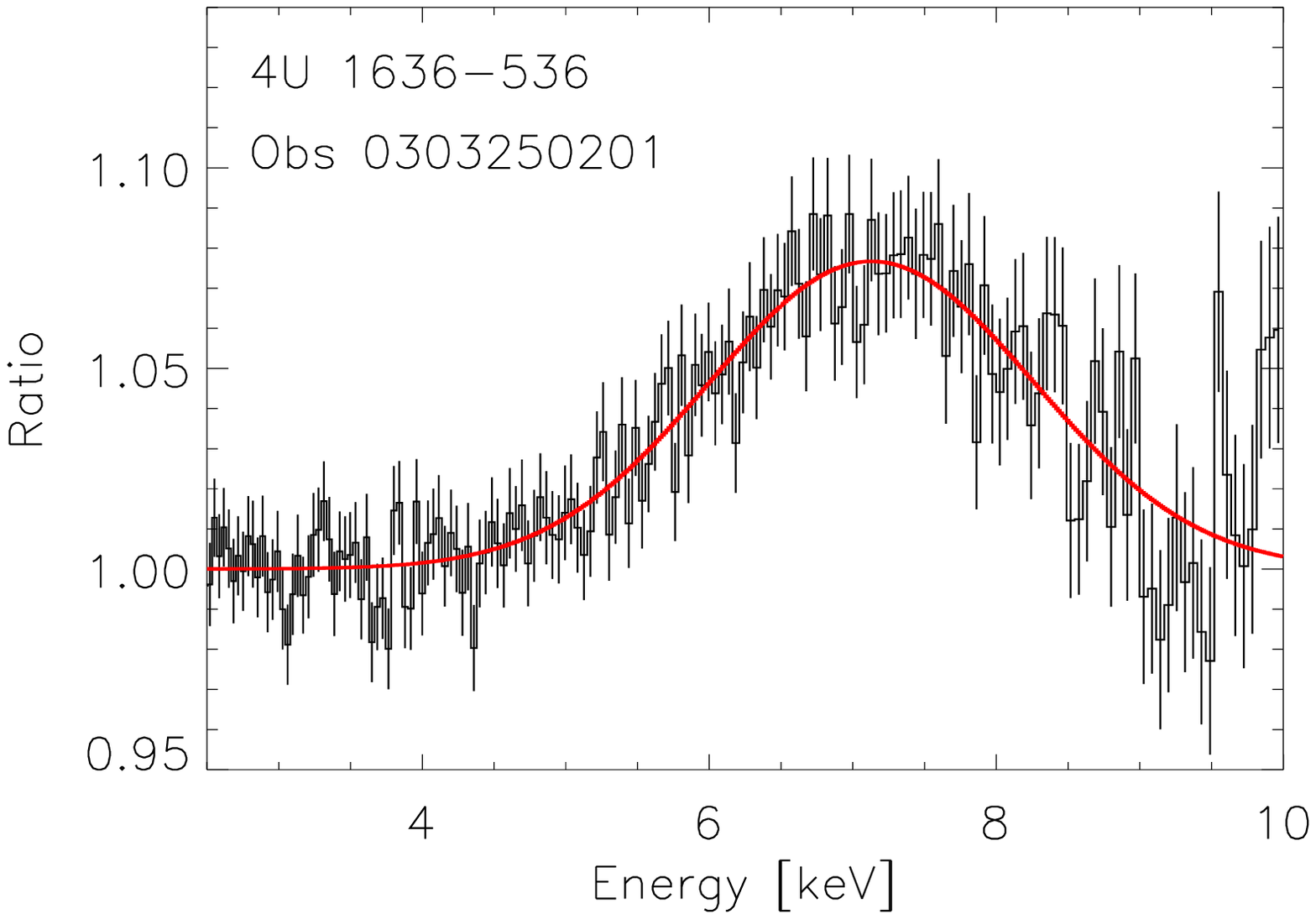}
\vspace{0.1cm}
\includegraphics[angle=0,width=0.33\textwidth]{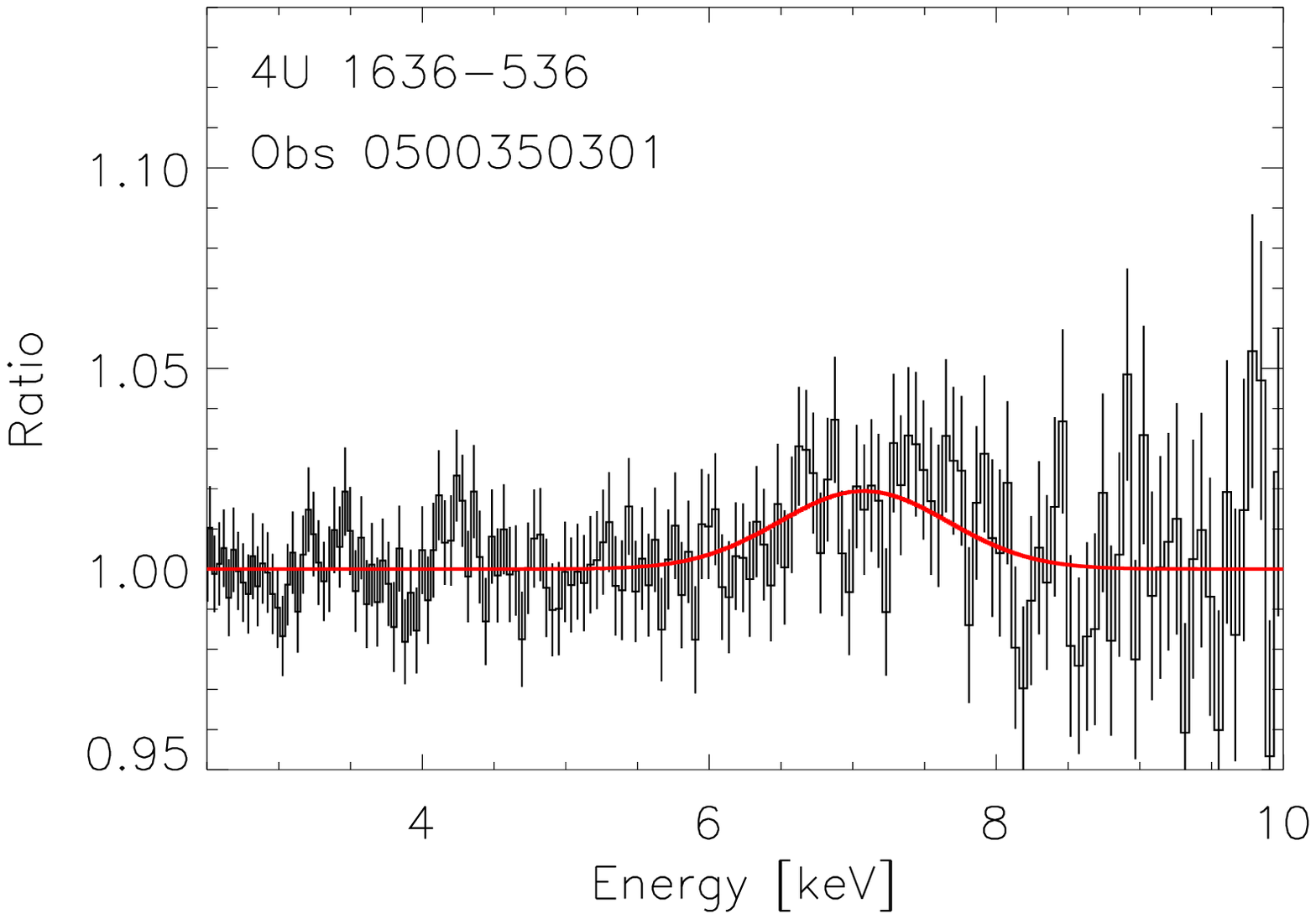}
\includegraphics[angle=0,width=0.33\textwidth]{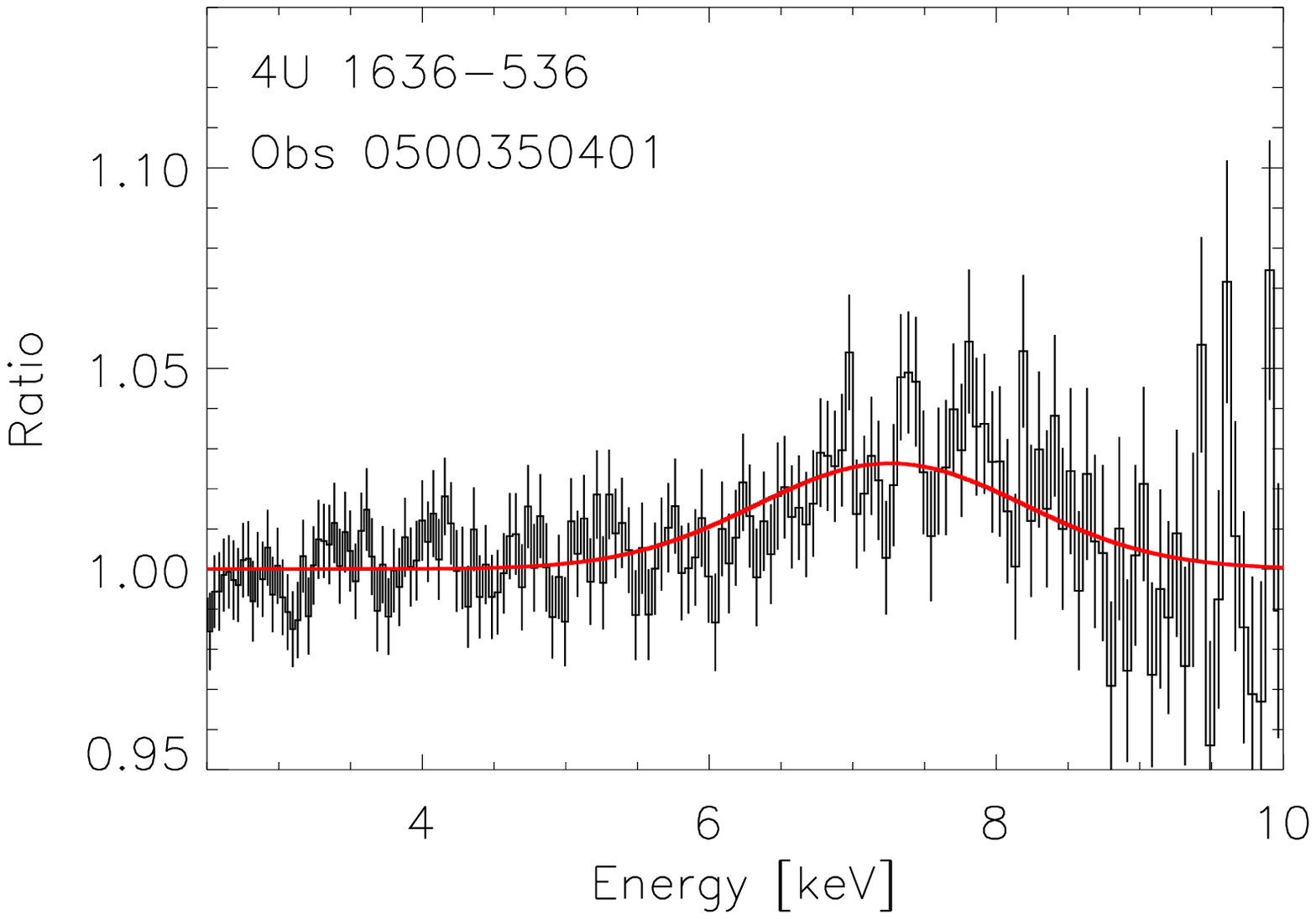}
\includegraphics[angle=0,width=0.33\textwidth]{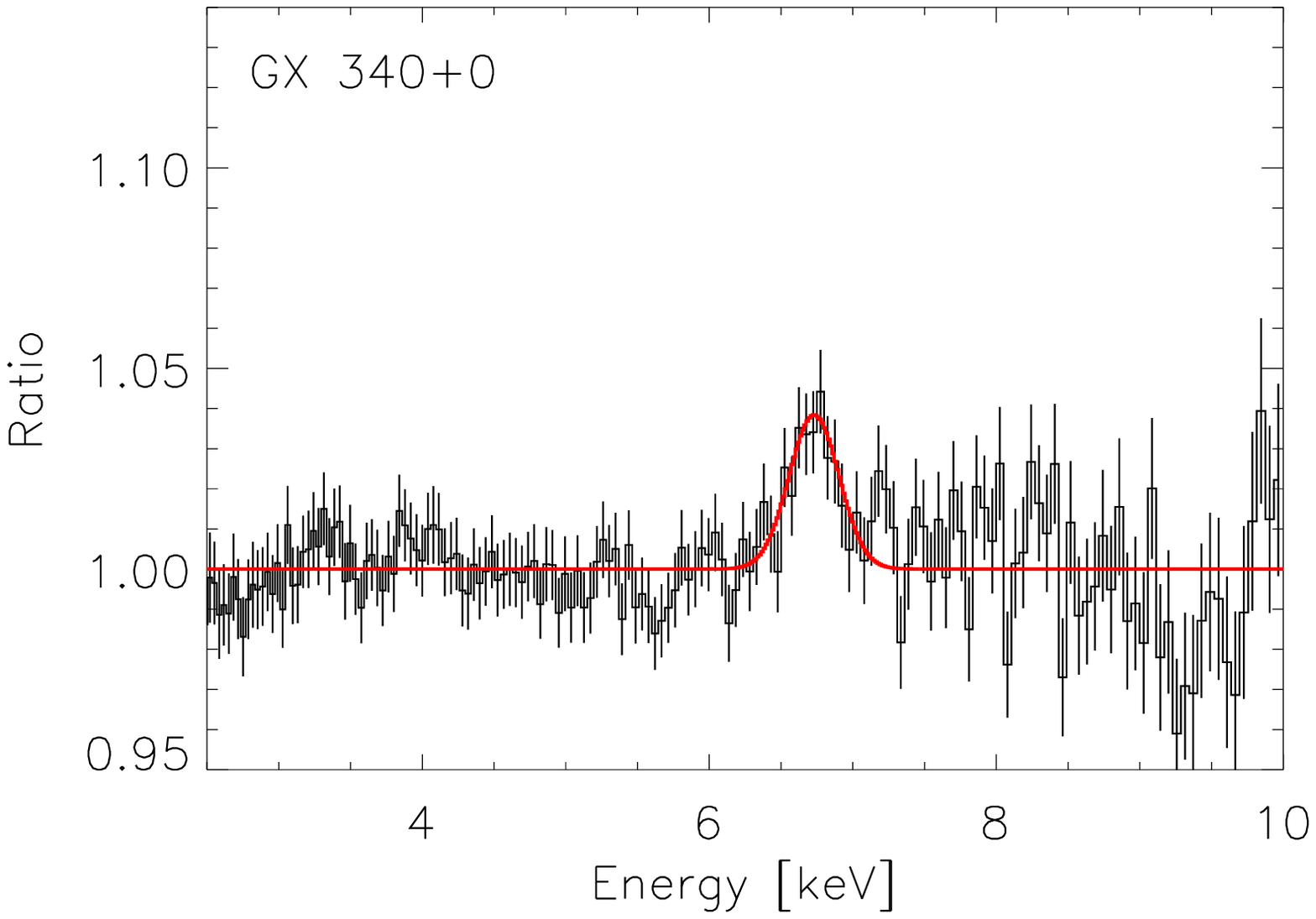}
\vspace{0.1cm}
\includegraphics[angle=0,width=0.33\textwidth]{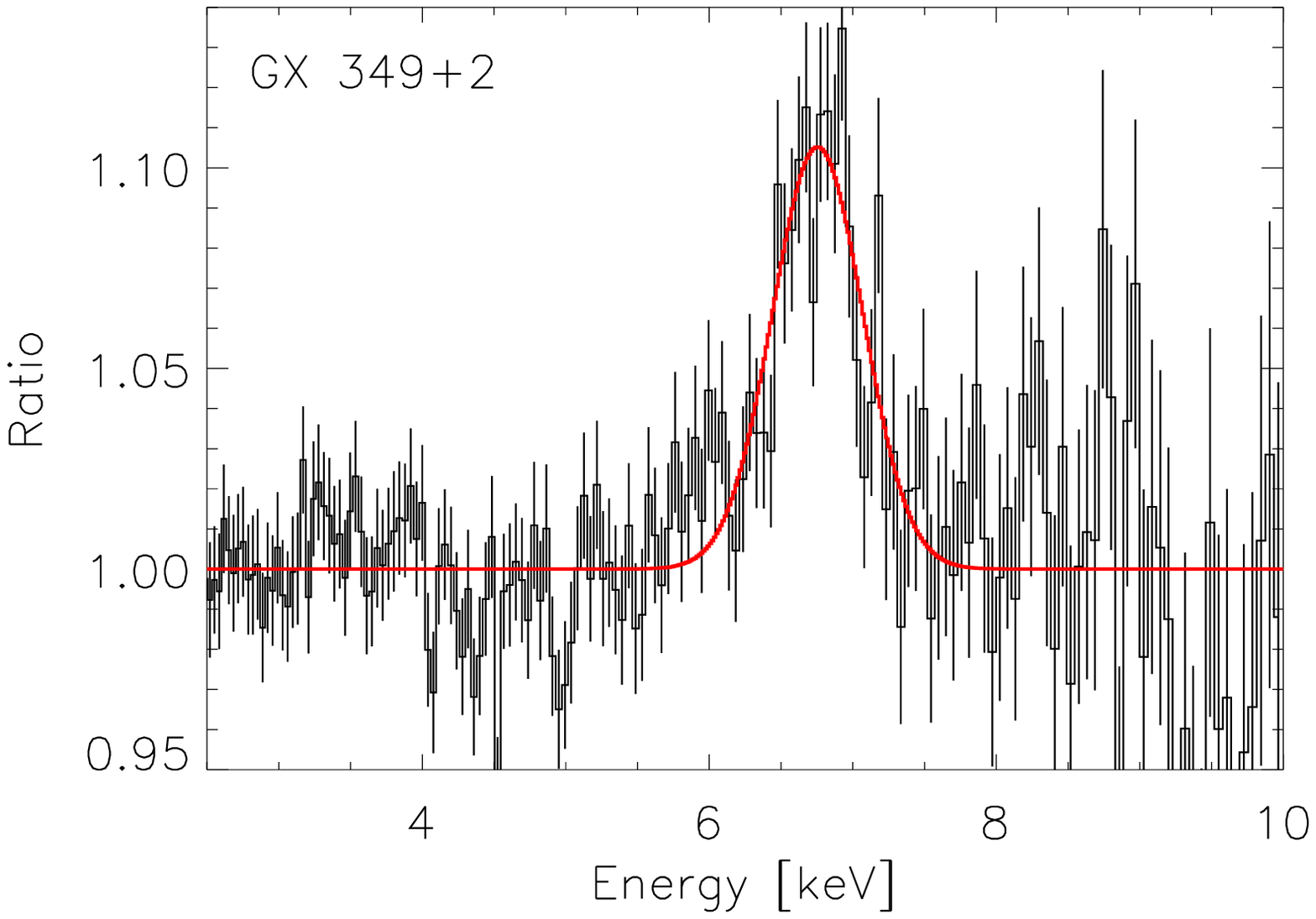}
\includegraphics[angle=0,width=0.33\textwidth]{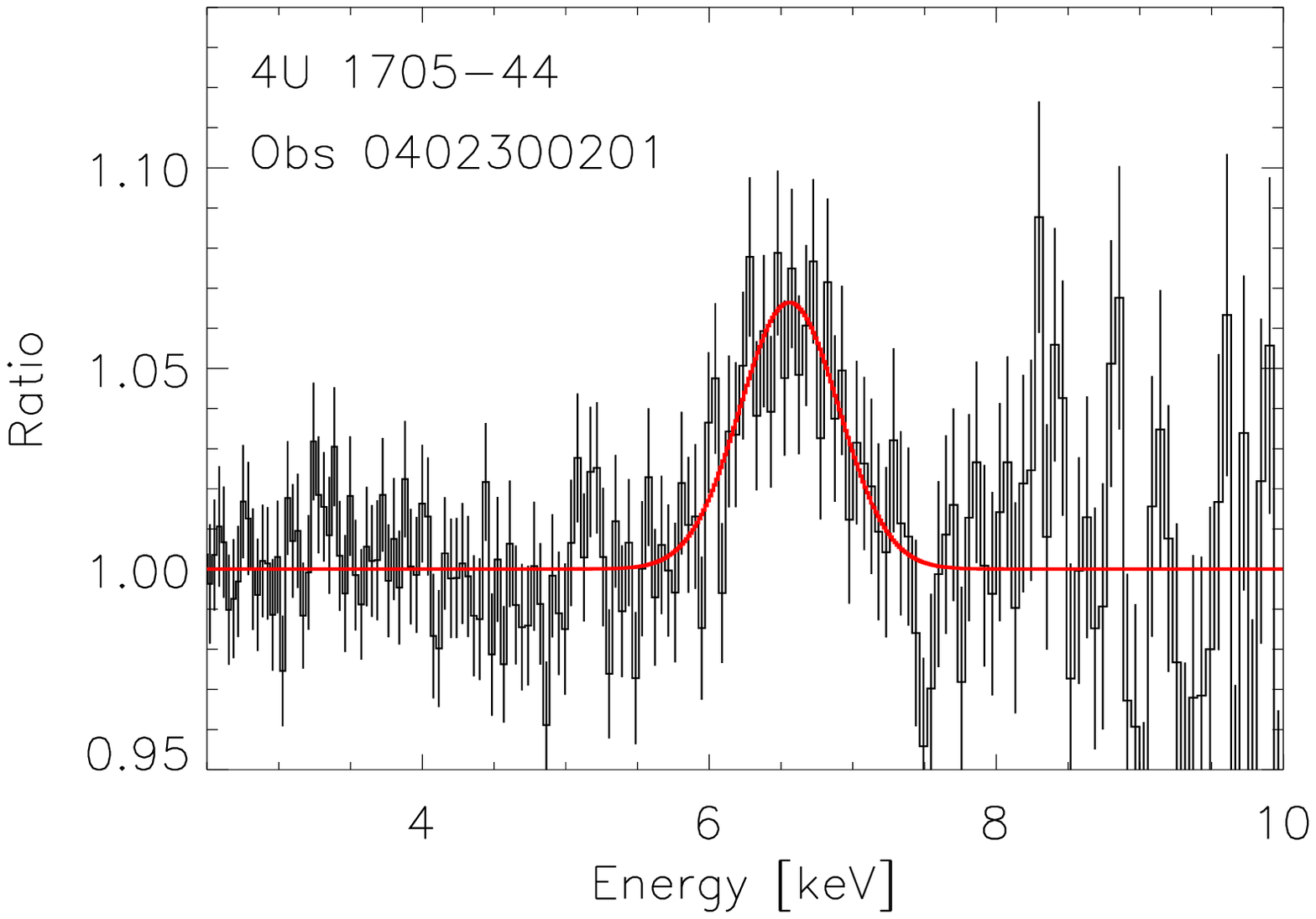}
\includegraphics[angle=0,width=0.33\textwidth]{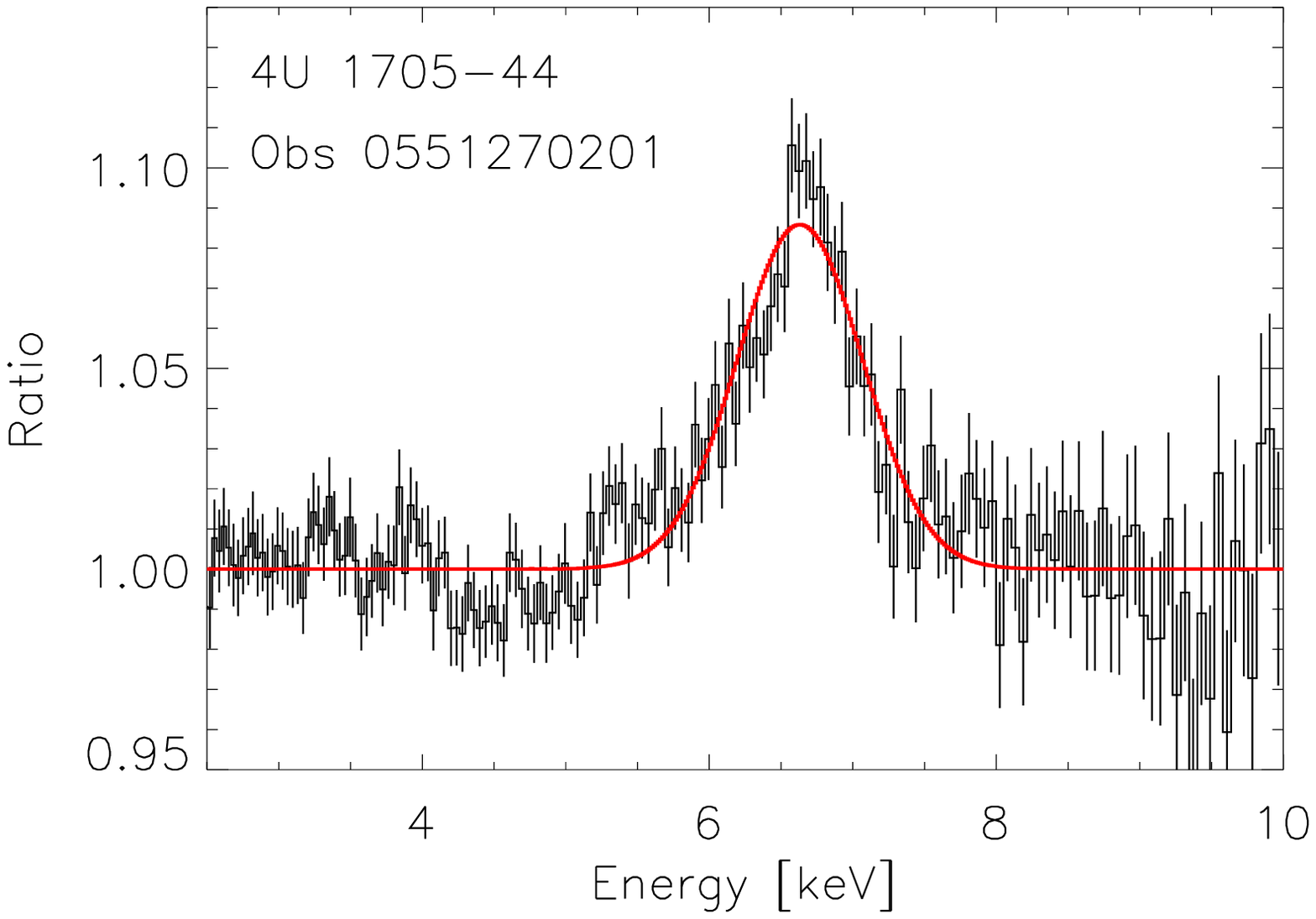}
\vspace{0.1cm}
\includegraphics[angle=0,width=0.33\textwidth]{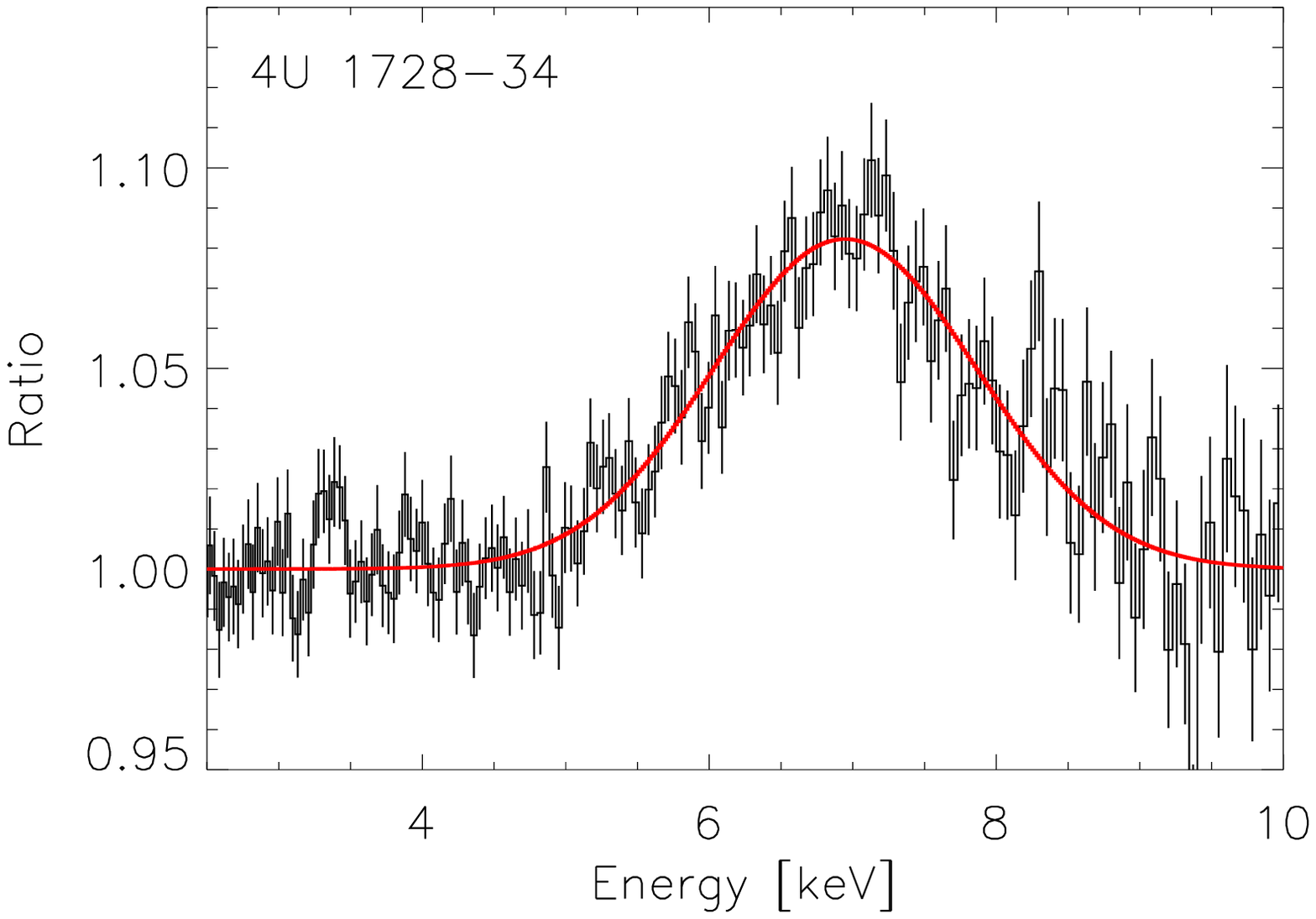}
\includegraphics[angle=0,width=0.33\textwidth]{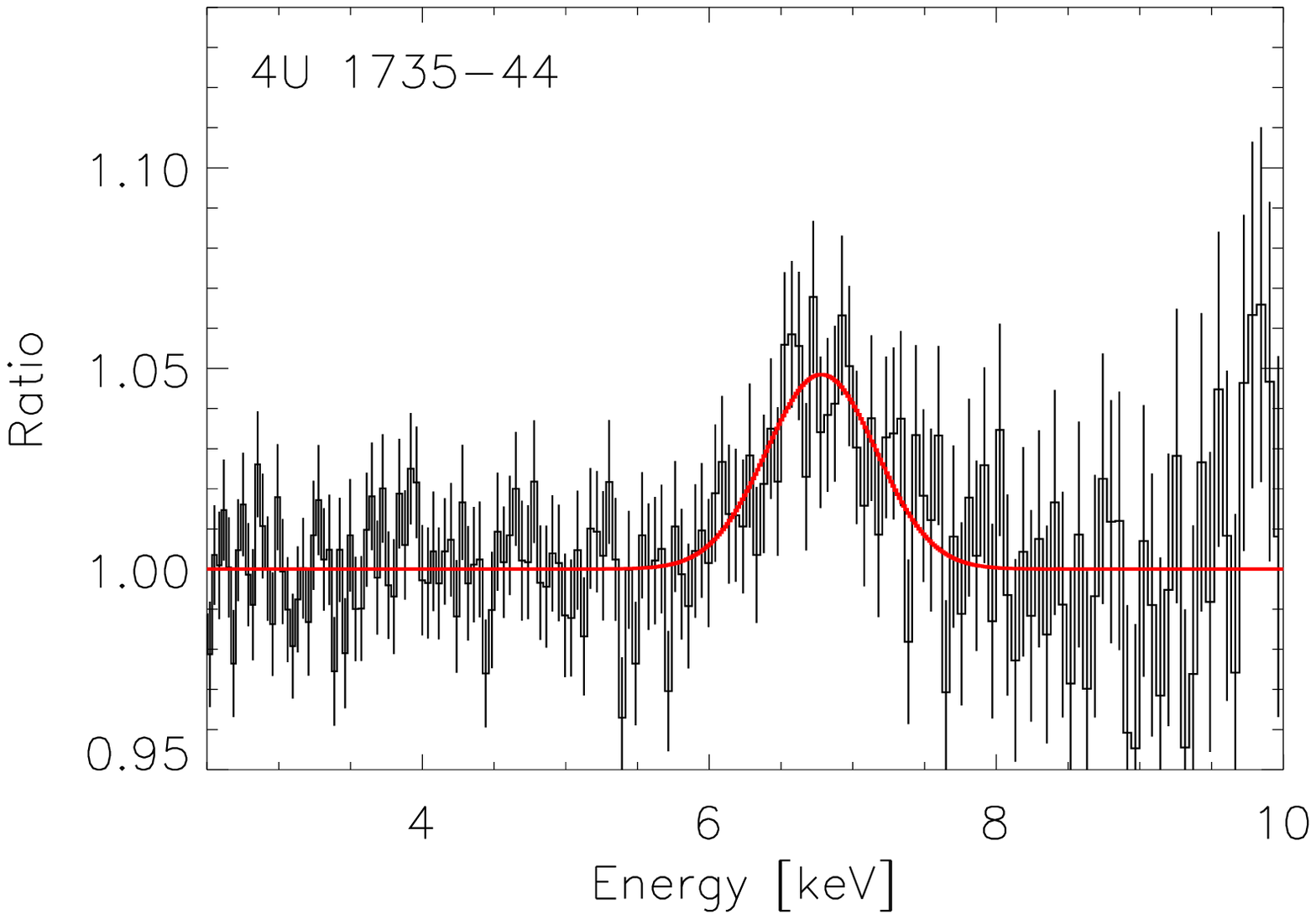}
\includegraphics[angle=0,width=0.33\textwidth]{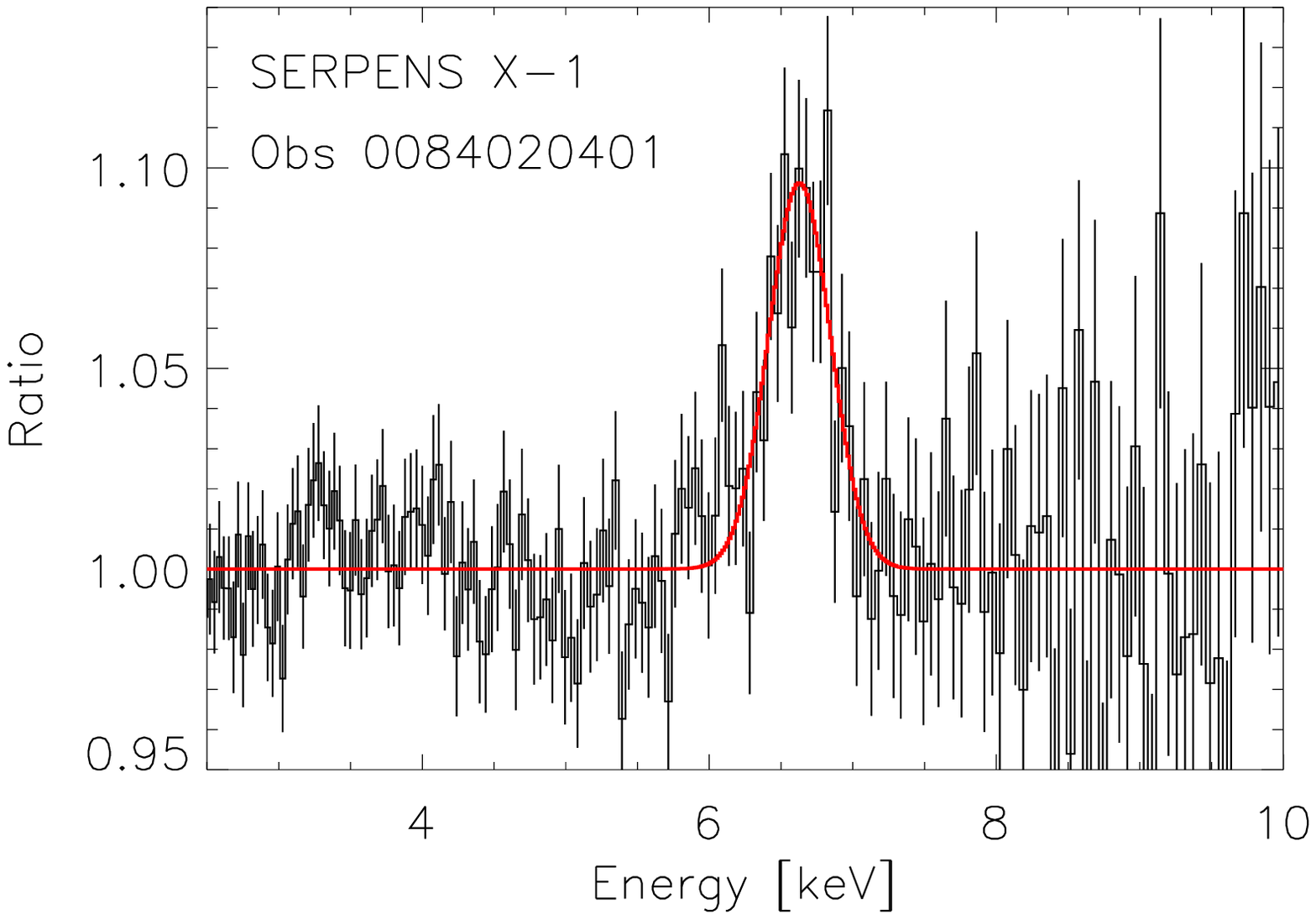}
\vspace{0.1cm}
\includegraphics[angle=0,width=0.33\textwidth]{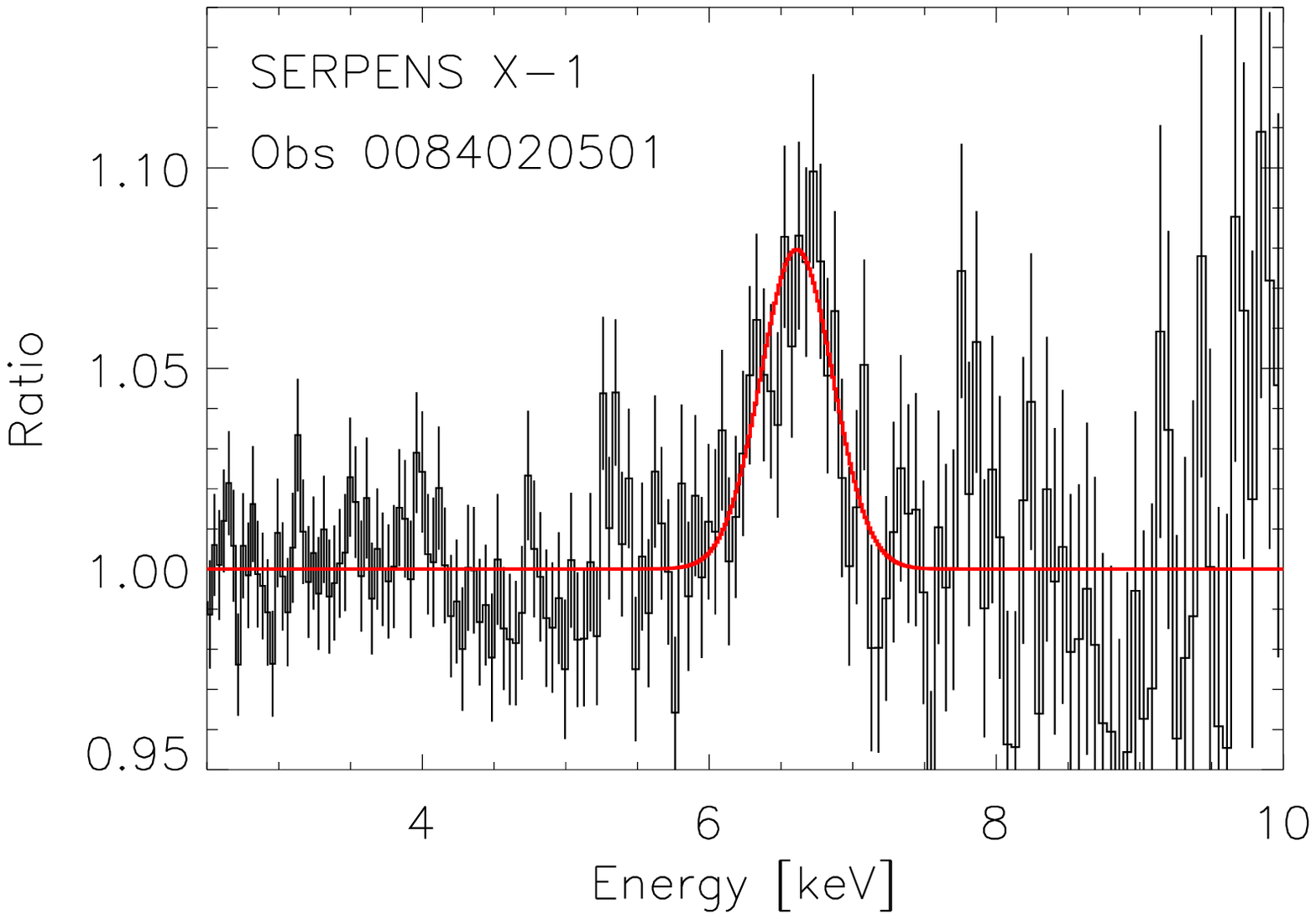}
\includegraphics[angle=0,width=0.33\textwidth]{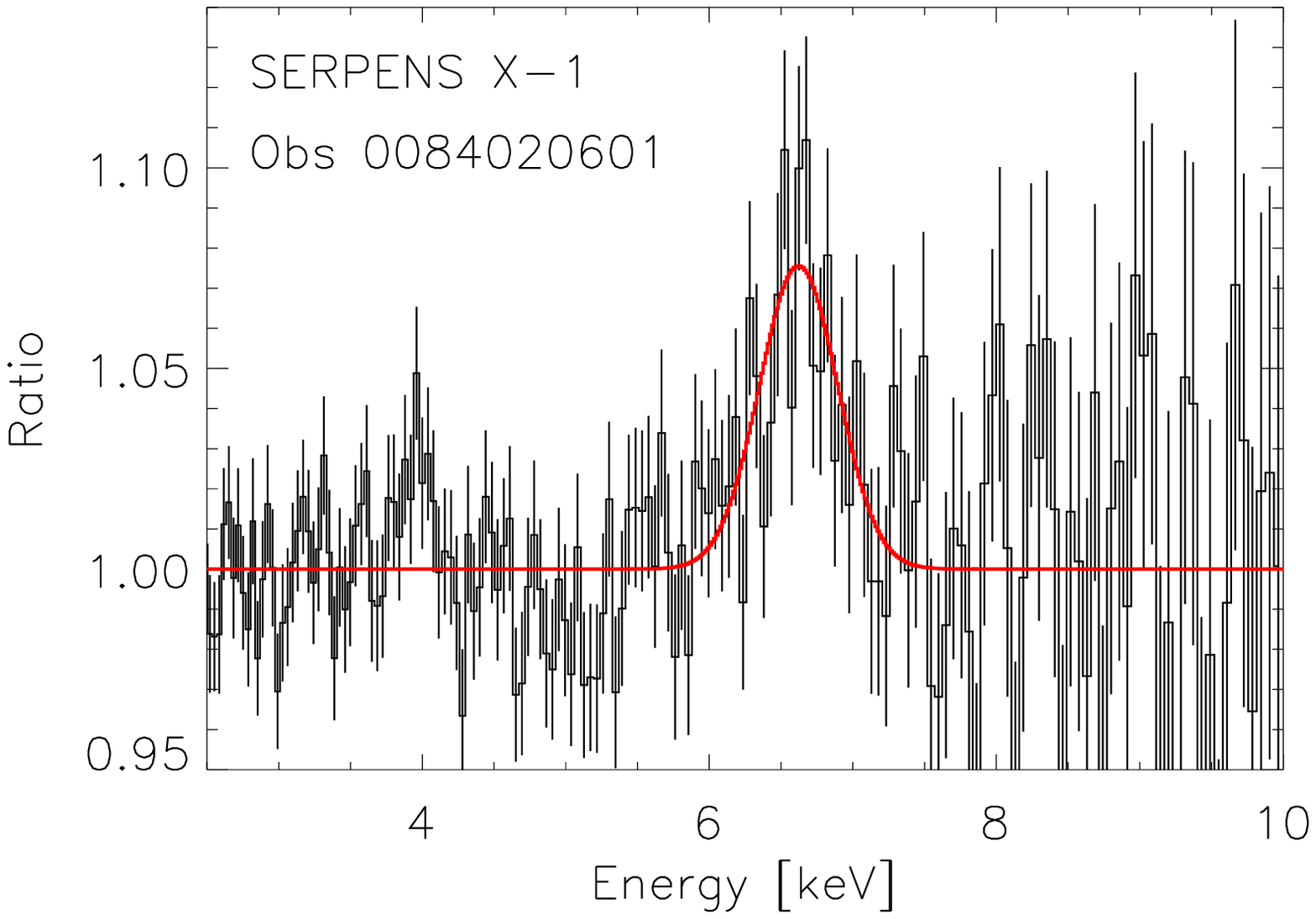}
\includegraphics[angle=0,width=0.33\textwidth]{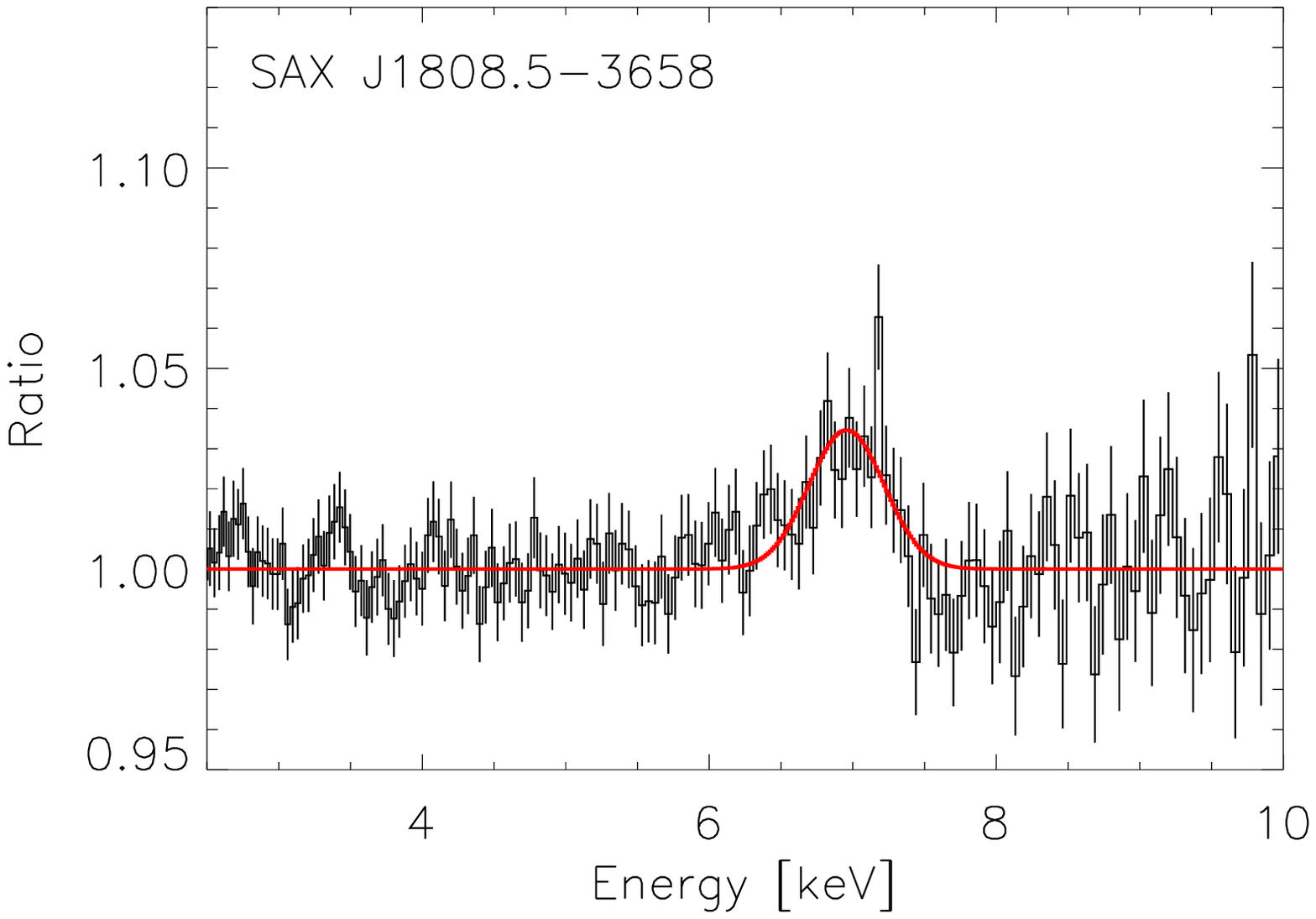}
\vspace{0.1cm}
\caption{Ratio of the data to the continuum model for all the neutron star LMXBs
analysed in this work for which significant Fe~K emission is detected. 
}
\label{fig:gallery}
\end{figure*}

Fig.~\ref{fig:gallery} shows the ratio between the data and the
best-fit model at the Fe~line energy band for all the sources with
significant emission when the Fe~line is not included. The fit of the
line with a Gaussian profile is shown in red.

\begin{landscape}
\begin{table}
\caption[]{Best-fits to the EPIC pn spectra for all the
  observations (except for \xtee, \sax, \osix\ and \sixteen\ Obs~0303250201, 
  which are shown in Table~\ref{tab:bestfit2}) 
  using
  Model~1a ({\tt
  tbabs*(bbrad+dbb+gau$_1$/edge$_1$+gau$_2$)+gau$_3$+gau$_4$}) or
  Model~2a ({\tt
  tbabs*(bbrad+dbb+gau$_1$/edge$_1$+laor)+gau$_3$+gau$_4$}). gau$_1$/edge$_1$
  models the emission feature at $\sim$1~keV with a Gaussian component 
  or an edge respectively. \ktbb\ and
  \ktdbb\ are the temperatures of the blackbody and disc blackbody components
  respectively. \kbb,
  \kdbb\ and \kgau\ are the normalizations of the blackbody,
  disc blackbody and gau$_1$ components in units of
  {(R$_{in}$/D$_{10}$)$^{2}$}, {(R$_{in}$/D$_{10}$)$^{2}$ cos$\theta$}
  and {10$^{-4}$ ph cm$^{-2}$ s$^{-1}$} respectively. \egau\ and $\sigma$ 
  represent the energy and width of
  the Gaussian feature. E$_{edge}$ is the edge energy
  and $\tau$ its optical depth. $F_9$ is the unabsorbed 2--10 keV total flux
  in units of 10$^{-9}$\small \ergcmsec. We show
  two rows per observation corresponding to fits with Models~1a and 2a, except
  for observations for which no significant emission was found at the 6~keV
  range, since in this case Models~1a and 2a are the
  same model, namely {\tt
  tbabs*(bbrad+dbb+gau$_1$/edge$_1$)+gau$_3$+gau$_4$}. We do not show
  the values of gau$_3$ and gau$_4$, since they are fixed calibration
  features. The parametres of gau$_2$ and {\tt laor} components are shown in
  Table~\ref{tab:bestfitlines}. The spectrum energy range is 1.5--10,
  1.8--10 and 2.2--10~keV for \seventeenof, \seventeentw\ and \gxtfz,
  respectively, and 0.7--10~keV otherwise (see text).} 
  \renewcommand{\tabcolsep}{1.5mm}
\begin{tabular}[t]{lccccccccccccc}
\hline \hline
Target & Observation & {\tt tbabs} & \multicolumn{2}{l}{\tt dbb}& \multicolumn{2}{l}{\tt bbrad} & \multicolumn{4}{l}{\tt {gau$_1$}} & $F_9$ & \rchisq (d.o.f.) & Model \\
 & & \nhabs\ & \ktdbb\ & \kdbb\ & \ktbb\ & \kbb\ & \egau\ & $\sigma$ & \kgau\ & EW & & \\
& & \small($10^{21}$ cm$^{-2}$) & {\small(keV)} &  & {\small(keV)} & & {\small(keV)} & {\small(keV)} & &  {\small(eV)} & \\
\hline
\sixteen\ & 0500350301 & 2.38\,$\pm$\,0.04 & 0.80\,$\pm$\,0.01          & 177\,$\pm$\,9      & 1.86\,$\pm$\,0.02 & 8.24\,$\pm$\,0.3          & 1.0 (f) & 0.31\,$\pm$\,0.02 & 252\,$\pm$\,30 & 63\,$\pm$\,7.5 & 1.34 & 0.71 (218) & 1a \\
& & 2.38\,$\pm$\,0.03 & 0.80\,$\pm$\,0.01 & 177\,$\pm$\,8 & 1.86\,$\pm$\,0.02 & 8.2\,$\pm$\,0.3 & 1.0 (f) & 0.31\,$\pm$\,0.02 & 253\,$_{-16}^{+32}$ & 62\,$_{-4}^{+8}$ & 1.34 & 0.70 (217) & 2a \\
& 0500350401 & 2.49\,$\pm$\,0.04 & 0.84\,$\pm$\,0.01          & 185\,$_{-4}^{+5}$      & 1.80\,$\pm$\,0.02 & 12.3\,$\pm$\,0.5 & 1.06\,$\pm$\,0.06 & 0.23\,$_{-0.06}^{+0.17}$ & 159\,$_{-55}^{+450}$ & 35\,$_{-12}^{+98}$  & 1.82 & 0.72 (217) & 1a \\
    &  & 2.50\,$_{-0.04}^{+0.09}$ & 0.83\,$\pm$0.01          & 196\,$\pm$\,7       & 1.80\,$\pm$\,0.01          & 12.7\,$\pm$\,0.4 & 1.08\,$_{-0.14}^{+0.05}$ & 0.20\,$_{-0.05}^{+0.09}$ & 123\,$_{-38}^{+135}$ & 27\,$_{-8}^{+30}$   & 1.82 & 0.80 (215) & 2a \\ \hline\noalign {\smallskip}
\gxtfz\   & 0505950101            & 65.0\,$\pm$\,1.9 & 0.96\,$_{-0.05}^{+0.07}$ & 757\,$_{-206}^{+235}$ & 1.64\,$\pm$\,0.03 & 147\,$_{-18}^{+13}$ & -                      & -                      & -                  & -               & 15.3 & 0.86 (163) & 1a \\
  &             & 65.2\,$_{-0.7}^{+1.8}$ & 0.95\,$\pm$\,0.05 & 787\,$_{-184}^{+228}$ & 1.64\,$\pm$\,0.02 & 150\,$_{-10}^{+13}$ &  -                      & -                      & -                  & -               & 15.3 & 0.88 (161) & 2a \\ \hline\noalign {\smallskip}
\gxtfn\   & 0506110101 & 7.0\,$\pm$\,0.1            & 0.92\,$\pm$\,0.03          & 465\,$_{-55}^{+59}$     & 1.57\,$\pm$\,0.02           &169\,$\pm$\,10 & 1.06\,$\pm$\,0.02 & 0.09\,$\pm$\,0.02          & 517\,$_{-101}^{+131}$& 31\,$_{-6}^{+8}$  & 12.1 & 1.15 (216) & 1a \\
&      & 7.0\,$\pm$\,0.1            & 0.93\,$\pm$\,0.04          & 443\,$\pm$\,56     & 1.58\,$\pm$\,0.02 & 164\,$_{-12}^{+8}$& 1.06\,$\pm$\,0.02 & 0.09\,$\pm$\,0.02          & 519\,$_{-101}^{+126}$ & 31\,$\pm$\,7  & 12.1 & 1.14 (214) & 2a \\ \hline\noalign {\smallskip}
\seventeenof\  & 0402300201 & 15.8\,$_{-0.4}^{+0.2}$ & 1.31\,$\pm$\,0.06 & 3.6\,$\pm$\,0.6 & 2.76\,$\pm$\,0.2 & 0.37\,$\pm$\,0.08 & -                      & -                      & -                  & -               & 0.25 & 0.99 (186) & 1a \\
  &            & 15.8\,$\pm$\,0.3 & 1.30\,$_{-0.03}^{+0.06}$ & 3.7\,$^{+0.6}_{-0.3}$ & 2.74\,$\pm$\,0.2 & 0.38\,$^{+0.05}_{-0.07}$ & -                      & -                      & -                  & -               & 0.25 & 1.01 (184) & 2a \\ 
& 0551270201 & 15.0\,$\pm$\,0.5 & 0.95\,$^{+0.07}_{-0.05}$ & 231\,$\pm$\,52 & 1.61\,$\pm$\,0.03 & 79\,$^{+6}_{-8}$ & - & - & - & - & 6.5 & 1.07 (187) & 1a \\
& & 14.7\,$^{+0.5}_{-0.3}$ & 1.01\,$\pm$\,0.06 & 184\,$^{+51}_{-32}$ & 1.64\,$\pm$\,0.03 & 71\,$\pm$\,6 & - & - & - & - & 6.5 & 1.03 (185) & 2a \\\hline\noalign {\smallskip}
\seventeensix\ & 0090340101 & 1.50\,$\pm$\,0.07 & 1.02\,$\pm$\,0.06 & 148\,$^{+31}_{-26}$ & 1.76\,$\pm$\,0.06 & 36\,$\pm$\,6 & 1.0 (f) & 0.20\,$\pm$\,0.07 & 213\,$\pm$\,100 & 29\,$\pm$\,14 & 4.55 & 0.99 (220) & 1a/2a \\ 
 & 0090340601 & 1.48\,$\pm$\,0.05          & 0.96\,$\pm$\,0.03 & 187\,$\pm$\,17            & 1.71\,$\pm$\,0.02          & 49\,$\pm$\,3 & 1.05\,$_{-0.05}^{+0.03}$ & 0.08\,$_{-0.03}^{+0.08}$ & 76\,$_{-23}^{+71}$   & 11\,$_{-3}^{+10}$ & 5.03 & 0.89 (219) & 1a/2a \\ \hline\noalign {\smallskip}
\seventeentw\  & 0149810101            & 27.0\,$\pm$\,0.5 & 1.9\,$\pm$\,0.3 & 3.5\,$_{-1.1}^{+1.5}$ & 3.8\,$_{-1.1}^{+8.4}$ & 0.3\,$_{-0.2}^{+0.9}$ & -                      & -                      & -              & -               & 0.97 & 1.02 (177) & 1a \\
 &             & 27.6\,$\pm$\,0.4 & 1.48\,$_{-0.10}^{+0.14}$ & 7.4\,$_{-0.6}^{+1.9}$ & 2.56\,$_{-0.06}^{+0.11}$ & 1.9\,$\pm$\,0.2 &  -                      & -                      & -              & -               & 0.97 & 1.07 (175) & 2a \\ \hline\noalign {\smallskip}
\seventeenth\ & 0090340201 & 1.85\,$_{-0.05}^{+0.10}$ & 0.93\,$\pm$\,0.02          & 171\,$_{-12}^{+15}$     & 1.89\,$\pm$\,0.02          & 33\,$\pm$\,2 & 1.04\,$_{-0.06}^{+0.02}$ & 0.15\,$_{-0.03}^{+0.06}$ & 218\,$_{-57}^{+142}$ & 36\,$_{-9}^{+23}$ & 4.60 & 0.68 (216) & 1a \\
&        & 1.85\,$_{-0.05}^{+0.08}$ & 0.93\,$\pm$\,0.02          & 172\,$_{-11}^{+17}$     & 1.89\,$\pm$\,0.02 & 33\,$\pm$\,1 & 1.04\,$_{-0.05}^{+0.03}$ & 0.15\,$\pm$\,0.04 & 220\,$_{-61}^{+91}$  & 36\,$_{-10}^{+15}$ & 4.60 & 0.78 (214) & 2a \\ \hline\noalign {\smallskip}
\ser\ & 0084020401  & 3.67\,$_{-0.06}^{+0.08}$ & 0.86\,$\pm$\,0.02          & 292\,$\pm$\,24     & 1.60\,$\pm$\,0.02          & 53\,$\pm$\,3 & 1.06\,$_{-0.04}^{+0.02}$ & 0.13\,$_{-0.02}^{+0.04}$ & 216\,$_{-45}^{+105}$ & 27\,$_{-6}^{+13}$ & 4.41 & 0.98 (216) & 1a \\
 & & 3.66\,$_{-0.04}^{+0.08}$ & 0.87\,$\pm$\,0.01          & 284\,$_{-21}^{+12}$     & 1.61\,$\pm$\,0.01          & 52\,$\pm$\,2 & 1.06\,$\pm$\,0.03 & 0.13\,$_{-0.02}^{+0.04}$ & 221\,$_{-38}^{+100}$ & 28\,$_{-5}^{+13}$ & 4.41 & 0.97 (214) & 2a \\ 
 & 0084020501  & 3.60\,$_{-0.05}^{+0.11}$ & 0.91\,$\pm$\,0.02          & 233\,$\pm$\,18     & 1.64\,$\pm$\,0.03 & 33.7\,$\pm$\,2.6 & 1.07\,$_{-0.07}^{+0.03}$ & 0.14\,$_{-0.03}^{+0.07}$ & 201\,$_{-46}^{+167}$ & 27\,$_{-6}^{+22}$ & 3.58 & 0.93 (216) & 1a \\
 &  & 3.59\,$_{-0.05}^{+0.07}$ & 0.92\,$\pm$\,0.02          & 230\,$_{-14}^{+7}$      & 1.65\,$\pm$\,0.03 & 33.0\,$_{-3.1}^{+2.6}$ & 1.07\,$_{-0.07}^{+0.02}$ & 0.14\,$_{-0.02}^{+0.06}$ & 201\,$_{-43}^{+155}$ & 27\,$_{-6}^{+21}$ & 3.58 & 0.92 (214) & 2a \\ 
 & 0084020601  & 3.68\,$_{-0.06}^{+0.10}$ & 0.87\,$\pm$\,0.02          & 277\,$_{-23}^{+25}$     & 1.61\,$\pm$\,0.02          & 44.2\,$\pm$\,3.0         & 1.06\,$_{-0.06}^{+0.03}$ & 0.14\,$_{-0.03}^{+0.05}$ & 203\,$_{-55}^{+134}$ & 26\,$_{-7}^{+17}$ & 3.97 & 1.04 (216) & 1a \\
 & & 3.69\,$_{-0.05}^{+0.08}$ & 0.87\,$\pm$0.02 & 277\,$_{-13}^{+21}$     & 1.61\,$\pm$\,0.02 & 44.0\,$_{-1.6}^{+2.5}$ & 1.06\,$_{-0.05}^{+0.03}$ & 0.14\,$_{-0.03}^{+0.05}$ & 201\,$_{-46}^{+122}$ & 25\,$_{-6}^{+15}$ & 3.97 & 1.04 (214) & 2a \\ \hline\noalign {\smallskip}
\aql\ & 0303220201 & 2.37\,$_{-0.09}^{+0.23}$ & 1.10\,$\pm$\,0.03          & 25\,$\pm$\,3  & 2.25\,$_{-0.05}^{+0.07}$ &  3.4\,$\pm$\,0.4         & 1.01\,$_{-0.09}^{+0.03}$ & 0.15\,$_{-0.03}^{+0.06}$ & 76\,$_{-20}^{+75}$   & 51\,$_{-14}^{+51}$ & 1.01 & 0.97 (221) & 1a/2a \\ \hline\noalign {\smallskip}
\hline\noalign {\smallskip}
 &  &  & & & & & \multicolumn{2}{l}{\tt {edge$_1$}} & & & & \\
 & & &  &  &  & & E$_{edge}$ & $\tau$ & & & & \\
& &  &  &  &  & & {\small(keV)} & & & & \\\hline\noalign {\smallskip}
\fifteen\ & 0061140201 & 0.71\,$\pm$\,0.04 & 0.640\,$\pm$\,0.005       & 140\,$\pm$\,5         & 1.539\,$\pm$\,0.006          & 6.99\,$_{-0.16}^{+0.14}$   & 0.842\,$_{-0.005}^{+0.008}$ & 0.20\,$\pm$\,0.02 & - & - & 0.47  & 0.97 (217) & 1a \\
&         & 0.70\,$\pm$\,0.04 & 0.642\,$^{+0.005}_{-0.008}$      & 137\,$\pm$\,5         & 1.541\,$\pm$\,0.006          & 6.94\,$\pm$\,0.12            & 0.841\,$_{-0.005}^{+0.008}$ & 0.20\,$\pm$\,0.02 & - & - & 0.47 & 0.96 (215) & 2a \\ \hline\noalign {\smallskip}
\label{tab:bestfit}
\end{tabular}
\end{table}
\end{landscape}

\begin{landscape}
\begin{table}
\caption[2]{Best-fits to the 0.7--10~keV EPIC pn
  spectra for \xtee, \sax\ and \osix\ using Model~1c ({\tt
  tbabs*(bbrad+po+gau$_1$+gau$_2$)+gau$_3$+gau$_4$}) or Model~2c ({\tt
  tbabs*(bbrad+po+gau$_1$+laor)+gau$_3$+gau$_4$}) and to the 0.7--10~keV EPIC pn
  spectra for \sixteen\ Obs~0303250201 with Model~1d ({\tt
  tbabs*(bbrad+dbb+po+gau$_1$+gau$_2$)+gau$_3$+gau$_4$}) or Model~2d ({\tt
  tbabs*(bbrad+dbb+po+gau$_1$+laor)+gau$_3$+gau$_4$}).
  The parameters of the models are the same
  as in Table~\ref{tab:bestfit}, except $\Gamma$ and \kpl, 
  index and normalization of the power law component, the latter
  is in units of {\small[ph keV$^{-1}$ cm$^{-2}$ s$^{-1}$]}. 
  We show one row for each observation
  whenever no line was found at the 6~keV range, since in this case the
  two models 1c and 2c or 1d and 2d are the same model. For \sax\ we show results for two spectra, which
  only differ in the pile-up treatment (see text).
  }
\renewcommand{\tabcolsep}{1.5mm}

\begin{tabular}[t]{lccccccccccccc}
\hline \hline
Target & Observation & {\tt tbabs} & \multicolumn{2}{l}{\tt bbrad}& \multicolumn{2}{l}{\tt po} & \multicolumn{4}{l}{\tt {gau$_1$}} & $F_9$ & \rchisq (d.o.f.) & M \\
 & & \nhabs\ & \ktbb\ & \kbb\ & $\Gamma$ & \kpl\ & \egau\ & $\sigma$ & \kgau\ & EW & & \\
& & \small($10^{21}$ cm$^{-2}$) & {\small(keV)} & & & & {\small(keV)} & {\small(keV)} & & {\small(eV)}  & \\
\hline
\xtee\ & 0157960101 & 5.27\,$\pm$\,0.14       & 0.69\,$\pm$\,0.02 & 12\,$\pm$\,2        & 1.79\,$\pm$\,0.03 & 0.043\,$\pm$\,0.002 & - &  - & - & - & 0.17 & 1.02 (224) & 1c/2c \\ \hline\noalign {\smallskip}
\sax\     & 0560180601 & 1.46\,$\pm$\,0.04           & 2.21\,$\pm$\,0.03          & 1.12\,$\pm$\,0.08          & 2.55\,$\pm$\,0.01           & 0.550\,$\pm$\,0.006           & - & - & - & - & 0.83 & 0.75 (220) & 1c \\
& & 1.43\,$\pm$\,0.04  & 2.26\,$\pm$\,0.05 & 1.01\,$_{-0.05}^{+0.13}$ & 2.54\,$\pm$\,0.02  & 0.546\,$\pm$\,0.006 & - & - & - & - & 0.83 & 0.71 (218) & 2c \\
\sax\     & 0560180601** & 1.52\,$\pm$\,0.05           & 2.08\,$\pm$\,0.03          & 1.4\,$\pm$\,0.1 & 2.59\,$\pm$\,0.02           & 0.540\,$\pm$\,0.007           & - & - & - & - & 0.80 & 0.86 (220) & 1c \\
   &  & 1.53\,$\pm$\,0.05  & 2.08\,$\pm$\,0.03 & 1.43\,$_{-0.12}^{+0.05}$ & 2.59\,$\pm$\,0.02 & 0.542\,$\pm$\,0.007 & - & - & - & - & 0.80 & 0.84 (218) & 2c \\
\hline\noalign {\smallskip}
\hline\noalign {\smallskip}
&  &  & & & & & \multicolumn{2}{l}{\tt {edge$_1$}} & & & & \\
& & &  &  &  & & E$_{edge}$ & $\tau$ & & & & \\
& &  &  &  &  & & {\small(keV)} & & & & \\\hline\noalign {\smallskip}
\osix\ & 0111040101 & 1.89\,$\pm$\,0.07 & 0.64\,$^{+0.07}_{-0.05}$ & 17\,$^{+10}_{-7}$ & 2.16\,$\pm$\,0.02 & 0.275\,$\pm$\,0.008 & 0.856\,$\pm$\,0.007 & 0.20\,$\pm$\,0.02 & - & - & 0.58 & 0.82 (217) & 1c \\
&  & 1.87\,$\pm$\,0.07  & 0.61\,$\pm$\,0.04 & 21\,$^{+10}_{-7}$ & 2.15\,$\pm$\,0.02 & 0.272\,$\pm$\,0.008 & 0.856\,$\pm$\,0.007 & 0.20\,$\pm$\,0.02 & - & - & 0.58 & 0.84 (215) & 2c \\
\hline\noalign {\smallskip}
\hline\noalign {\smallskip}
&  &  & & & & & \multicolumn{4}{l}{\tt {gau$_1$}} & & \\
 & &  &  &  &  &  & \egau\ & $\sigma$ & \kgau\ & EW & & \\
&  &  & & & & & {\small(keV)} & {\small(keV)} & & {\small(eV)} & & \\
\hline\noalign {\smallskip}
\sixteen\ & 0303250201 & 2.77\,$\pm$\,0.03 & 1.78\,$^{+0.45}_{-0.18}$ & 1.0\,$^{+0.5}_{-0.3}$ & 1.86\,$\pm$\,0.02 & 0.204\,$_{-0.002}^{+0.001}$ & 1.02\,$\pm$\,0.03 & 0.22\,$^{+0.05}_{-0.03}$ & 83\,$\pm$\,3 & 38\,$\pm$\,1 & 0.76 & 0.76 (216) & 1d \\
& & 2.61\,$^{+0.02}_{-0.05}$ & 1.74\,$_{-0.08}^{+0.02}$ & 1.29\,$_{-0.27}^{+0.25}$ & 1.820$_{-0.022}^{+0.004}$ & 0.187\,$_{-0.004}^{0.005}$ & 0.88\,$_{-0.03}^{+0.02}$ & 0.314\,$_{-0.040}^{+0.025}$ & 183$_{-24}^{+35}$ & 69\,$^{+13}_{-9}$ & 0.76 & 0.78 (214) & 2d \\
& & & \multicolumn{2}{l}{\tt dbb} & \\
& & & \ktdbb\ & \kdbb\ & \\
& & & {\small(keV)} & & \\
& & & 0.42\,$\pm$\,0.04 & 63\,$^{+32}_{-29}$ & & & & & & & & & 1d \\
& & & 0.58\,$_{-0.02}^{+0.2}$ & 29.9\,$_{-5.5}^{+4.0}$ & & & & & & & & & 2d \\
\hline\noalign {\smallskip}
\label{tab:bestfit2}
\end{tabular}
\end{table}
\end{landscape}

\begin{landscape}
\begin{table}
\begin{center}
\begin{small}
\caption[]{Parameters of the Gaussian (Models~1a, 1b, 1c and 1d) and {\tt
  laor} (Models~2a, 2b, 2c and 2d) components used to model the
  Fe~emission at $\sim$6~keV. \egau\ and $\sigma$ represent the
  energy and width of the Gaussian feature. The parameters of the 
  {\tt laor} component E$_{laor}$,
  $\beta$, r$_{in}$, $i$ and $k_{laor}$ account for energy, emissivity
  index, inner radius (in units of r$_g$), inclination and
  normalization. $\beta$ is constrained to be $<$\,5. The inner radius
  is constrained to be $>$\,1.235~r$_g$ ($GM/c^2$). The inclination is
  constrained to be $<$\,70\deg, since none of the sources is a known
  dipper or eclipsing system. The outer radius has been fixed to
  400~$r_g$ for all sources, since it is not well constrained. The
  normalization is in units of {\small(10$^{-4}$ ph cm$^{-2}$
  s$^{-1}$)} for both components. Parameters that are completely
  unconstrained when the errors are calculated have been fixed at
  their best value and are indicated in the table with ``(f)''. A star
  (*) means that the error has reached the maximum or minimum allowed value of
  the parameter. For \sax\ we show results for two spectra, which
  only differ in the pile-up treatment (see text).}
\begin{tabular}{lccccccccccc}
\hline \hline\noalign{\smallskip}
Target & Observation & \multicolumn{4}{l}{\tt gau$_2$} & \multicolumn{6}{l}{\tt {laor}} \\
& & \egau\ & $\sigma$ & \kgau\ & EW & E$_{laor}$ & $\beta$ & r$_{in}$ & $i$ & k$_{laor}$ & EW \\
& & {\small(keV)} &  {\small(keV)} &  & {\small(eV)} & {\small(keV)} & & & \deg\ &  & {\small(eV)} \\
\hline\noalign{\smallskip\hrule\smallskip}
\osix\ & 0111040101 & 6.79\,$\pm$\,0.16 & 0.94\,$_{-0.2}^{+0.3}$ & 8\,$^{+5}_{-3}$ & 185\,$_{-69}^{+116}$ & 6.50\,$\pm$\,0.10 & 2.1\,$\pm$\,0.3 & 4.0$_{-3.9*}^{+5.2}$ & 70\,$_{-7}^{+0*}$ & 7\,$_{-2}^{+1}$ & 160\,$_{-46}^{+23}$  \\\hline\noalign {\smallskip}
\fifteen\   & 0061140201 & 6.77\,$_{-0.16}^{+0.14}$ & 0.33\,$\pm$\,0.13          & 0.97\,$_{-0.39}^{+0.44}$ & 23\,$_{-9}^{+11}$  & 6.56\,$_{-0.10}^{+0.41*}$ & 3.6\,$_{-2.1}^{+1.4*}$& 64\,$_{-54}^{+37}$  &63\,$_{-39}^{+7*}$ & 1.21\,$\pm$\,0.35 & 32\,$\pm$\,9   \\\hline\noalign {\smallskip}

\sixteen\ & 0303250201 & 6.78\,$\pm$\,0.01 & 1.15\,$\pm$\,0.17 & 14.6\,$\pm$\,9 & 210\,$\pm$\,129 & 6.40\,$^{+0.12}_{-0*}$ & 2.3\,$\pm$\,0.2 & $<$\,6.25 & 70\,$^{+0*}_{-1}$ & 9\,$\pm$\,1 & 130\,$\pm$\,14 \\
& 0500350301 & 6.97\,$^{+0*}_{-0.16}$  & 0.6\,$\pm$\,0.3 & 3\,$\pm$\,2 & 28\,$\pm$\,19 & 6.97 (f) & 2.2\,$_{-1.1}^{+2.8*}$ & $<$\,64 & 50\,$_{-8}^{+20*}$ & 3.5\,$_{-1.7}^{+3.3}$ & 36\,$^{+34}_{-16}$ \\
& 0500350401 & 6.97\,$_{-0.08}^{+0.00*}$ & 0.94\,$_{-0.26}^{+0.37}$ & 8.3\,$_{-2.4}^{+2.0}$    & 59\,$_{-17}^{+15}$ & 6.95\,$_{-0.10}^{+0.02*}$ & 4.8 (f) & 205.8 (f)                  & 14 (f)             & 0.9\,$\pm$\,0.5   & 6\,$\pm$\,3 \\\hline\noalign {\smallskip}
\gxtfz\  & 0505950101            & 6.72\,$\pm$\,0.06 & 0.17\,$_{-0.08}^{+0.14}$ & 22.0\,$_{-7.5}^{+11.8}$ & 17\,$_{-6}^{+9}$ & 6.82\,$_{-0.13}^{+0.07}$ & 2.56 (f) & $>$\,16 & $<$\,34 & 19.7\,$_{-5.4}^{+5.6}$ & 15\,$\pm$\,5 \\\hline\noalign {\smallskip}
\gxtfn\   & 0506110101          & 6.72\,$\pm$\,0.06 & 0.32\,$_{-0.07}^{+0.11}$ & 96\,$_{-18}^{+26}$  & 82\,$_{-15}^{+22}$  & 6.94\,$_{-0.22}^{+0.03*}$ & 3.1\,$_{-1.0}^{+0.7}$ & 11\,$_{-5}^{+7}$    & 17\,$\pm$\,9          &104\,$_{-15}^{+21}$& 96\,$_{-14}^{+20}$ \\\hline\noalign {\smallskip}
\seventeenof\  & 0402300201         &6.53\,$\pm$\,0.07           & 0.34\,$^{+0.05}_{-0.07}$ & 1.4\,$\pm$\,0.4          & 57\,$\pm$\,16            & 6.51\,$\pm$\,0.08 & $<$\,3.1 & $<$\,110 & 45\,$\pm$\,12 & 1.4\,$\pm$\,0.4 & 58\,$\pm$\,16 \\ 
& 0551270201 & 6.56\,$\pm$\,0.05 & 0.44\,$\pm$\,0.06 & 61\,$^{+15}_{-5}$ & 92\,$^{+23}_{-8}$ & 6.45\,$^{+0.05}_{-0.05*}$ & 1.8\,$^{+0.1}_{-0.2}$ & $<$\,9.1 & 65\,$^{+5*}_{-5}$ & 83\,$^{+6}_{-13}$ & 135\,$^{+10}_{-21}$ \\\hline\noalign {\smallskip}
\ser\ & 0084020401 & 6.61\,$\pm$\,0.05          & 0.22\,$_{-0.05}^{+0.08}$ & 22\,$_{-5}^{+6}$ & 52\,$_{-11}^{+14}$ & 6.79\,$_{-0.2}^{+0.1}$ & 2.84\,$_{-0.8}^{+0.9}$ & 13.3\,$_{-7}^{+19}$ & $<$\,33 & 25.2\,$\pm$\,4.8          & 63\,$\pm$\,12        \\
 & 0084020501 & 6.58\,$\pm$\,0.07 & 0.25\,$\pm$0.08 & 15.7\,$_{-3.9}^{+4.5}$ & 50\,$_{-12}^{+14}$ & 6.6\,$\pm$\,0.1 & 2.03 (f) & 11.15 (f) & 28\,$\pm$\,9 & 17.5\,$\pm$\,4  & 59\,$\pm$\,13.5 \\
 & 0084020601& 6.59\,$_{-0.09}^{+0.07}$ & 0.27\,$_{-0.10}^{+0.14}$ & 18.7\,$_{-5.6}^{+7.7}$   & 51\,$_{-15}^{+21}$ & 6.73\,$_{-0.23}^{+0.24*}$& 2.6\,$_{-1.1}^{+2.4*}$ & 11\,$_{-10*}^{+26}$  & $<$\,25 & 19.4\,$_{-5.8}^{+6.2}$  & 57\,$\pm$\,18 \\\hline\noalign {\smallskip}
\seventeentw\ & 0149810101          & 6.72\,$\pm$\,0.10 & 0.98\,$_{-0.16}^{+0.19}$ & 21\,$_{-9}^{+12}$ & 189\,$_{-81}^{+108}$ & 6.53\,$_{-0.06}^{+0.11}$ & 2.6\,$\pm$\,0.4 & 20\,$_{-7}^{+5}$ & 70\,$_{-16}^{+0*}$ & 7.7\,$_{-1.2}^{+1.6}$ & 80\,$_{-15}^{+22}$ \\\hline\noalign {\smallskip}
\seventeenth\ & 0090340201          & 6.74\,$\pm$\,0.10          & 0.38\,$\pm$\,0.11 & 22\,$\pm$\,7   & 46\,$\pm$\,14        & 6.73\,$_{-0.18}^{+0.09}$ & $<$\,2.94 & 5.41 (f)                   & 44\,$_{-19}^{+16}$  & 20\,$\pm$\,6  & 43\,$\pm$\,13 \\\hline\noalign {\smallskip}
\sax\ & 0560180601 & 6.94\,$_{-0.12}^{+0.03*}$ & 0.26\,$_{-0.08}^{+0.12}$ & 1.5\,$_{-0.4}^{+0.6}$ & 23\,$_{-7}^{+9}$ & 6.50\,$_{-0.10*}^{+0.42}$ & 3\,$_{-0.8}^{+2*}$ & 6.3\,$_{-5.1*}^{+26}$ & 46\,$_{-13}^{+7}$    & 3.1\,$_{-1.0}^{+1.2}$ & 51\,$_{-16}^{+20}$ \\
& 0560180601** & 6.77\,$\pm$\,0.13 & 0.39\,$_{-0.11}^{+0.14}$ & 1.9\,$\pm$\,0.7 & 30\,$\pm$\,11 & 6.96\,$_{-0.24}^{+0.01*}$ & 5.0\,$_{-2.6}^{+0*}$ & 33\,$_{-18}^{+60}$ & 26\,$_{-4}^{+15}$ & 1.8\,$\pm$\,0.6 & 31\,$\pm$\,10 \\\hline\noalign {\smallskip}
\noalign{\smallskip\hrule\smallskip}
\label{tab:bestfitlines}
\end{tabular}
\end{small}
\end{center}
\end{table}
\end{landscape}

\begin{table*}
\begin{center}
\caption[]{\rchisq\ of spectral fits to (a) {\tt diskbb+bbodyrad}, (b)
{\tt diskbb+po} and (c) {\tt bbodyrad+po}
components before inclusion of the iron line component. Columns (d)
and (e) show the \rchisq\ of the fit after including a Gaussian (d)
and a {\tt laor} (e) component to the best-fit model.}
\begin{tabular}{ccccccc}
\hline \noalign {\smallskip}
\hline \noalign {\smallskip}
Source  & Observation & \multicolumn{5}{c}{\tt \rchisq} \\
& ID   & a & b & c & d & e \\
\hline \noalign {\smallskip}
\osix\ & 0111040101 & 2.25 (220) & 1.21 (220) & 1.19 (220) & 0.83 (217) & 0.85 (216) \\
\fifteen\ & 0061140201 & 1.11 (220) & 12.5 (224) & 4.28 (220) & 0.97 (217) & 0.96 (215) \\
\sixteen\ & 0303250201 & 6.46 (224) & 32.9 (224) & 48.7 (224) & 0.76 (216) & 0.78 (214) \\
& 0500350301 & 0.76 (220) & 4.14 (220) & 4.15 (220) & 0.71 (217) & 0.70 (215)\\
& 0500350401 & 0.82 (220) & 4.84 (224) & 6.00 (224) & 0.72 (217) & 0.80 (215)\\
\gxtfz\ & 0505950101 & 1.15 (166) & 2.04 (166) & 1.83 (166) & 0.86 (163) & 0.88 (161) \\
\gxtfn\ & 0506110101 & 1.97 (219) & 3.10 (219) & 3.75 (219) & 1.15 (216) & 1.14 (214) \\
\seventeenof\ & 0402300201 & 1.34 (189) & 1.56 (189) & 1.58 (189) & 1.01 (186) & 1.03 (184) \\
& 0551270201 & 3.05 (189) & 5.88 (189) & 4.10 (189) & 1.07 (187) & 1.03 (185) \\
\seventeensix\ & 0090340101 & 0.99 (219) & 3.58 (221) & 1.26 (219) & - & - \\
& 0090340601 & 0.91 (219) & 1.27 (221) & 2.13 (221) & - & - \\
\seventeentw\ & 0149810101 & 1.55 (180) & 1.82 (180) & 1.80 (180) & 1.02 (177) & 1.07 (175)\\
\seventeenth\ & 0090340201 & 0.99 (219) & 1.82 (219) & 1.98 (219) & 0.68 (216) & 0.78 (214) \\
\ser\ & 0084020401 & 1.47 (219) & 2.11 (219) & 3.04 (219) & 0.98 (216) & 0.97 (214) \\
& 0084020501 & 1.24 (219) & 2.07 (219)  & 3.04 (219) & 0.93 (216) & 0.92 (214) \\
& 0084020601 & 1.31 (219) & 1.95 (219)  & 2.92 (219) & 1.04 (216) & 1.04 (214) \\
\aql\ & 0303220201 & 0.97 (221) & 1.22 (221) & 1.35 (221) &  - & - \\
\xtee\ & 0157960101 & 1.20 (224) & 1.00 (224) & 1.02 (224)  & - & - \\
\sax\ & 0560180601 & 30.2 (224) & 1.30 (223) & 0.98 (223) & 0.75 (220) & 0.71 (218) \\
& 056018060** & 33.4 (224) & 1.31 (223) & 1.00 (223) & 0.86 (220) & 0.84 (218)\\
\noalign {\smallskip} \hline \label{tab:chisq}
\end{tabular}
\end{center}
\end{table*}

\section{Properties of the sample}
\label{sec:properties}

\subsection{Fe line emission}

\begin{table*}
\begin{center}
\caption[]{2--10~keV luminosities in units of 10$^{36}$ \ergsec\ and corresponding
Eddington ratios (L$_{2-10~keV}$/L$_{Eddington}$) for all the observations in the sample. 
}
\begin{tabular}{ccccc}
\hline \noalign {\smallskip}
\hline \noalign {\smallskip}
Source  & Observation & d & L$_{36}$ & R$_{Edd}$ \\
 & ID   &  [kpc] & \ergsec\ & \\
        &  \\
\hline \noalign {\smallskip}
\osix\ & 0111040101 & 3.2\,$\pm$\,0.5 \citep{0614:kuulkers09aa} & 0.73 & 0.004 \\
\fifteen\ & 0061140201 &  7 \citep{1543:wang04apjl} & 2.87 & 0.016 \\
\sixteen\ & 0303250201 & 6.0$\pm$0.1 \citep{galloway08apjs} & 3.41 & 0.019 \\
& 0500350301 & & 6.01 & 0.033 \\
& 0500350401 & & 8.17 & 0.045 \\
\gxtfz\ & 0505950101 & 11 \citep{grimm02aa} & 230.72 & 1.264 \\
\gxtfn\ & 0506110101 & 9.2 \citep{grimm02aa} & 127.64 & 0.701 \\
\seventeenof\ & 0402300201 & 5.8$\pm$0.2 \citep{galloway08apjs} & 1.05 & 0.006 \\
& 0551270201 & & 27.25 & 0.150\\
\seventeensix\ & 0090340101 & 4.4 \citep{grimm02aa} & 10.98 & 0.060  \\
& 0090340601 & & 12.14 & 0.067\\
\seventeentw\ & 0149810101 & 4.0\,$\pm$\,0.4 \citep{galloway08apjs} & 0.48 & 0.003 \\
\seventeenth\ & 0090340201& 6.6\,$\pm$\,1.0  \citep{galloway08apjs} &  24.97 &  0.137 \\
\ser\ & 0084020401 & 7.7\,$\pm$\,0.9 \citep{galloway08apjs} & 32.59 & 0.179 \\
& 0084020501 &  & 26.45 & 0.145 \\
& 0084020601 & & 29.33 & 0.161 \\
\aql\ & 0303220201 & 3.9\,$\pm$\,0.7 \citep{galloway08apjs} & 1.91 & 0.010 \\
\xtee\ & 0157960101 & 8 (assumed) & 1.36 & 0.007 \\
\sax\ & 0560180601 & 2.77\,$\pm$\,0.11 \citep{galloway08apjs} &  0.79 &  0.004  \\
\noalign {\smallskip} \hline \label{tab:lumin}
\end{tabular}
\end{center}
\end{table*}

Examination of the spectral fits
(Tables~\ref{tab:bestfit} to \ref{tab:bestfitlines} and
Figs.~\ref{fig:spectra}--\ref{fig:eeuf2}) shows that we are able to
successfully model the emission in the Fe~K band with a simple
Gaussian component. Fits of the excess at the
Fe~K band with the more complicated {\tt laor} component do not improve significantly
the goodness of the fit for any of the observations presented here (see Table~\ref{tab:chisq}). 
\begin{figure*}[!ht]
\includegraphics[angle=0,width=0.33\textwidth]{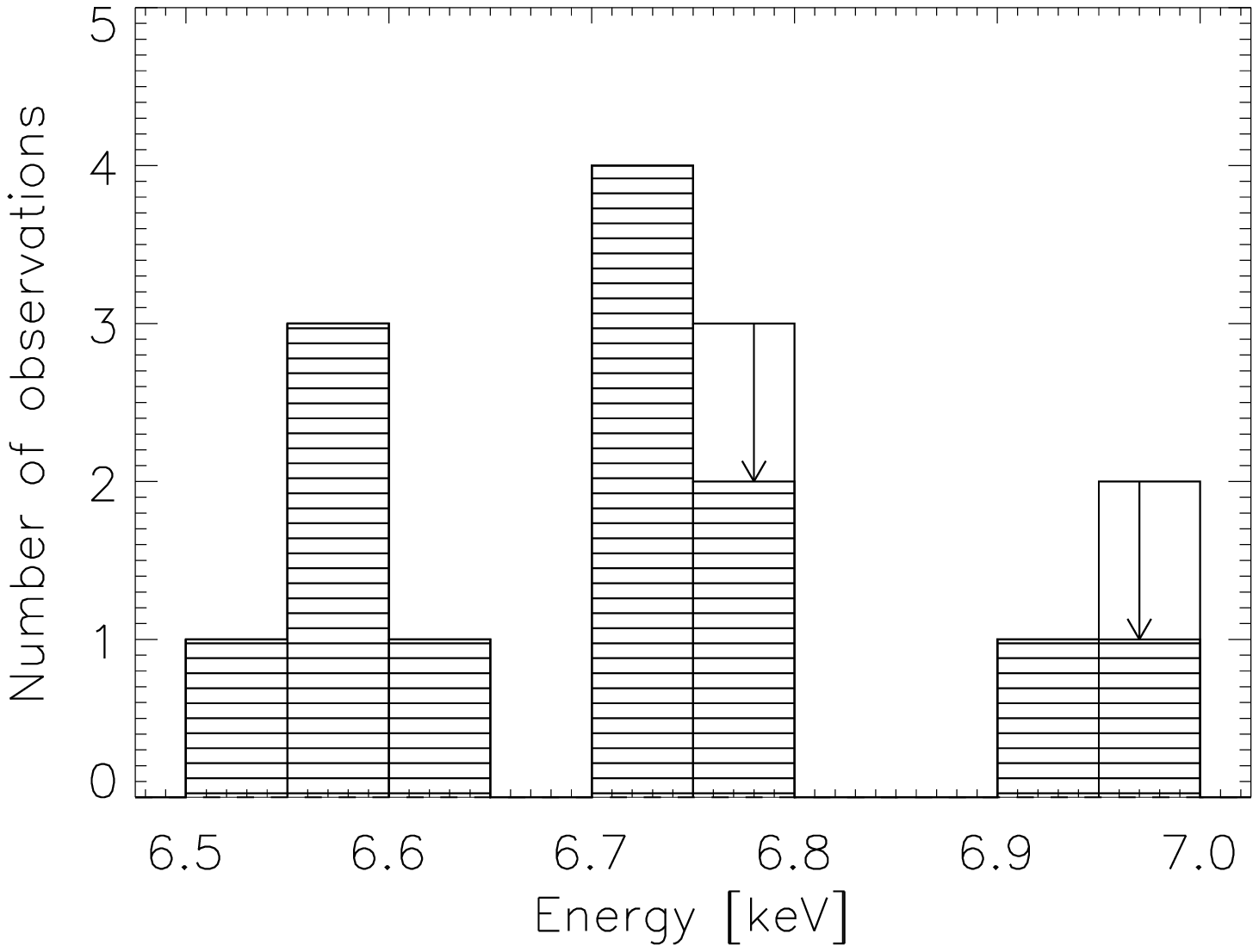}
\includegraphics[angle=0,width=0.33\textwidth]{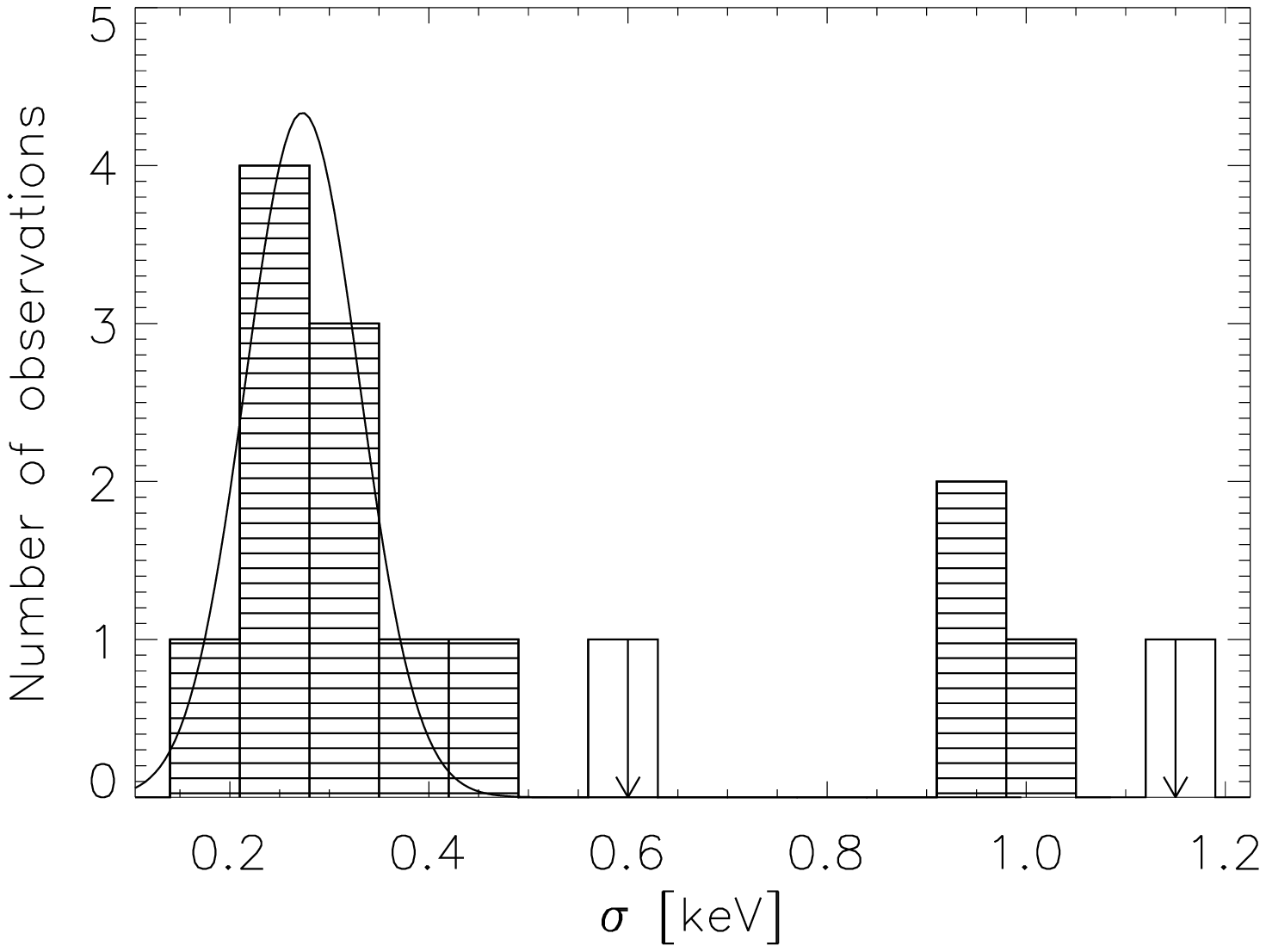}
\includegraphics[angle=0,width=0.33\textwidth]{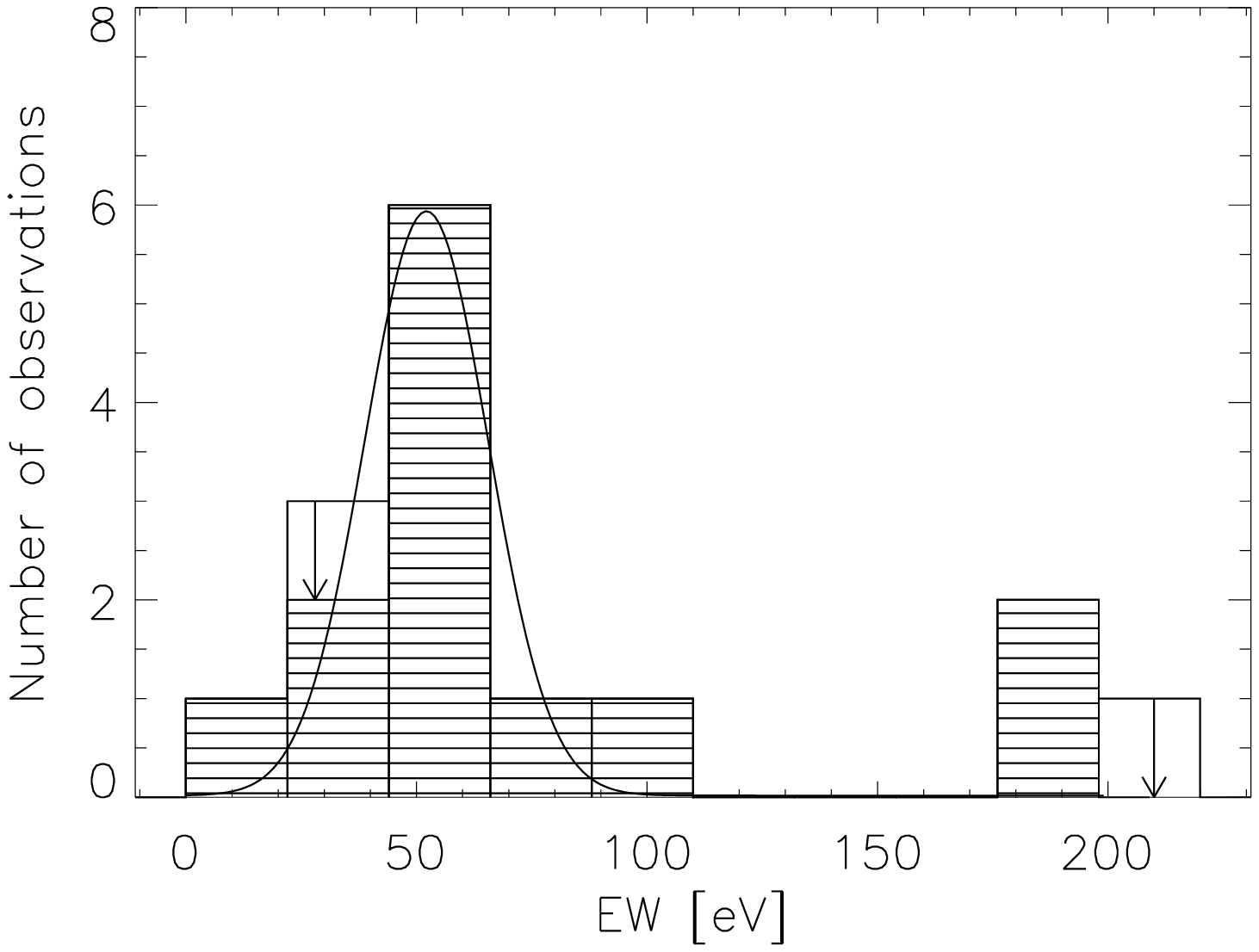}
\caption{Histogram of the parameters of the Fe line when fitted with a Gaussian
component for the full sample considered in this work. The white rectangles
marked with an arrow correspond to the lines which are detected below a
significance of 3~$\sigma$ (see Fig.~\ref{fig:fe_properties}).
The width (middle panel) and the \ew\ (right panel) have a
well defined ``Gaussian-like'' distribution around the weighted
average value, while the outliers have large errors associated to their $\sigma$ and \ew\ values.}

\label{fig:histo}
\end{figure*}

Considering the whole sample, the Fe line has a weighted average
energy of 6.67\,$\pm$\,0.02~keV, a width of 0.33\,$\pm$\,0.02~keV and
an \ew\ of 42\,$\pm$\,3~eV. The statistical distribution of these
values is shown in Fig.~\ref{fig:histo}. The width and the \ew\ have a
well defined ``Gaussian-like'' distribution around the weighted
average value. The outliers of the histograms in
Fig.~\ref{fig:histo} have values with large errors and therefore do
not contribute significantly to the weighted average. The energy
distribution peaks at $\sim$6.7~keV, in agreement with the value of
the weighted average. This is consistent with emission from
\fetfive. Clearly, the distribution has values consistent with
emission from highly ionised species of iron, from \fettwo\ to 
\fetsix, and it never shows a value consistent with neutral iron.
\begin{figure}[!ht]
\includegraphics[angle=0,width=0.5\textwidth]{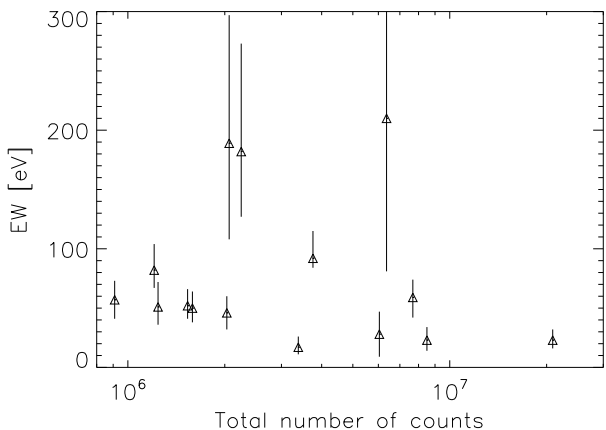}
\caption{
\ew\ of the Fe line when fitted with a Gaussian
component versus the total number of counts of each spectrum. 
}
\label{fig:fe_properties}
\end{figure}
\begin{figure*}[!ht]
\includegraphics[angle=-90,width=0.33\textwidth]{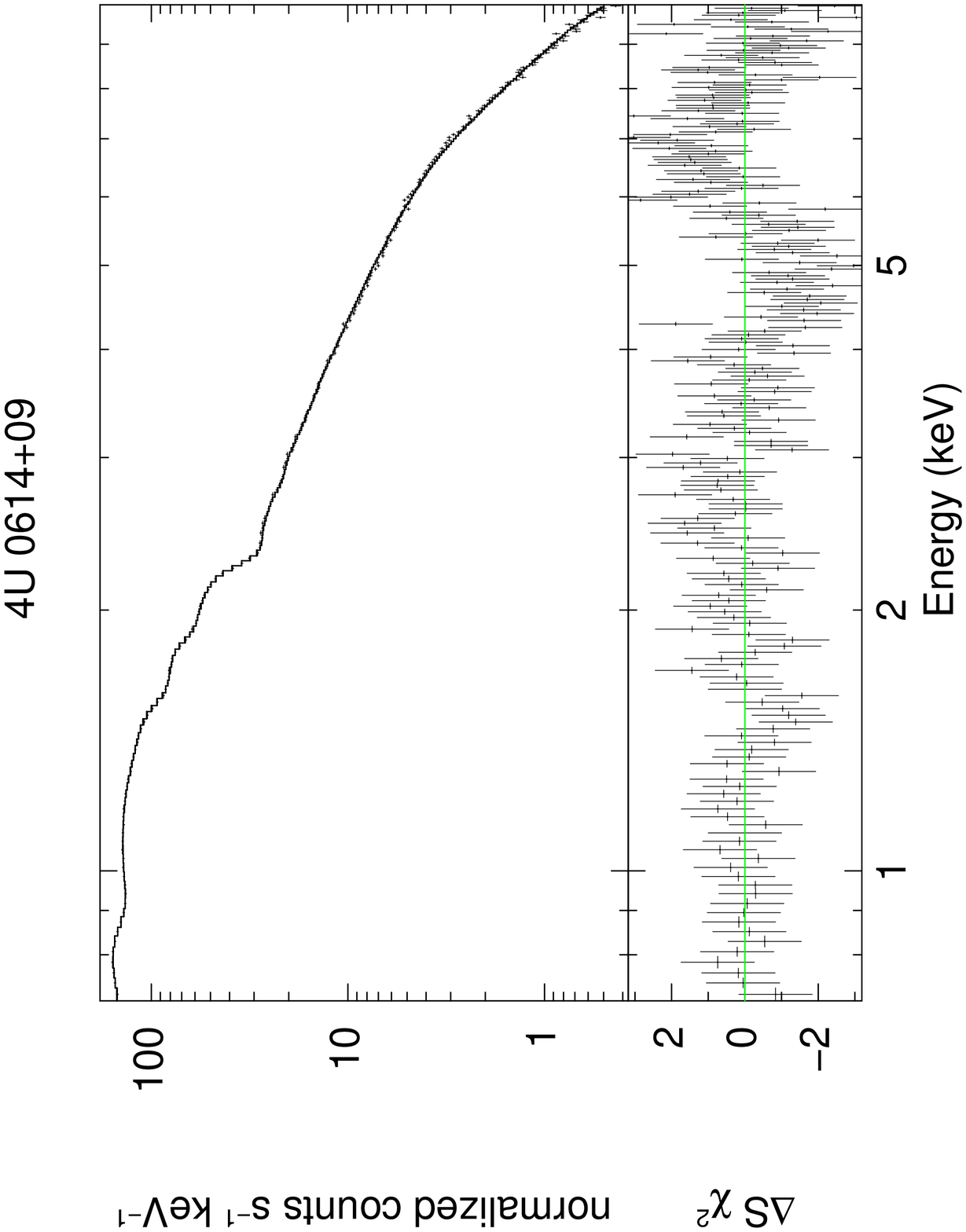}
\includegraphics[angle=-90,width=0.33\textwidth]{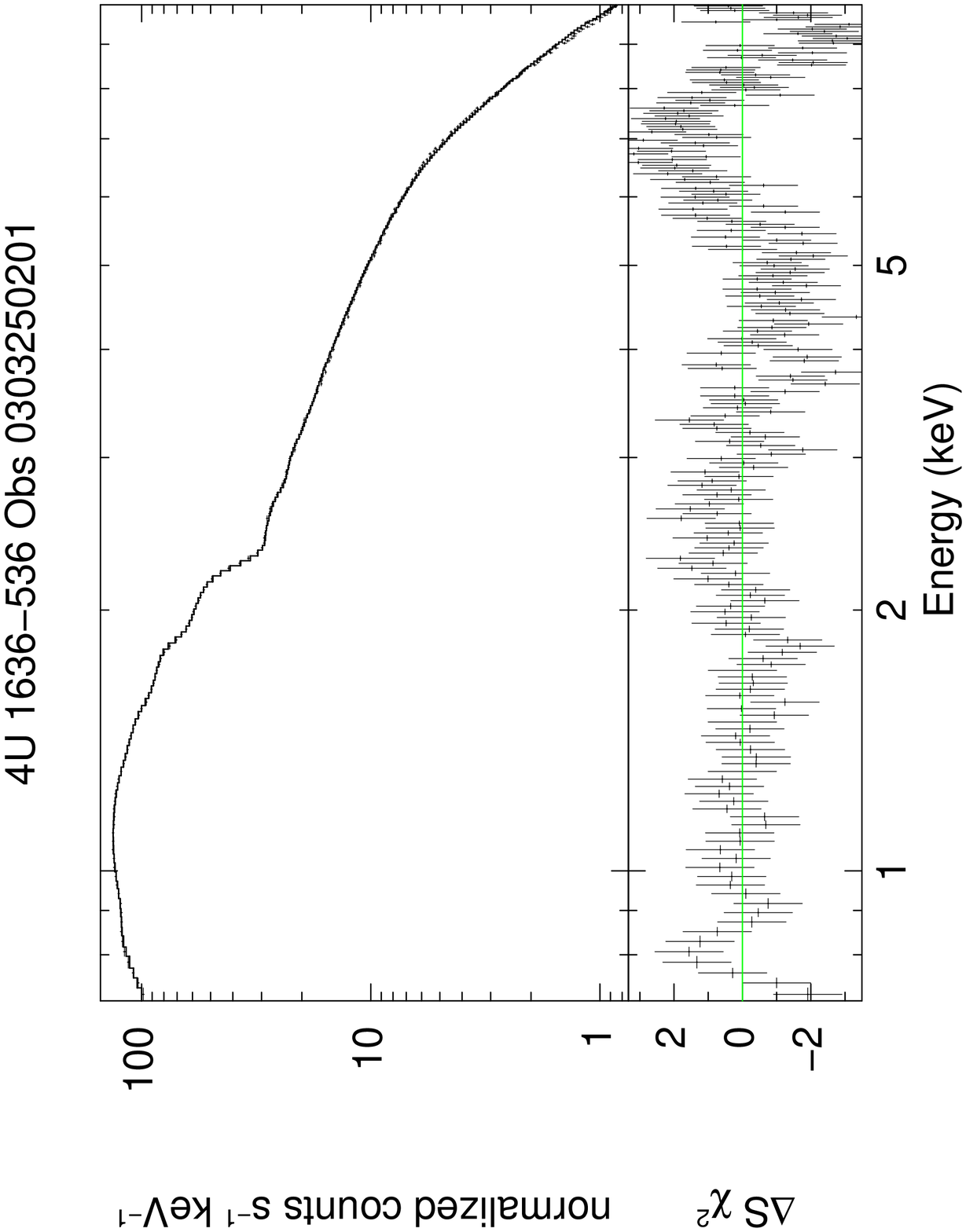}
\includegraphics[angle=-90,width=0.33\textwidth]{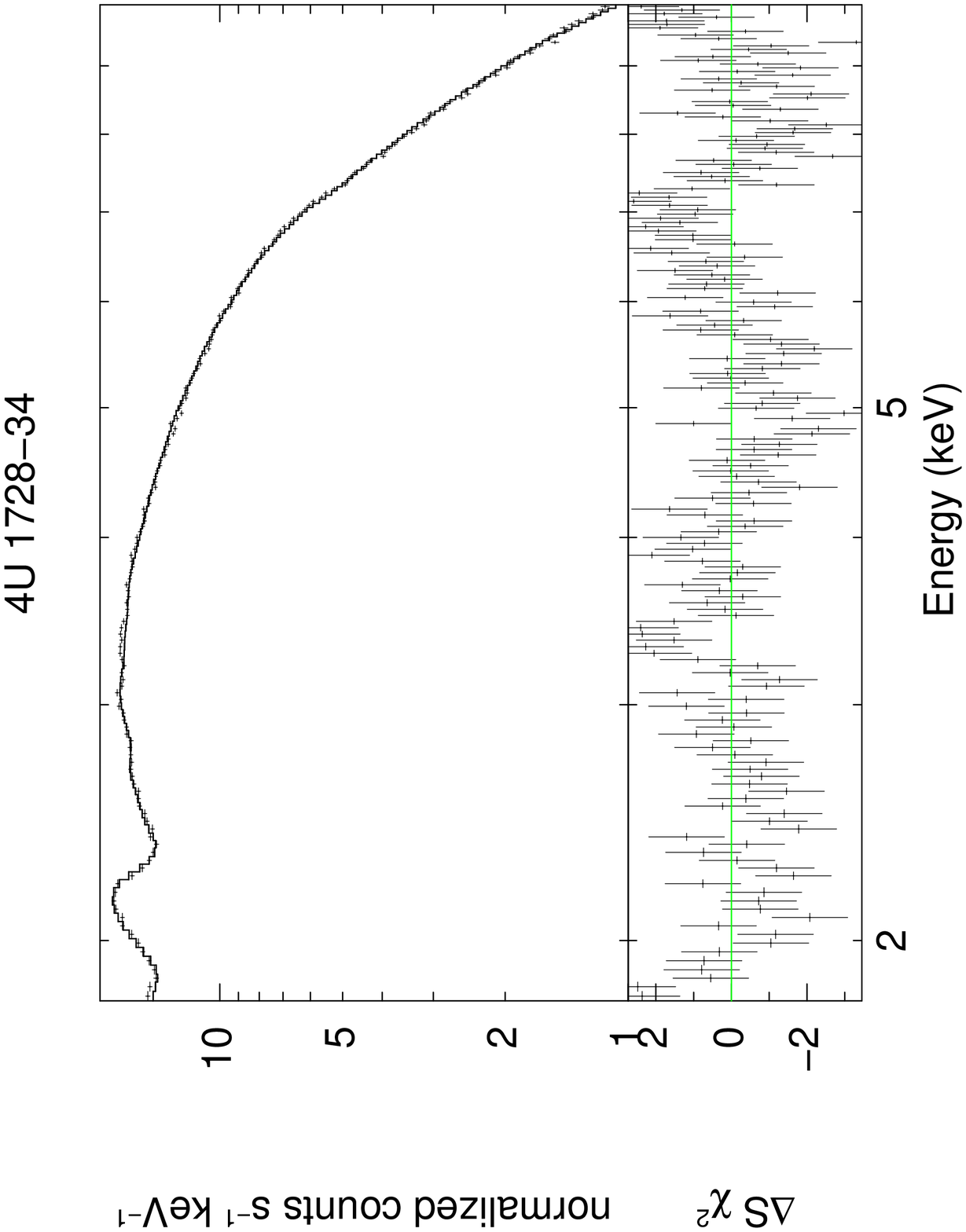}
\caption{Residuals in units of standard
deviation from the best-fit model to the data for \osix, \sixteen\ 
Obs~0303250201 
and \seventeentw\ before including the Fe~line. 
}
\label{fig:residuals}
\end{figure*}

In order to find out the reason for the large errors in the
lines with an \ew\ above 30~eV,
we plotted the \ew\ of the lines as a function of the total number
of counts in the spectrum (see
Fig.~\ref{fig:fe_properties}).  
The lines which show large errors are not associated to the spectra with
less statistics, indicating that the
detection of the lines within our sample is not limited by statistics.
Therefore, we examined in detail the three cases for which the errors on the 
parameters of the line are very large (corresponding to \osix, \sixteen\ 
Obs~0303250201 
and \seventeentw) to determine the reason for
such large errors. 
As indicated in Sect.~\ref{sec:spectra}, fits to \seventeentw\ 
spectra of \chandra\ and RXTE or BeppoSAX \citep{1728:dai06aa}
showed equally good fits when modelling the emission at $\sim$6.5~keV with
a broad line or with two absorption edges. Including one edge in
the \xmm\ spectrum at 7.6~keV and no Fe line yields a \rchisq\ of 
 1.32 (178). Although the \rchisq\ is higher than when including the
Fe line, the residuals at the Fe band are reduced considerably. 
The fits to the \xmm\ spectrum have
significantly different continuum parameters when fitting the residuals
at the Fe band with Models~1a or 1b (see Table~\ref{tab:bestfit}), in contrast to all other observations. 
This indicates that after inclusion of the line, the parameters
of the fit change significantly and thus the line fit may not be realistic.
Interestingly, \osix\ and \seventeentw\ show the lowest luminosities of the
full sample. Therefore, it is plausible that such observations have a 
significantly different, more complex, spectrum compared to higher luminosity 
observations. For comparison, the MOS spectra of the observation of \osix\ analysed in this
work were previously fitted with an absorbed power-law and an emission 
feature at 0.65~keV \citep{0614:mendez02} and did not require a line at $\sim$6.5~keV.
In the case of \sixteen\ Obs~0303250201, although
the luminosity is higher than for other sources in our sample
fitted with the ``standard'' Model 1a/1b, we needed three continuum components
to obtain an acceptable fit. This points
again to a complex spectrum.
Fig.~\ref{fig:residuals} shows the residuals of the fit to
the continuum before including the Fe~line for the three observations discussed.
The residuals at the Fe energy band are very different compared to those shown
in Fig.~\ref{fig:gallery}, indicating that the lines fitted for these
observations may not be realistic and that such fits may need e.g. the inclusion
of absorption edges. For these reasons we regard the
properties of these lines with caution.

\begin{figure*}[!ht]
\includegraphics[angle=0,width=0.33\textwidth]{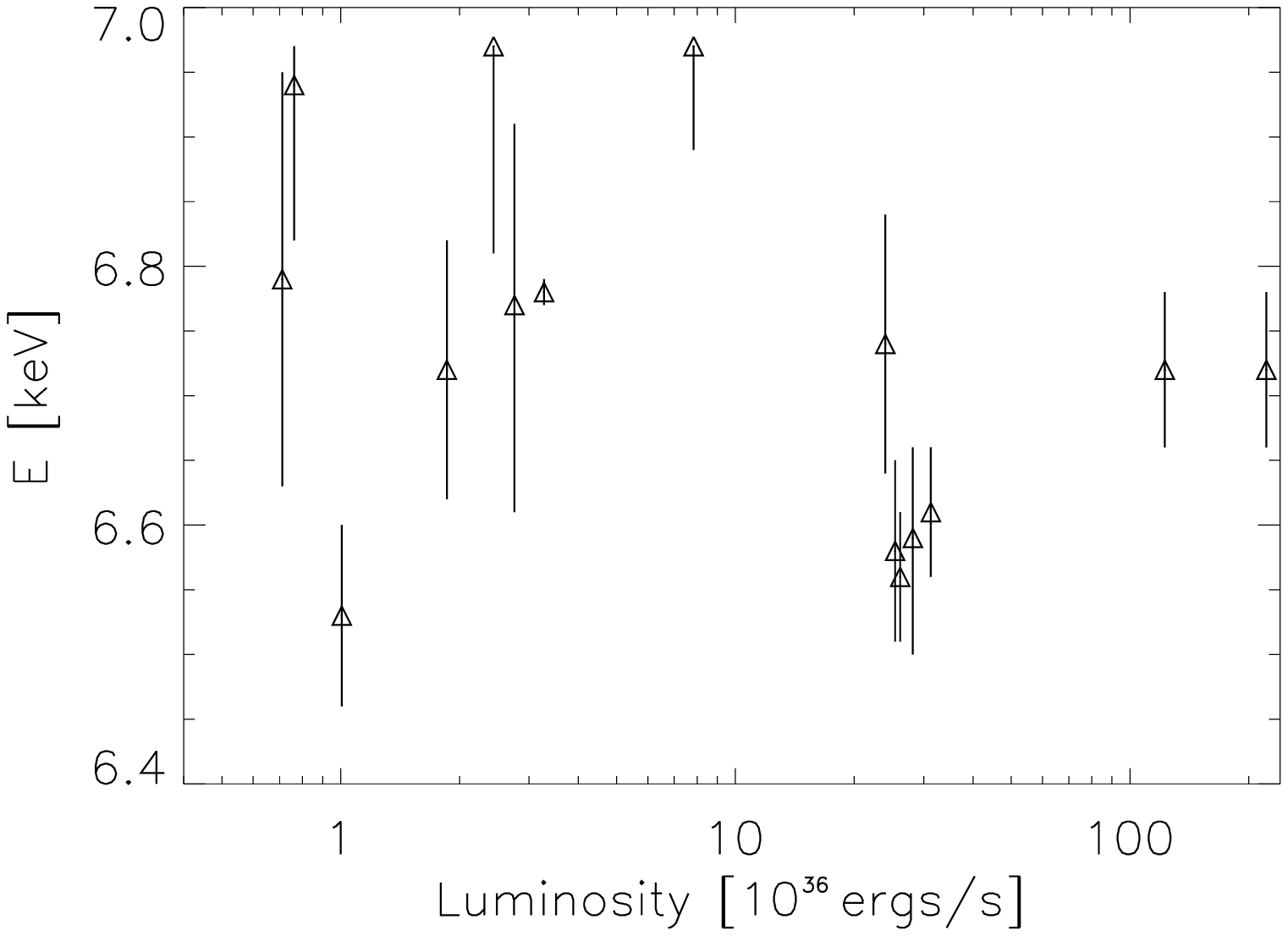}
\includegraphics[angle=0,width=0.33\textwidth]{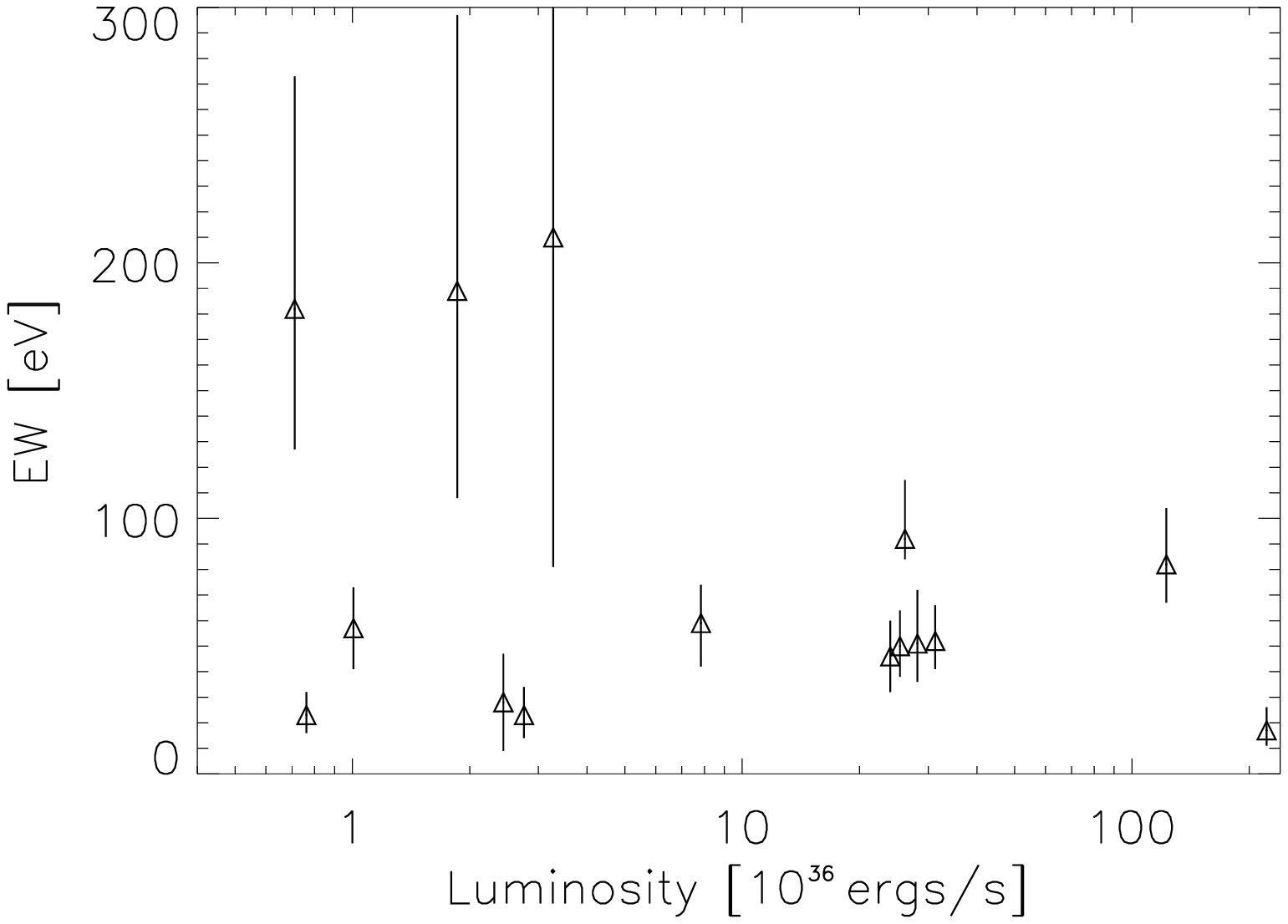}
\includegraphics[angle=0,width=0.33\textwidth]{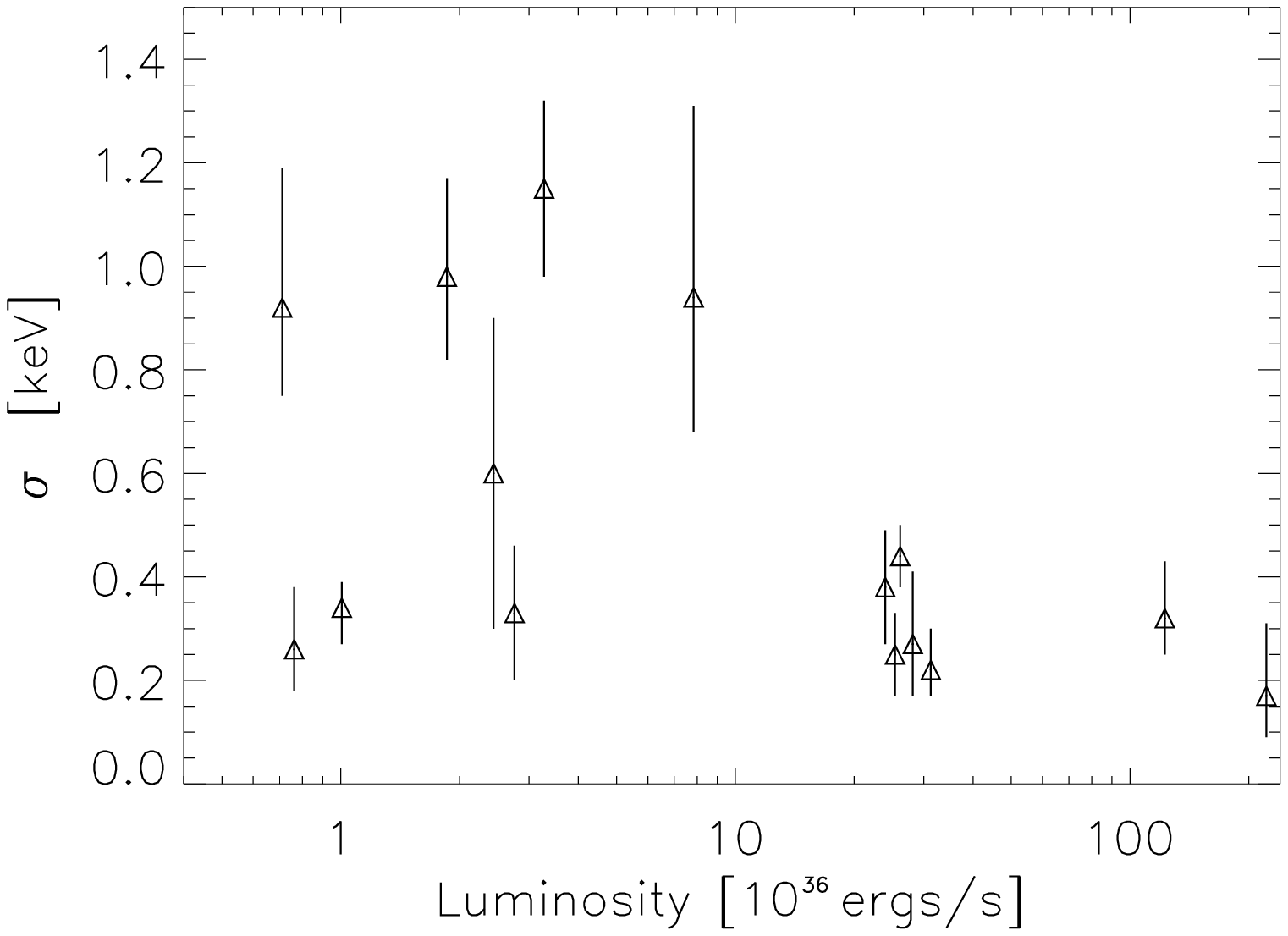}
\caption{Energy, \ew\ and $\sigma$ of the Fe line versus luminosity of the source
in the 2--10~keV energy band. The luminosity has been calculated based on
the distances reported in Table~\ref{tab:lumin}. 
}
\label{fig:fe_properties1}
\end{figure*}

It is
interesting to compare the properties of the Fe~\ka\ line with general
properties of the sources such as luminosity or temperatures of the
blackbody and disc blackbody components, with the aim to constrain the
origin of the line emission region. For example, we may expect a
correlation between the source accretion rate and the breadth of the
line, if the latter is caused by relativistic effects.
Fig.~\ref{fig:fe_properties1} shows the
properties of the Fe~\ka\ line (width and \ew) as a function
of the luminosity calculated in Table~\ref{tab:lumin}. 

We do not find any clear correlation between the energy centroid, the
width or the \ew\ of the Fe~line with the source luminosity (see
Fig.~\ref{fig:fe_properties1}). The source luminosities range over 2
orders of magnitude, covering the full range of Eddington ratios from
0.003 to 1.26, while the line centroid, width and \ew\ do not show any
systematic trend. Similary, we do not find any clear correlation between
the \ew\ of the Fe~line and the temperature or flux of the blackbody 
or disc blackbody components.

\subsection{Continuum emission}
\label{sec:properties2}

We examined the continuum properties of the sample for all the sources shown
in Table~\ref{tab:bestfit}. 

\begin{figure*}[!ht]
\includegraphics[angle=0,width=0.33\textwidth]{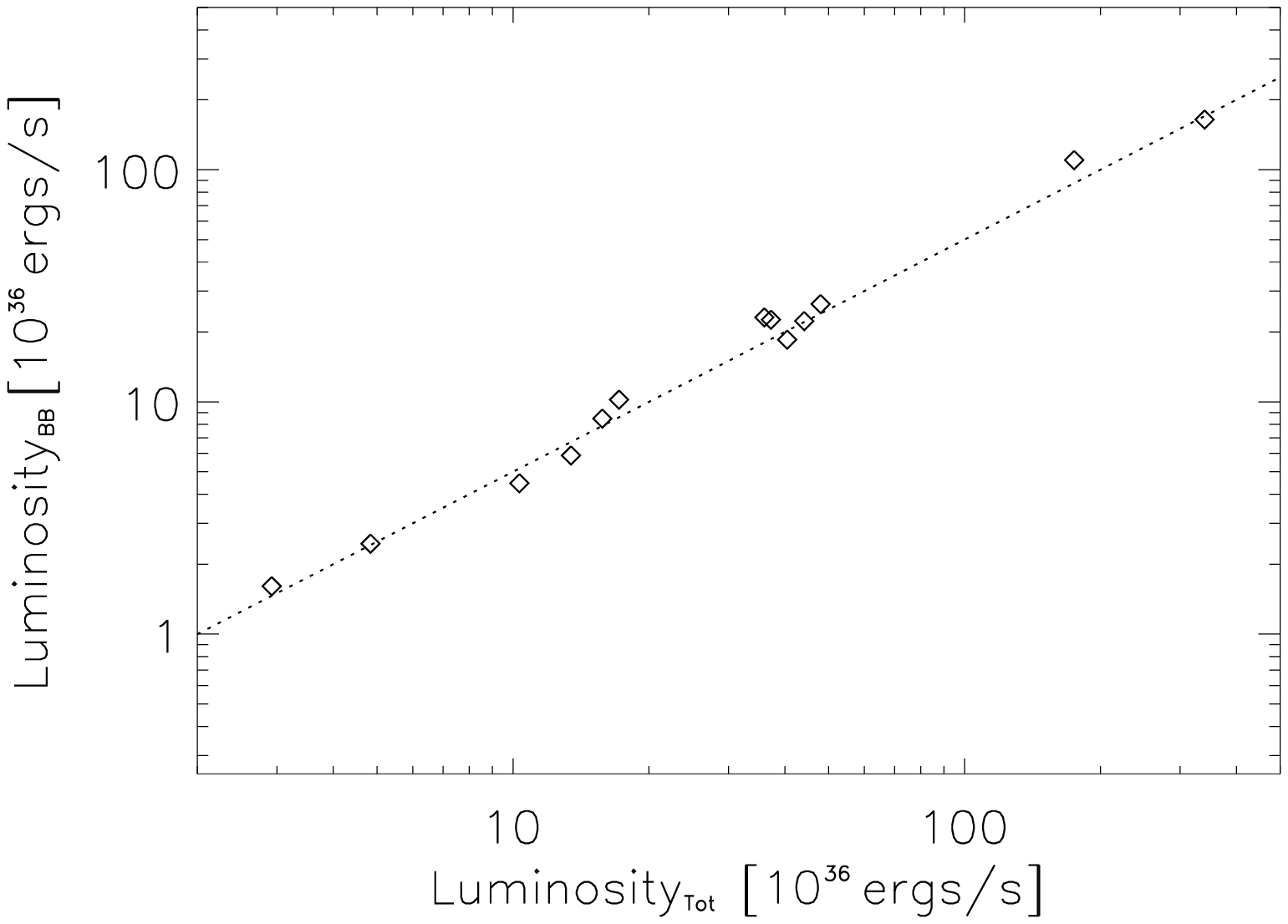}
\includegraphics[angle=0,width=0.33\textwidth]{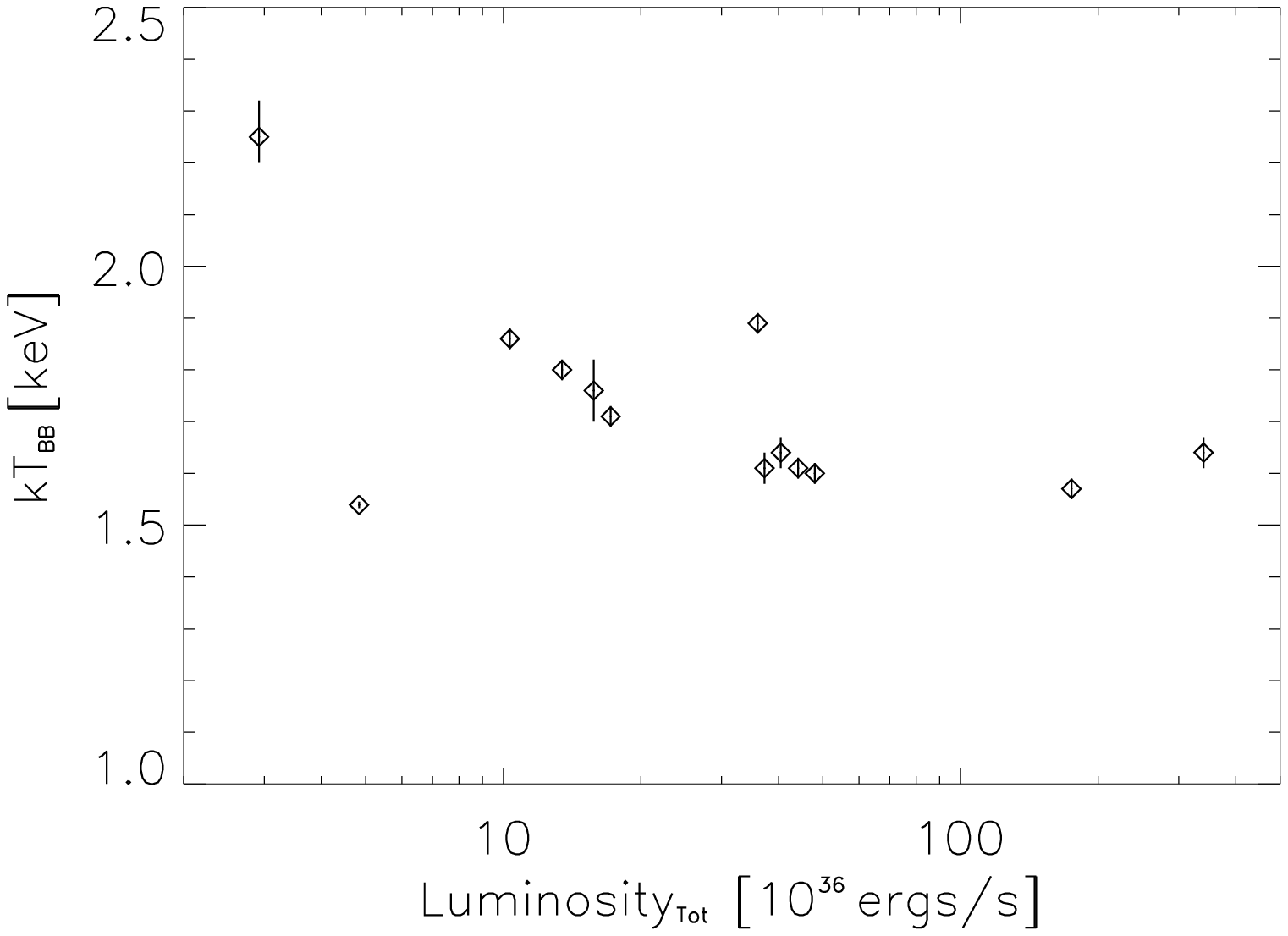}
\includegraphics[angle=0,width=0.33\textwidth]{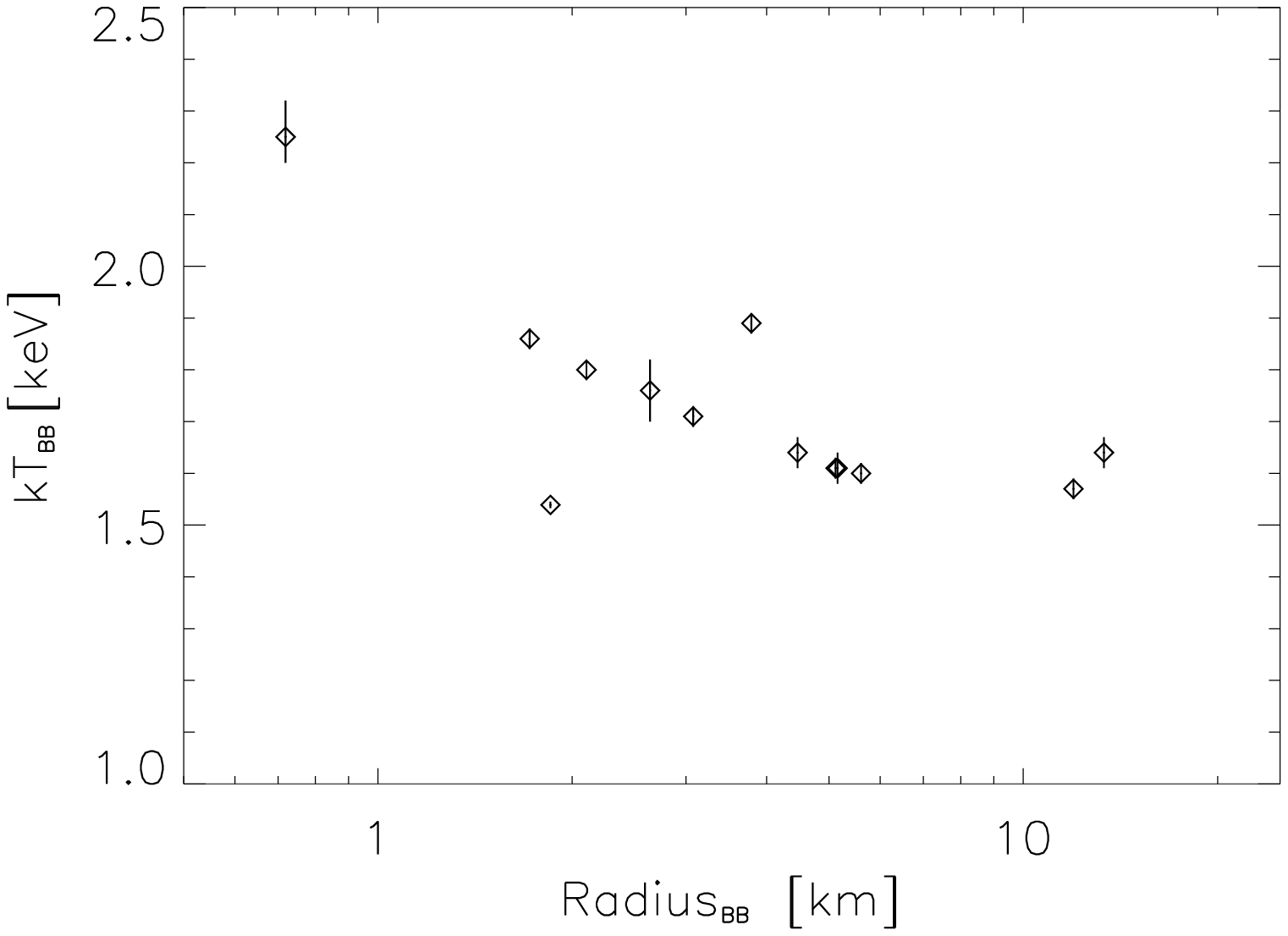}
\caption{{\it Left panel:} Variation of blackbody luminosity in the band 0.5--30~keV
with total luminosity in the same band. The dashed line shows the 50\% of the total
luminosity. 
{\it Middle panel:} Temperature of the blackbody component versus the total luminosity
in the 0.5--30~keV energy band. 
{\it Right panel:} Temperature of the blackbody component versus the 
radius of the blackbody emission. The radius has been inferred from the normalisation of the blackbody component.
 See Sect.~\ref{sec:properties2} for the observations included in this figure.}

\label{fig:continuum_properties1}
\end{figure*}

Figure~\ref{fig:continuum_properties1} shows the variation of the blackbody luminosity
in the band 0.5--30~keV as a function of the total luminosity in the same band. In this
figure, we excluded the two 
AMSP \xtee\ and \sax, \osix\ and Obs~0303250201 of \sixteen. The reason was
that these observations were fitted with a continuum different than the standard
one of blackbody and disc blackbody components and are therefore indicative of
different properties. We further excluded the two dim observations corresponding to 
\seventeenof\ (Obs~0402300201) and \seventeentw, since for such dim sources
Compton scattering is expected to play an important role and was not included in our model.

The blackbody luminosity is very near the value of 50\% of total luminosity expected 
from simple energy considerations for all the sources of the sample.
Interestingly, the temperature of the blackbody component, \ktbb, is nearly independent
upon the global luminosity at luminosities $\approxgt$2$\times$10$^{36}$\ergsec\
(Fig.~\ref{fig:continuum_properties1}). Ignoring the point with the highest temperature 
since it has associated a significantly larger error than the other points, \ktbb\ falls by  
$\sim$45\% while the total luminosity increases by more than two orders of magnitude and the 
blackbody radius by more than a factor of 40.

\section{Discussion}
\label{sec:discussion}

We performed a systematic analysis of 26 \xmm\ observations corresponding to
all the NS LMXBs observed with EPIC~pn Timing mode and publicly available up 
until the 30th of September 2009 to establish the characteristics
of the Fe~K line emission in these objects.

In seven observations we did not detect the source significantly. For the remaining nineteen
observations we extracted one spectrum per observation. We payed special attention 
to the effects of pile-up and background subtraction. 

\subsection{Fe line emission}

We detected Fe~\ka\ line emission in 80\% of the observations where the source 
was significantly detected.

The energy of the line has values between 6.53 and 6.97~keV,
consistent with highly ionised iron, from \fettwo\ to \fetsix. The
width has values between 0.17 and 1.15~keV and the \ew\ between 17 and 190~eV. 
Only in four cases out of the fifteen lines reported in this work did we find
a width near or above 1~keV and an \ew\ above 100~eV. Interestingly,
there is a gap in the distributions of $\sigma$ and \ew\ between these 
high values and the rest of the sample. 
Lines with high values of $\sigma$ or \ew\ have also associated large errors. 
\citet{vaughan08mnras} interpreted
a similar finding in the context of detection of narrow lines from AGN 
as the lines with large \ews\ being most likely false detections. In 
the cases studied here the lines with widths near 1~keV are also in the limit of
detectability of 3$\sigma$ and the large errors most likely point to an 
inappropriate modelling of the continuum for \osix, \seventeentw, and \sixteen.

Considering the whole sample, the Fe line has a weighted average
energy of 6.67\,$\pm$\,0.02~keV, a width ($\sigma$) of 0.33\,$\pm$\,0.02~keV and
an \ew\ of 42\,$\pm$\,3~eV. 

Recently, broad skewed Fe~K emission
lines from the disc were reported from LMXBs containing a
NS \citep[e.g.][]{serx1:bhattacharyya07apjl,
gx349:iaria09aa, gx340:dai09apjl}. All the \xmm\ observations for which
a broad skewed Fe~line has been reported were included in our sample
(see Sect.~\ref{sec:properties}).
However, in contrast to previous analyses, we {\it do not} need 
to invoke relativistic effects to explain its width, which
could be due to mechanisms such as Compton broadening. 
The lines are {\it equally well} fitted with the 
relativistic {\tt laor} component or with a simpler Gaussian component. Further,
the profiles shown in Fig.~\ref{fig:gallery} do not show an asymmetric shape, as
expected if the lines are emitted close to the NS and shaped by relativistic effects.
The line profile is instead symmetric, similarly to the one found in dipping sources
\citep{ionabs:diaz06aa}.
The major difference between the analysis 
presented in this work and previous works is a careful treatment
of pile-up effects, common in the observations of bright LMXBs, and
a different modelling of the continuum in some cases, which has a strong effect in the 
\ew\ of the lines. 

As shown by \citet{brandt94mnras}, in a scenario where the iron line
is generated at the inner disc, we would expect the centroid and the
width of the iron line to decrease as the accretion rate diminishes
and the disc recedes. The behaviour of the equivalent width is
somewhat more complex: the line equivalent width is expected to first
increase with \xil\ up to a certain value at which it starts to
decrease (see Fig.~5b from \citet{brandt94mnras}). This is due to the
emission properties of iron atoms as a function of the ionisation
parameter (Matt et al. 1994). When the most abundant ions are
\fetfour--\fetsix, line emission is more intense than in the neutral
case, due to the greater fluorescent yield and the smaller
photoabsorption cross-section at the line energy. However, if \xil\
increases further, the ion distribution becomes dominated by fully
stripped atoms and the line emission decreases. In the observations
presented in this paper, we do not observe a correlation between the
centroid and the width of the iron line with luminosity, in agreement
with a similar systematic analysis based on ASCA observations \citep{asai00apjs}. 
In contrast, the three cases for which a large width and equivalent
width are measured occur at relatively low luminosities and are actually 
associated to large errors. 
This seems
to be at odds with the behaviour predicted by \citet{brandt94mnras}
for lines generated at the disc and suggests instead that 
the physical condition of the line-emitting region is rather similar among the LMXBs.

An interpretation of the line in the context of disc emission would
imply that the line is produced far from the neutron star in order to
explain its symmetric shape. However, in such a case it is difficult
to justify the variations of the line as the disc inner radius
changes.

Alternatively, the origin of the line broadening could be Compton scattering in a
corona. Based on a systematic analysis of twenty~NS LMXBs with ASCA, 
\citet{asai00apjs} concluded that the iron lines were likely produced through
the radiative recombination of photoionised plasma, being the line
width a result of the combination of line blending, Doppler broadening
and Compton scattering. Combining such effects, \citet{kallman89apj} 
estimated that \ews\ of up to 100 eV were attainable from an ADC using 
standard parameters. This in agreement with the values obtained in
this work. 

Finally, the line could originate in a partially ionised wind as a
result of illumination by the central source continuum photons and
broadened by electron downscattering in the wind environment
\citep[e.g.,][]{laurent07apj, titarchuk09apj}. The line profiles
obtained via such mechanism are again asymmetric, similar to the ones
produced by relativistic effects. Since the line profiles obtained
here are highly symmetric, we do not attribute the line origin to the
outflow based on these observations.

\subsection{Continuum emission}

The X-ray continua for 80\% of the observations were well fitted by a
model consisting of a blackbody and disc-blackbody components absorbed
by neutral material.  For 15\% of the observations, a model consisting
of a blackbody and power-law components absorbed by neutral material
was preferred to the combination of blackbody and disc-blackbody
components.  For one observation a fit with three components, namely,
blackbody, disc-blackbody and power-law was required in order to get
an acceptable \rchisq.

For the fits with blackbody and disc-blackbody components the
temperature of the blackbody component had values between 1.5 and
2.8~keV, except for \seventeentw, for which a temperature of
3.8\,$^{+8.4}_{-1.1}$~keV was found. The temperature of the disc
blackbody component had lower values between 0.6 and 1.3~keV, except
for \seventeentw, for which a temperature of 1.9\,$\pm$\,0.3~keV was
found. This is consistent with the interpretation of the blackbody
component as the boundary layer between the neutron star and the inner
disc, which shows smaller radii and higher temperatures.
\citet{church01aa}, based on an ASCA survey, found a broad dependence 
of the blackbody
luminosity on the total luminosity not previously known for LMXBs in
general, but while they found that the blackbody luminosity was
falling more rapidly than the total luminosity as the mass accretion
rate decreases, we find a very tight correlation where the blackbody
luminosity is very near the value of 50\% of total luminosity at all
luminosities. This may be due to the different continuum model used, blackbody
and cutoff power-law components \citep{church01aa} versus blackbody and disc
blackbody components (this work). Similarly to \citet{church01aa}, we find that the major
change in blackbody emission results primarily from changes in the emitting
area, not the temperature. The independence of the temperature of this component on the global value of
Eddington ratio lends support to the theoretical suggestion that the boundary layer is 
radiation pressure supported \citep{inogamov99,revnivtsev06aa}.

For the sources fitted with blackbody and power-law components, the
temperature of the blackbody component had values between 0.6 and
2.2~keV and the index of the power-law was between 1.8 and 2.6.

We detected an excess of emission at $\sim$1~keV in {\it all} the
observations where we could use the spectrum down to 0.7~keV, except
for the two accreting millisecond pulsars \xtee\ and \sax. We modelled
such emission with a Gaussian component centred between 1.0 and
1.15~keV in 12 observations and as an edge between 0.84 and 0.86~keV
in 2 observations. Provided the feature has an astrophysical origin,
its appearance always at similar energy points to line emission, since
a soft component due to a e.g. blackbody emission is expected to
change its temperature for different sources. Its energy is consistent
with emission from highly ionised Fe, from \fettwenty\ to \fetfour,
and/or \neten. We did not find a clear correlation between the 
width of the line emission at $\sim$1~keV and the Fe~\ka\ band. 
However, the fact that the same feature is not observed in 
simultaneous EPIC and RGS exposures \citep[e.g.][]{1323:boirin05aa}
indicates that the feature is likely blended emission from various
ions and additionally broadened by the same mechanism that the Fe~\ka\
line.

In this work, we show that effects of pile-up can significantly affect
the width, and therefore the physical interpretation, of the iron line
in NS X-ray binaries. Furthermore, parallel works on \xmm\ and \suzaku\
observations of the black hole X-ray binary GX~339-4 
demonstrated that, taking into account pile-up effects,
the breadth of the line was consistent with a truncation of the disc \citep{gx339:done10mnras}, 
or with a black hole with moderate spin \citep{gx339:yamada09apjl}, 
in contrast to previous claims \citep[e.g.][]{gx339:miller06apj,gx339:miller08apjl}.
Therefore, we urge caution in using piled-up data for detailed
spectral analysis.


\begin{acknowledgements}
Based on observations obtained with XMM-Newton, an ESA science
mission with instruments and contributions directly funded by ESA
member states and the USA (NASA). We thank the anonymous referee for 
helpful comments. We also thank M. Guainazzi and
E. Kuulkers for a careful reading of this manuscript and suggestions
that helped to improve it. We acknowledge support from the 
Faculty of the European Space Astronomy Centre (ESAC).
\end{acknowledgements}


\bibliographystyle{aa}
\bibliography{felines}

\begin{thebibliography}{65}
\expandafter\ifx\csname natexlab\endcsname\relax\def\natexlab#1{#1}\fi

\bibitem[{{Arnaud}(1996)}]{arnaud96conf}
{Arnaud}, K.~A. 1996, in ASP Conf. Ser. 101: Astronomical Data Analysis
  Software and Systems V, 17

\bibitem[{{Asai} {et~al.}(2000){Asai}, {Dotani}, {Nagase}, \&
  {Mitsuda}}]{asai00apjs}
{Asai}, K., {Dotani}, T., {Nagase}, F., \& {Mitsuda}, K. 2000, \apjs, 131, 571

\bibitem[{{Bhattacharyya} \& {Strohmayer}(2007)}]{serx1:bhattacharyya07apjl}
{Bhattacharyya}, S. \& {Strohmayer}, T. 2007, \apjl, 664, 103B

\bibitem[{{Boirin} {et~al.}(2005){Boirin}, {M\'endez}, {Parmar}, \&
  {Kaastra}}]{1323:boirin05aa}
{Boirin}, L., {M\'endez}, M.~{D{\'i}az Trigo}, M., {Parmar}, A.~N., \&
  {Kaastra}, J. 2005, \aap, 436, 195

\bibitem[{{Boirin} \& {Parmar}(2003)}]{1254:boirin03aa}
{Boirin}, L. \& {Parmar}, A.~N. 2003, \aap, 407, 1079

\bibitem[{{Boirin} {et~al.}(2004){Boirin}, {Parmar}, {Barret}, {Paltani}, \&
  {Grindlay}}]{1916:boirin04aa}
{Boirin}, L., {Parmar}, A.~N., {Barret}, D., {Paltani}, S., \& {Grindlay},
  J.~E. 2004, \aap, 418, 1061

\bibitem[{{Brandt} \& {Matt}(1994)}]{brandt94mnras}
{Brandt}, W.~M. \& {Matt}, G. 1994, \mnras, 268, 1051

\bibitem[{{Cackett} {et~al.}(2009{\natexlab{a}}){Cackett}, {Altamirano},
  {Patruno}, {Miller}, {Reynolds}, {Linares}, \&
  {Wijnands}}]{sax1808:cackett09apjl}
{Cackett}, E., {Altamirano}, D., {Patruno}, A., {et~al.} 2009{\natexlab{a}},
  \apjl, 694, L21

\bibitem[{{Cackett} {et~al.}(2009{\natexlab{b}}){Cackett}, {Miller},
  {Ballantyne}, {Barret}, {Bhattacharyya}, {Boutelier}, {Miller}, {Strohmayer},
  \& {Wijnands}}]{cackett09apj}
{Cackett}, E.~M., {Miller}, J.~M., {Ballantyne}, D.~R., {et~al.}
  2009{\natexlab{b}}, \apj, Subm., arxiv:0908.1098

\bibitem[{{Cackett} {et~al.}(2008){Cackett}, {Miller}, {Bhattacharyya},
  {Grindlay}, {Homan}, {van der Klis}, {Miller}, {Strohmayer}, \&
  {Wijnands}}]{cackett08apj}
{Cackett}, E.~M., {Miller}, J.~M., {Bhattacharyya}, S., {et~al.} 2008, \apj,
  674, 415

\bibitem[{{Casares} {et~al.}(2006){Casares}, {Cornelisse}, {Steeghs},
  {Charles}, {Hynes}, {O'Brien}, \& {Strohmayer}}]{1636:casares06mnras}
{Casares}, J., {Cornelisse}, R., {Steeghs}, D., {et~al.} 2006, \mnras, 373,
  1235

\bibitem[{{Church} \& {Baluci{\'n}ska-Church}(2001)}]{church01aa}
{Church}, M.~J. \& {Baluci{\'n}ska-Church}, M. 2001, \aap, 369, 915

\bibitem[{{D'A\`{\i}} {et~al.}(2006){D'A\`{\i}}, {Di Salvo}, {Iaria},
  {M\'endez}, {Burderi}, {Lavagetto}, {Lewin}, {Robba}, {Stella}, \& {van der
  Klis}}]{1728:dai06aa}
{D'A\`{\i}}, A., {Di Salvo}, T., {Iaria}, R., {et~al.} 2006, \aap, 448, 817

\bibitem[{{D'A\`{\i}} {et~al.}(2009){D'A\`{\i}}, {Di Salvo}, {Matt}, \&
  {Robba}}]{gx340:dai09apjl}
{D'A\`{\i}}, A., {Di Salvo}, T., {Matt}, G., \& {Robba}, N.~R. 2009, \apjl,
  693, L1

\bibitem[{{Den Herder} {et~al.}(2001){Den Herder}, {Brinkman}, {Kahn},
  {Branduardi-Raymont}, {Thomsen}, {Aarts}, {Audard}, {Bixler}, {den Boggende},
  {Cottam}, {Decker}, {Dubbeldam}, {Erd}, {Goulooze}, {G{\" u}del},
  {Guttridge}, {Hailey}, {Janabi}, {Kaastra}, {de Korte}, {van Leeuwen},
  {Mauche}, {McCalden}, {Mewe}, {Naber}, {Paerels}, {Peterson}, {Rasmussen},
  {Rees}, {Sakelliou}, {Sako}, {Spodek}, {Stern}, {Tamura}, {Tandy}, {de
  Vries}, {Welch}, \& {Zehnder}}]{xmm:denherder01aa}
{Den Herder}, J.~W., {Brinkman}, A.~C., {Kahn}, S.~M., {et~al.} 2001, \aap,
  365, L7

\bibitem[{{Di Salvo} {et~al.}(2009){Di Salvo}, {D'A\`{\i}}, {Iaria}, {Burderi},
  {Dov{\v c}iak}, {Karas}, {Matt}, {Papitto}, {Piraino}, {Riggio}, {Robba}, \&
  {Santangelo}}]{1705:disalvo09mnras}
{Di Salvo}, T., {D'A\`{\i}}, A., {Iaria}, R., {et~al.} 2009, \mnras, 398, 2022

\bibitem[{{D{\'i}az Trigo} {et~al.}(2006){D{\'i}az Trigo}, {Parmar}, {Boirin},
  {M\'endez}, \& {Kaastra}}]{ionabs:diaz06aa}
{D{\'i}az Trigo}, M., {Parmar}, A.~N., {Boirin}, L., {M\'endez}, M., \&
  {Kaastra}, J. 2006, \aap, 445, 179

\bibitem[{{Done} \& {D{\'i}az Trigo}(2009)}]{gx339:done10mnras}
{Done}, C. \& {D{\'i}az Trigo}, M. 2009, \mnras, Subm., arxiv:0911.3243

\bibitem[{{Done} {et~al.}(2007){Done}, {Sobolewska}, {Gierli{\'n}ski}, \&
  {Schurch}}]{done07mnras}
{Done}, C., {Sobolewska}, M.~A., {Gierli{\'n}ski}, M., \& {Schurch}, N.~J.
  2007, \mnras, 374, L15

\bibitem[{{Fabian} \& {Miniutti}(2005)}]{fabian05}
{Fabian}, A. \& {Miniutti}, G. 2005, in {\it The Kerr Spacetime}, edited by
  Wiltshire, D.~L., Visser, M., Scott, S.~M.

\bibitem[{{Fabian} {et~al.}(1989){Fabian}, {Rees}, {Stella}, \&
  {White}}]{cygx1:fabian89mnras}
{Fabian}, A.~C., {Rees}, M.~J., {Stella}, L., \& {White}, N.~E. 1989, \mnras,
  238, 729

\bibitem[{{Galloway} {et~al.}(2008){Galloway}, {Muno}, {Hartman}, {Psaltis}, \&
  {Chakrabarty}}]{galloway08apjs}
{Galloway}, D.~K., {Muno}, M.~P., {Hartman}, J.~M., {Psaltis}, D., \&
  {Chakrabarty}, D. 2008, \apjs, 179, 360

\bibitem[{{Grimm} {et~al.}(2002){Grimm}, {Gilfanov}, \& {Sunyaev}}]{grimm02aa}
{Grimm}, H.~J., {Gilfanov}, M., \& {Sunyaev}, R. 2002, \aap, 391, 923

\bibitem[{{Guainazzi} {et~al.}(2006){Guainazzi}, {Bianchi}, \&
  {Dovciak}}]{guainazzi06an}
{Guainazzi}, M., {Bianchi}, S., \& {Dovciak}, M. 2006, Astronomische
  Nachrichten, 327, 1032

\bibitem[{{Hasinger} {et~al.}(1990){Hasinger}, {van der Klis}, {Ebisawa},
  {Dotani}, \& {Mitsuda}}]{cygx2:hasinger90aa}
{Hasinger}, G., {van der Klis}, M., {Ebisawa}, K., {Dotani}, T., \& {Mitsuda},
  K. 1990, \aap, 235, 131

\bibitem[{{Hiemstra} {et~al.}(2009){Hiemstra}, {Soleri}, {M{\'e}ndez},
  {Belloni}, {Mostafa}, \& {Wijnands}}]{1753:hiemstra09mnras}
{Hiemstra}, B., {Soleri}, P., {M{\'e}ndez}, M., {et~al.} 2009, \mnras, 394,
  2080

\bibitem[{{Hirano} {et~al.}(1987){Hirano}, {Hayakawa}, {Nagase}, {Masai}, \&
  {Mitsuda}}]{hirano87pasj}
{Hirano}, T., {Hayakawa}, S., {Nagase}, F., {Masai}, K., \& {Mitsuda}, K. 1987,
  \pasj, 39, 619

\bibitem[{{Iaria} {et~al.}(2009){Iaria}, {D'A\`{\i}}, {Di Salvo}, {Robba},
  {Riggio}, {Papitto}, \& {Burderi}}]{gx349:iaria09aa}
{Iaria}, R., {D'A\`{\i}}, A., {Di Salvo}, T., {et~al.} 2009, \aap, 505, 1143

\bibitem[{{Iaria} {et~al.}(2004){Iaria}, {Di Salvo}, {Robba}, {Burderi},
  {Stella}, {Frontera}, \& {van der Klis}}]{gx349:iaria04apj}
{Iaria}, R., {Di Salvo}, T., {Robba}, N.~R., {et~al.} 2004, \apj, 600, 358

\bibitem[{{Inogamov} \& {Sunyaev}(1999)}]{inogamov99}
{Inogamov}, N.~A. \& {Sunyaev}, R.~A. 1999, Astronomy Letters, 25, 269

\bibitem[{{Jansen} {et~al.}(2001){Jansen}, {Lumb}, {Altieri}, {Clavel}, {Ehle},
  {Erd}, {Gabriel}, {Guainazzi}, {Gondoin}, {Much}, {Munoz}, {Santos},
  {Schartel}, {Texier}, \& {Vacanti}}]{xmm:jansen01aa}
{Jansen}, F., {Lumb}, D., {Altieri}, B., {et~al.} 2001, \aap, 365, L1

\bibitem[{{Kallman} \& {White}(1989)}]{kallman89apj}
{Kallman}, T. \& {White}, N.~E. 1989, \apj, 341, 955

\bibitem[{{Kuulkers} {et~al.}(2009){Kuulkers}, {in 't Zand}, {Atteia},
  {Levine}, {Brandt}, {Smith}, {Linares}, {Falanga}, {Sanchez-Fernandez},
  {Markwardt}, {Strohmayer}, {Cumming}, \& {Suzuki}}]{0614:kuulkers09aa}
{Kuulkers}, E., {in 't Zand}, J.~J.~M., {Atteia}, J.~L., {et~al.} 2009, \aap

\bibitem[{{Laming} \& {Titarchuk}(2004)}]{laming04apj}
{Laming}, J.~M. \& {Titarchuk}, L. 2004, \apjl, 615, L121

\bibitem[{{Laurent} \& {Titarchuk}(2007)}]{laurent07apj}
{Laurent}, P. \& {Titarchuk}, L. 2007, \apj, 656, 1056

\bibitem[{{Mason} {et~al.}(2001){Mason}, {Breeveld}, {Much}, {Carter},
  {Cordova}, {Cropper}, {Fordham}, {Huckle}, {Ho}, {Kawakami}, {Kennea},
  {Kennedy}, {Mittaz}, {Pandel}, {Priedhorsky}, {Sasseen}, {Shirey}, {Smith},
  \& {Vreux}}]{xmm:mason01aa}
{Mason}, K.~O., {Breeveld}, A., {Much}, R., {et~al.} 2001, \aap, 365, L36

\bibitem[{{Matt}(2006)}]{matt06an}
{Matt}, G. 2006, Astronomische Nachrichten, 327, 949

\bibitem[{{M\'endez} {et~al.}(2002){M\'endez}, {Cottam}, \&
  {Paerels}}]{0614:mendez02}
{M\'endez}, M., {Cottam}, J., \& {Paerels}, F. 2002, in New Visions of the
  X-ray Universe in the XMM-Newton and Chandra Era, ed. ESA, SP--488

\bibitem[{{Miller} {et~al.}(2006){Miller}, {Homan}, {Steeghs}, {Rupen},
  {Hunstead}, {Wijnands}, {Charles}, \& {Fabian}}]{gx339:miller06apj}
{Miller}, J.~M., {Homan}, J., {Steeghs}, D., {et~al.} 2006, \apj, 653, 525

\bibitem[{{Miller} {et~al.}(2008){Miller}, {Reynolds}, {Fabian}, {Cackett},
  {Miniutti}, {Raymond}, {Steeghs}, {Reis}, \& {Homan}}]{gx339:miller08apjl}
{Miller}, J.~M., {Reynolds}, C.~S., {Fabian}, A.~C., {et~al.} 2008, \apjl, 679,
  L113

\bibitem[{{Miller} {et~al.}(2009){Miller}, {Reynolds}, {Fabian}, {Miniutti}, \&
  {Gallo}}]{miller09apj}
{Miller}, J.~M., {Reynolds}, C.~S., {Fabian}, A.~C., {Miniutti}, G., \&
  {Gallo}, L.~C. 2009, \apj, 697, 900

\bibitem[{{Nandra} {et~al.}(1997){Nandra}, {George}, {Mushotzky}, {Turner}, \&
  {Yaqoob}}]{nandra97apj}
{Nandra}, K., {George}, I.~M., {Mushotzky}, R.~F., {Turner}, T.~J., \&
  {Yaqoob}, T. 1997, \apj, 477, 602

\bibitem[{{Nandra} {et~al.}(2007){Nandra}, {O'Neill}, {George}, \&
  {Reeves}}]{nandra07mnras}
{Nandra}, K., {O'Neill}, P.~M., {George}, I.~M., \& {Reeves}, J.~N. 2007,
  \mnras, 382, 194

\bibitem[{Pandel {et~al.}(2008)Pandel, Kaaret, \& Corbel}]{1636:pandel08apj}
Pandel, D., Kaaret, P., \& Corbel, S. 2008, \apj, 688, 1288

\bibitem[{{Papitto} {et~al.}(2009){Papitto}, {Di Salvo}, {D'A\`{\i}}, {Iaria},
  {Burderi}, {Riggio}, {Menna}, \& {Robba}}]{1808:papitto09aa}
{Papitto}, A., {Di Salvo}, T., {D'A\`{\i}}, A., {et~al.} 2009, \aap, 493, L39

\bibitem[{{Patruno} {et~al.}(2009){Patruno}, {Rea}, {Altamirano}, {Linares},
  {Wijnands}, \& {van der Klis}}]{1808:patruno09mnras}
{Patruno}, A., {Rea}, N., {Altamirano}, D., {et~al.} 2009, \mnras, 396, L51

\bibitem[{{Pozdnyakov} {et~al.}(1979){Pozdnyakov}, {Sobol}, \&
  {Sunyaev}}]{pozdnyakov79aa}
{Pozdnyakov}, L.~A., {Sobol}, I.~M., \& {Sunyaev}, R.~A. 1979, \aap, 75, 214

\bibitem[{{Reis} {et~al.}(2009){Reis}, {Fabian}, {Ross}, \&
  {Miller}}]{1655:reis09mnras}
{Reis}, R.~C., {Fabian}, A.~C., {Ross}, R.~R., \& {Miller}, J.~M. 2009, \mnras,
  395, 1257

\bibitem[{{Revnivtsev} \& {Gilfanov}(2006)}]{revnivtsev06aa}
{Revnivtsev}, M.~G. \& {Gilfanov}, M.~R. 2006, \aap, 453, 253

\bibitem[{{Reynolds} \& {Nowak}(2003)}]{reynolds03ps}
{Reynolds}, C.~S. \& {Nowak}, M.~A. 2003, Physics Reports, 377, 389

\bibitem[{{Ross} \& {Fabian}(2007)}]{ross07mnras}
{Ross}, R.~R. \& {Fabian}, A.~C. 2007, \mnras, 381, 1697

\bibitem[{{Sidoli} {et~al.}(2001){Sidoli}, {Oosterbroek}, {Parmar}, {Lumb}, \&
  {Erd}}]{1658:sidoli01aa}
{Sidoli}, L., {Oosterbroek}, T., {Parmar}, A.~N., {Lumb}, D., \& {Erd}, C.
  2001, \aap, 379, 540

\bibitem[{{Str{\" u}der} {et~al.}(2001){Str{\" u}der}, {Briel}, {Dennerl},
  {Hartmann}, {Kendziorra}, {Meidinger}, {Pfeffermann}, {Reppin}, {Aschenbach},
  {Bornemann}, {Br{\" a}uninger}, {Burkert}, {Elender}, {Freyberg}, {Haberl},
  {Hartner}, {Heuschmann}, {Hippmann}, {Kastelic}, {Kemmer}, {Kettenring},
  {Kink}, {Krause}, {M{\" u}ller}, {Oppitz}, {Pietsch}, {Popp}, {Predehl},
  {Read}, {Stephan}, {St{\" o}tter}, {Tr{\" u}mper}, {Holl}, {Kemmer},
  {Soltau}, {St{\" o}tter}, {Weber}, {Weichert}, {von Zanthier},
  {Carathanassis}, {Lutz}, {Richter}, {Solc}, {B{\" o}ttcher}, {Kuster},
  {Staubert}, {Abbey}, {Holland}, {Turner}, {Balasini}, {Bignami}, {La
  Palombara}, {Villa}, {Buttler}, {Gianini}, {Lain{\' e}}, {Lumb}, \&
  {Dhez}}]{xmm:struder01aa}
{Str{\" u}der}, L., {Briel}, U., {Dennerl}, K., {et~al.} 2001, \aap, 365, L18

\bibitem[{{Sunyaev} \& {Titarchuk}(1980)}]{sunyaev80aa}
{Sunyaev}, R.~A. \& {Titarchuk}, L.~G. 1980, \aap, 86, 121

\bibitem[{{Tanaka} {et~al.}(1995){Tanaka}, {Nandra}, {Fabian}, {Inoue},
  {Otani}, {Dotani}, {Hayashida}, {Iwasawa}, {Kii}, {Kunieda}, {Makino}, \&
  {Matsuoka}}]{tanaka95nat}
{Tanaka}, Y., {Nandra}, K., {Fabian}, A.~C., {et~al.} 1995, \nat, 375, 659

\bibitem[{{Titarchuk} {et~al.}(2003){Titarchuk}, {Kazanas}, \&
  {Becker}}]{titarchuk03apj}
{Titarchuk}, L., {Kazanas}, D., \& {Becker}, P.~A. 2003, \apj, 598, 411

\bibitem[{{Titarchuk} {et~al.}(2009){Titarchuk}, {Laurent}, \&
  {Shaposhnikov}}]{titarchuk09apj}
{Titarchuk}, L., {Laurent}, P., \& {Shaposhnikov}, N. 2009, \apj, 700, 1831

\bibitem[{{Turner} {et~al.}(2001){Turner}, {Abbey}, {Arnaud}, {Balasini},
  {Barbera}, {Belsole}, {Bennie}, {Bernard}, {Bignami}, {Boer}, {Briel},
  {Butler}, {Cara}, {Chabaud}, {Cole}, {Collura}, {Conte}, {Cros}, {Denby},
  {Dhez}, {Di Coco}, {Dowson}, {Ferrando}, {Ghizzardi}, {Gianotti}, {Goodall},
  {Gretton}, {Griffiths}, {Hainaut}, {Hochedez}, {Holland}, {Jourdain},
  {Kendziorra}, {Lagostina}, {Laine}, {La Palombara}, {Lortholary}, {Lumb},
  {Marty}, {Molendi}, {Pigot}, {Poindron}, {Pounds}, {Reeves}, {Reppin},
  {Rothenflug}, {Salvetat}, {Sauvageot}, {Schmitt}, {Sembay}, {Short},
  {Spragg}, {Stephen}, {Str{\" u}der}, {Tiengo}, {Trifoglio}, {Tr{\" u}mper},
  {Vercellone}, {Vigroux}, {Villa}, {Ward}, {Whitehead}, \&
  {Zonca}}]{xmm:turner01aa}
{Turner}, M.~J.~L., {Abbey}, A., {Arnaud}, M., {et~al.} 2001, \aap, 365, L27

\bibitem[{{Vaughan} \& {Uttley}(2008)}]{vaughan08mnras}
{Vaughan}, S. \& {Uttley}, P. 2008, \mnras, 390, 421

\bibitem[{{Vrtilek} {et~al.}(1993){Vrtilek}, {Soker}, \&
  {Raymond}}]{vrtilek93apj}
{Vrtilek}, S.~D., {Soker}, N., \& {Raymond}, J.~C. 1993, \apj, 404, 696

\bibitem[{{Wang} \& {Chakrabarty}(2004)}]{1543:wang04apjl}
{Wang}, Z. \& {Chakrabarty}, D. 2004, \apjl, 616, L139

\bibitem[{{White} \& {Holt}(1982)}]{white82apj}
{White}, N.~E. \& {Holt}, S.~S. 1982, \apj, 257, 318

\bibitem[{{White} {et~al.}(1986){White}, {Peacock}, {Hasinger}, {Mason},
  {Manzo}, {Taylor}, \& {Branduardi-Raymont}}]{white86mnras}
{White}, N.~E., {Peacock}, A., {Hasinger}, G., {et~al.} 1986, \mnras, 218, 129

\bibitem[{{Wilms} {et~al.}(2000){Wilms}, {Allen}, \& {McCray}}]{wilms00apj}
{Wilms}, J., {Allen}, A., \& {McCray}, R. 2000, \apj, 542, 914

\bibitem[{{Yamada} {et~al.}(2009){Yamada}, {Makishima}, {Uehara}, {Nakazawa},
  {Takahashi}, {Dotani}, {Ueda}, {Ebisawa}, {Kubota}, \&
  {Gandhi}}]{gx339:yamada09apjl}
{Yamada}, S., {Makishima}, K., {Uehara}, Y., {et~al.} 2009, \apjl, 707, L109

\end{thebibliography}

\appendix

\section{Examples of pile-up removal}
\label{sec:app1}

\begin{figure*}[!ht]
\includegraphics[angle=0,width=0.24\textwidth]{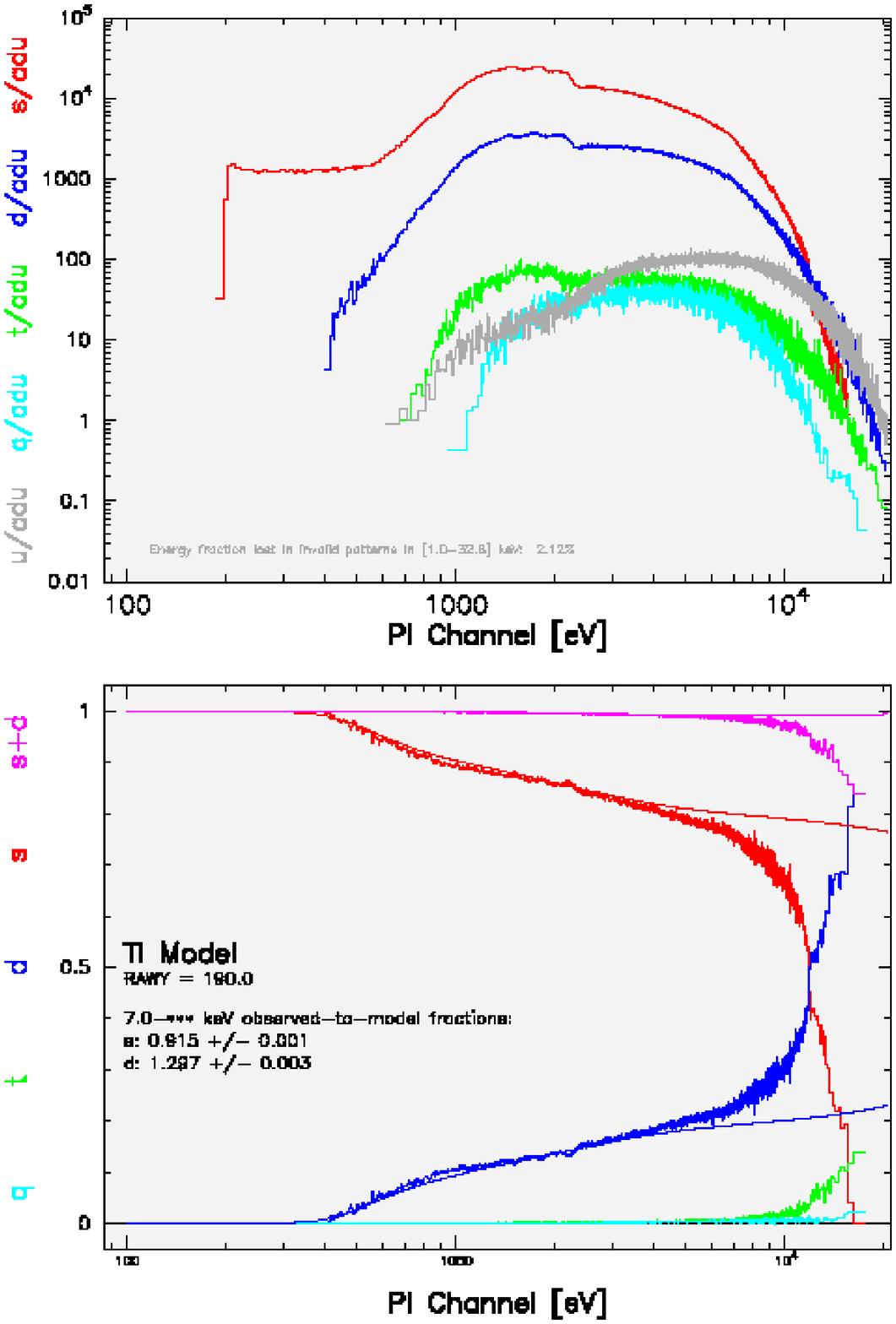}
\includegraphics[angle=0,width=0.24\textwidth]{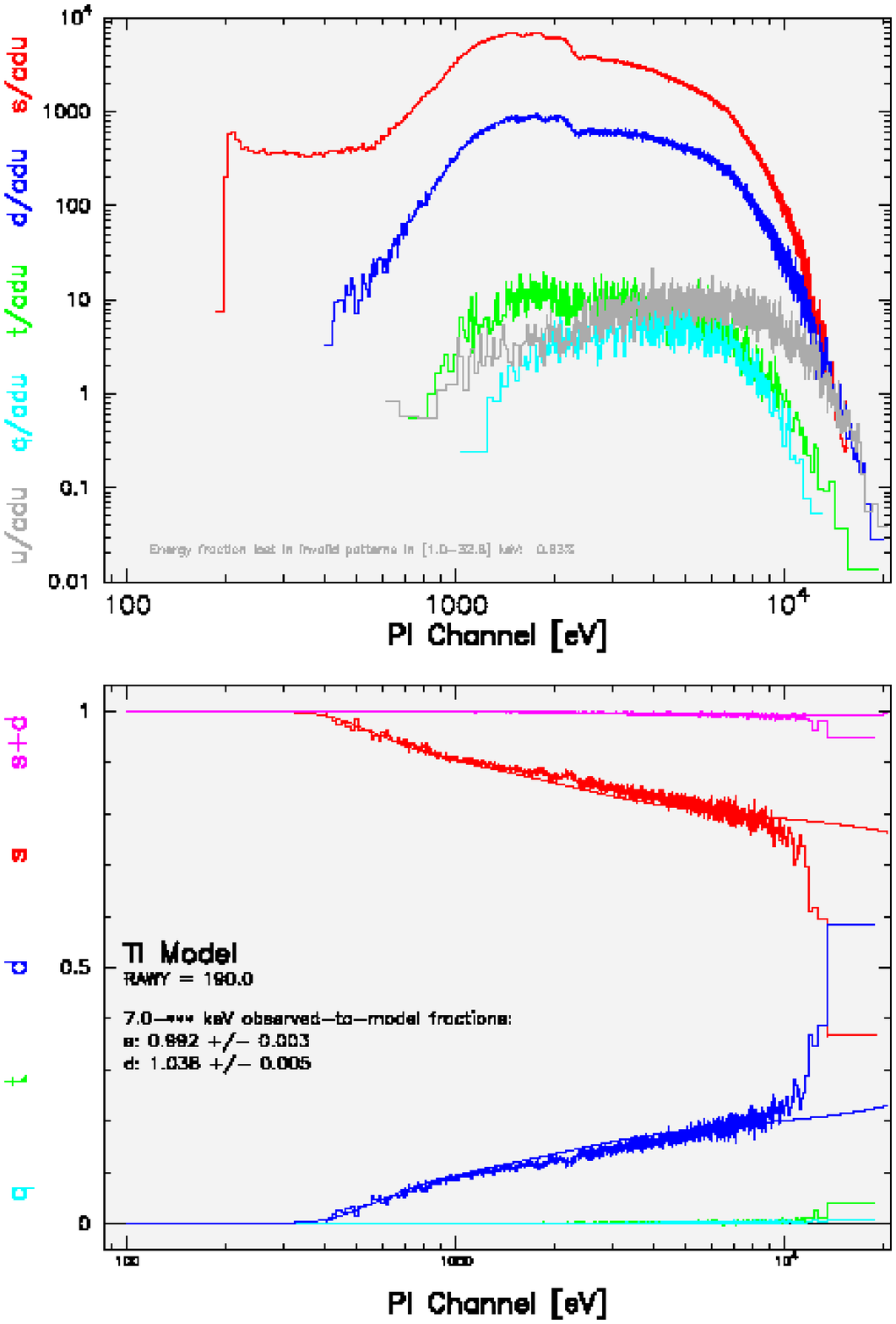}
\includegraphics[angle=0,width=0.24\textwidth]{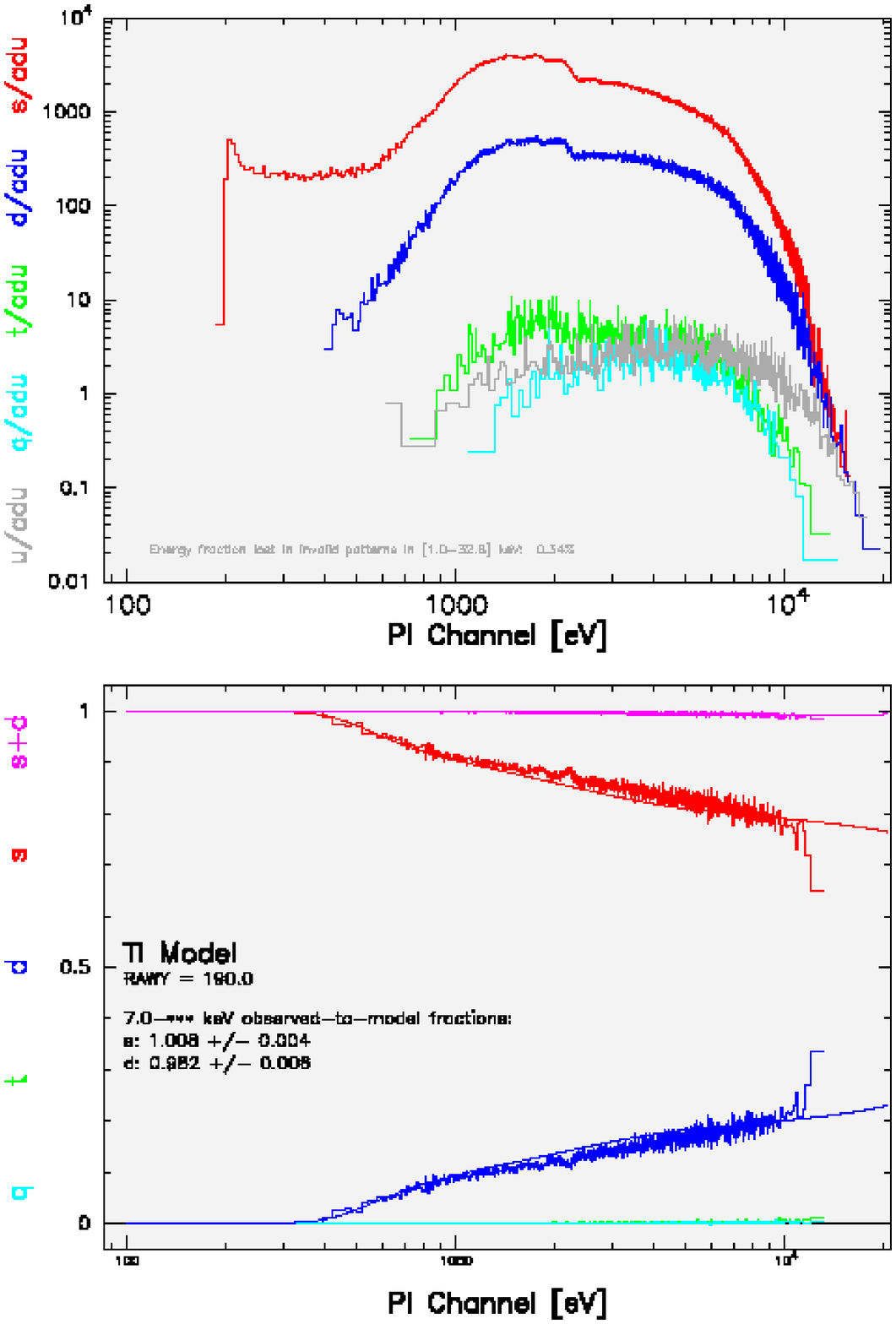}
\includegraphics[angle=0,width=0.24\textwidth]{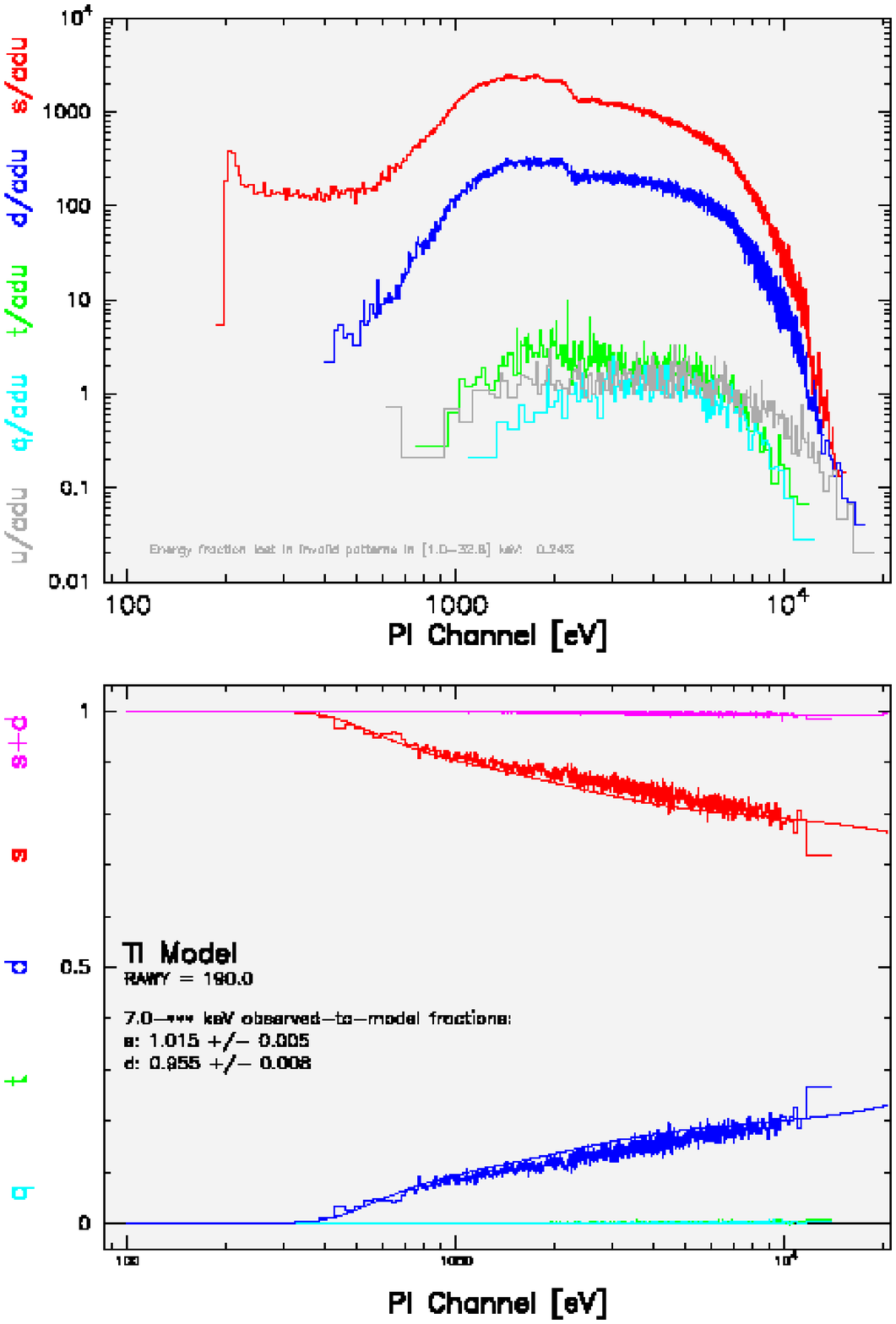}
\caption{Measure of pile-up in the EPIC pn Timing Mode using the SAS task {\tt epatplot} for the bright source 
\gxtfn. 
Source events were extracted from a 64\arcsec\
(16 columns) wide box centred on the source position (left panel). Middle-left, middle-right and right panels 
show the same
region as in the left panel but after exclusion of the neighbouring 4, 6 and 8 columns from the centre of 
the box.
}
\label{fig:epatplots}
\end{figure*}

\begin{figure*}[!ht]
\includegraphics[angle=0,width=0.24\textwidth]{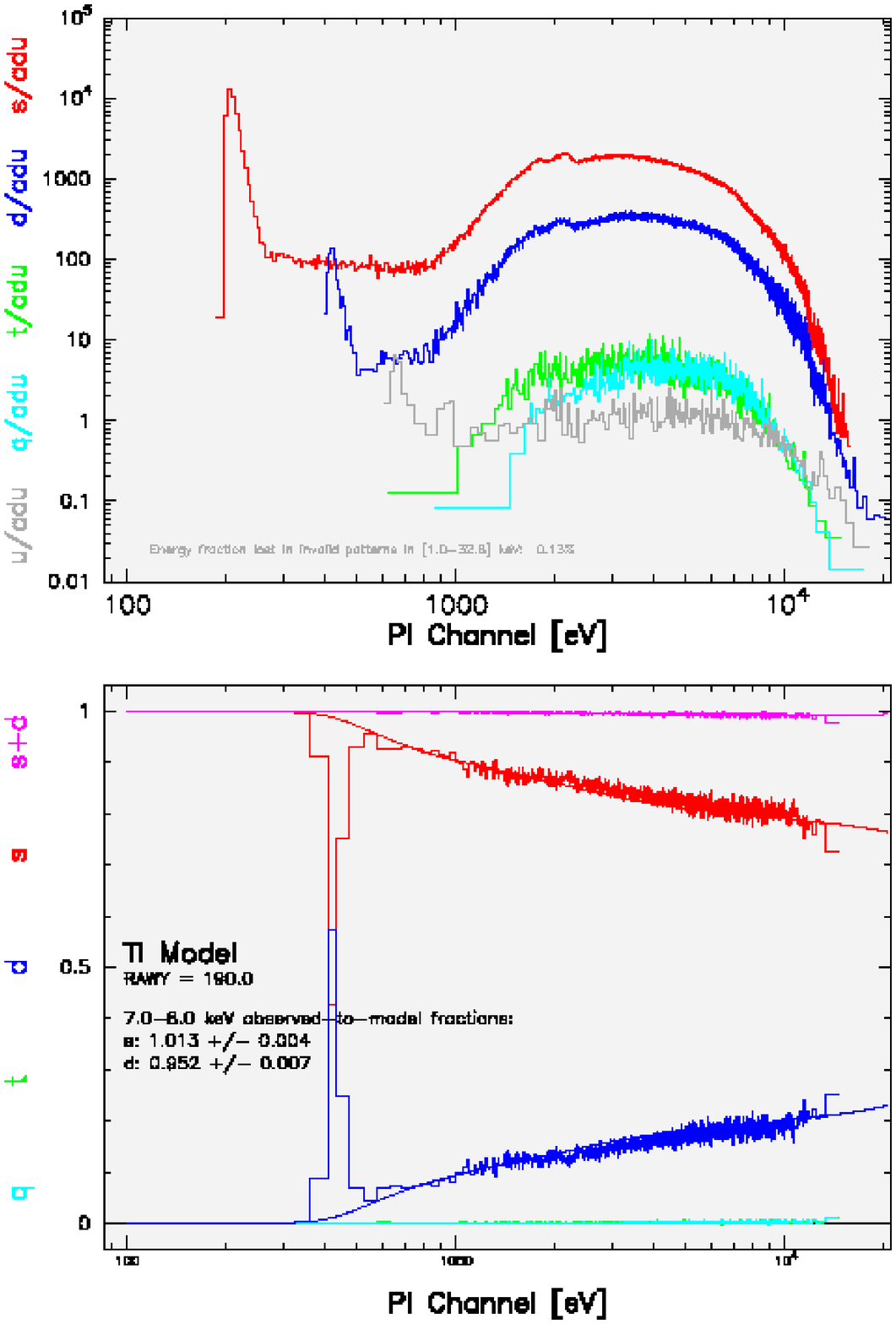}
\includegraphics[angle=0,width=0.24\textwidth]{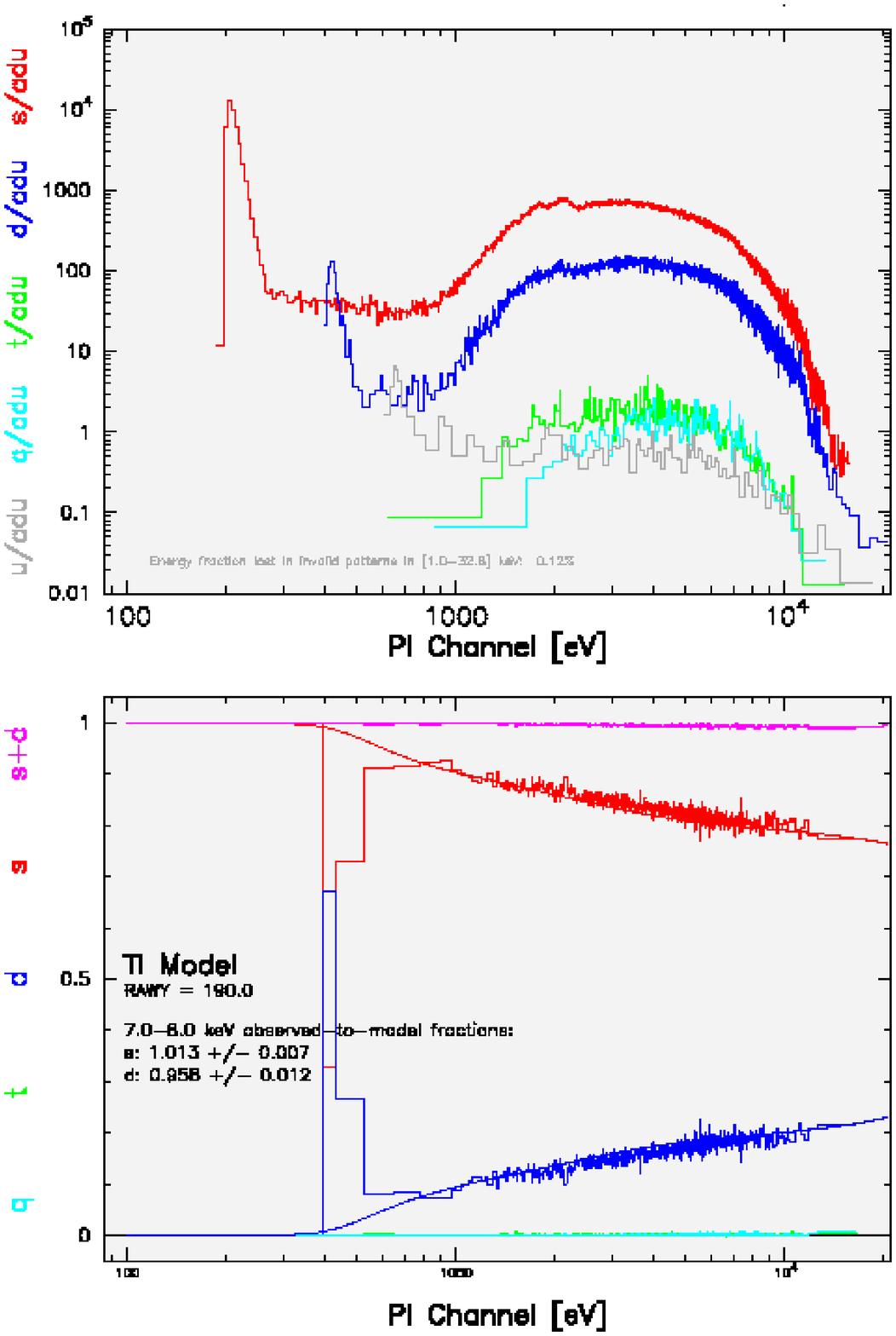}
\includegraphics[angle=0,width=0.24\textwidth]{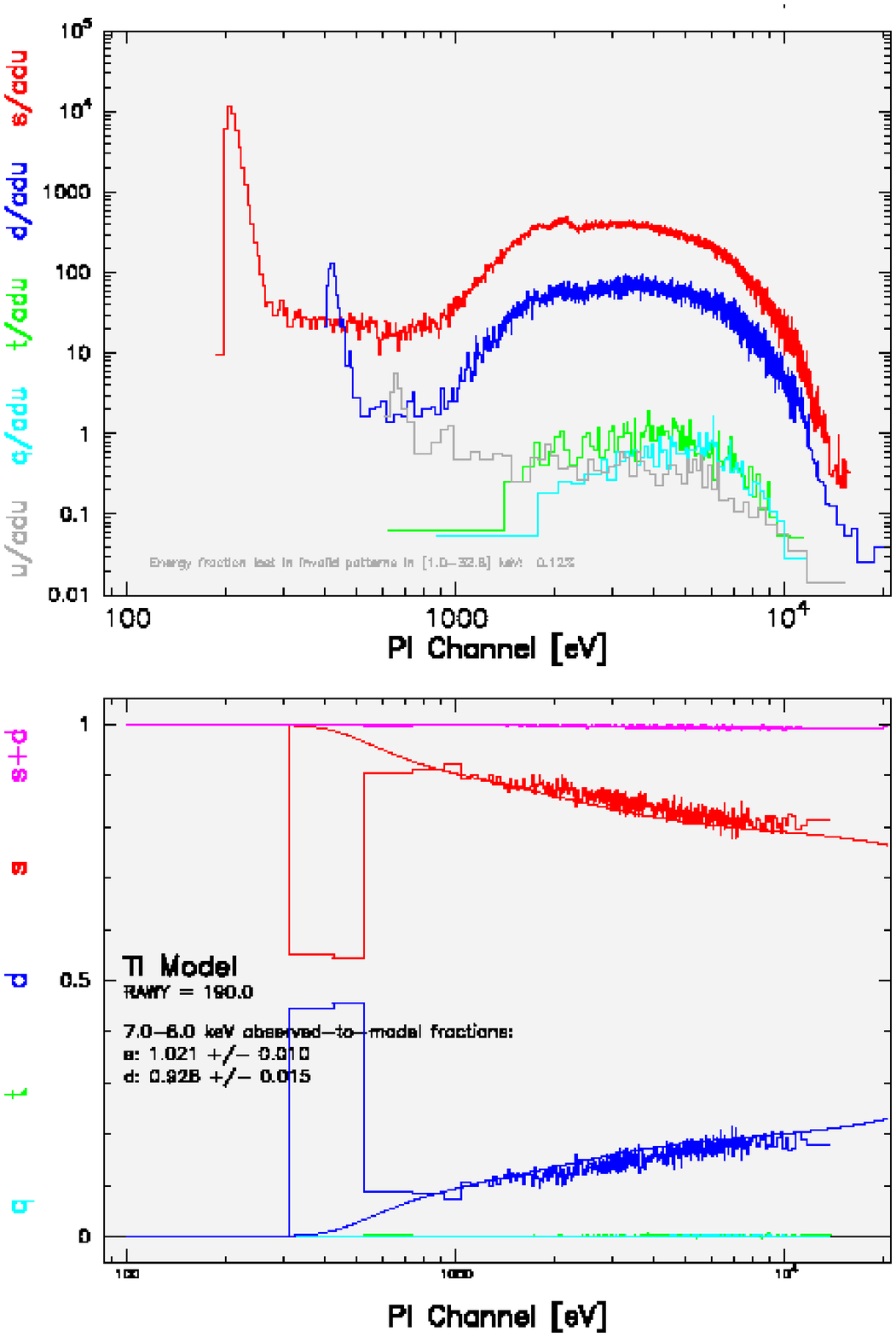}
\includegraphics[angle=0,width=0.24\textwidth]{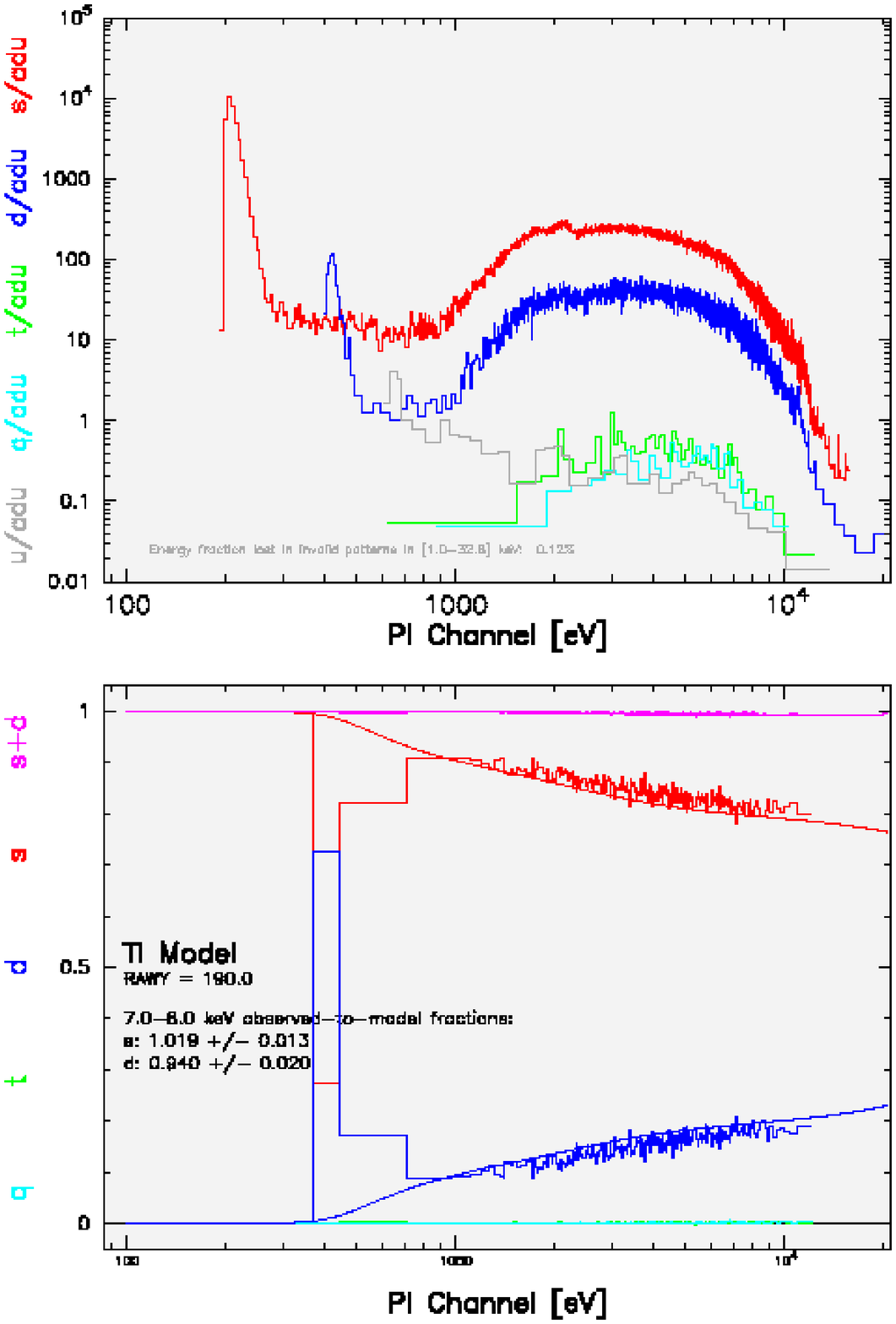}
\caption{Same as Fig.~\ref{fig:epatplots} but for the dim source \seventeentw. 
}
\label{fig:epatplots2}
\end{figure*}

Fig.~\ref{fig:epatplots} and ~\ref{fig:epatplots2} show the effects of pile-up 
in the {\tt epatplot} in the bright source \gxtfn\ and the
dim source \seventeentw, respectively.

For \gxtfn\ we observe a significant energy-dependent deviation
between the data and the model $\approxgt$~4 keV when the full 
PSF region is used (Fig.~\ref{fig:epatplots}, upper left).

As we remove the inner columns from the PSF the effects of pile-up 
are mitigated. After removal of 6 columns, there is still a small deviation above 9 keV.
After removal of 8 columns, there is not 
an energy-dependent deviation anymore (Fig.~\ref{fig:epatplots}, right panel).

For \seventeentw\ we do not observe any energy-dependent deviation
between the data and the model when the full 
PSF region is used (Fig.~\ref{fig:epatplots2}, left panel).
Similarly, when we excise the core of the PSF the pattern ratios do not change
with respect to the case where the core of the PSF is included.
This is an indication that pile-up is not present in this source.

\section{Comparison with previous results}

A number of the sources analysed in this paper were previously
published by other authors \citep[e.g.][]{serx1:bhattacharyya07apjl,
gx349:iaria09aa, gx340:dai09apjl}.
In what follows we
compare the properties of the Fe~\ka\ line found in previous analyses
with the ones inferred in this paper.  

For most of the sources we obtained significantly different parameters
for the Fe line when fitted with a {\tt laor} component, compared to
previous analyses (we discuss the discrepancies in detail below).

Therefore, in order to obtain a quantitative 
measurement  of the difference between the Fe line obtained by other authors and
in this work we performed fits to the spectra with the best-fit model shown in
Sect.~\ref{sec:spectra} but fixing all the parameters of the {\tt laor} component to the 
values obtained in previous analyses except the normalisation, which was left
free. The \rchisq\ of these fits is shown in Table~\ref{tab:line_comparison}. For
comparison, we show in the same table the \rchisq\ obtained when the line is 
fitted both with a Gaussian and a {\tt laor} component and leaving all the parameters
free, as in Tables~\ref{tab:bestfit}-\ref{tab:bestfitlines}. We note that we compared
the results in this work with the results obtained by \citet{cackett09apj} only in Table~\ref{tab:line_comparison}
and not in the subsections that follow, since the latter work is in refereeing process at the moment of acceptance 
of this paper, and their parameters may still change in their final version. 

We found that the \rchisq\ of the fits to the lines with the parameters of the 
literature yielded higher values of \rchisq\ for eight of the observations analysed 
in this work (see Table~\ref{tab:line_comparison}).
For the three observations of \sixteen, the difference in \rchisq\ is not statistically 
significant. Indeed the lines fitted in this work have a width $\sigma$ near 1~keV, similarly
to previous analyses. However, we note that two of the three lines are below the level of 3~$\sigma$ 
detectability significance in this work and the third one is only slightly above such limit 
(see Sect.~\ref{sec:properties}). We also note that in the case of \sixteen\ Obs~0500350401, 
we obtained a lower, though not significant, \rchisq\ when using the parameters of the line from previous
analyses. This is due to the fact that we did not allow in our analysis an inclination higher than 70\deg, since
the source shows neither dips nor eclipses. In contrast, the value of the inclination in the {\tt laor} component in previous
analyses had a large value of $>$\,81\deg\ \citep[e.g.,][]{1636:pandel08apj,cackett09apj}.
For the observations of \ser, \sax\ and \gxtfz, the increase of the \rchisq\ when we use the parameters
of the lines in the literature is not statistically significant, since the \rchisq\ is already well below 1 in the fits 
in this work.
However, we note that since the \rchisq\ is already below 1 for fits with the simple Gaussian model there is no reason
to consider a {\tt laor} component for these fits.

\begin{table*}
\begin{center}
\caption[]{\rchisq\ of spectral fits for sources for which asymmetric Fe lines 
showing relativistic effects have been reported in the literature. Columns (a)
and (b) show the \rchisq\ of the fit after including a Gaussian and a {\tt laor} 
component, respectively, to the best-fit model. Column (c) shows the \rchisq\ of 
the fit after including a {\tt laor} 
component with all the parameters fixed to the values obtained by previous authors, except
the normalisation, to the best-fit model. 
}
\begin{tabular}{ccccl}
\hline \noalign {\smallskip}
\hline \noalign {\smallskip}
Source  & Observation & \multicolumn{3}{c}{\tt \rchisq} \\
& ID   & a & b & c \\
\hline \noalign {\smallskip}

\sixteen\ & 0303250201 & 0.76 (216) & 0.78 (214) & 0.85 (218)$^1$; 0.84 (218)$^2$ \\
& 0500350301  & 0.71 (217) & 0.70 (215) & 0.71 (219)$^1$; 0.70 (219)$^2$  \\
& 0500350401 & 0.72 (217) & 0.80 (215)& 0.69 (219)$^1$; 0.69 (219)$^2$ \\
\gxtfz\ & 0505950101 & 0.86 (163) & 0.88 (161) & 0.98 (165)$^3$  \\
\gxtfn\ & 0506110101 & 1.15 (216) & 1.14 (214) & 1.37 (218)$^1$; 1.30 (218)$^4$  \\
\seventeenof\ & 0402300201  & 1.01 (186) & 1.03 (184) & 1.10 (188)$^1$  \\
& 0551270201 & 1.07 (187) & 1.03 (185) & 1.37 (188)$^5$  \\
\ser\ & 0084020401  & 0.98 (216) & 0.97 (214) & 1.00 (218)$^1$; 1.04 (218)$^6$  \\
& 0084020501  & 0.93 (216) & 0.92 (214) & 0.95 (218)$^1$; 0.97 (218)$^6$  \\
& 0084020601  & 1.04 (216) & 1.04 (214) & 1.04 (218)$^1$; 1.05 (218)$^6$  \\
\sax\ & 0560180601 & 0.86 (220) & 0.84 (218)& 0.88 (222)$^7$; 0.92 (222)$^8$; 0.90 (222)$^9$ \\

\noalign {\smallskip} \hline \label{tab:line_comparison}
\end{tabular}
\end{center}
\footnotetext{}{$^1$\citet{cackett09apj}; $^2$ \citet{1636:pandel08apj}; $^3$ \citet{gx340:dai09apjl};
$^4$ \citet{gx349:iaria09aa}; $^5$ \citet{1705:disalvo09mnras}; $^6$ \citet{serx1:bhattacharyya07apjl};
$^7$ \citet{sax1808:cackett09apjl}; $^8$ \citet{1808:patruno09mnras}; $^9$\citet{1808:papitto09aa}
 }
\end{table*}

\subsection{\ser}

\citet{serx1:bhattacharyya07apjl} first reported the presence of a
skewed iron \ka\ line with a moderately extended red wing in a NS LMXB
based on the three \xmm\ observations of \ser\ that we re-analysed in
this paper. They fitted the line with a {\tt laor} component and found
\ews\ for the lines between 86\,$^{+9}_{-11}$ and 105\,$^{+7}_{-8}$~eV,
inner radii between 4.04\,$^{+2.14}_{-0.68}$ and
16.19\,$^{+8.77}_{-3.19}$~r$_g$ and an energy centroid of
6.4\,$^{+0.08}_{-0.0*}$~keV. On average we found smaller \ews, between
57\,$\pm$\,18 and 63\,$\pm$\,12~eV and larger energy centroids, 
between 6.60\,$\pm$\,0.1 and
6.79\,$^{+0.1}_{-0.2}$~keV, when fitting the line with the same
component and \ews\ between 50\,$^{+14}_{-12}$ and 52\,$^{+14}_{-11}$ and
an average energy centroid of 6.60\,$\pm$\,0.04~keV when fitting the line with a
Gaussian component. We also inferred a smaller average inclination for
the source from the {\tt laor} fit: between $<$25 and 28\,$\pm$\,9\deg, 
compared to between 39.7$^{+1.4}_{-1.5}$ and 50.2$^{+8.8}_{-5.4}$\deg\ found by
\citet{serx1:bhattacharyya07apjl}. The inner radius was not well constrained when
fitting the line with a {\tt laor} component, with a value of 13.3\,$^{+19.3}_{-6.9}$~r$_g$
for the best constrained observation (Obs~0084020401). In our analysis the
fit with the {\tt laor} component was not statistically preferred to the one with the
Gaussian component.
 
\begin{figure}[!ht]
\includegraphics[angle=-90,width=0.5\textwidth]{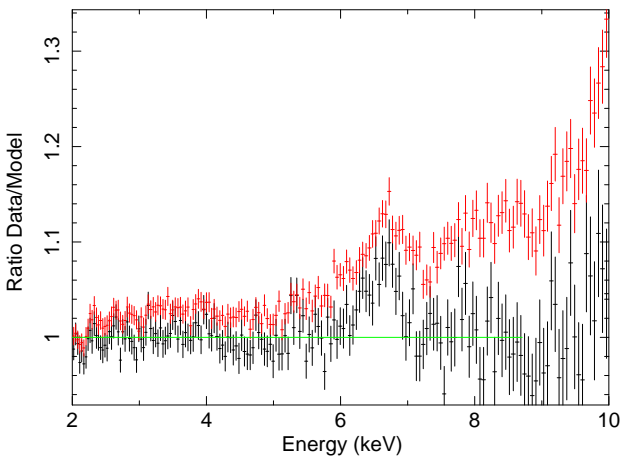}
\caption{Ratio of the EPIC pn spectrum of \ser\ (Obs~0084020501) to its best fit continuum model 
for the spectrum free of pile-up used in this paper (black). The red points show the ratio of the
piled-up spectrum of the same observation, obtained when the full PSF is used, to the best
fit continuum model of the spectrum free of pile-up. The pile-up shows up mainly as a significant
hardening of the spectrum at energies $\approxgt$5~keV. This adds an 'artificial' red-wing to the
iron line due to the different curvature of the spectrum.
}
\label{fig:pileup}
\end{figure}

We explain the discrepancies in the
results as due mainly to two differences in the analysis: first, while
\citet{serx1:bhattacharyya07apjl} considered pile-up to be insignificant
for these observations, we found that the {\tt epatplot} shows a
significant degree of pile-up before we remove the central 4 columns of
the PSF. Fig.~\ref{fig:pileup} shows the difference in the spectra extracted
including the core of the PSF (red), as done by
\citet{serx1:bhattacharyya07apjl}, and excluding the 4 central columns
to remove pile-up effects (black), as in this paper. The pile-up
effects show up as an excess of photons from as low as 3~keV and
becomes especially prominent above 5~keV in the
red spectrum.
This, combined with a improper continuum modelling (see below), may
create an excess of photons below 6~keV which could be misinterpreted
as a red wing of the Fe~line.

The second difference in the analysis is precisely that we used a
different continuum model to fit the spectrum. As shown in
Table~\ref{tab:chisq} we got a significantly worse \rchisq\ using a
combination of {\tt diskbb} and {\tt comptt} components to fit the
continuum, as done by \citet{serx1:bhattacharyya07apjl}, when compared
to {\tt diskbb} and {\tt bbody} components, as used 
in this paper.

Therefore, for this particular source, the main contribution to the
broad Fe~line detected in the first analysis by
\citet{serx1:bhattacharyya07apjl} is likely related to residuals from
the fit to the continuum, showing that modelling the spectra with
different continua may modify significantly the parameters of the
Fe~line.

\subsection{\gxtfz}

\citet{gx340:dai09apjl} reported the presence of a broad asymmetric
emission line in the \xmm\ spectrum of \gxtfz. Due to the variability
of the source as revealed by the light curve (see Fig.~\ref{fig:lc}),
they divided the observation in five segments and analysed each
segment separately. They modelled the spectral continuum with {\tt
diskbb} and {\tt bbody} components and found a significant improvement
in the fit when using the {\tt diskline} component to model the
Fe~\ka\ emission compared to a Gaussian component. The line had an
average energy of 6.69\,$\pm$\,0.02~keV, inner radius of
13$\pm$3~r$_g$, and \ew\ of 41\,$\pm$\,3~eV. In spite of using the
same continuum to model the spectrum, we obtained equally good fits when
modelling the Fe~line with a Gaussian or with a {\tt laor}
component. The line had an average energy of
6.82\,$^{+0.07}_{-0.13}$~keV, an inner radius $>$16~r$_g$, and a
smaller \ew\ of 15\,$\pm$\,5~eV. We inferred an average inclination for
the source of $<$\,34\deg, compared to 35$\pm$1\deg\ found by
\citet{gx340:dai09apjl}. When fitted with a Gaussian profile, the line
had an energy of 6.76\,$\pm$\,0.02~keV and width of
0.24\,$\pm$\,0.02~keV \citep{gx340:dai09apjl}, consistent with our
values of 6.72\,$\pm$\,0.06~keV and width of
0.17\,$^{+0.14}_{-0.08}$~keV.

The main difference bewteen our analysis and theirs is again
the treatment of pile-up. While they considered only the initial
$\sim$7~ks of the observation (interval 5) to be affected by
pile-up and excised the 2 central columns to remove its effects, we
found that the full observation was strongly affected by pile-up and
removed the central 8 columns. This is consistent with the high count
rate of the source, with peaks up to $\approxgt$1100~\countsec and a
minimum of 650~\countsec\ along the observation, as well as the hardness of the
spectrum (see Sect.~\ref{sec:pileup}). Note that the light curve in this paper
(see Fig.~\ref{fig:lc}) shows a count rate twice as large as the one
shown in Fig.~1 from \citet{gx340:dai09apjl}. A potential explanation of this difference 
in the light curves could be
that \citet{gx340:dai09apjl} averaged intervals of real data with those for which
no data were recorded, due to telemetry saturation.

\subsection{\gxtfn}

\citet{gx349:iaria09aa} reported the presence of several emission
features, consistent with the transitions of L-shell
\fettwo-\fetthree, \ssixteen, \areighteen, \canineteen\ and highly
ionised Fe in the \xmm\ spectrum of \gxtfn. While the first four
features could be fitted equally well with Gaussian features or the relativistic {\tt
diskline} component, they found that the Fe~\ka\ feature was better fitted
using the {\tt diskline} component at 6.76~keV or two {\tt diskline}
components at 6.7 and 6.97~keV. The line had an energy of
6.80$\pm$0.02~keV, width ($\sigma$) of 0.28$^{+0.03}_{-0.04}$~keV and \ew\
of 49$^{+6}_{-7}$~eV when modelled with a Gaussian component. The
same line had an energy of 6.76$\pm$0.02~keV, inner radius
6.2$^{+19.1}_{-0.2}~$r$_g$ and \ew\ of 61$\pm$9~eV when modelled with
a {\tt diskline} component.

In spite of using the same continuum to model the spectrum, we obtained
equally good fits when modelling the Fe~line with a Gaussian or with a
{\tt laor} component and a significantly larger \ew\ in any of the
fits, compared to \citet{gx349:iaria09aa}. In this paper, the line has
an energy of 6.72$\pm$0.06~keV, width ($\sigma$) of
0.32$^{+0.11}_{-0.07}$~keV and \ew\ of 82$^{+22}_{-15}$~eV when
modelled with a Gaussian component and energy of
6.94$^{+0.03}_{-0.22}$~keV, inner radius of 11$^{+7}_{-5}$~r$_g$
and \ew\ of 96$^{+20}_{-14}$~eV when modelled with a {\tt laor}
component. We obtained a significantly different inclination for the
source, 17\,$\pm$\,9\deg, compared to 41.4\,$^{+1}_{-2.1}$\deg\
\citep{gx349:iaria09aa}. The line parameters, apart from the inclination and the
\ew, were not significantly different within the errors. We attribute
the differences in some properties of the line to the PSF
extraction regions used. While
\citet{gx349:iaria09aa} excluded only the 2 central pixels of the PSF before spectral
extraction, their Fig.~4 shows an excess of double events and
a deficit of single events above $\sim$7.5~keV compared to the
predicted model. This is a clear indication of residual pile-up in
their spectra (see {\it XMM-Newton Users Handbook}). In contrast we found that 
rejection of the inner 8
columns of the PSF was necessary in order to remove the pile-up
effects completely. 

We found a lower
temperature for the {\tt diskbb} and {\tt bbody} components,
0.92\,$\pm$\,0.03 and 1.57\,$\pm$\,0.02~keV, compared to
1.05\,$^{+0.02}_{-0.03}$ and 1.792\,$^{+0.006}_{-0.019}$~keV
\citep{gx349:iaria09aa}. 
Thus the spectrum extracted in
this paper is significantly softer compared to the analysis by \citet{gx349:iaria09aa}. Again, this
is expected if pile-up effects have not been completely removed from
the spectra in their analysis.

\subsection{\seventeenof}

\seventeenof\ was observed twice by \xmm\ with the EPIC pn camera in 
Timing mode.

The first observation had a relatively low flux,
2.5$\times$10$^{-10}$\ergcmsec, and was at least one order of
magnitude below the typical count rate where pile-up effects become
important.
In 2008 \xmm\ observed \seventeenof\ at a significantly higher flux,
642~\countsec\ compared to 28~\countsec\ in 2006 (see
Table~\ref{tab:obslog}). \citet{1705:disalvo09mnras} reported the presence
of four emission features and one absorption edge in the spectral
residuals after modelling the continuum with a {\tt bbody} and a {\tt
comptt} components, which were identified with emission of \ssixteen,
\areighteen, \canineteen\ and \fetfive, and (redshifted) absorption
from \fetfive. The lines were broad with $\sigma$ between 120 and
260~eV. They found a significant improvement of the fit when the
\fetfive\ feature was modelled with a {\tt diskline} component
compared to a Gaussian component and reported values of
6.66\,$\pm$\,0.01~keV, 14\,$\pm$\,2~r$_g$, 39\,$\pm$\,1 and
56\,$\pm$\,2~eV for the energy, inner radius, inclination and \ew\ for
the fit with a {\tt diskline} component. The absorption edge appeared
to be smeared (width of 0.7 keV) and redshifted with an energy of
8.3\,$\pm$\,0.1~ keV with respect to the rest frame energy of 8.83~keV.
We found values of 6.45\,$\pm$\,0.05~keV, $<$\,9~r$_g$,
$<$\,34\deg and 135\,$^{+10}_{-21}$~eV for the energy, inner
radius, inclination and \ew\ for the fit with a {\tt laor} component,
i.e. significantly different to those reported by
\citet{1705:disalvo09mnras}. When fitted with a Gaussian component the centroid was
6.56\,$\pm$\,0.05~keV and the \ew\ 92\,$^{+23}_{-8}$~eV.

The difference between both analyses is related again to both the pile-up
treatment and the modelling of the continuum. We found that it 
was necessary to excise the 7 central columns of the PSF in order to remove
completely pile-up effects, while \citet{1705:disalvo09mnras} did not remove any
column. In addition, \citet{1705:disalvo09mnras} used a
continuum of {\tt bbody} and {\tt comptt} to fit the spectrum, while
we used {\tt bbody} and {\tt diskbb}.

\subsection{\sax}

\sax\ was observed by \xmm\ in 2008 and its spectrum modelled with a
combination of {\tt diskbb}, {\tt blackbody} and {\tt power law}
components \citep{1808:papitto09aa,
1808:patruno09mnras,sax1808:cackett09apjl}.  The residual emission at
the Fe~\ka\ band is modelled with a {\tt diskline} profile in all the
analyses, although \citet{1808:papitto09aa} obtained only a slight
improvement in the fit when using a {\tt diskline} profile compared to
a Gaussian profile. The parameters of the {\tt diskline} component
were in general consistent in the three analyses, with some exceptions
such as the inclination derived for the source, due probably to small
differences in the analyses such as including or not the RGS in the
simultaneous fit or considering slightly different energy bands. For
example, \citet{1808:papitto09aa} \citep{sax1808:cackett09apjl} used
the 1.4--11~keV (1.2--11~keV) band for spectral fitting since they
found large residuals below 1.4 (1.2)~keV, respectively.

As explained in Sect.~\ref{sec:spectra} we extracted for this
observation two different spectra. Spectrum~1 was extracted including
the core of the PSF, as done by the authors above. Despite using the
same extraction region, we found a significantly smaller line, with an
\ew\ of 51\,$^{+20}_{-16}$~eV when fitting the
line with a {\tt laor} component and 23\,$^{+9}_{-7}$~eV in the fits
with a Gaussian component. In contrast, in the previous analyses with
a {\tt laor} component, the \ew\ was found to be as large as
121\,$^{+20}_{-16}$~eV
\citep{1808:papitto09aa}, 118\,$\pm$\,10 \citep{sax1808:cackett09apjl}
and 97.7\,$\pm$\,31.4~eV \citep{1808:patruno09mnras}. 

This may be related to the different continuum used to fit the
spectrum. While all the previous analyses of the \xmm\ observation of
\sax\ required 3 continuum components to achieve an acceptable fit, we
obtain already a \rchisq\ of 0.98 (223) with a continuum consistent of
blackbody and power-law components. 

\subsection{\sixteen}

\citet{1636:pandel08apj} found evidence for the presence of
relativistic lines from iron in different ionization states in the
three \xmm\ observations of \sixteen\ presented here.
They reported \ews\ of 215, 98 and 140 eV for the
iron \ka\ line present in Obs~0303250201, 0500350301 and
0500350401 respectively, which they modelled with a {\tt diskline} component. 
In contrast we found smaller \ews\ of
130\,$\pm$\,14, 36\,$^{+34}_{-16}$ and 6\,$\pm$\,3 eV when fitting the line with
a {\tt laor} component and of 210\,$\pm$\,129, 28\,$\pm$\,19 and 59\,$^{+15}_{-17}$~eV 
when using a Gaussian component instead. We note that when fitting the line with a Gaussian component, the lines from Obs~0303250201 and 0500350301 are below the 3~$\sigma$ significance and should therefore not considered as detections. 
\citet{1636:pandel08apj} regarded the high inclination, $>$\,81\deg, 
obtained for Obs~0303250201 and 0500350401 as unrealistic (given the values
of the inclination of 36-74\deg determined 
for this source \citet{1636:casares06mnras}) and interpreted it as an indication
that the excess at the Fe~\ka\ band was a blend of at least two lines.
In contrast, we found only a high inclination for Obs~0303250201 and an upper value
of 70\deg\ (as used for all the fits in this sample) gave already an acceptable
fit with a \rchisq\ of 0.76 (216) for this observation.
We obtained equally good fits when
modelling the Fe~line with a Gaussian or with a {\tt laor}
component. 

We attribute the differences in some properties of the line to the PSF
extraction regions used. \citet{1636:pandel08apj} considered that none
of the three \xmm\ observations suffer from pile-up effects based on
the count rate limit for the pn Timing mode. In contrast we found that
rejection of the inner 1(3) columns of the PSF was necessary in order
to remove the pile-up effects completely for Obs~0500350301
(0500350401). The residual pile-up effects in their spectra are most
likely responsible for the harder spectrum. They reported a difference
of their spectra with respect to the simultaneous RXTE PCA spectra
both of flux and slope. The PCA spectra show a flux excess of
$\sim$30\% at 3~keV and $\sim$10-15\% at 10~keV with respect to their
EPIC pn spectra. Qualitatively, this is what we expect if the \xmm\
spectra are affected by pile-up: photons will be lost so that in
average the flux is lower and in addition soft photons will be counted
as hard photons, so that in average the PCA will show a softer
spectrum.

\section{Spectra}
\label{sec:app2}

The best-fits of each individual spectrum together with the residuals of the fits are shown in Figs.~\ref{fig:spectra}-\ref{fig:spectra2}. Figs.~\ref{fig:eeuf}-\ref{fig:eeuf2} show the unfolded spectra for each observation. 

\begin{figure*}[!ht]
\includegraphics[angle=-90,width=0.33\textwidth]{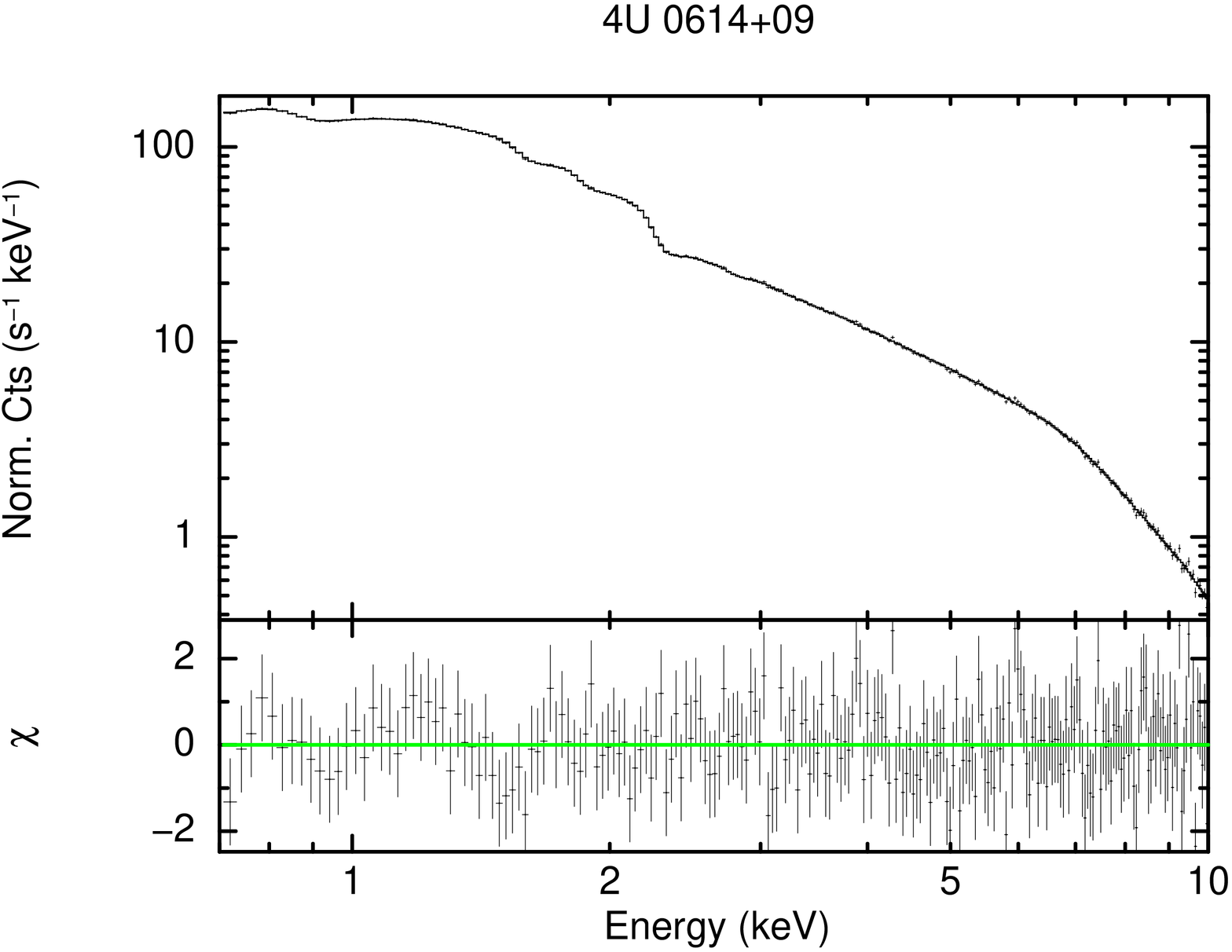}
\includegraphics[angle=-90,width=0.33\textwidth]{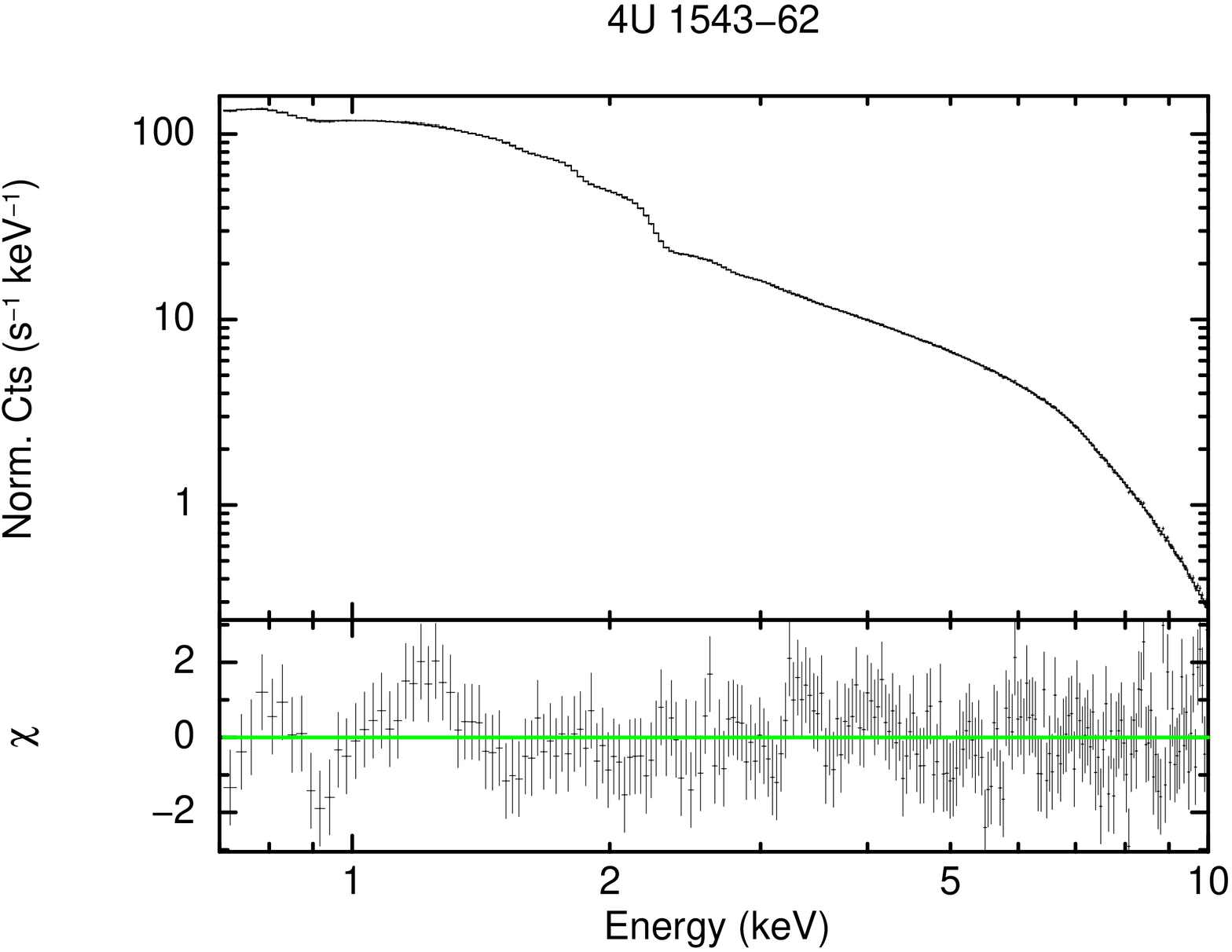}
\includegraphics[angle=-90,width=0.33\textwidth]{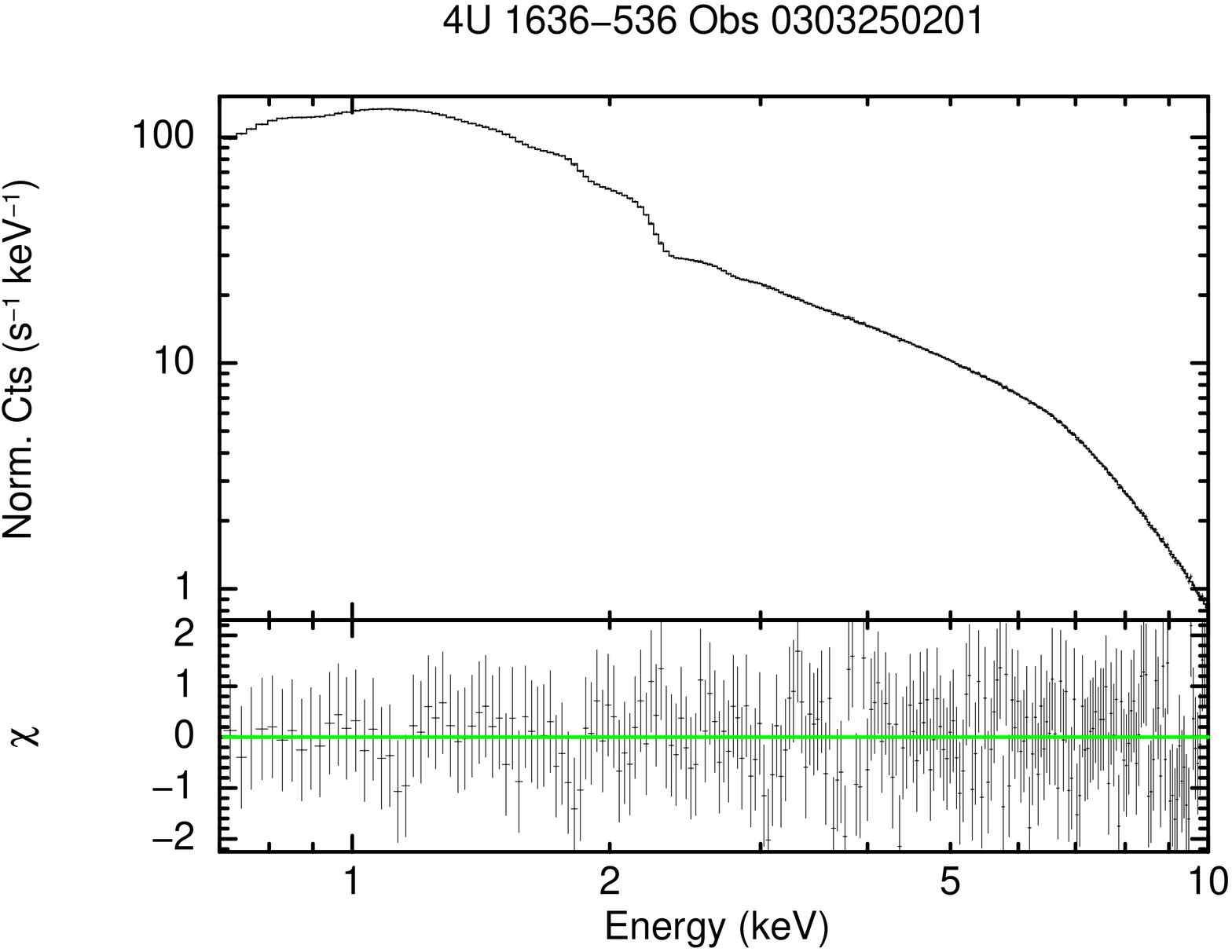}
\includegraphics[angle=-90,width=0.33\textwidth]{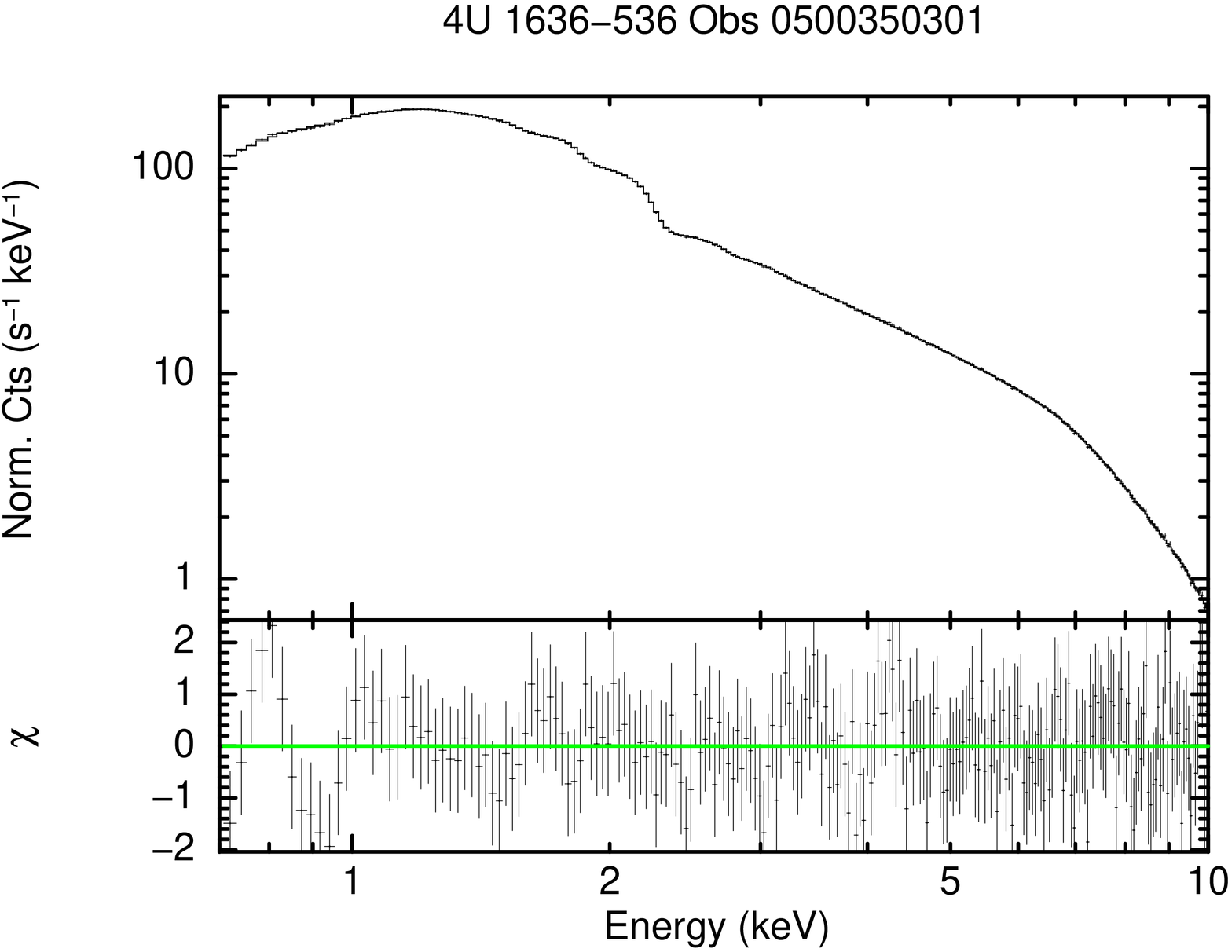}
\includegraphics[angle=-90,width=0.33\textwidth]{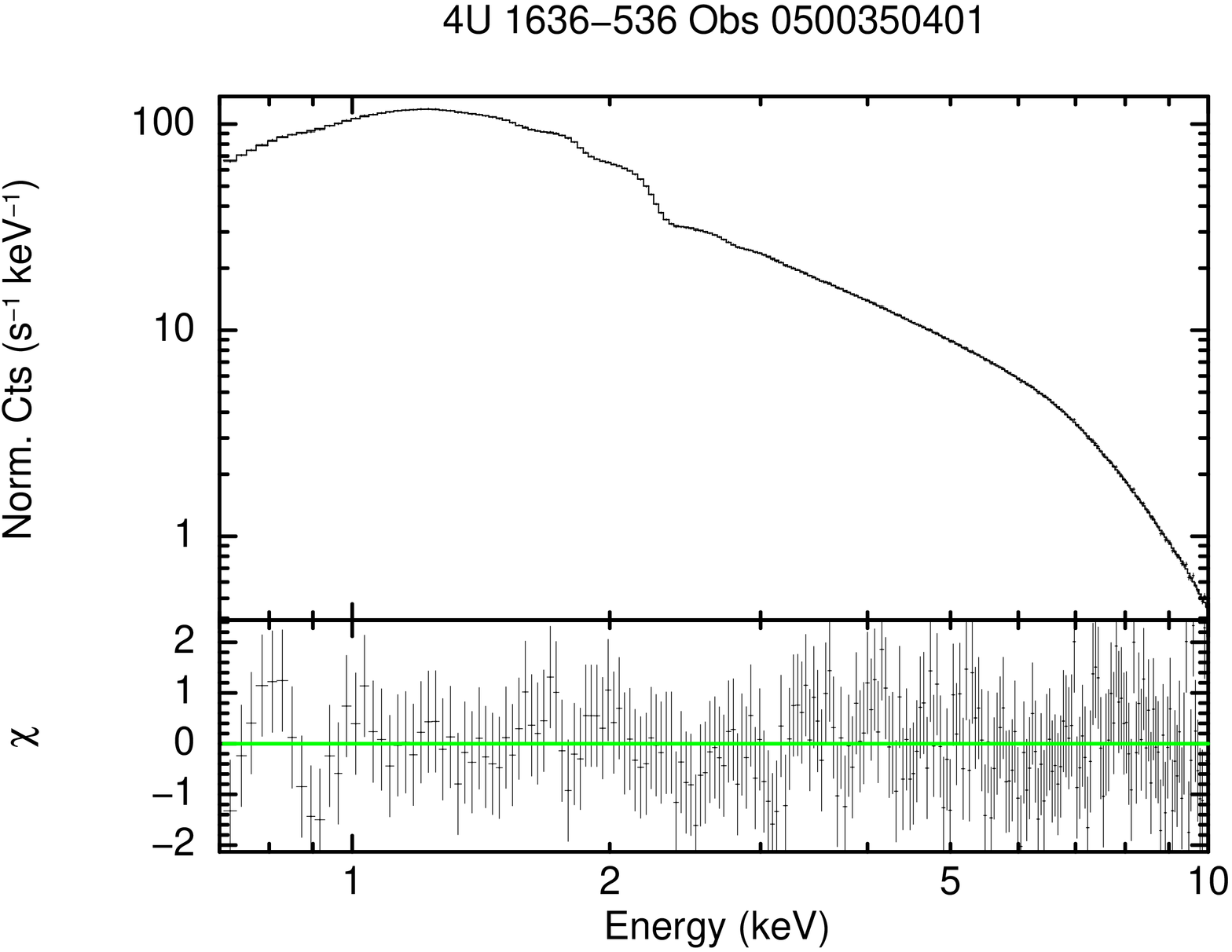}
\includegraphics[angle=-90,width=0.33\textwidth]{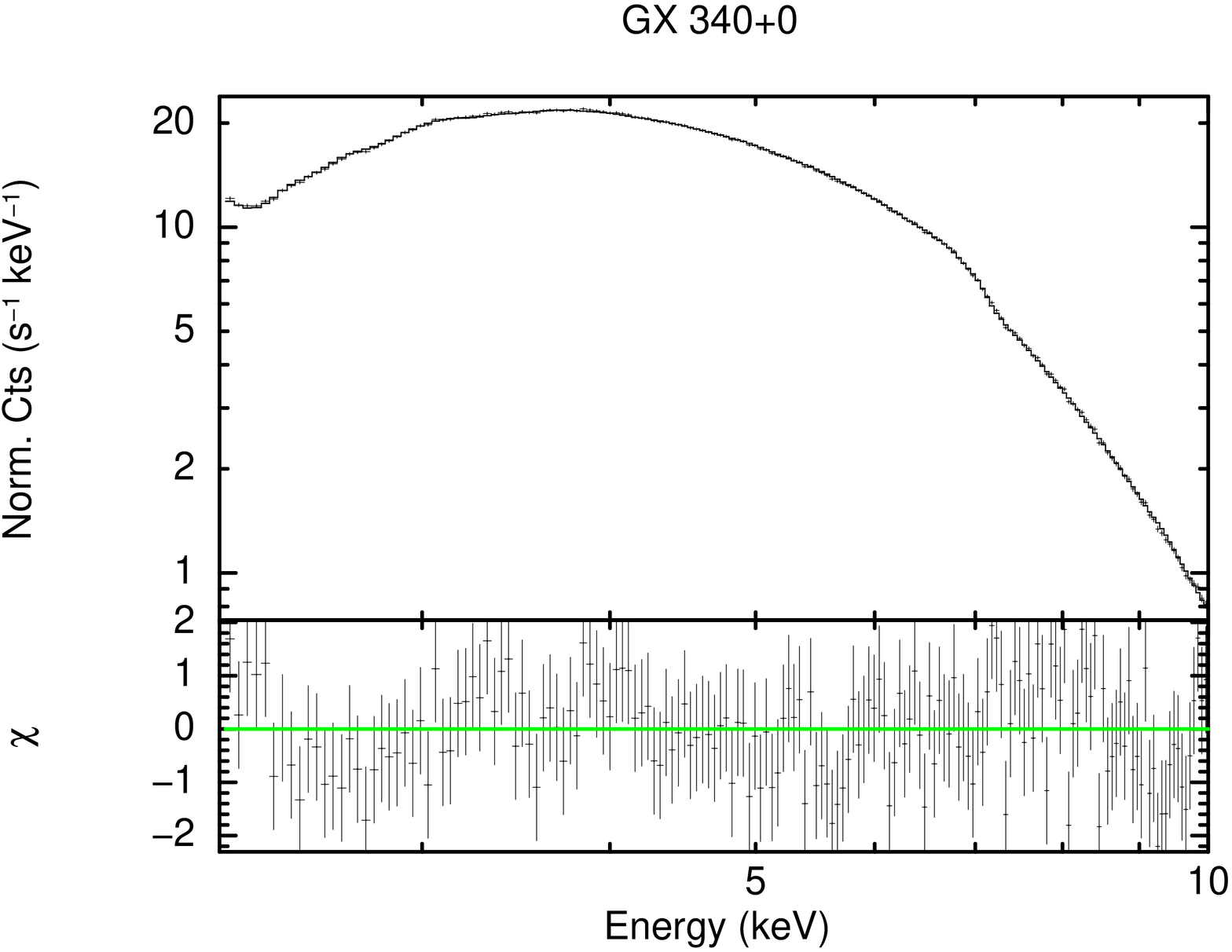}
\includegraphics[angle=-90,width=0.33\textwidth]{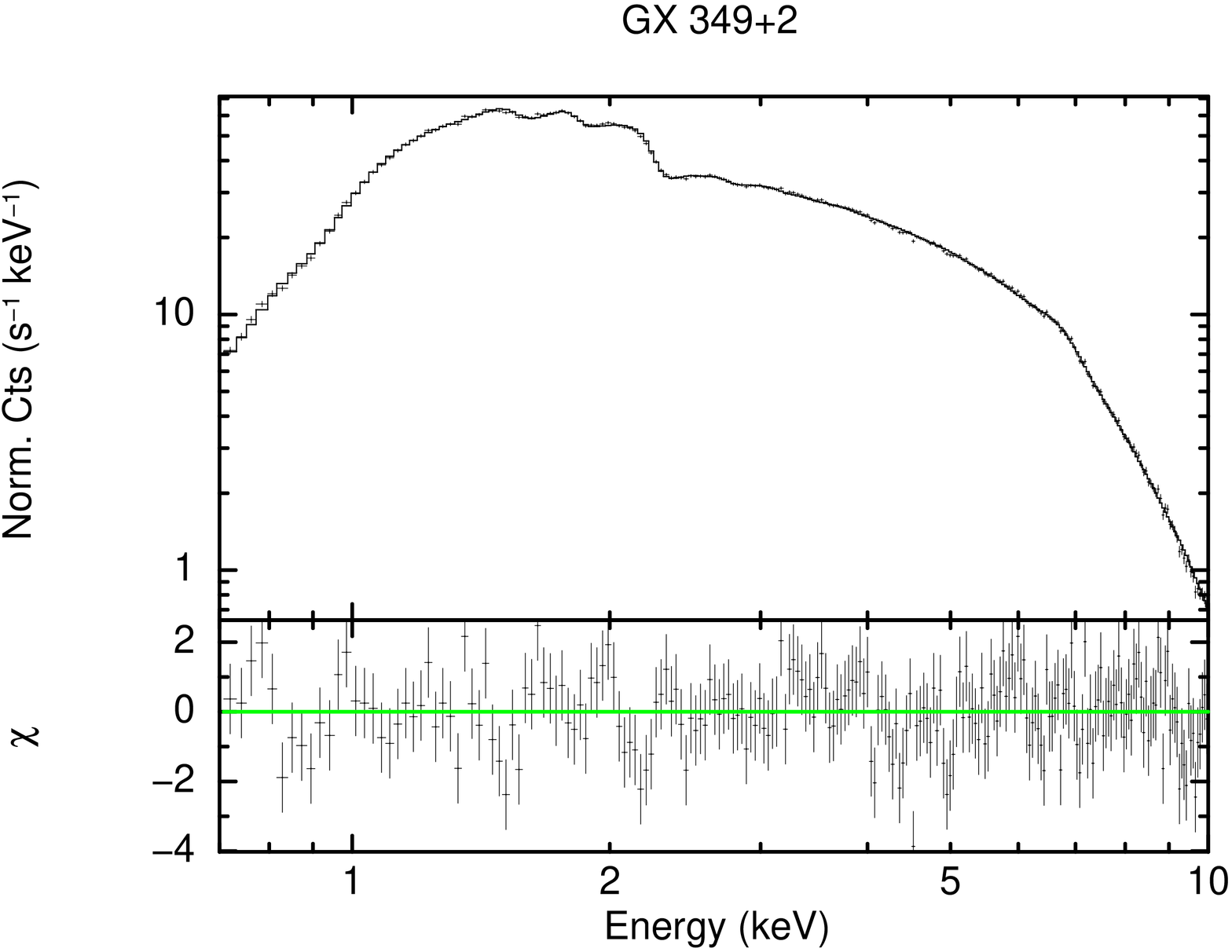}
\includegraphics[angle=-90,width=0.33\textwidth]{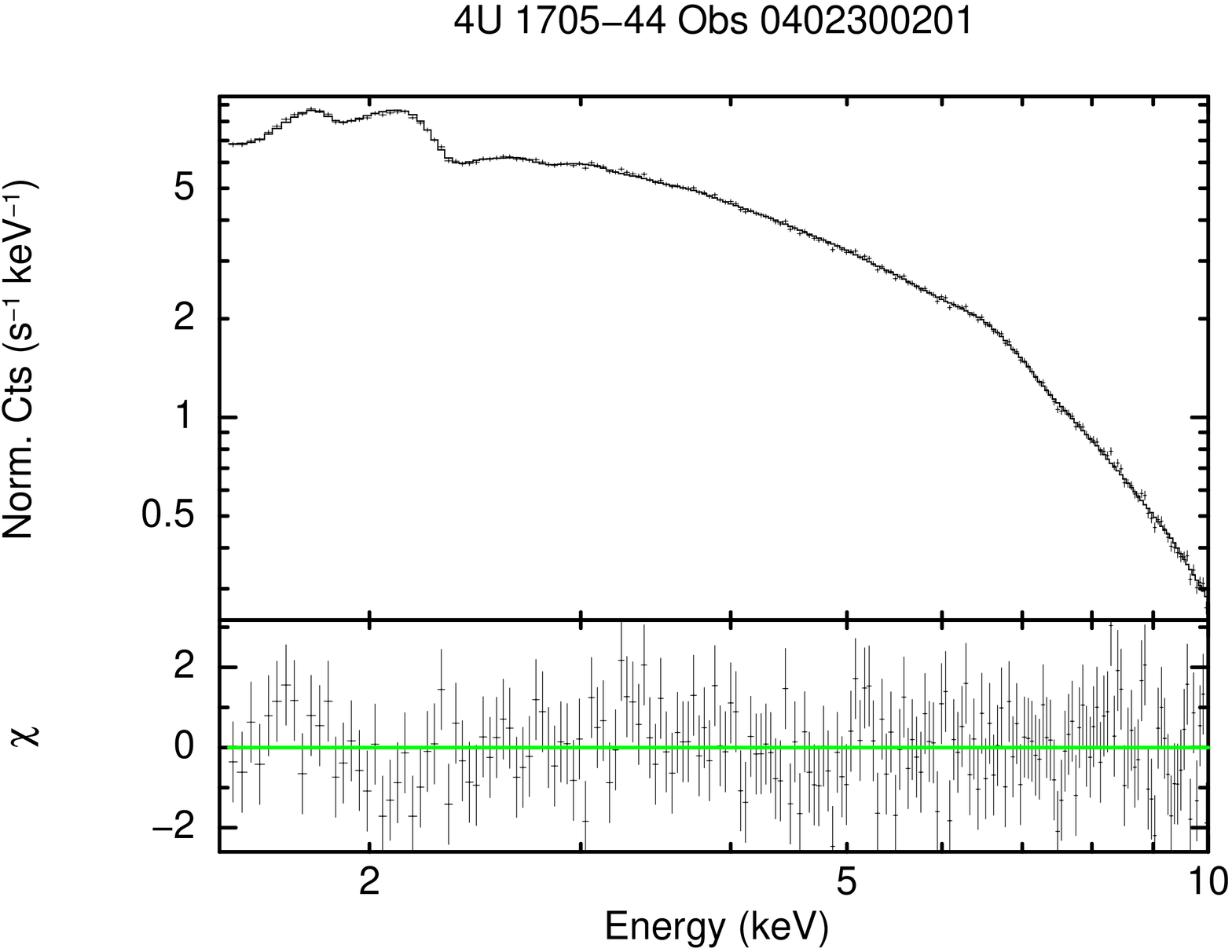}
\includegraphics[angle=-90,width=0.33\textwidth]{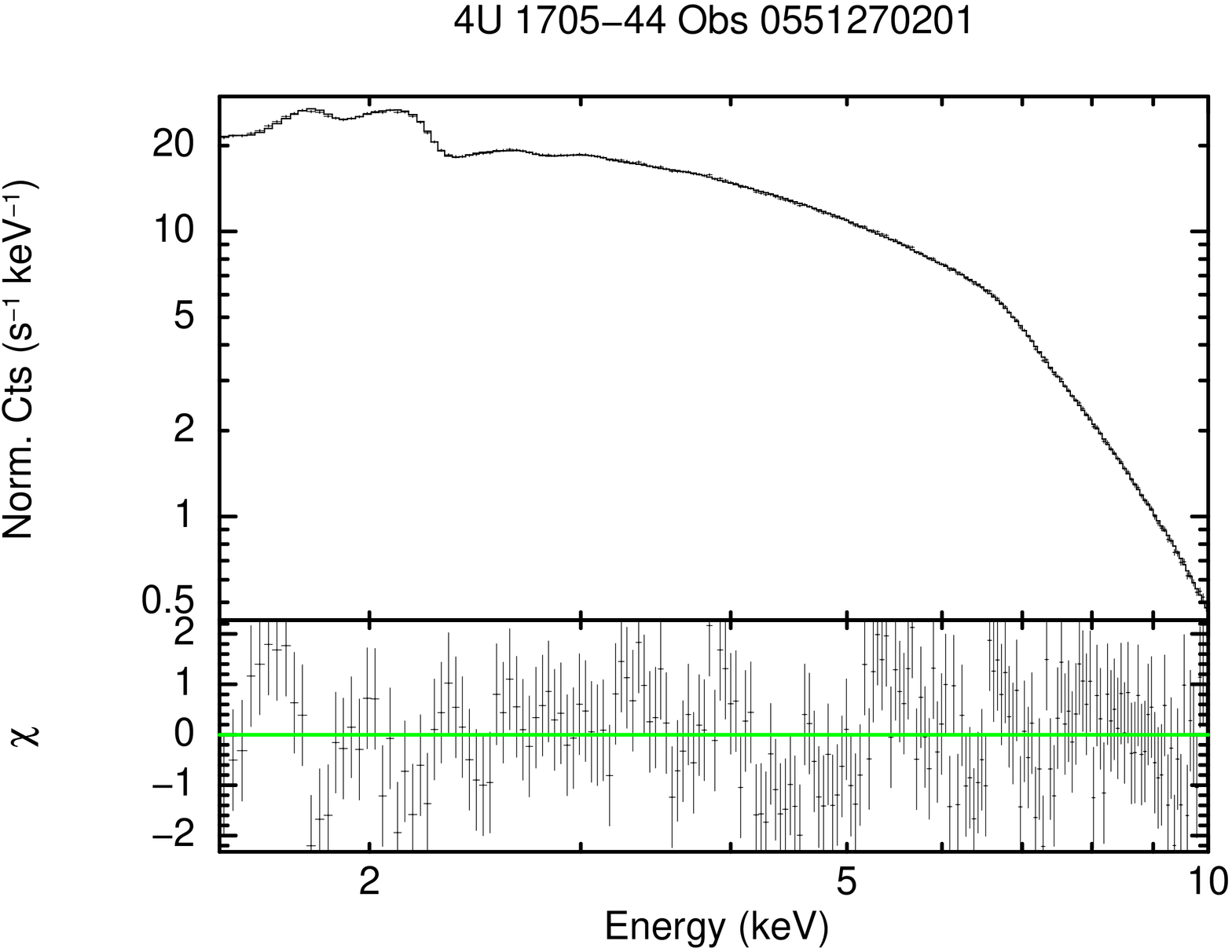}
\includegraphics[angle=-90,width=0.33\textwidth]{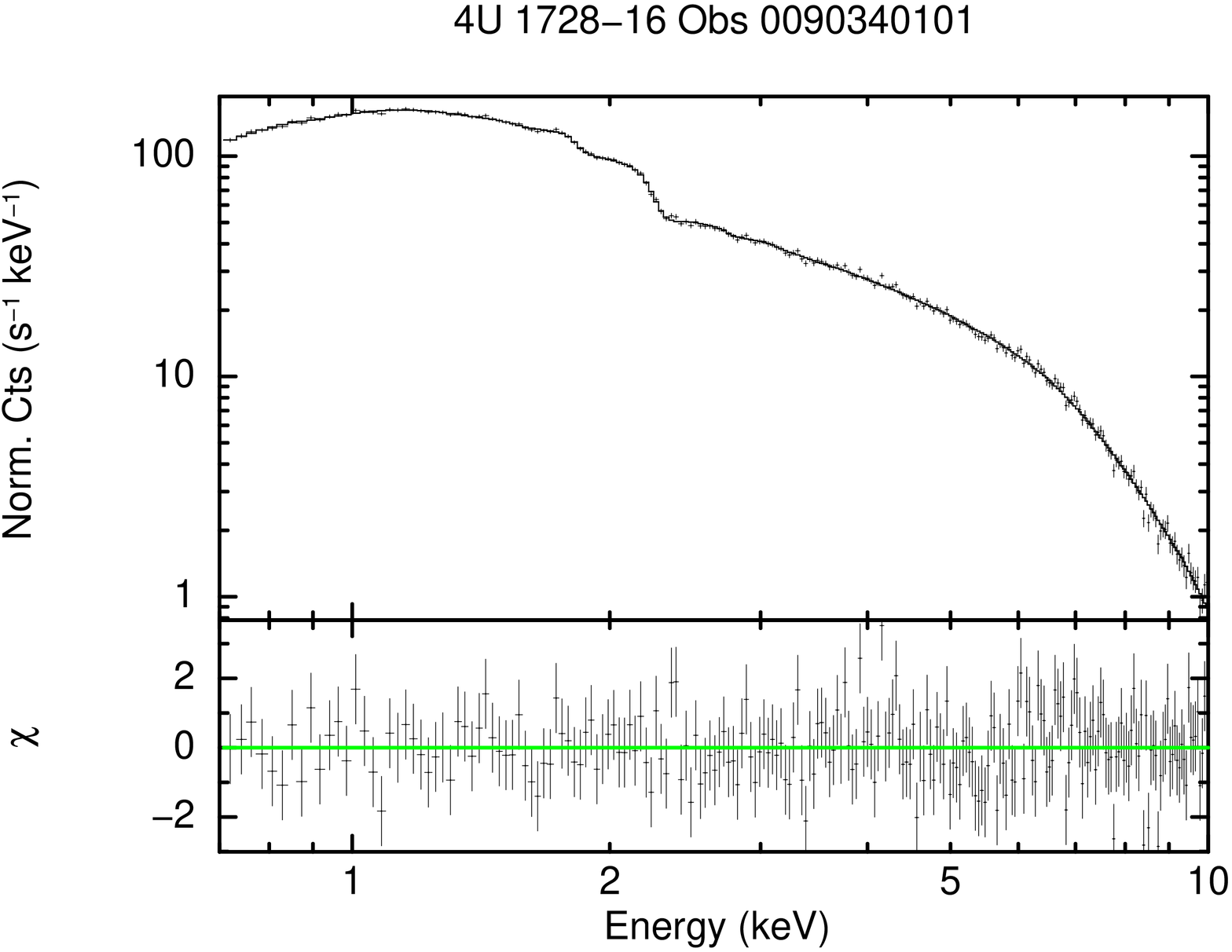}
\includegraphics[angle=-90,width=0.33\textwidth]{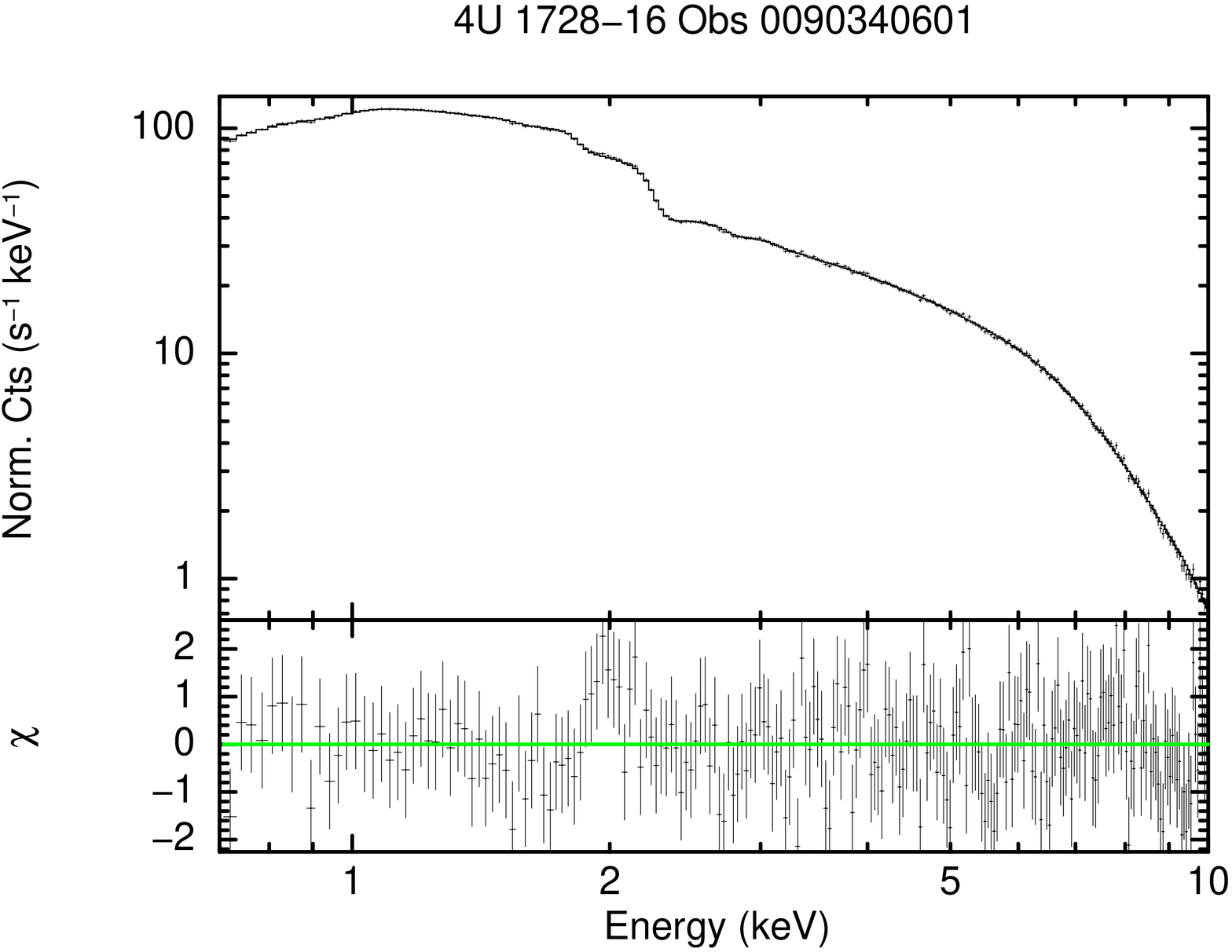}
\includegraphics[angle=-90,width=0.33\textwidth]{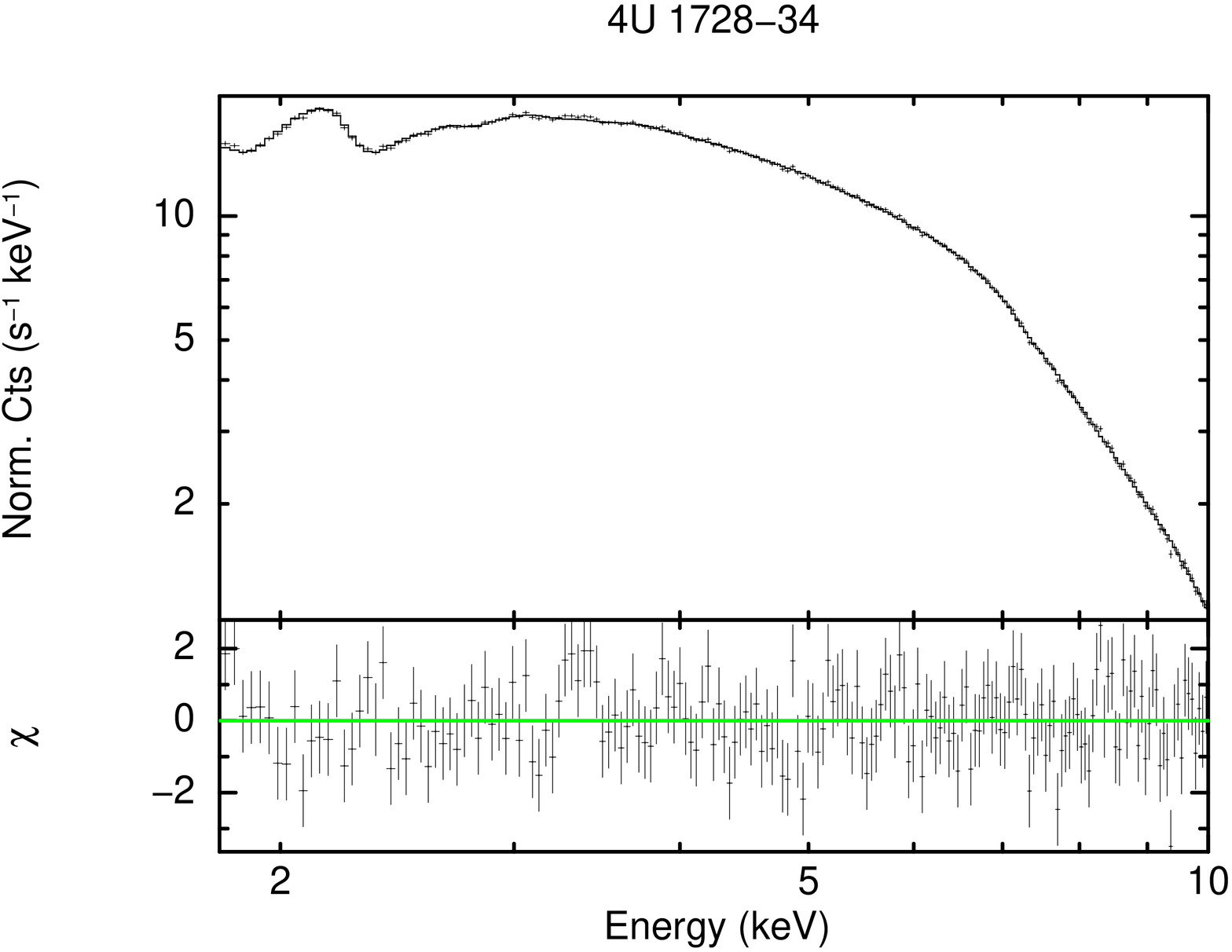}
\includegraphics[angle=-90,width=0.33\textwidth]{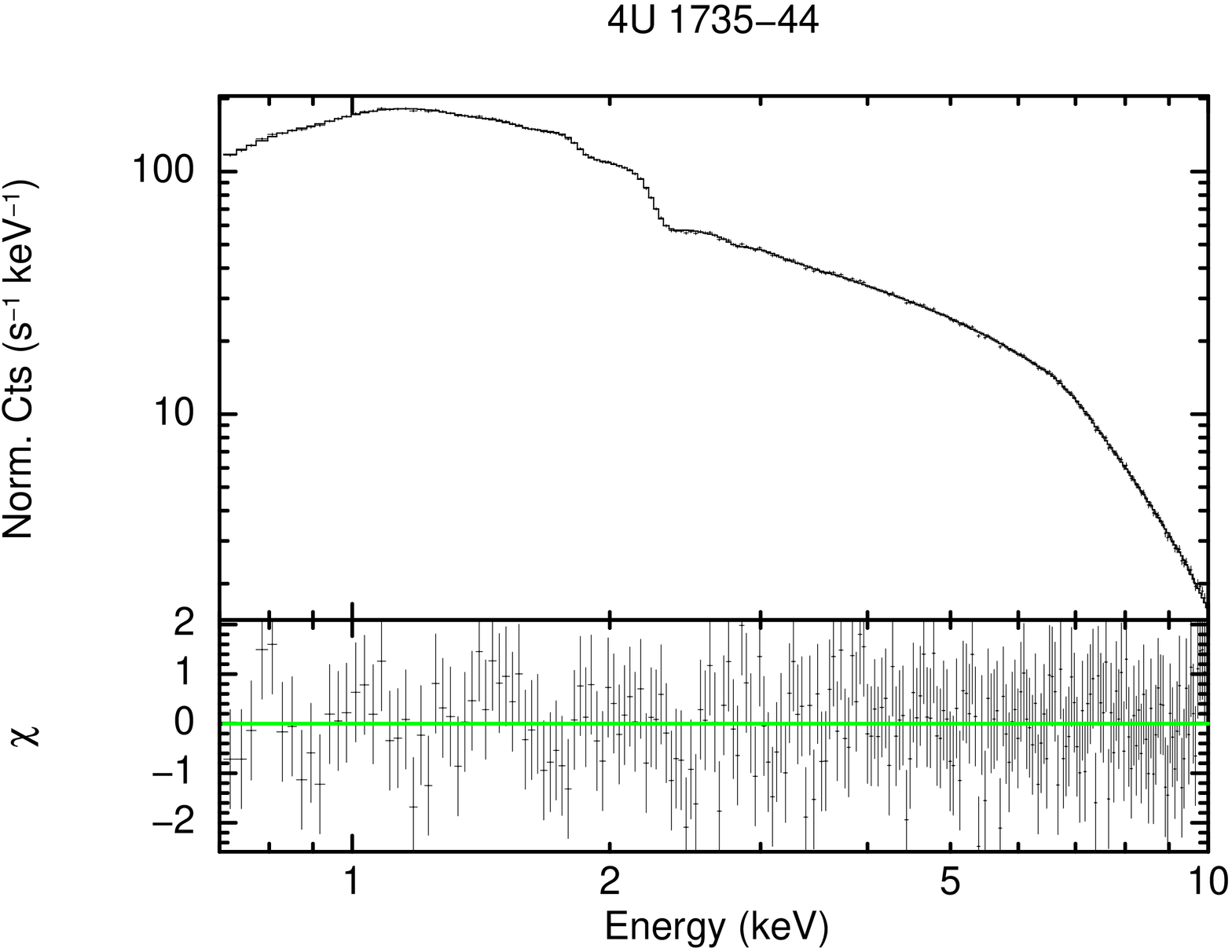}
\includegraphics[angle=-90,width=0.33\textwidth]{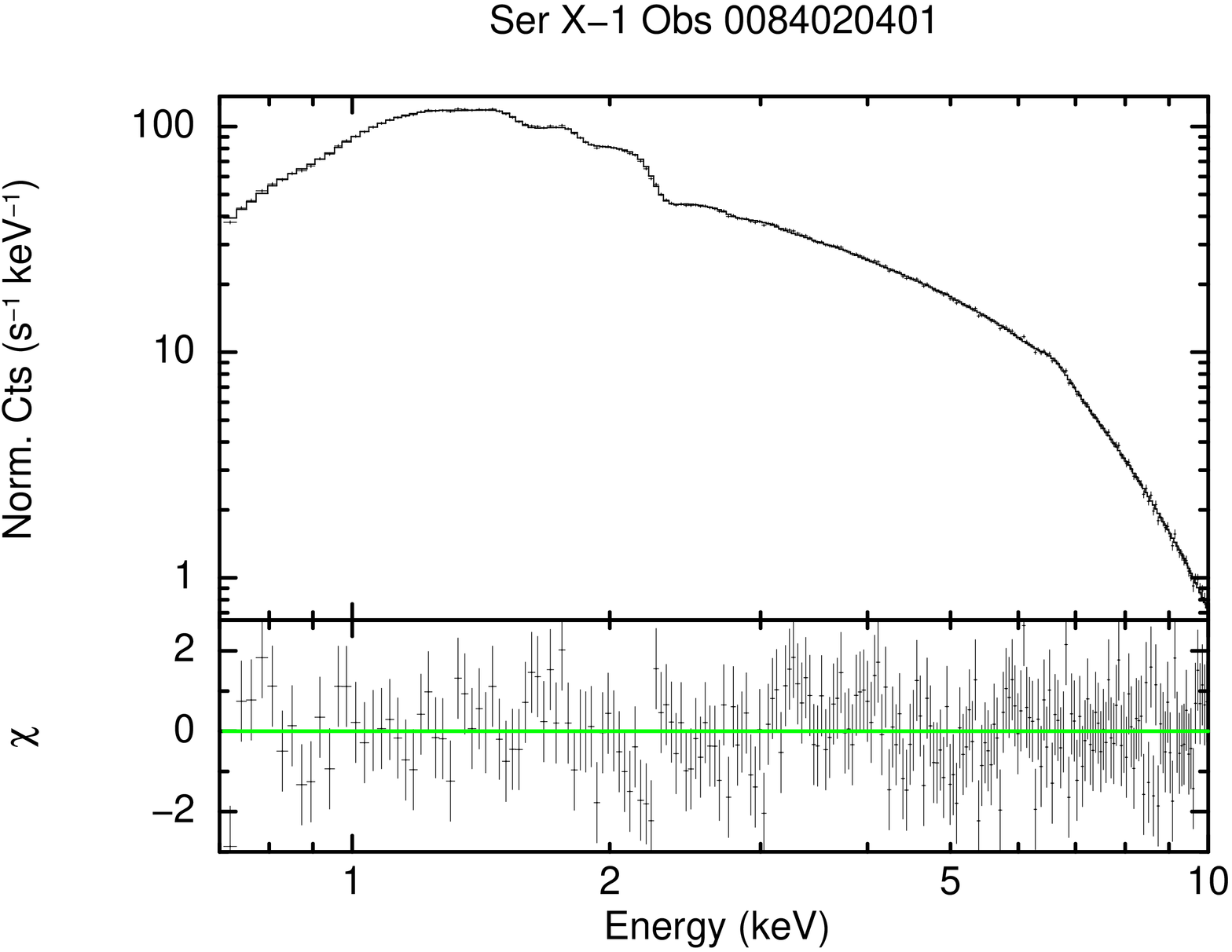}
\includegraphics[angle=-90,width=0.33\textwidth]{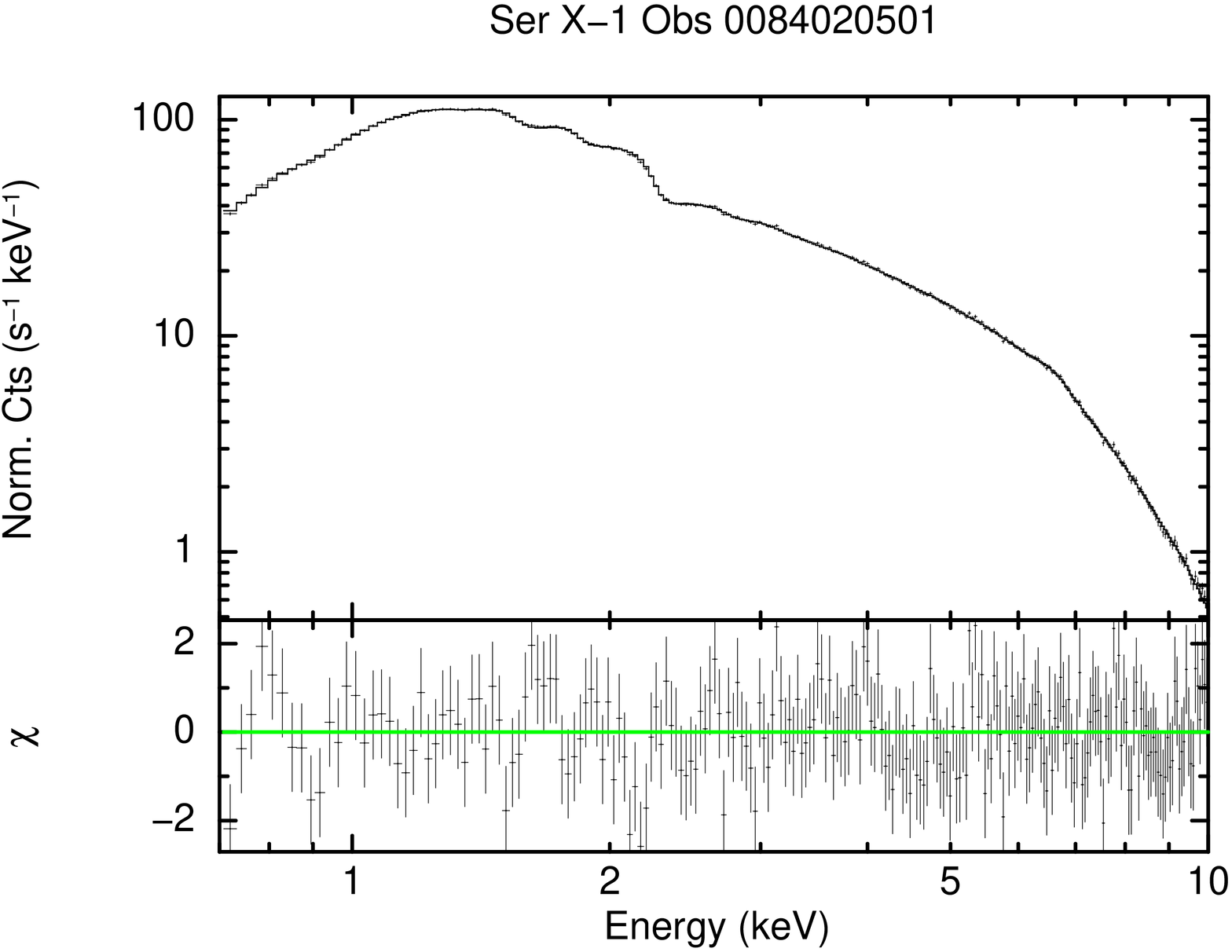}
\caption{{\it Upper panels:} 0.7--10 keV EPIC pn (black) spectra fit with 
Models~1a--1d. For each source the best-fit 
is shown (see Tables~\ref{tab:bestfit}-\ref{tab:bestfit2} and text). 
{\it Lower panels:} Residuals in units of standard
deviation from the corresponding model.
}
\label{fig:spectra}
\end{figure*}

\begin{figure*}[!ht]
\includegraphics[angle=-90,width=0.33\textwidth]{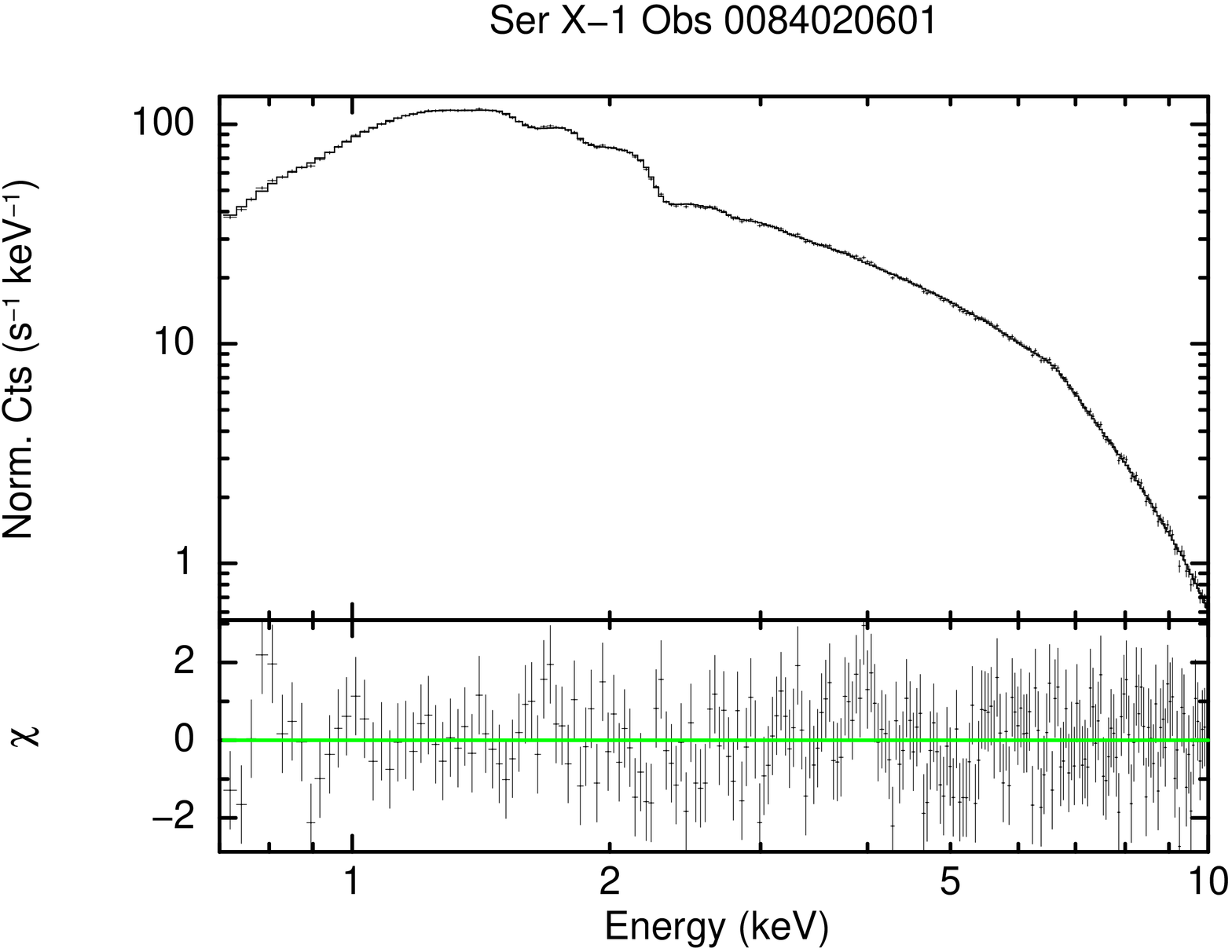}
\includegraphics[angle=-90,width=0.33\textwidth]{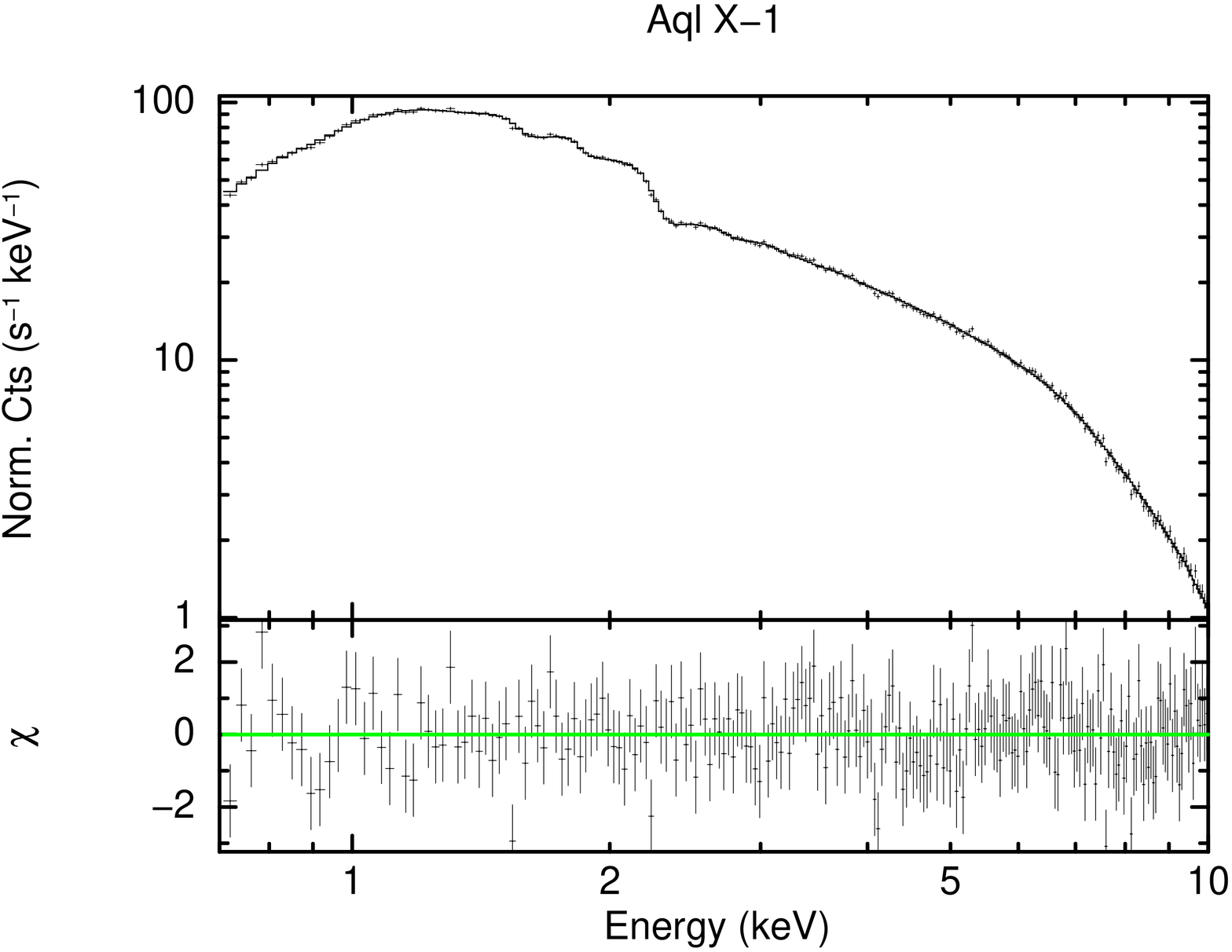}
\includegraphics[angle=-90,width=0.33\textwidth]{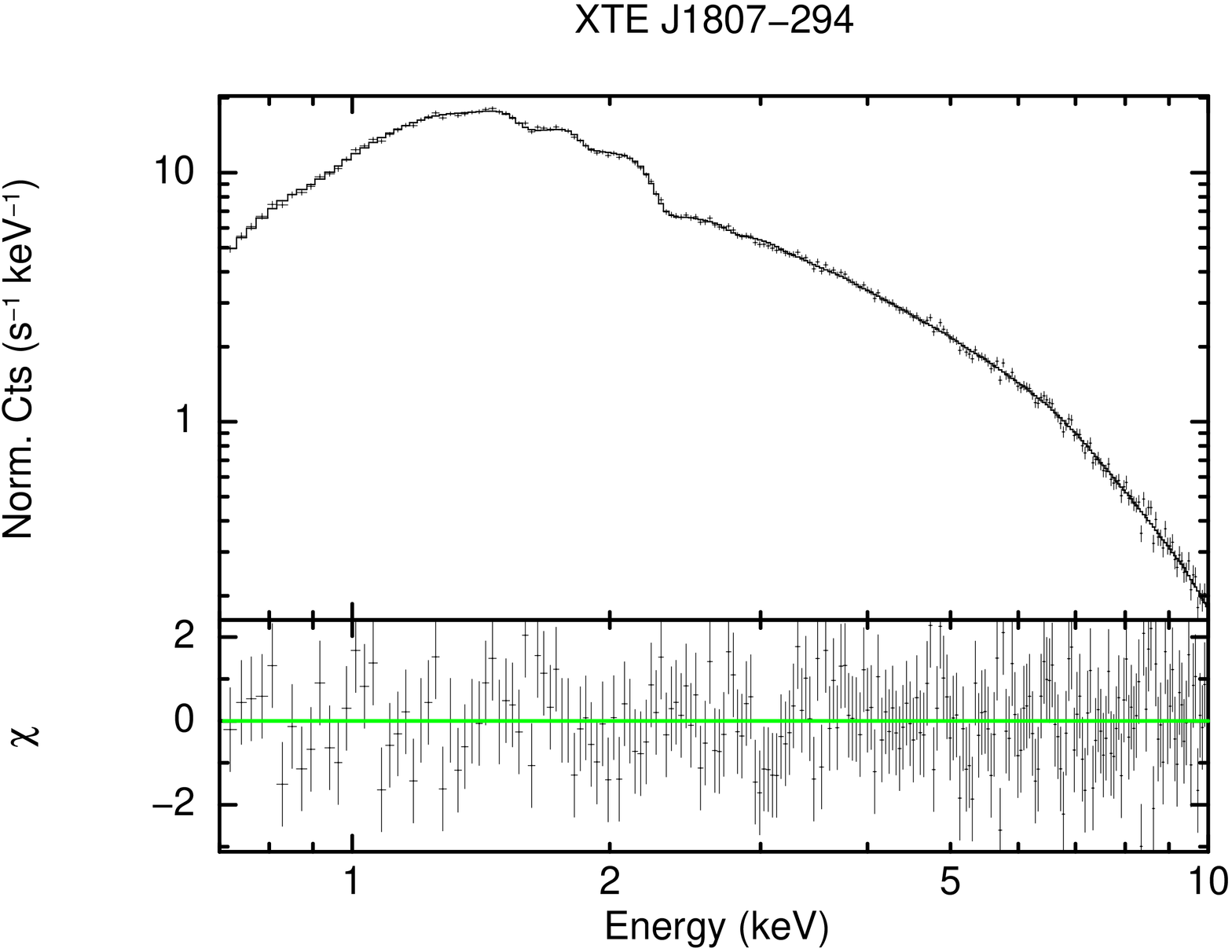}
\vspace{0.1cm}
\includegraphics[angle=-90,width=0.33\textwidth]{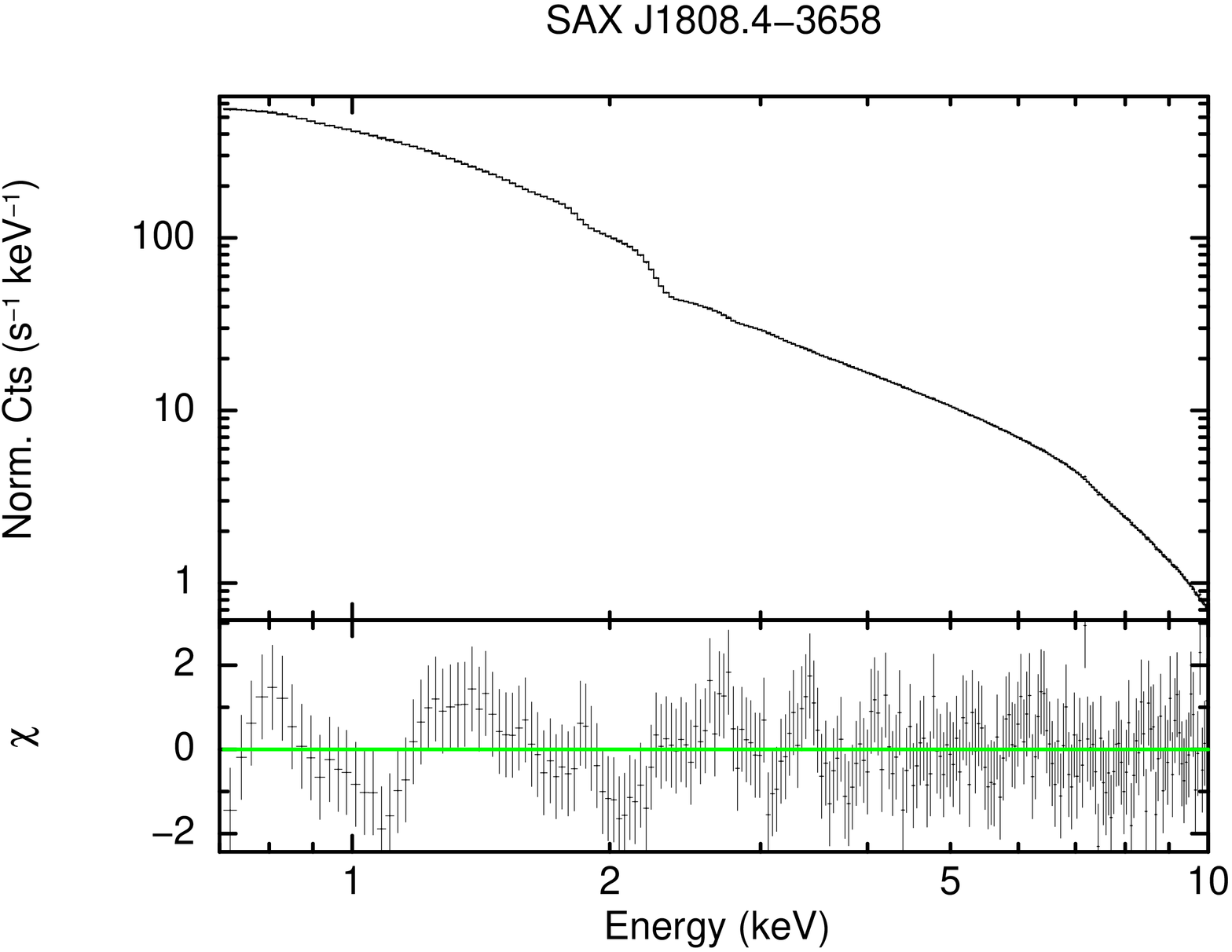}
\caption{Continued from Fig.~\ref{fig:spectra}}
\label{fig:spectra2}
\end{figure*}

\begin{figure*}[!ht]
\includegraphics[angle=-90,width=0.33\textwidth]{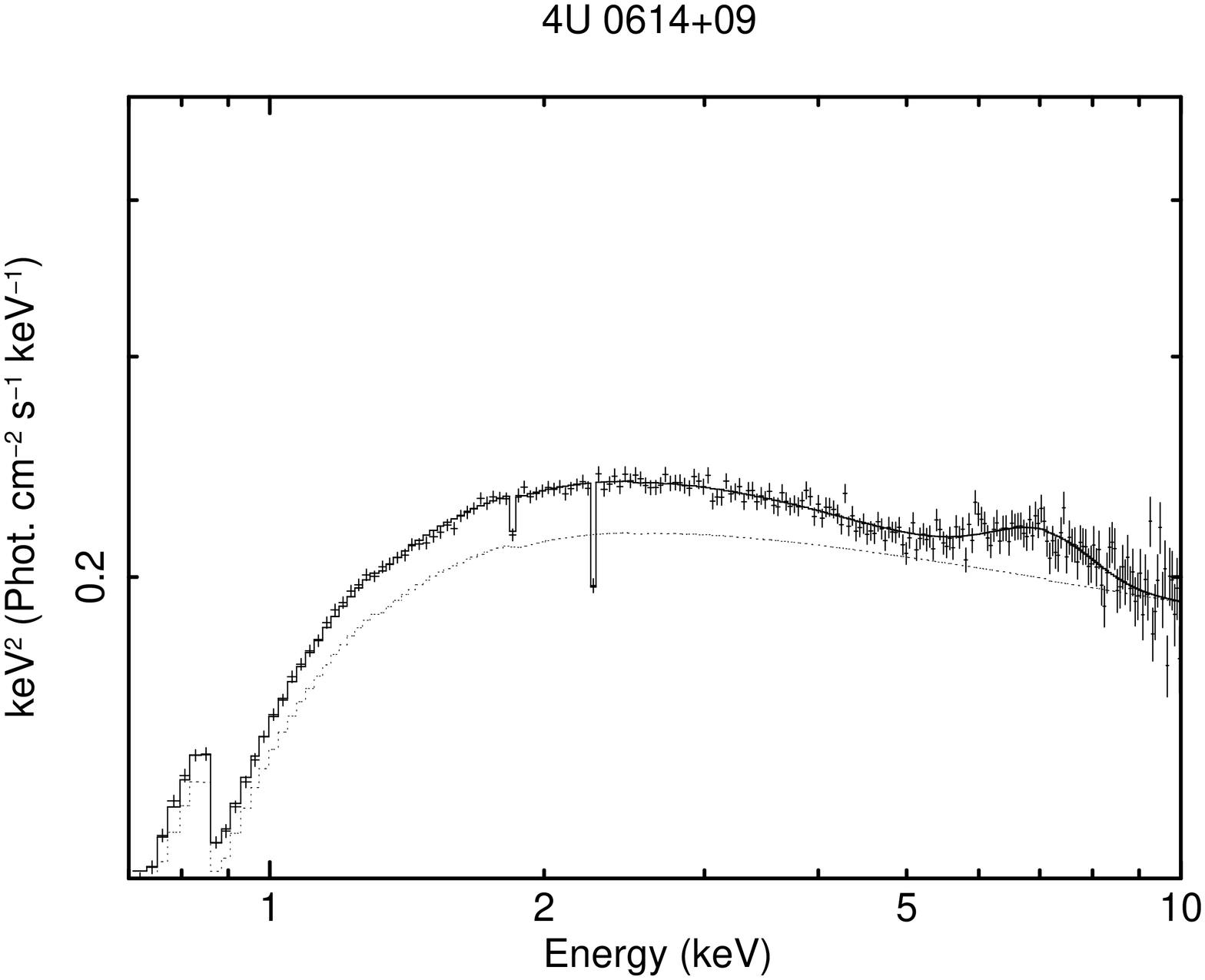}
\includegraphics[angle=-90,width=0.33\textwidth]{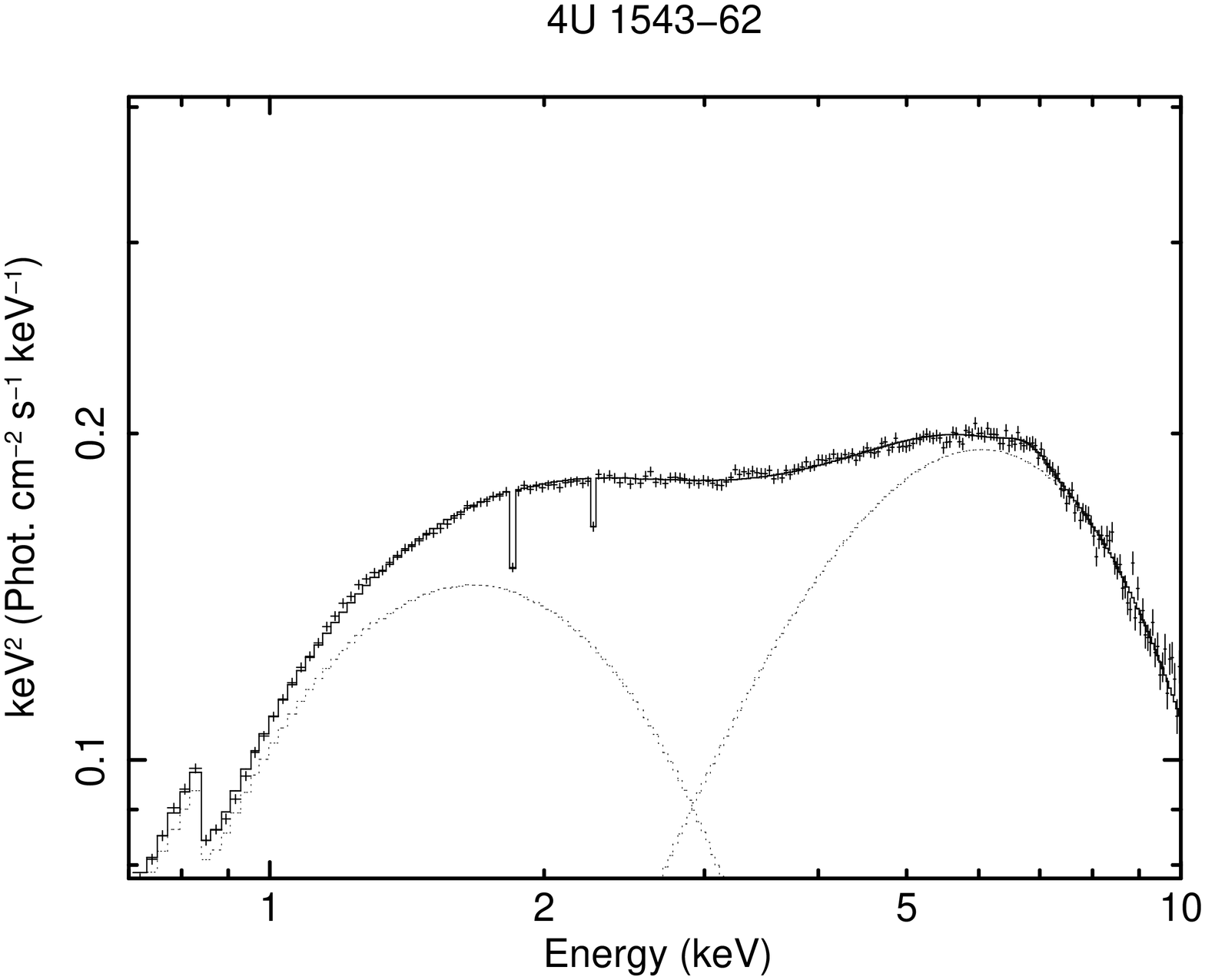}
\includegraphics[angle=-90,width=0.33\textwidth]{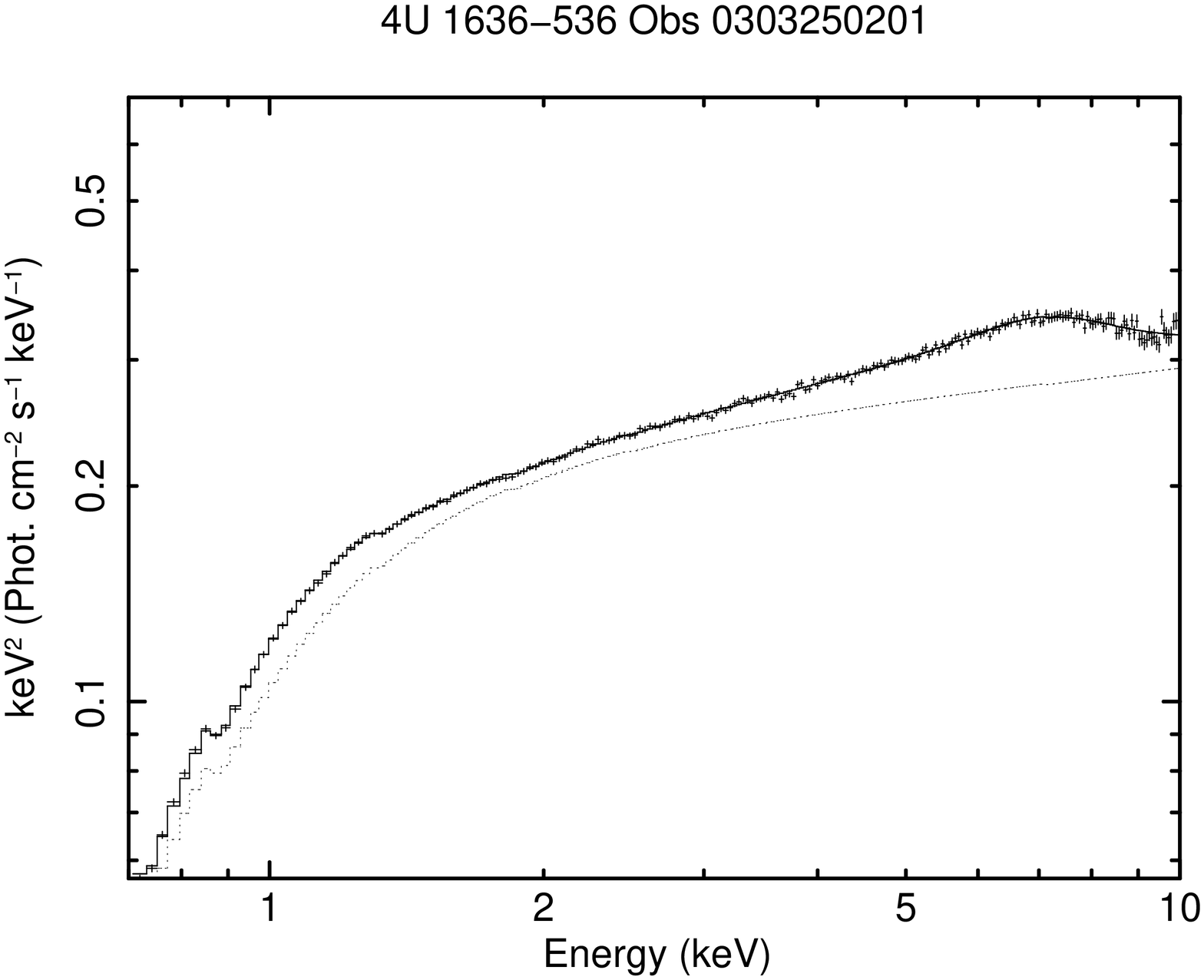}
\includegraphics[angle=-90,width=0.33\textwidth]{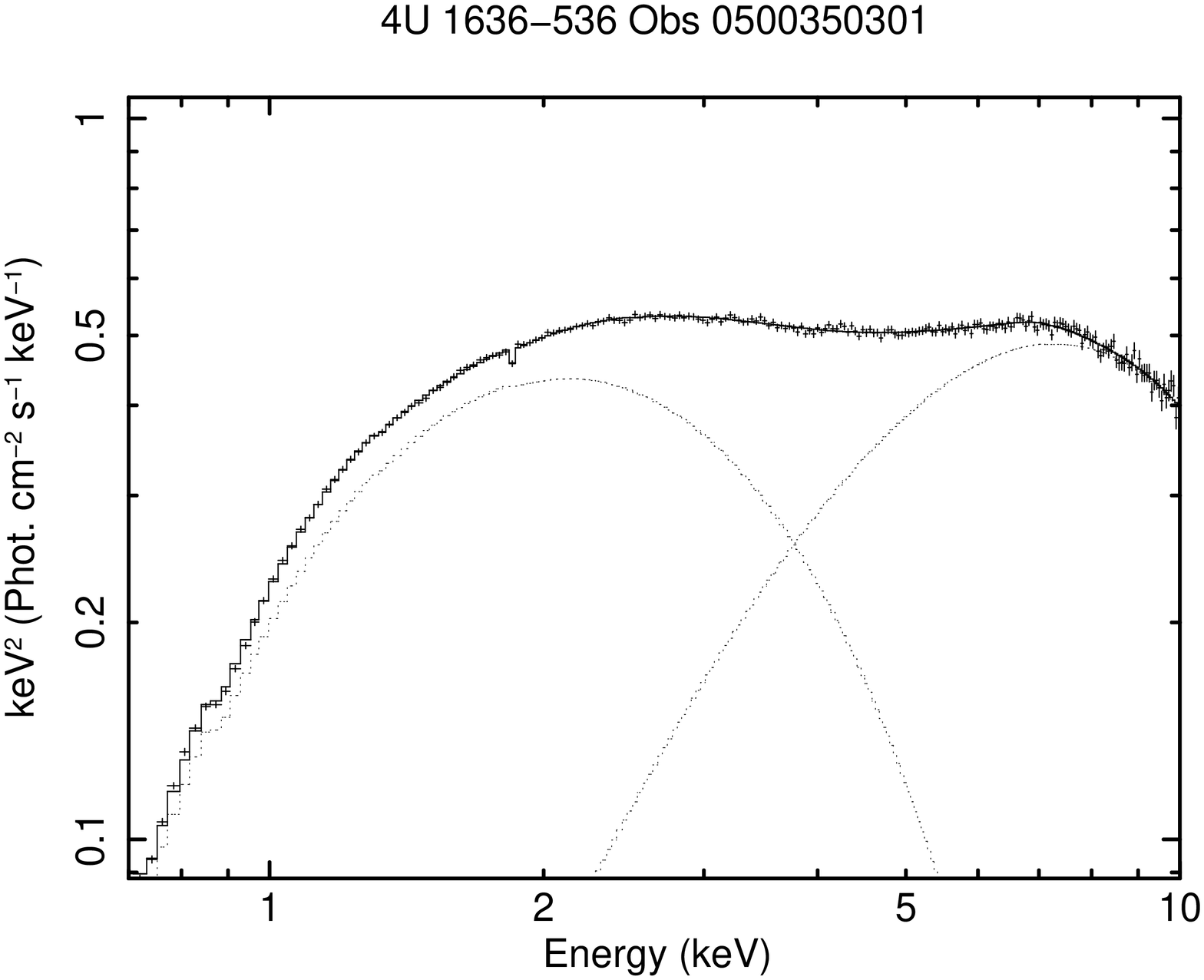}
\includegraphics[angle=-90,width=0.33\textwidth]{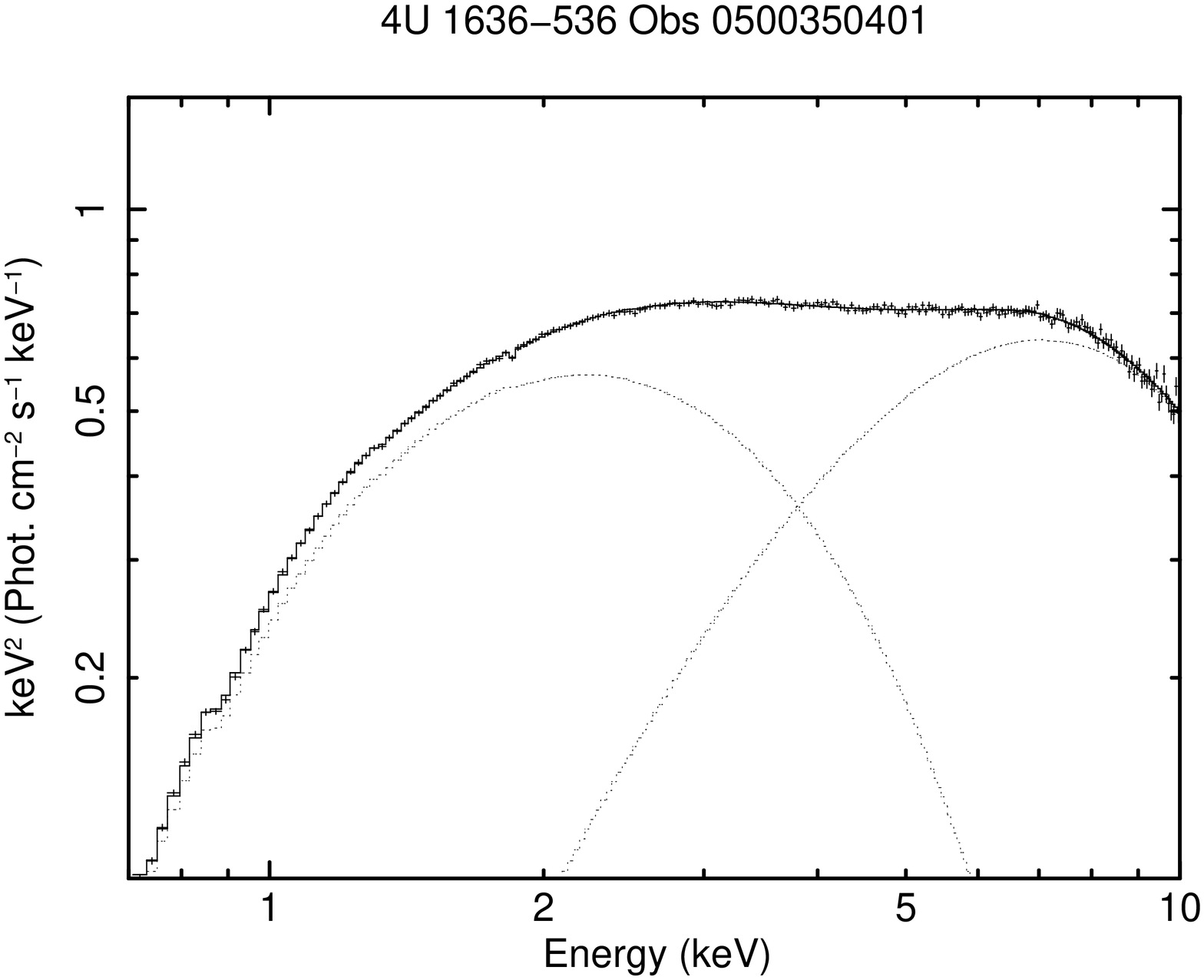}
\includegraphics[angle=-90,width=0.33\textwidth]{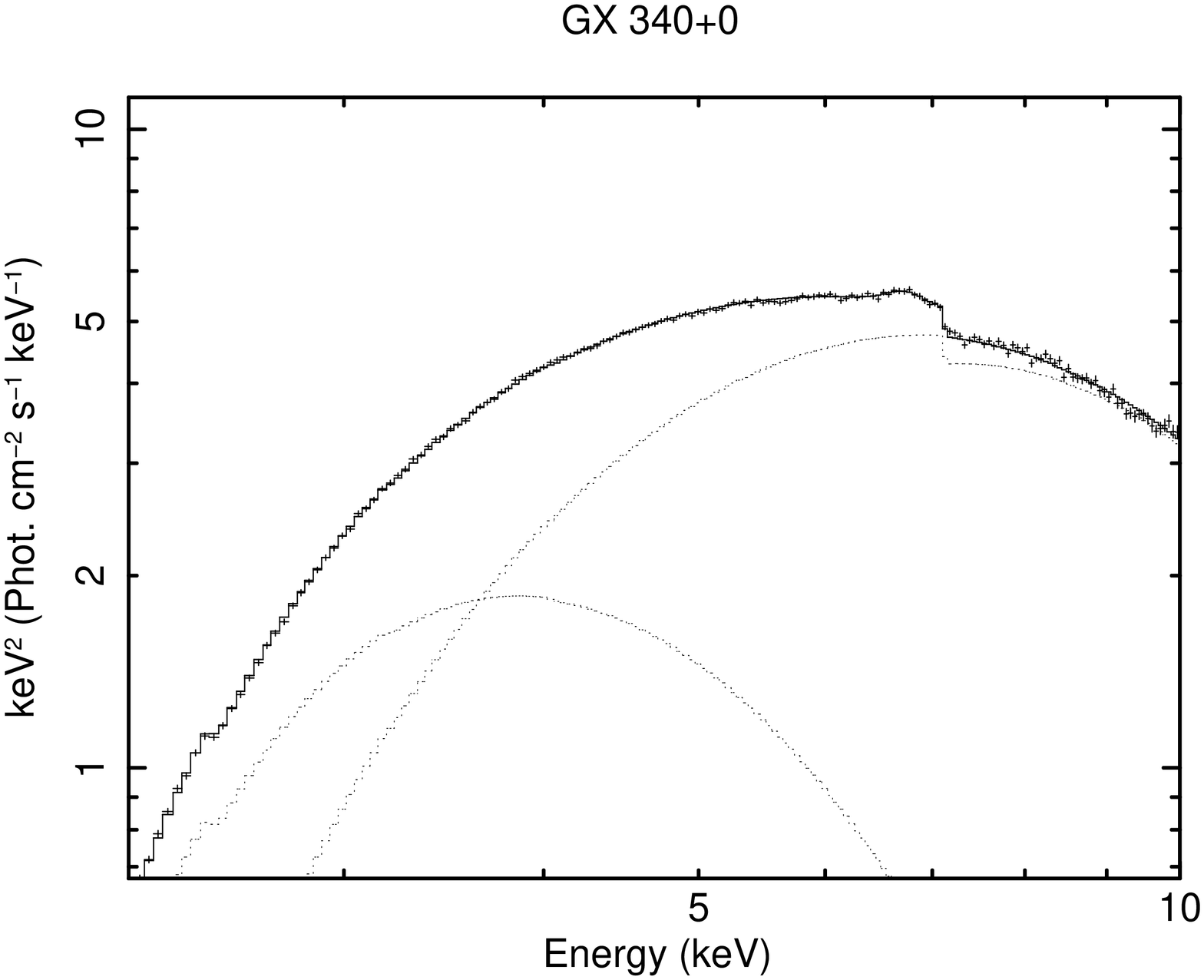}
\includegraphics[angle=-90,width=0.33\textwidth]{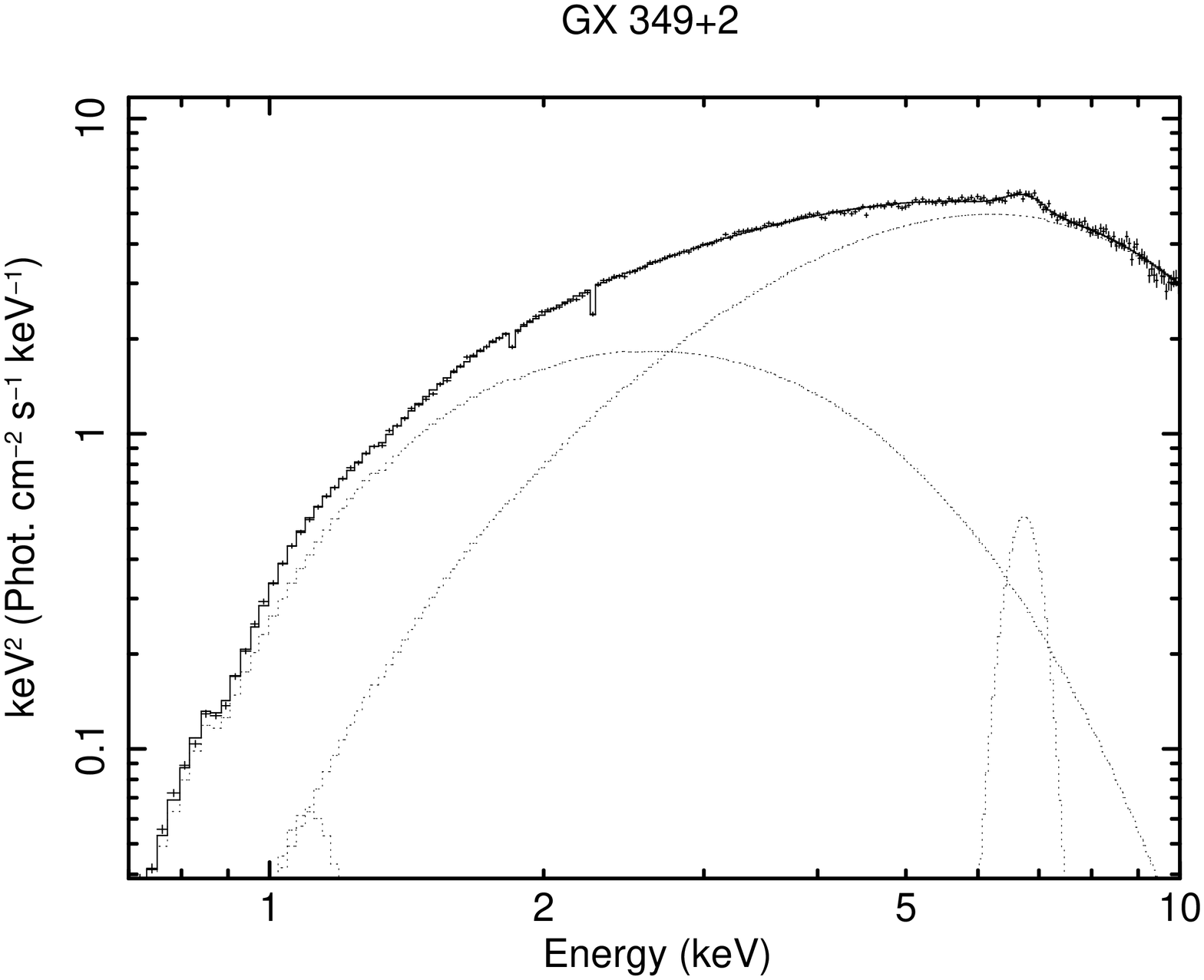}
\includegraphics[angle=-90,width=0.33\textwidth]{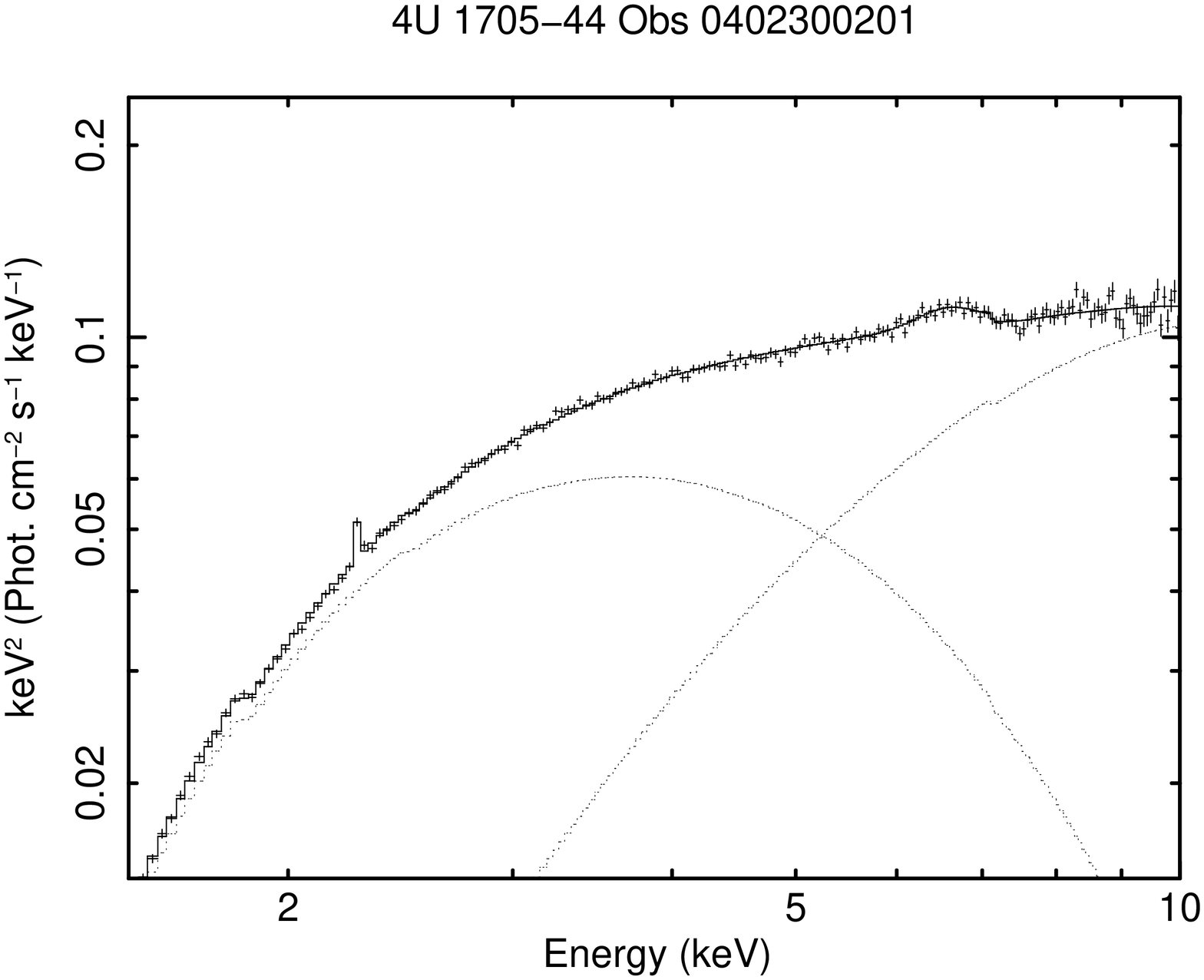}
\includegraphics[angle=-90,width=0.33\textwidth]{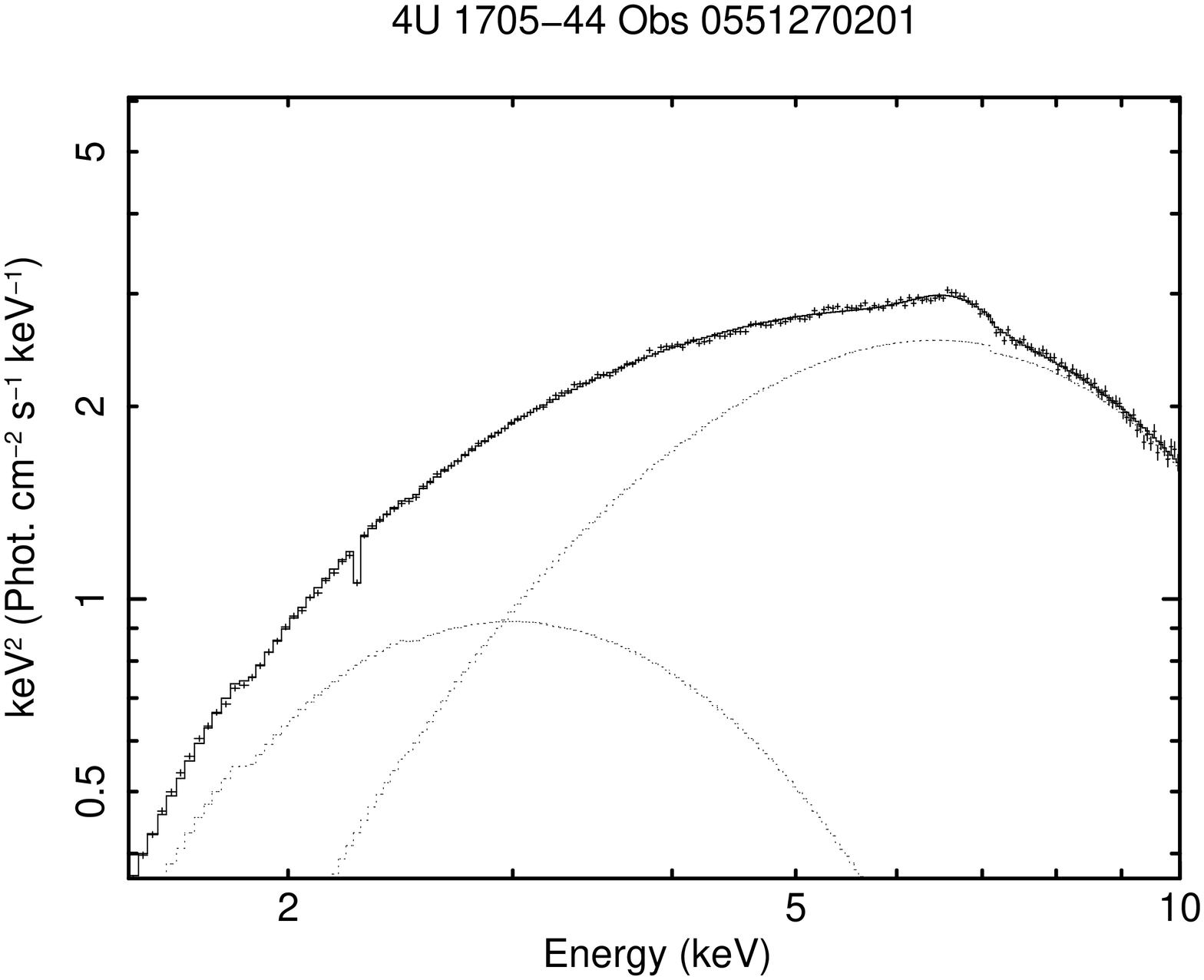}
\includegraphics[angle=-90,width=0.33\textwidth]{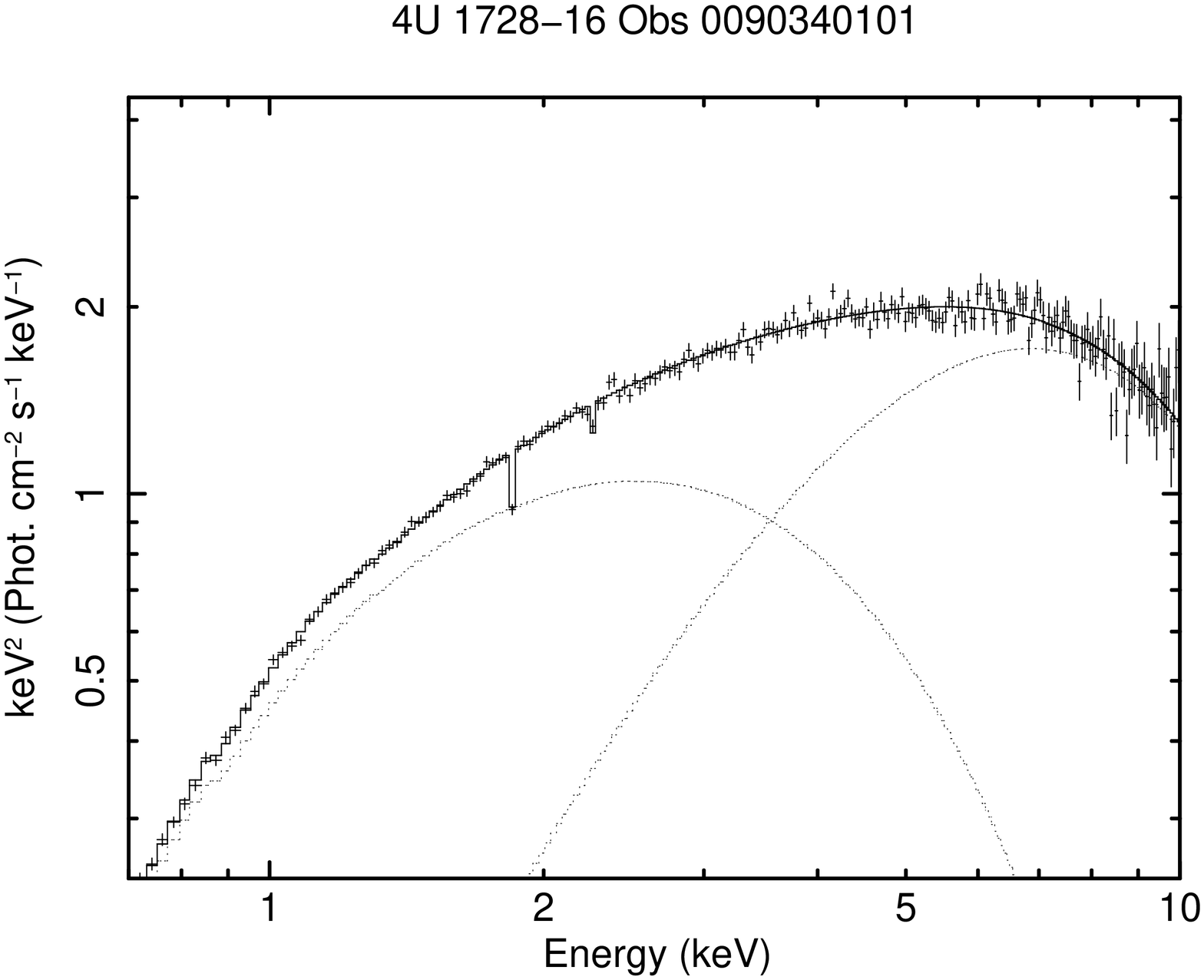}
\includegraphics[angle=-90,width=0.33\textwidth]{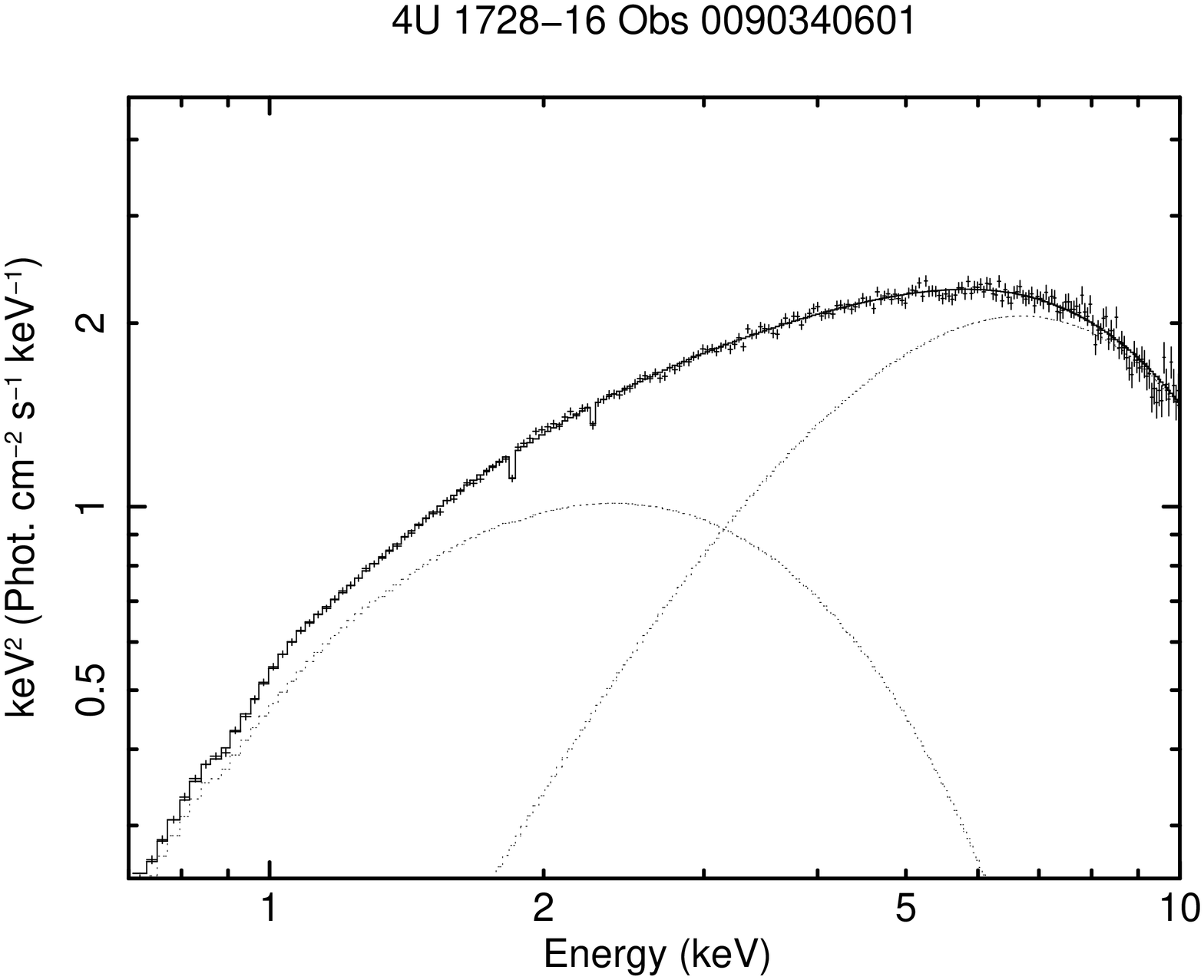}
\includegraphics[angle=-90,width=0.33\textwidth]{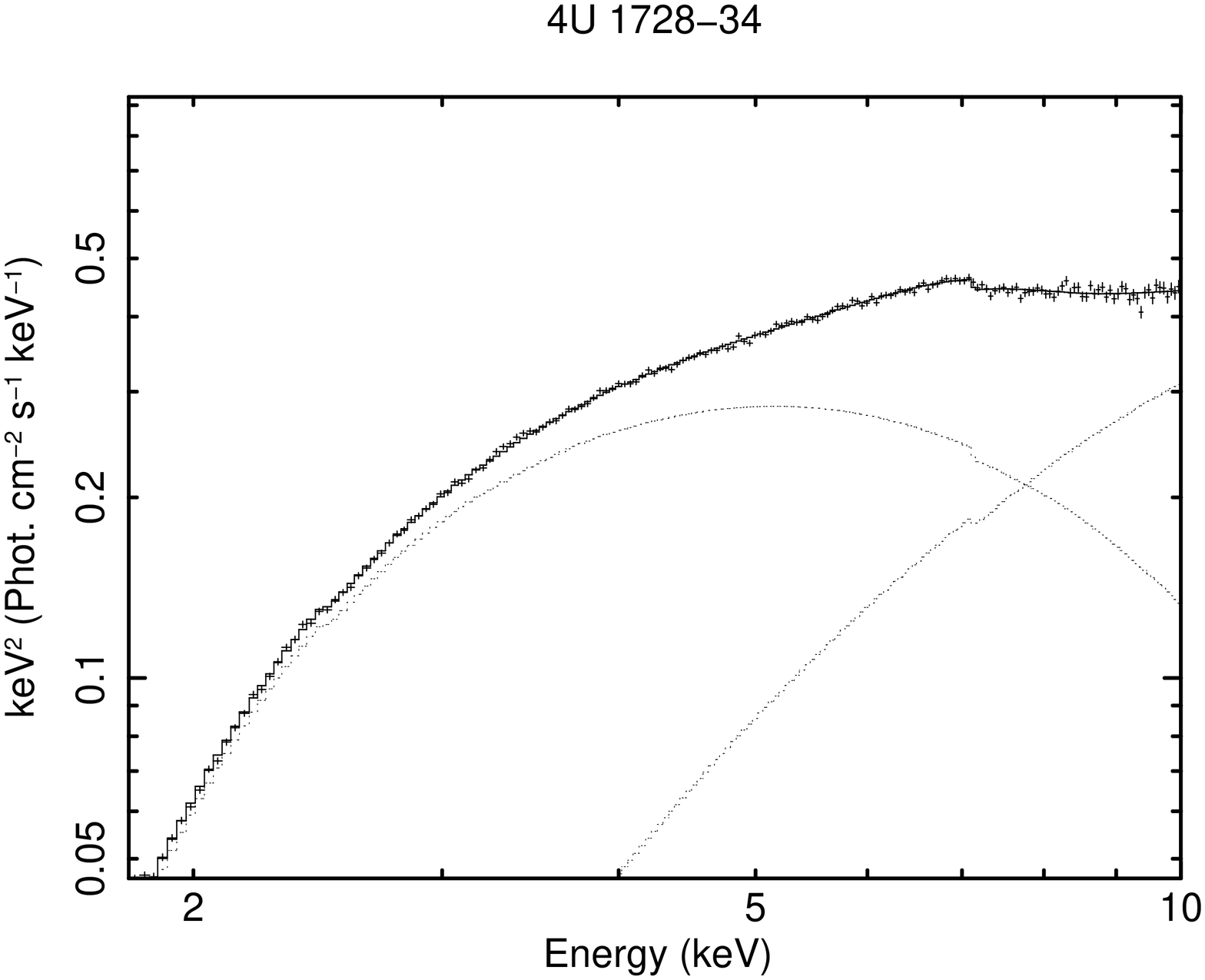}
\includegraphics[angle=-90,width=0.33\textwidth]{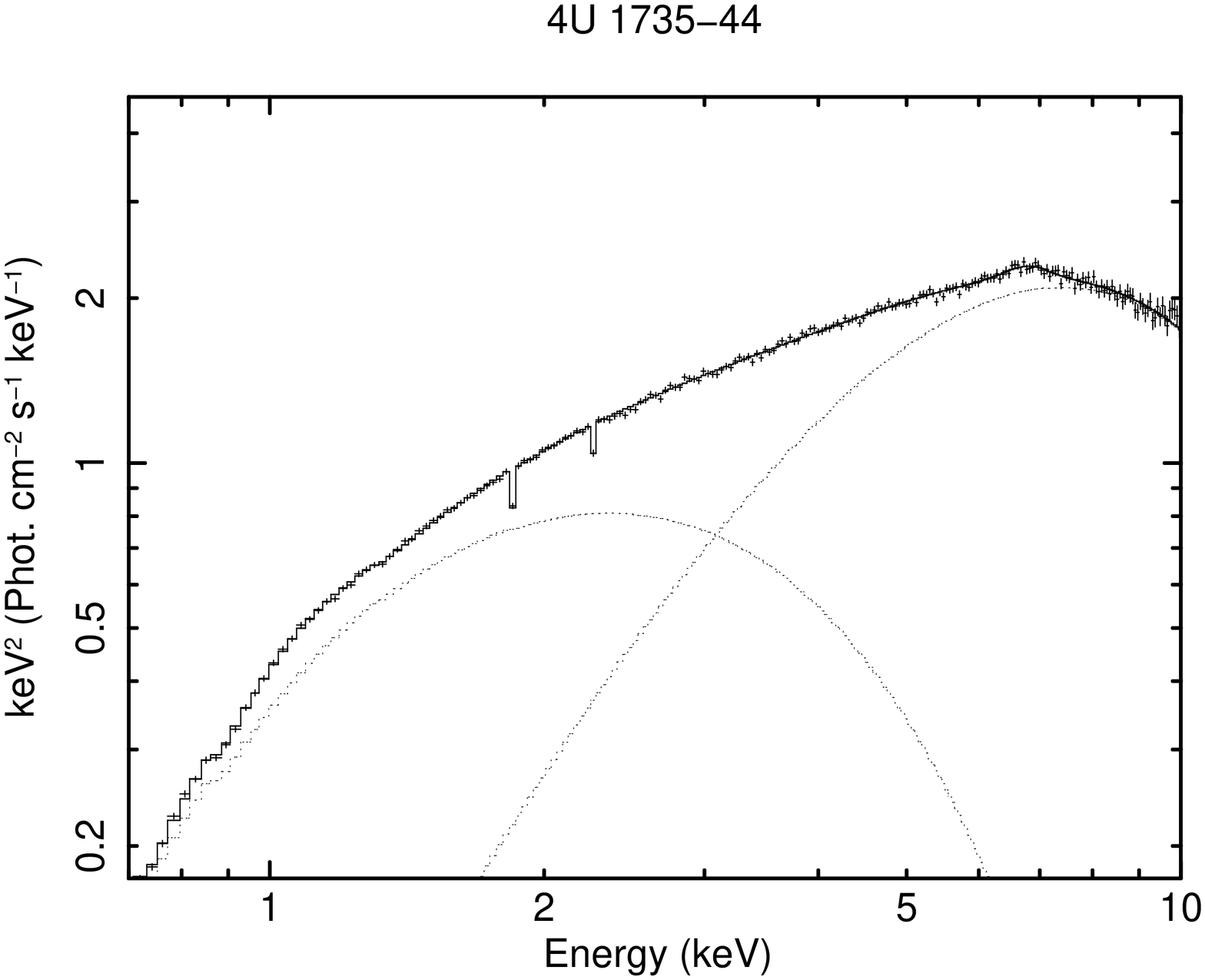}
\includegraphics[angle=-90,width=0.33\textwidth]{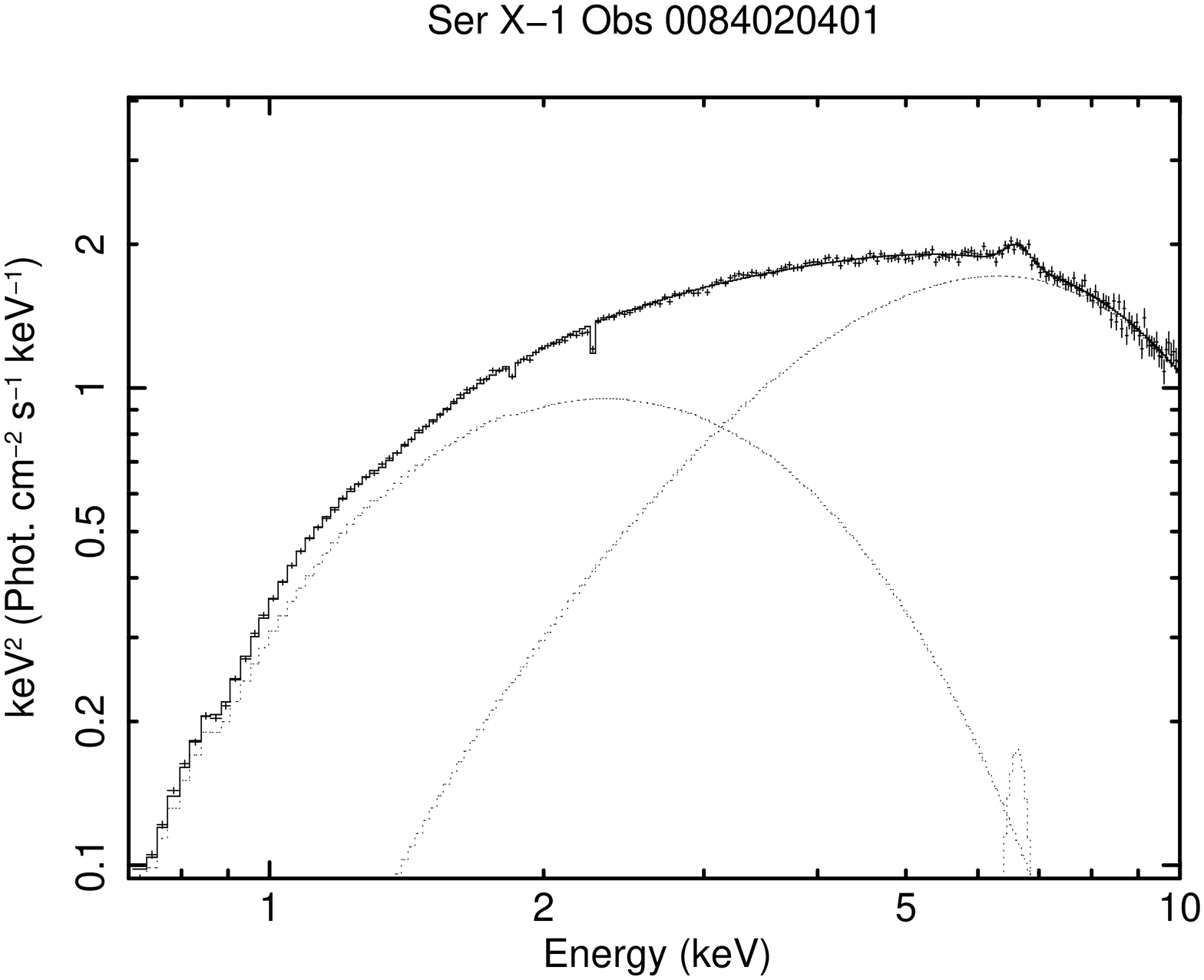}
\includegraphics[angle=-90,width=0.33\textwidth]{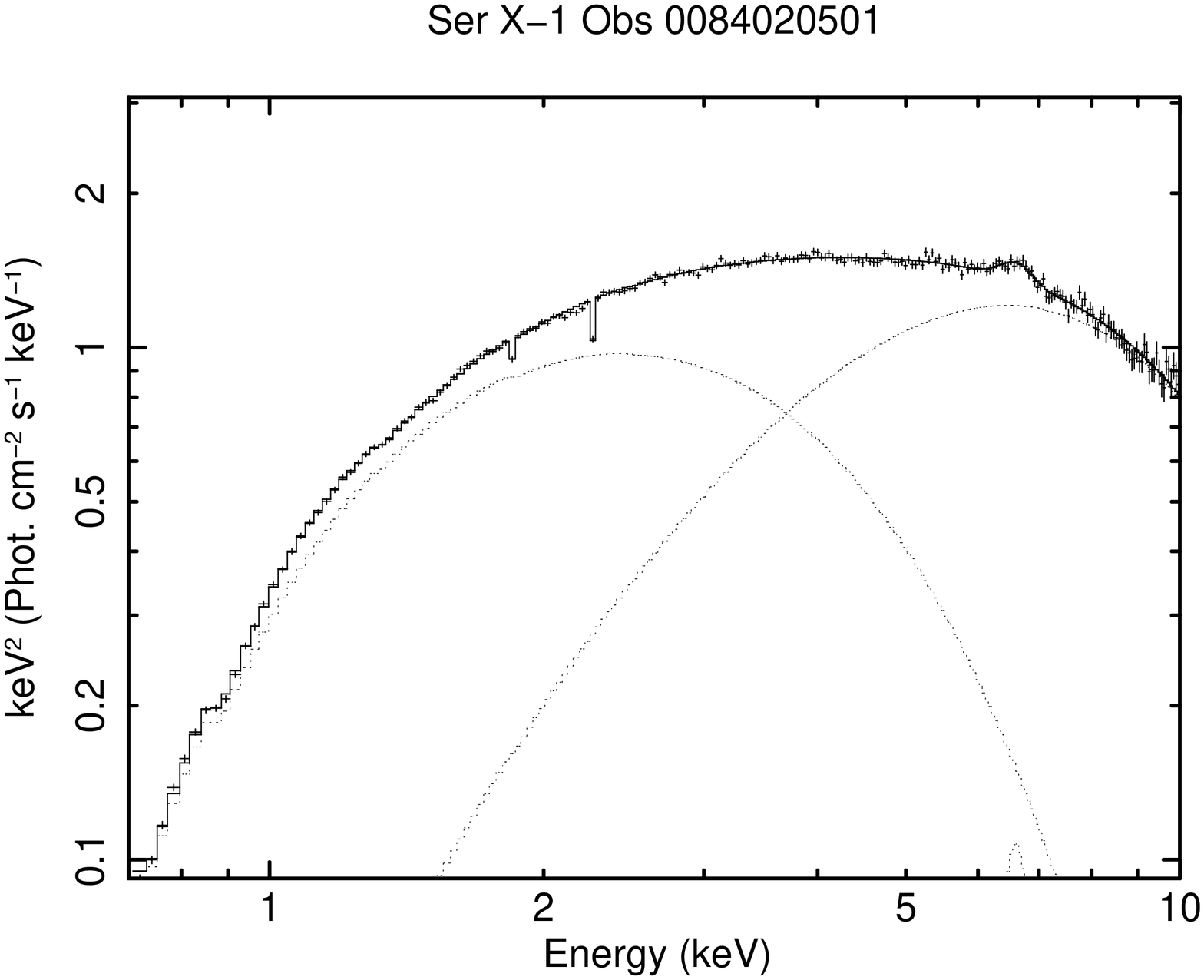}
\caption{0.7--10 keV EPIC pn (black) spectra fit with 
their best-fit model (see Tables~\ref{tab:bestfit}-\ref{tab:bestfit2}) shown in flux units.  
}
\label{fig:eeuf}
\end{figure*}

\begin{figure*}[!ht]
\includegraphics[angle=-90,width=0.33\textwidth]{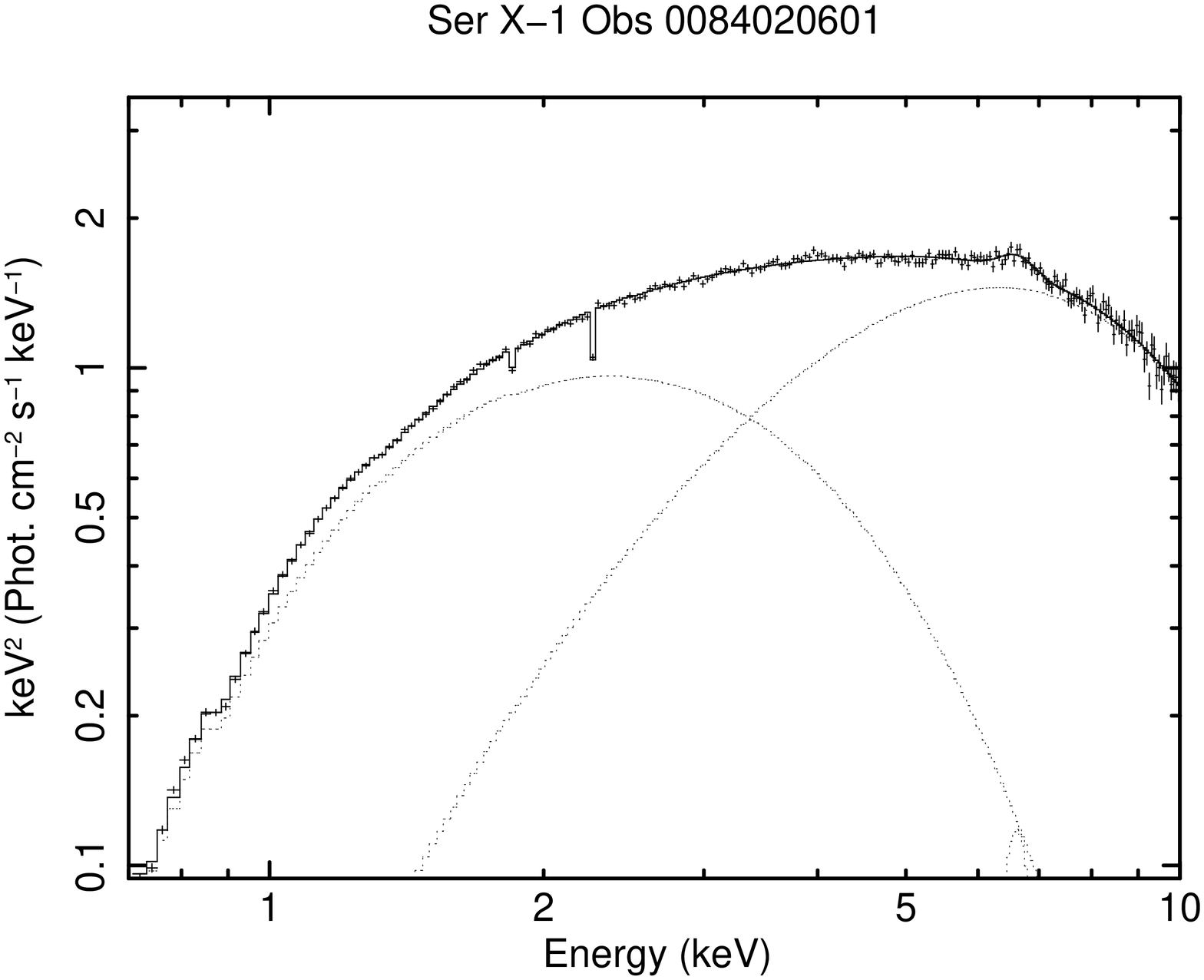}
\includegraphics[angle=-90,width=0.33\textwidth]{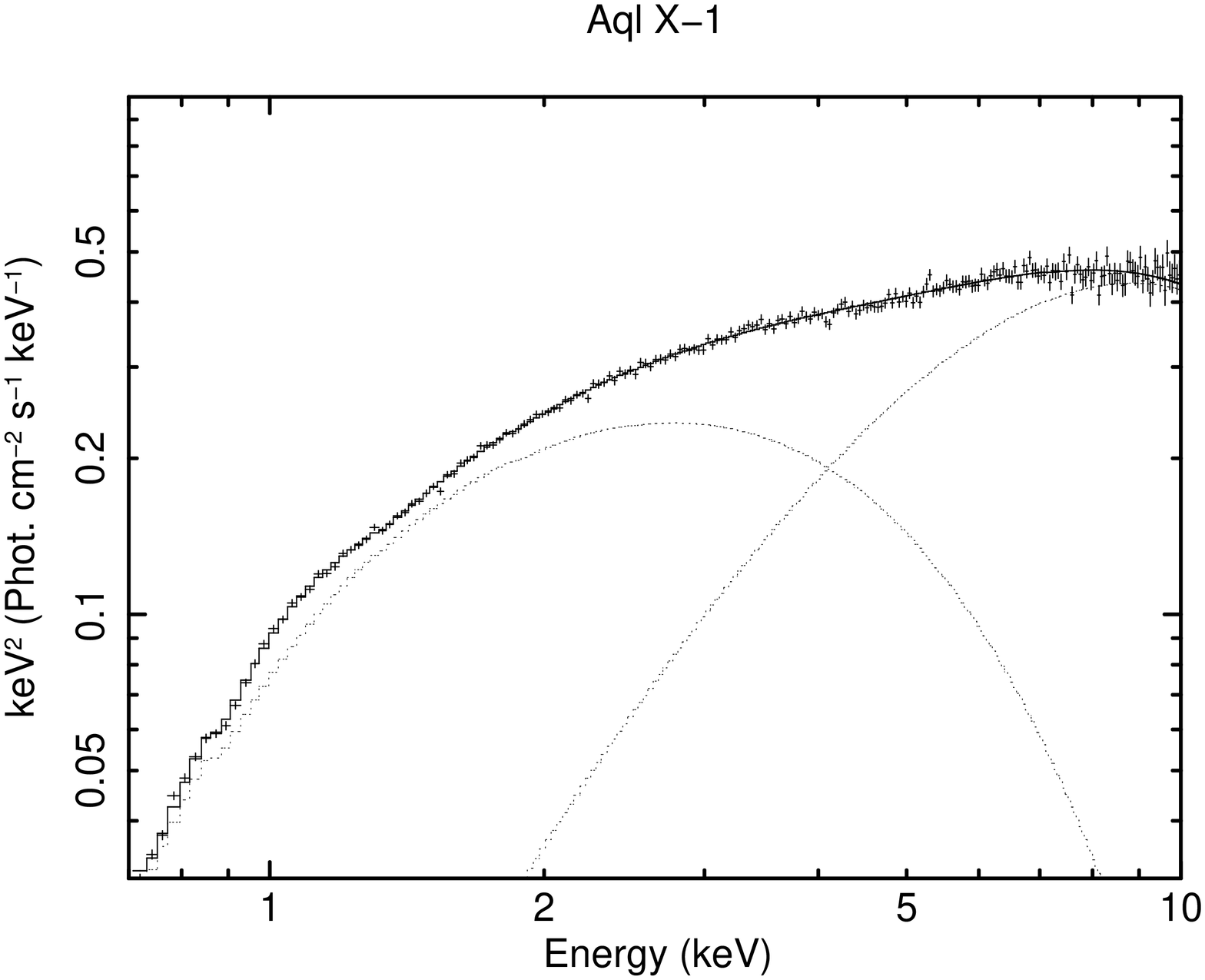}
\includegraphics[angle=-90,width=0.33\textwidth]{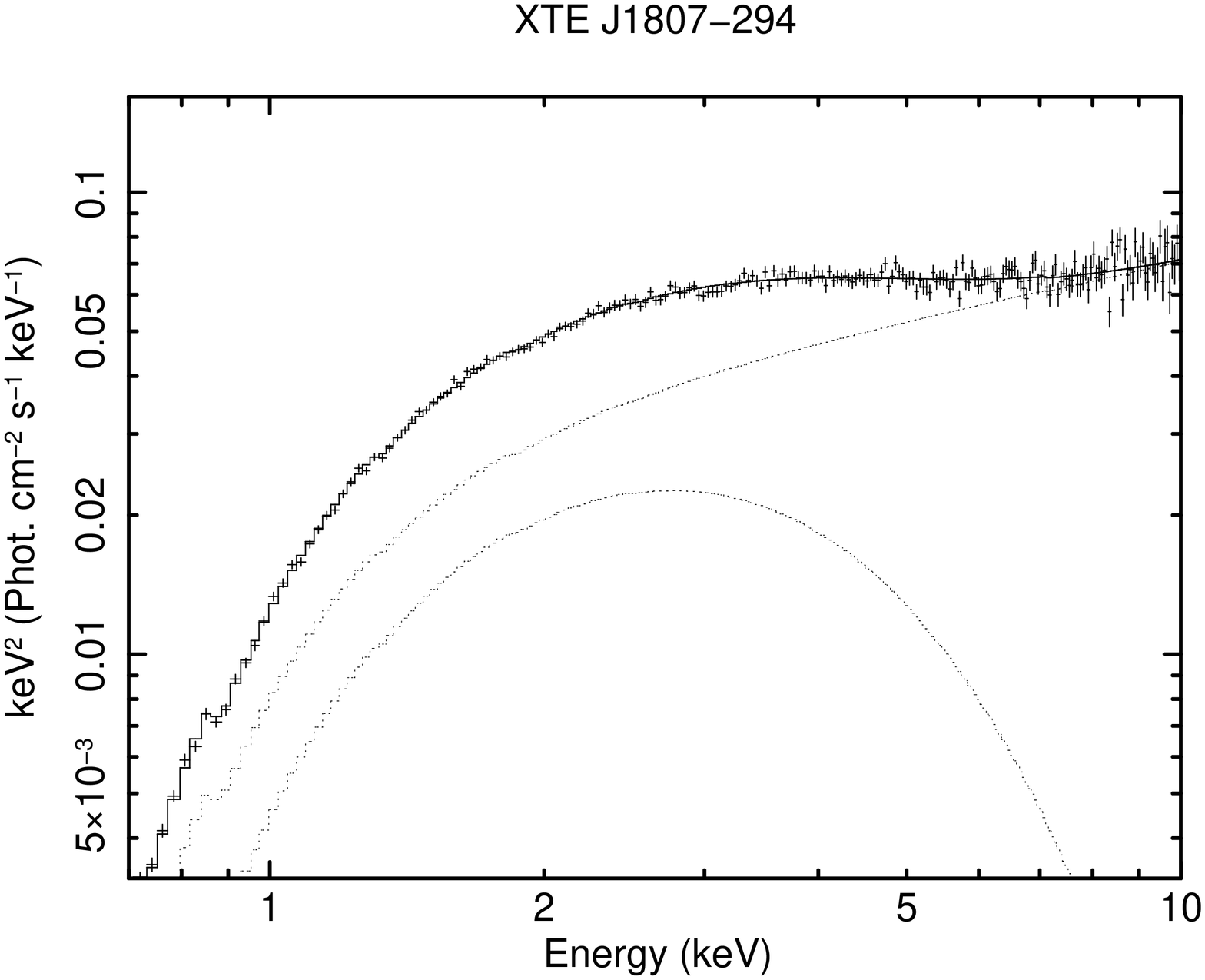}
\includegraphics[angle=-90,width=0.33\textwidth]{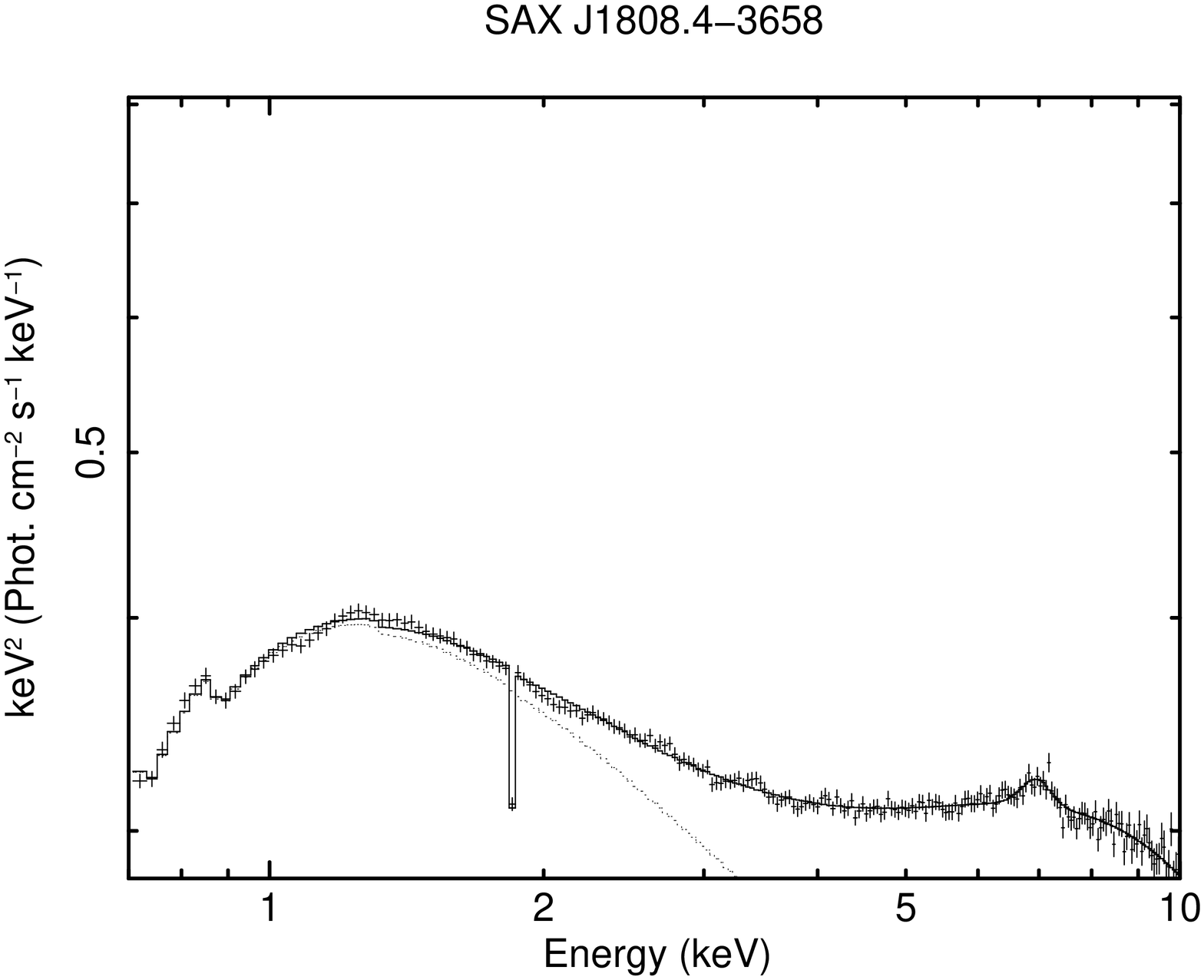}
\caption{Continued from Fig.~\ref{fig:eeuf}  
}
\label{fig:eeuf2}
\end{figure*}

\end{document}